\documentstyle[12pt,twoside]{article}

\let\ced=\c			
\let\sll=\l			
\let\Sll=\L			

\def\a{\alpha}
\def\b{\beta}
\def\c{\chi}
\def\d{\delta}
\def\e{\epsilon}		
\def\f{\phi}			
\def\g{\gamma}
\def\h{\eta}

\def\j{\psi}
\def\k{\kappa}
\def\l{\lambda}
\def\m{\mu}
\def\n{\nu}
\def\o{\omega}
\def\p{\pi}			
\def\q{\theta}			
\def\r{\rho}			
\def\s{\sigma}			
\def\t{\tau}

\def\x{\xi}
\def\z{\zeta}
\def\D{\Delta}
\def\F{\Phi}
\def\G{\Gamma}
\def\J{\Psi}
\def\L{\Lambda}
\def\O{\Omega}
\def\P{\Pi}
\def\Q{\Theta}
\def\S{\Sigma}
\def\U{\Upsilon}
\def\X{\Xi}

\def\ca{{\cal A}}
\def\cb{{\cal B}}
\def\cc{{\cal C}}
\def\cd{{\cal D}}

\def\cg{{\cal G}}
\def\ch{{\cal H}}

\def\cj{{\cal J}}

\def\cl{{\cal L}}
\def\cm{{\cal M}}
\def\cn{{\cal N}}
\def\co{{\cal O}}
\def\cp{{\cal P}}
\def\cq{{\cal Q}}

\def\ct{{\cal T}}

\def\cv{{\cal V}}

\catcode`@=11

\def\un#1{\relax\ifmmode\@@underline#1\else $\@@underline{\hbox{#1}}$\relax\fi}

{\catcode`\'=\active \gdef'{{}^\bgroup\prim@s}}

\def\magstep#1{\ifcase#1 \@m\or 1200\or 1440\or 1728\or 2074\or 2488\or
	2986\fi\relax}

\font\twfvmi=cmmi10\@magscale5
    \skewchar\twfvmi='177
\font\twfvsy=cmsy10\@magscale5
    \skewchar\twfvsy='60
\font\twfvly=lasy10\@magscale5
\font\thtyrm=cmr10\@magscale6

\def\vpt{\textfont\z@\fivrm
  \scriptfont\z@\fivrm \scriptscriptfont\z@\fivrm
\textfont\@ne\fivmi \scriptfont\@ne\fivmi \scriptscriptfont\@ne\fivmi
\textfont\tw@\fivsy \scriptfont\tw@\fivsy \scriptscriptfont\tw@\fivsy
\textfont\thr@@\tenex \scriptfont\thr@@\tenex \scriptscriptfont\thr@@\tenex
\def\prm{\fam\z@\fivrm}%
\def\unboldmath{\everymath{}\everydisplay{}\@nomath
  \unboldmath\fam\@ne\@boldfalse}\@boldfalse
\def\boldmath{\@subfont\boldmath\unboldmath}%
\def\pit{\@getfont\pit\itfam\@vpt{cmti5}}%
\def\psl{\@subfont\sl\it}%
\def\pbf{\@getfont\pbf\bffam\@vpt{cmbx5}}%
\def\ptt{\@subfont\tt\rm}%
\def\psf{\@subfont\sf\rm}%
\def\psc{\@subfont\sc\rm}%
\def\ly{\fam\lyfam\fivly}\textfont\lyfam\fivly
    \scriptfont\lyfam\fivly \scriptscriptfont\lyfam\fivly
\@setstrut\rm}

\def\@vpt{}

\def\vipt{\textfont\z@\sixrm
  \scriptfont\z@\sixrm \scriptscriptfont\z@\sixrm
\textfont\@ne\sixmi \scriptfont\@ne\sixmi \scriptscriptfont\@ne\sixmi
\textfont\tw@\sixsy \scriptfont\tw@\sixsy \scriptscriptfont\tw@\sixsy
\textfont\thr@@\tenex \scriptfont\thr@@\tenex \scriptscriptfont\thr@@\tenex
\def\prm{\fam\z@\sixrm}%
\def\unboldmath{\everymath{}\everydisplay{}\@nomath
  \unboldmath\@boldfalse}\@boldfalse
\def\boldmath{\@subfont\boldmath\unboldmath}%
\def\pit{\@subfont\it\rm}%
\def\psl{\@subfont\sl\rm}%
\def\pbf{\@getfont\pbf\bffam\@vipt{cmbx6}}%
\def\ptt{\@subfont\tt\rm}%
\def\psf{\@subfont\sf\rm}%
\def\psc{\@subfont\sc\rm}%
\def\ly{\fam\lyfam\sixly}\textfont\lyfam\sixly
    \scriptfont\lyfam\sixly \scriptscriptfont\lyfam\sixly
\@setstrut\rm}

\def\@vipt{}

\def\xxxpt{\textfont\z@\thtyrm
  \scriptfont\z@\twfvrm \scriptscriptfont\z@\twtyrm
\textfont\@ne\twfvmi \scriptfont\@ne\twfvmi \scriptscriptfont\@ne\twtymi
\textfont\tw@\twfvsy \scriptfont\tw@\twfvsy \scriptscriptfont\tw@\twtysy
\textfont\thr@@\tenex \scriptfont\thr@@\tenex \scriptscriptfont\thr@@\tenex
\def\unboldmath{\everymath{}\everydisplay{}\@nomath\unboldmath
        \textfont\@ne\twfvmi \textfont\tw@\twfvsy \textfont\lyfam\twfvly
        \@boldfalse}\@boldfalse
\def\boldmath{\@subfont\boldmath\unboldmath}%
\def\prm{\fam\z@\thtyrm}%
\def\pit{\@subfont\it\rm}%
\def\psl{\@subfont\sl\rm}%
\def\pbf{\@getfont\pbf\bffam\@xxxpt{cmbx10\@magscale6}}%
\def\ptt{\@subfont\tt\rm}%
\def\psf{\@subfont\sf\rm}%
\def\psc{\@subfont\sc\rm}%
\def\ly{\fam\lyfam\twfvly}\textfont\lyfam\twfvly
   \scriptfont\lyfam\twfvly \scriptscriptfont\lyfam\twtyly
\@setstrut \rm}

\def\@xxxpt{}

\def\Huge{\@setsize\Huge{36pt}\xxxpt\@xxxpt}

\font\thtymi=cmmi10\@magscale6
    \skewchar\thtymi='177
\font\thtysy=cmsy10\@magscale6
    \skewchar\thtysy='60
\font\thtyly=lasy10\@magscale6
\font\thsirm=cmr12\@magscale6

\def\xxxvipt{\textfont\z@\thsirm
  \scriptfont\z@\thtyrm \scriptscriptfont\z@\twfvrm
\textfont\@ne\thtymi \scriptfont\@ne\thtymi \scriptscriptfont\@ne\twfvmi
\textfont\tw@\thtysy \scriptfont\tw@\thtysy \scriptscriptfont\tw@\twfvsy
\textfont\thr@@\tenex \scriptfont\thr@@\tenex \scriptscriptfont\thr@@\tenex
\def\unboldmath{\everymath{}\everydisplay{}\@nomath\unboldmath
        \textfont\@ne\thtymi \textfont\tw@\thtysy \textfont\lyfam\thtyly
        \@boldfalse}\@boldfalse
\def\boldmath{\@subfont\boldmath\unboldmath}%
\def\prm{\fam\z@\thsirm}%
\def\pit{\@subfont\it\rm}%
\def\psl{\@subfont\sl\rm}%
\def\pbf{\@getfont\pbf\bffam\@xxxpt{cmss12\@magscale6}}%
\def\ptt{\@subfont\tt\rm}%
\def\psf{\@subfont\sf\rm}%
\def\psc{\@subfont\sc\rm}%
\def\ly{\fam\lyfam\thtyly}\textfont\lyfam\thtyly
   \scriptfont\lyfam\thtyly \scriptscriptfont\lyfam\twfvly
\@setstrut \rm}

\def\@xxxvipt{}

\def\HUGE{\@setsize\HUGE{43pt}\xxxvipt\@xxxvipt}

\catcode`@=12

\font\titlefont=cmbx10 scaled 3000
\font\tenex=cmex10 scaled 1200
\font\go=eufm10 scaled 1200			
\def\sc{\go}
\def\Go#1{\hbox{\go #1}}	
\def\Sc#1{\hbox{\sc #1}}	
\def\Sf#1{\hbox{\sf #1}}	

\def\bo{{\raise.05ex\hbox{\large$\Box$}\:}}		
\def\cbo{{\,\raise-.15ex\Sc [\,}}			
\def\pa{\partial}					
\def\de{\nabla}						
\def\su{\sum}						
\def\pr{\prod}						
\def\iff{\leftrightarrow}				
\def\conj{{\hbox{\large *}}}				
\def\TH{{\raise.2ex\hbox{$\displaystyle \bigodot$}\mskip-4.7mu \llap H \;}}
\def\Face{{\raise.2ex\hbox{$\displaystyle \bigodot$}\mskip-2.2mu \llap {$\ddot
	\smile$}}}					
\def\dg{\sp\dagger}					
\def\Lhat{{\bf\rlap{\kern-.09em$\hat{\phantom L}$}L}}
\def\Lcheck{{\bf\rlap{\kern-.09em$\check{\phantom L}$}L}}

\def\sp#1{{}^{#1}}				
\def\sb#1{{}_{#1}}				
\def\sl#1{\rlap{\hbox{$\mskip 1 mu /$}}#1}	
\def\Sl#1{\rlap{\hbox{$\mskip 3 mu /$}}#1}	
\def\SL#1{\rlap{\hbox{$\mskip 4.5 mu /$}}#1}	
\def\Tilde#1{\widetilde{#1}}			
\def\Hat#1{\widehat{#1}}			
\def\Bar#1{\overline{#1}}			
\def\bra#1{\Big\langle #1\Big|}			
\def\ket#1{\Big| #1\Big\rangle}			
\def\VEV#1{\Big\langle #1\Big\rangle}		
\def\sbra#1{\left\langle #1\right|}		
\def\sket#1{\left| #1\right\rangle}		
\def\leftrightarrowfill{$\mathsurround=0pt \mathord\leftarrow \mkern-6mu
	\cleaders\hbox{$\mkern-2mu \mathord- \mkern-2mu$}\hfill
	\mkern-6mu \mathord\rightarrow$}
\def\dvec#1{\vbox{\ialign{##\crcr
	\leftrightarrowfill\crcr\noalign{\kern-1pt\nointerlineskip}
	$\hfil\displaystyle{#1}\hfil$\crcr}}}		
\def\dt#1{{\buildrel {\hbox{\LARGE .}} \over {#1}}}	
\def\ddt#1{{\buildrel {\hbox{\LARGE .\kern-2pt.}} \over {#1}}}
\def\der#1{{\pa \over \pa {#1}}}		
\def\fder#1{{\d \over \d {#1}}}			

\def\frac#1#2{{\textstyle{#1\over\vphantom2\smash{\raise.20ex
	\hbox{$\scriptstyle{#2}$}}}}}			
\def\ha{\frac12}					
\def\sfrac#1#2{{\vphantom1\smash{\lower.5ex\hbox{\small$#1$}}\over
	\vphantom1\smash{\raise.4ex\hbox{\small$#2$}}}}	
\def\bfrac#1#2{{\vphantom1\smash{\lower.5ex\hbox{$#1$}}\over
	\vphantom1\smash{\raise.3ex\hbox{$#2$}}}}	
\def\afrac#1#2{{\vphantom1\smash{\lower.5ex\hbox{$#1$}}\over#2}}    
\def\on#1#2{\mathop{\null#2}\limits^{#1}}	
\def\oover#1{\on\circ{#1}}				

\def\boxes#1{
	\newcount\num
	\num=1
	\newdimen\downsy
	\downsy=-1.64ex
	\mskip-7.8mu
	\bo
	\loop
	\ifnum\num<#1
	\llap{\raise\num\downsy\hbox{$\bo$}}
	\advance\num by1
	\repeat}
\def\boxup#1#2{\newcount\numup
	\numup=#1
	\advance\numup by-1
	\newdimen\upsy
	\upsy=.82ex
	\mskip7.8mu
	\raise\numup\upsy\hbox{$#2$}}

\newskip\humongous \humongous=0pt plus 1000pt minus 1000pt
\def\caja{\mathsurround=0pt}

\newif\ifdtup
\def\panorama{\global\dtuptrue \openup2\jot \caja
	\everycr{\noalign{\ifdtup \global\dtupfalse
	\vskip-\lineskiplimit \vskip\normallineskiplimit
	\else \penalty\interdisplaylinepenalty \fi}}}
\def\li#1{\panorama \tabskip=\humongous				
	\halign to\displaywidth{\hfil$\displaystyle{##}$
	\tabskip=0pt&$\displaystyle{{}##}$\hfil
	\tabskip=\humongous&\llap{$##$}\tabskip=0pt
	\crcr#1\crcr}}

\def\NP{Nucl. Phys. B}
\def\PL{Phys. Lett. }
\def\PR{Phys. Rev. Lett. }
\def\PRD{Phys. Rev. D}
\def\ref#1{$\sp{#1]}$}

\parskip=\medskipamount			
\def\baselinestretch{1.255}	
\thicklines			    
\def\start#1{\pagestyle{myheadings}\begin{document}\thispagestyle{myheadings}
	\setcounter{page}{#1}}

\catcode`@=11

\def\ps@myheadings{\def\@oddhead{\hbox{}\footnotesize\bf\rightmark \hfil
	\thepage}\def\@oddfoot{}\def\@evenhead{\footnotesize\bf 
	\thepage\hfil\leftmark\hbox{}}\def\@evenfoot{}
	\def\sectionmark##1{}\def\subsectionmark##1{}
	\topmargin=-.35in\headheight=.17in\headsep=.35in}
\def\ps@acidheadings{\def\@oddhead{\hbox{}\rightmark\hbox{}}
	\def\@oddfoot{\rm\hfil\thepage\hfil}
	\def\@evenhead{\hbox{}\leftmark\hbox{}}\let\@evenfoot\@oddfoot
	\def\sectionmark##1{}\def\subsectionmark##1{}
	\topmargin=-.35in\headheight=.17in\headsep=.35in}

\catcode`@=12

\def\secty#1#2\par{\bigskip\medskip\par\penalty-3000\noindent{\large\bf{#1}}\nobreak\par%
	\nobreak\medskip\markright{#1}\nobreak\par\nobreak\bookmark0#2}
\def\sect#1{\secty{#1}{#1}}
\def\chscy#1#2#3#4{\newpage\pagestyle{myheadings}\thispagestyle{myheadings}\phantom 	m\vskip.5in\noindent{\LARGE\bf{#1}}\par\vskip.75in
	\noindent{\large\bf{#2}}\par\medskip\markboth{#1}{#2}
	\bookmark{#4}{#1}\bookmark0{#3}}
\def\chsc#1#2#3{\chscy{#1}{#2}{#2}{#3}}
\def\Chsc#1#2#3#4#5{\newpage\pagestyle{myheadings}\thispagestyle{myheadings}\phantom 	m\vskip.5in\noindent\halign{\LARGE\bf##&
	\LARGE\bf##\hfil\cr{#1}&{#2}\cr\noalign{\vskip8pt}&{#3}\cr}\par\vskip
	.75in\noindent{\large\bf{#4}}\par\medskip\markboth{{#1}{#2}{#3}}{#4}
	\bookmark{#5}{#1#2#3}\bookmark0{#4}}
\def\chap#1#2{\newpage\pagestyle{myheadings}\thispagestyle{myheadings}\phantom 	m\vskip.5in\noindent{\LARGE\bf{#1}}\par\vskip.75in
	\markboth{#1}{#1}\bookmark{#2}{#1}}
\def\refs\par{\bigskip\medskip\goodbreak\noindent{\large\bf{REFERENCES}}\nobreak\par%
	\bigskip\markboth{REFERENCES}{REFERENCES}\frenchspacing\nobreak%
	\parskip=0pt\parindent=21pt\renewcommand{\baselinestretch}{1}\small\par\nobreak
	\bookmark0{References}}
\def\unrefs{\normalsize \nonfrenchspacing \parskip=medskipamount}
\def\Item{\par\hang\textindent}
\def\Itemitem{\par\indent \hangindent2\parindent \textindent}
\def\makelabel#1{\hfil #1}
\def\topic{\par\noindent \hangafter1 \hangindent20pt}
\def\Topic{\par\noindent \hangafter1 \hangindent60pt}

\def\postscript#1{\special{" #1}}	
\def\textures{\gdef\postscript##1{\special{postscript ##1}}}	
\def\prelim{\postscript{
	gsave
	initgraphics
	-12 -15 translate
	.5 setgray
	150 100 moveto
	/Helvetica-Bold findfont 100 scalefont setfont
	55 rotate
	(PRELIMINARY) show
	grestore}}
\def\pdfinit{\postscript{
	/bd {bind def} bind def
	/fsd {findfont exch scalefont def} bd
	/sms {setfont moveto show} bd
	/ms {moveto show} bd
	/pdfmark where
	{pop} {userdict /pdfmark /cleartomark load put} ifelse
	[ /PageMode /UseOutlines
	/DOCVIEW pdfmark}}
\def\bookmark#1#2{\postscript{
	[ /Title (#2) /Count #1
	/OUT pdfmark}}
\newcount\countdp \newcount\countwd \newcount\countht 
\def\pdfklink#1#2{\hskip-.25em\setbox0=\hbox{#1}%
		\countdp=\dp0 \countwd=\wd0 \countht=\ht0%
		\divide\countdp by65536 \divide\countwd by65536%
			\divide\countht by65536%
		\advance\countdp by1 \advance\countwd by1%
			\advance\countht by1%
		\def\linkdp{\the\countdp} \def\linkwd{\the\countwd}%
			\def\linkht{\the\countht}%
	\postscript{
		[ /Rect [ -1.5 -\linkdp.0 0\linkwd.0 0\linkht.5 ] 
		/Border [ 0 0 1 ]
		/Color [ 1 0 0 ]
		/Action << /Subtype /URI /URI (#2) >>
		/Subtype /Link
		/ANN pdfmark}{#1}}
\def\pdflink#1{\pdfklink{#1}{#1}}
\def\xxxlink#1{\pdfklink{#1}{http://xxx.lanl.gov/abs/#1}}

\pdfinit

\def\lline#1{\noindent{#1}\\}
\def\iline#1#2{\hbox to\hsize{\indent{#1}\hfill{#2}\quad}}
\def\niline#1#2{\hbox to\hsize{{#1}\hfill{#2}\quad}}

\begin{document}

\phantom m\vskip1in
\begin{center}

{\titlefont INTRODUCTION to\\[.2in] STRING FIELD THEORY}

\vskip.5in

{\Large\bf Warren Siegel}

{\it University of Maryland\\[-.03in] College Park, Maryland}

\end{center}

\vfill

\noindent Present address:  State University of New York, Stony Brook\\
\pdflink{mailto:warren@wcgall.physics.sunysb.edu}\\
\pdfklink{http://insti.physics.sunysb.edu/\~{}siegel/plan.html}
{http://insti.physics.sunysb.edu/\noexpand~siegel/plan.html}

\newpage

\baselineskip=.91\normalbaselineskip

\twocolumn[\hbox to\textwidth{\large\bf\hfill CONTENTS\hfill}\vskip.3in]

\lline{Preface}
\lline{1. Introduction}
\iline{1.1. Motivation}{1}
\iline{1.2. Known models (interacting)}{3}
\iline{1.3. Aspects}{4}
\iline{1.4. Outline}{6}
\lline{2. General light cone}
\iline{2.1. Actions}{8}
\iline{2.2. Conformal algebra}{10}
\iline{2.3. Poincar\'e algebra}{13}
\iline{2.4. Interactions}{16}
\iline{2.5. Graphs}{19}
\iline{2.6. Covariantized light cone}{20}
\iline{Exercises}{23}
\lline{3. General BRST}
\indent 3.1. Gauge invariance and\\
\hbox to\hsize{\hfill constraints \hfill 25\quad}
\iline{3.2. IGL(1)}{29}
\iline{3.3. OSp(1,1$|$2)}{35}
\iline{3.4. From the light cone}{38}
\iline{3.5. Fermions}{45}
\iline{3.6. More dimensions}{46}
\iline{Exercises}{51}
\lline{4. General gauge theories}
\iline{4.1. OSp(1,1$|$2)}{52}
\iline{4.2. IGL(1)}{62}
\iline{4.3. Extra modes}{67}
\iline{4.4. Gauge fixing}{68}
\iline{4.5. Fermions}{75}
\iline{Exercises}{79}
\lline{5. Particle}
\iline{5.1. Bosonic}{81}
\iline{5.2. BRST}{84}
\iline{5.3. Spinning}{86}
\iline{5.4. Supersymmetric}{95}
\iline{5.5. SuperBRST}{110}
\iline{Exercises}{118}
\lline{6. Classical mechanics}
\iline{6.1. Gauge covariant}{120}
\iline{6.2. Conformal gauge}{122}
\iline{6.3. Light cone}{125}
\iline{Exercises}{127}
\lline{7. Light-cone quantum mechanics}
\iline{7.1. Bosonic}{128}
\iline{7.2. Spinning}{134}
\iline{7.3. Supersymmetric}{137}
\iline{Exercises}{145}
\lline{8. BRST quantum mechanics}
\iline{8.1. IGL(1)}{146}
\iline{8.2. OSp(1,1$|$2)}{157}
\iline{8.3. Lorentz gauge}{160}
\iline{Exercises}{170}
\lline{9. Graphs}
\iline{9.1. External fields}{171}
\iline{9.2. Trees}{177}
\iline{9.3. Loops}{190}
\iline{Exercises}{196}
\niline{10. Light-cone field theory}{197}
\iline{Exercises}{203}
\lline{11. BRST field theory}
\iline{11.1. Closed strings}{204}
\iline{11.2. Components}{207}
\iline{Exercises}{214}
\lline{12. Gauge-invariant interactions}
\iline{12.1. Introduction}{215}
\iline{12.2. Midpoint interaction}{217}
\iline{Exercises}{228}
\niline{References}{230}
\niline{Index}{241}

\onecolumn

\baselineskip=.9\normalbaselineskip

\chap{PREFACE}0
\pagestyle{empty}\thispagestyle{empty}

First, I'd like to explain the title of this book.  I always hated books
whose titles began ``Introduction to...''  In particular, when I was a
grad student, books titled ``Introduction to Quantum Field Theory'' were
the most difficult and advanced textbooks available, and I always feared 
what a quantum field theory book which was not introductory would look like.
There is now a standard reference on relativistic string theory by
Green, Schwarz, and Witten, {\it Superstring Theory} [0.1], which consists
of two volumes, is over 1,000 pages long, and yet admits to having some
major omissions.  Now that I see, from an author's point of view, how
much effort is necessary to produce a non-introductory text, the words
``Introduction to'' take a more tranquilizing character.  (I have worked
on a one-volume, non-introductory text on another topic, but that was in
association with three coauthors.)  Furthermore, these words leave me
the option of omitting topics which I don't understand, or at least
being more heuristic in the areas which I haven't studied in detail yet.

The rest of the title is ``String Field Theory.''  This is the newest
approach to string theory, although the older approaches are
continuously developing new twists and improvements.  The main
alternative approach is the quantum mechanical 
(/analog-model/path-integral/interacting-string-picture/Polyakov/conformal-
``field-theory'') one, which necessarily treats a fixed number of 
fields, corresponding to homogeneous equations in the field theory.  
(For example, there is no analog in the mechanics approach of even the 
nonabelian gauge transformation of the field theory, which includes such
fundamental concepts as general coordinate invariance.)
It is also an S-matrix approach, and can thus calculate only quantities 
which are gauge-fixed (although limited background-field techniques 
allow the calculation of 1-loop effective actions with only some 
coefficients gauge-dependent).  In the old S-matrix
approach to field theory, the basic idea was to start with the S-matrix,
and then analytically continue to obtain quantities which are off-shell
(and perhaps in more general gauges).  However, in the long run, it
turned out to be more practical to work directly with field theory
Lagrangians, even for semiclassical results such as spontaneous symmetry
breaking and instantons, which change the meaning of ``on-shell'' by
redefining the vacuum to be a state which is not as obvious from
looking at the unphysical-vacuum S-matrix.  Of course, S-matrix methods are 
always valuable for perturbation theory, but even in perturbation theory it is
far more convenient to start with the field theory in order to determine
which vacuum to perturb about, which gauges to use, and what
power-counting rules can be used to determine divergence structure
without specific S-matrix calculations.  (More details on this
comparison are in the Introduction.)

Unfortunately, string field theory is in a rather primitive state right
now, and not even close to being as well understood as ordinary
(particle) field theory.  Of course, this is exactly the reason why the
present is the best time to do research in this area.  (Anyone who can
honestly say, ``I'll learn it when it's better understood,'' should mark
a date on his calendar for returning to graduate school.)  It is
therefore simultaneously the best time for someone to read a book on the
topic and the worst time for someone to write one.  I have tried to
compensate for this problem somewhat by expanding on the more
introductory parts of the topic.  Several of the early chapters are
actually on the topic of general (particle/string) field theory, but
explained from a new point of view resulting from insights gained from
string field theory.  (A more standard course on quantum field theory is
assumed as a prerequisite.)
This includes the use of a universal method for
treating free field theories, which allows the derivation of a single,
simple, free, local, Poincar\'e-invariant, gauge-invariant action that 
can be applied directly to any field.  (Previously, only some special
cases had been treated, and each in a different way.)  As a result, even
though the fact that I have tried to make this book self-contained
with regard to string theory in general
means that there is significant overlap with other treatments, within
this overlap the approaches are sometimes quite different, and perhaps
in some ways complementary.  (The treatments of ref.\ [0.2] are also
quite different, but for quite different reasons.)

Exercises are given at the end of each chapter (except the introduction) to 
guide the reader to examples which illustrate the ideas in the chapter, 
and to encourage him to perform calculations which have been omitted to 
avoid making the length of this book diverge. 

This work was done at the University of Maryland, with partial support
from the National Science Foundation.  It is partly based on courses I
gave in the falls of 1985 and 1986.  I received valuable comments from
Aleksandar Mikovi\'c, Christian Preitschopf, Anton van de Ven, and 
Harold Mark Weiser.  I especially thank Barton Zwiebach, who collaborated
with me on most of the work on which this book was based.


\vskip.1in

\hbox to\hsize{June 16, 1988\hfil Warren Siegel}

\vfill

\noindent Originally published 1988 by World Scientific Publishing Co Pte Ltd.\\
ISBN 9971-50-731-5, 9971-50-731-3 (pbk)

\noindent {\bf July 11, 2001:}  liberated, corrected, bookmarks added (to pdf)

%
%

 \baselineskip=\normalbaselineskip

\chsc{1. INTRODUCTION}{1.1. Motivation}4
\setcounter{page}{1}

The experiments which gave us quantum theory and 
general relativity are now quite old, but a satisfactory theory which 
is consistent with both of them has yet to be found.  Although the 
importance of such a theory is undeniable, the urgency of finding it 
may not be so obvious, since the quantum effects of gravity are not 
yet accessible to experiment.  
However, recent progress in the problem has indicated that the 
restrictions imposed by quantum mechanics on a field theory of 
gravitation are so stringent as to {\it require} that it also be a 
unified theory of all interactions, and thus quantum gravity would 
lead to predictions for other interactions which can be subjected to 
present-day experiment.  Such indications were given by supergravity 
theories [1.1], where finiteness was found at some higher-order loops as a 
consequence of supersymmetry, which requires the presence of matter 
fields whose quantum effects cancel the ultraviolet divergences of the 
graviton field.  Thus, quantum consistency led to higher symmetry 
which in turn led to unification.  However, even this symmetry was 
found insufficient to guarantee finiteness at all loops [1.2] (unless 
perhaps the graviton were found to be a bound-state of a truly finite 
theory).  Interest then returned to theories which had already 
presented the possibility of consistent quantum gravity theories as a 
consequence of even larger (hidden) symmetries: theories of 
relativistic strings [1.3-5].  Strings thus offer a possibility of
consistently describing all of nature.  However, even if strings eventually
turn out to disagree with nature, or to be too intractable to be useful
for phenomenological applications, they are still the only consistent
toy models of quantum gravity (especially for the theory of the graviton
as a bound state), so their study will still be useful for discovering
new properties of quantum gravity.

The fundamental difference between a particle and a string is that a 
particle is a 0-dimensional object in space, with a 1-dimensional 
world-line describing its trajectory in spacetime, while a string is a 
(finite, open or closed) 1-dimensional object in space, which sweeps 
out a 2-dimensional world-sheet as it propagates through spacetime:

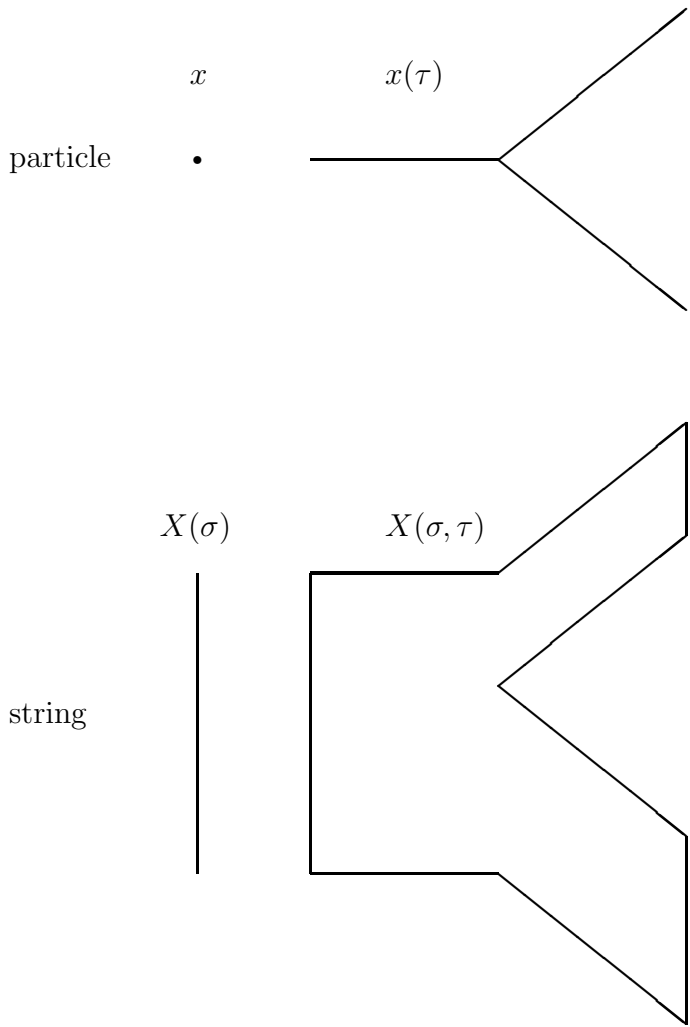
\setlength{\unitlength}{1mm}
\begin{picture}(90,135)
\put(24,125){$x$}
\put(50,125){$x( \t )$}
\put(0,114){particle}
\put(25,115){\circle*{1}}
\put(40,115){\line(1,0){25}}
\put(65,115){\line(5,4){25}}
\put(65,115){\line(5,-4){25}}
\put(20,65){$X( \s )$}
\put(50,65){$X( \s , \t )$}
\put(0,40){string}
\put(25,20){\line(0,1){40}}
\put(40,20){\line(0,1){40}}
\put(40,60){\line(1,0){25}}
\put(65,60){\line(5,4){25}}
\put(40,20){\line(1,0){25}}
\put(65,20){\line(5,-4){25}}
\put(65,45){\line(5,4){25}}
\put(65,45){\line(5,-4){25}}
\put(90,65){\line(0,1){15}}
\put(90,0){\line(0,1){25}}
\end{picture}

\noindent The nontrivial topology of the coordinates describes interactions.  
A string can be either open or closed, depending on
whether it has 2 free ends (its boundary) or is a continuous ring (no
boundary), respectively.  The corresponding spacetime figure is then
either a sheet or a tube (and their combinations, and topologically more
complicated structures, when they interact).

Strings were originally intended to describe hadrons directly, since 
the observed spectrum and high-energy behavior of hadrons (linearly 
rising Regge trajectories, which in a perturbative framework implies 
the property of hadronic duality) seems realizable only in a string 
framework.  After a quark structure for hadrons became generally 
accepted, it was shown that confinement would naturally lead to a 
string formulation of hadrons, since the topological expansion which 
follows from using $1/N\sb{color}$ 
as a perturbation parameter (the only dimensionless one in massless 
QCD, besides $1/N\sb{flavor}$), after summation in the other 
parameter (the gluon coupling, 
which becomes the hadronic mass scale after dimensional 
transmutation), is the same perturbation expansion as occurs in 
theories of fundamental strings [1.6].  Certain string theories can thus be 
considered alternative and equivalent formulations of QCD, just as 
general field theories can be equivalently formulated either in terms 
of ``fundamental'' particles or in terms of the particles which arise 
as bound states.  However, in practice certain criteria, in particular 
renormalizability, can be simply formulated only in one formalism:  
For example, QCD is easier to use than a theory where gluons are 
treated as bound states of self-interacting quarks, the latter being a 
nonrenormalizable theory which needs an unwieldy criterion 
(``asymptotic safety'' [1.7]) to restrict the available infinite number of 
couplings to a finite subset.  On the other hand, atomic physics is 
easier to use as a theory of electrons, nuclei, and photons 
than a formulation in terms of fields describing self-interacting 
atoms whose excitations lie on Regge trajectories (particularly since 
QED is not confining).  Thus, the choice 
of formulation is dependent on the dynamics of the particular theory, 
and perhaps even on the region in momentum space for that particular 
application: perhaps quarks for large transverse momenta and strings 
for small.  In particular, the running of the gluon coupling may lead 
to nonrenormalizability problems for small transverse momenta [1.8] (where 
an infinite number of arbitrary couplings may show up as 
nonperturbative vacuum values of operators of arbitrarily high 
dimension), and thus QCD may be best considered as an effective theory 
at large transverse momenta (in the same way as a perturbatively 
nonrenormalizable theory at low energies, like the Fermi theory of weak 
interactions, unless asymptotic safety is 
applied).  Hence, a string formulation, where mesons are the fundamental fields
(and baryons appear as skyrmeon-type solitons [1.9]) may be unavoidable.
Thus, strings may be important for hadronic physics as well as 
for gravity and unified theories; however, the presently known string models 
seem to apply only to the latter, since they contain massless 
particles and have (maximum) spacetime dimension $D = 10$ (whereas 
confinement in QCD occurs for $D \le 4$).

\sect{1.2. Known models (interacting)}

Although many string theories have been invented which are consistent 
at the tree level, most have problems at the one-loop level.  (There 
are also theories which are already so complicated at the free level 
that the interacting theories have been too difficult to formulate to 
test at the one-loop level, and these will not be discussed here.)  
These one-loop problems generally show up as anomalies.  It turns out 
that the anomaly-free theories are exactly the ones which are finite.  
Generally, topological arguments based on reparametrization 
invariance (the ``stretchiness'' of the string world sheet) show that 
any multiloop string graph can be represented as a tree graph with 
many one-loop insertions [1.10], so all divergences should be representable as 
just one-loop divergences.  The fact that one-loop divergences 
should generate overlapping divergences then implies that one-loop 
divergences cause anomalies in reparametrization invariance, since the 
resultant multi-loop divergences are in conflict with the 
one-loop-insertion structure implied by the invariance.  Therefore, 
finiteness should be a necessary requirement for string theories (even 
purely bosonic ones) in order to avoid anomalies in reparametrization 
invariance.  Furthermore, the absence of anomalies in such global 
transformations determines the dimension of spacetime, which in all 
known nonanomalous theories is $D=10$.  (This is also known as the
``critical,'' or maximum, dimension, since some of the dimensions can be
compactified or otherwise made unobservable, although the number of
degrees of freedom is unchanged.)    

In fact, there are only four such theories:
\halign{\hskip.2in#\hfill&\quad#\hfill\cr
	I:	&N=1 supersymmetry, SO(32) gauge group, open [1.11]\cr
	IIA,B:	&N=2 nonchiral or chiral supersymmetry [1.12]\cr
	heterotic:&N=1 supersymmetry, SO(32) or E$\sb 8\otimes$E$\sb 8$ 
			[1.13]\cr
	&{\it or} broken N=1 supersymmetry, 
			SO(16)$\otimes$SO(16) [1.14]\cr}
\noindent All except the first describe only closed strings; the first
describes open strings, which produce closed strings as bound states.
(There are also many cases of each of these theories due to the 
various possibilities for compactification of the extra dimensions 
onto tori or other manifolds, including some which have tachyons.)  
However, for simplicity we will first 
consider certain inconsistent theories: the bosonic string, which has 
global reparametrization anomalies unless $D=26$ (and for which the 
local anomalies described above even for $D=26$ have not yet been 
explicitly derived), and the spinning string, which is nonanomalous 
only when it is truncated to the above strings.  The heterotic
strings are actually closed strings for which modes 
propagating in the clockwise direction are nonsupersymmetric and 
26-dimensional, while the counterclockwise ones are $N=1$ (perhaps-broken)
supersymmetric and 10-dimensional, or vice versa.  

\sect{1.3. Aspects}

There are several aspects of, or approaches to, string theory which can
best be classified by the spacetime dimension in which they work: $D = 2, 
4, 6, 10$.  The 2D approach is the method of first-quantization in
the two-dimensional world sheet swept out by the string as it propagates, 
and is applicable solely to (second-quantized) perturbation theory, for
which it is the only tractable method of calculation.  Since it 
discusses only the properties of individual graphs, it can't discuss 
properties which involve an unfixed number of string fields: gauge 
transformations, spontaneous symmetry breaking, semiclassical solutions 
to the string field equations, etc.  Also, it can describe only the
gauge-fixed theory, and only in a limited set of gauges.  (However, by
introducing external particle fields, a limited amount of information on
the gauge-invariant theory can be obtained.)  Recently most of the
effort in this area has been concentrated on applying this approach to
higher loops.  However, in particle field theory, particularly for
Yang-Mills, gravity, and supersymmetric theories (all of which are
contained in various string theories), significant (and sometimes
indispensable) improvements in higher-loop calculations have required
techniques using the gauge-invariant field theory action.  Since such
techniques, whose string versions have not yet been derived, could
drastically affect the S-matrix techniques of the 2D approach, we do not
give the most recent details of the 2D approach here, but some of the 
basic ideas, and the
ones we suspect most likely to survive future reformulations,  will be
described in chapters 6-9.  

The 4D approach is concerned with the
phenomenological applications of the low-energy effective theories
obtained from the string theory.  Since these theories are still very
tentative (and still too ambiguous for many applications), they will 
not be discussed here. (See [1.15,0.1].)

The 6D approach describes the compactifications (or equivalent
eliminations) of the 6 additional dimensions which must shrink from
sight in order to obtain the observed dimensionality of the macroscopic 
world.  Unfortunately, this approach has several problems which inhibit
a useful treatment in a book:  (1) So far, no justification has been
given as to why the compactification occurs to the desired models, or to
4 dimensions, or at all; (2) the style of compactification (Ka\sll
u\.za-Klein, Calabi-Yau, toroidal, orbifold, fermionization, etc.)
deemed most promising changes from year to year; and (3) the string 
model chosen to compactify (see previous section) also changes every few
years.  Therefore, the 6D approach won't be discussed here, either 
(see [1.16,0.1]).  

What is discussed here is
primarily the 10D approach, or second quantization, which seeks to
obtain a more systematic understanding of string theory that would allow
treatment of nonperturbative as well as perturbative aspects, and
describe the enlarged hidden gauge symmetries which give string theories
their finiteness and other unusual properties.  In particular, it would
be desirable to have a formalism in which all the symmetries (gauge,
Lorentz, spacetime supersymmetry) are manifest, finiteness follows from simple
power-counting rules, and all possible models (including possible 4D
models whose existence is implied by the $1/N$ expansion of QCD and
hadronic duality) can be straightforwardly classified.  In ordinary
(particle) supersymmetric field theories [1.17], such a formalism ({\it
superfields} or {\it superspace}) has resulted in much simpler rules for
constructing general actions, calculating quantum corrections ({\it
supergraphs}), and explaining all finiteness properties (independent
from, but verified by, explicit supergraph calculations).  The
finiteness results make use of the background field gauge, which can be
defined only in a field theory formulation where all symmetries are
manifest, and in this gauge divergence cancellations are automatic,
requiring no explicit evaluation of integrals.

\sect{1.4. Outline}

String theory can be considered a particular kind of particle theory, in
that its modes of excitation correspond to different particles.  
All these particles, which differ in spin and other quantum numbers, are
related by a symmetry which reflects the properties of the string.  As
discussed above, quantum field theory is the most complete framework
within which to study the properties of particles.  Not only is this
framework not yet well understood for strings, but the study of string
field theory has brought attention to aspects which are not well
understood even for general types of particles.  (This is another
respect in which the study of strings resembles the study of
supersymmetry.)  We therefore devote chapts.\ 2-4 to a general study of
field theory.  Rather than trying to describe strings in the language of
old quantum field theory, we recast the formalism of field theory in a
mold prescribed by techniques learned from the study of strings.  This
language clarifies the relationship between physical states and gauge
degrees of freedom, as well as giving a general and straightforward
method for writing free actions for arbitrary theories.  

In chapts.\ 5-6
we discuss the mechanics of the particle and string.  As mentioned
above, this approach is a useful calculational tool for evaluating
graphs in perturbation theory, including the interaction vertices
themselves.  The quantum mechanics of the string is developed in chapts.\
7-8, but it is primarily discussed directly as an operator algebra for the
field theory, although it follows from quantization of the classical
mechanics of the previous chapter, and vice versa.  In general, the 
procedure of first-quantization of a relativistic system serves only 
to identify its constraint algebra, which directly corresponds to both 
the field equations and gauge transformations of the free field theory.
However, as described in chapts.\
2-4, such a first-quantization procedure does not exist for general
particle theories, but the constraint system can be derived by other
means.  The free gauge-covariant theory then follows in a
straightforward way.  String perturbation theory is discussed in chapt.\ 9.  

Finally, the methods of chapts.\ 2-4 are applied to strings in 
chapts.\ 10-12, where string field theory is discussed.  These chapters
are still rather introductory, since many problems still remain in
formulating interacting string field theory, even in the light-cone
formalism.  However, a more complete understanding of the extension of
the methods of chapts.\ 2-4 to just particle field theory should help in
the understanding of strings.

Chapts.\ 2-5 can be considered almost as an independent book, an attempt
at a general approach to all of field theory.  For those few high energy
physicists who are not intensely interested in strings (or do not have
high enough energy to study them), it can be read as a new introduction to
ordinary field theory, although familiarity with quantum field theory as
it is usually taught is assumed.  Strings can then be left for later as
an example.  On the other hand, for those who want just a brief
introduction to strings, a straightforward, though less
elegant, treatment can be found via the light cone in chapts.\ 6,7,9,10
(with perhaps some help from sects.\ 2.1 and 2.5).  These chapters
overlap with most other treatments of string theory.  The remainder of the
book (chapts.\ 8,11,12) is basically the synthesis of these two topics.

%
%

\chsc{2. GENERAL LIGHT CONE}{2.1. Actions}7

Before discussing the string
we first consider some general properties of gauge theories and field
theories, starting with the light-cone formalism.

In general, light-cone field theory [2.1] looks like {\it non}relativistic 
field theory.  Using light-cone notation, for vector indices $a$ and the
Minkowski inner product $A \cdot B = \h \sp {ab} A \sb b B \sb a =
A\sp a B\sb a$,
$$ a = ( + , - , i ) \quad , \quad A \cdot B = A \sb + B \sb - +
A \sb - B \sb + + A \sb i B \sb i \quad , \eqno(2.1.1)$$
we interpret $x \sb +$ as being the ``time'' coordinate (even though it
points in a lightlike direction), in terms of which the evolution of the
system is described.  The metric can be diagonalized by $A \sb \pm
\equiv 2 \sp{-1/2} ( A \sb 1 \mp A \sb 0 )$.  For positive energy
$E$($=p\sp 0 =-p\sb 0$), we have on shell $p\sb +\ge 0$ and $p\sb -\le 0$
(corresponding to paths with $\D x\sb +\ge 0$ and $\D x\sb -\le 0$),
with the opposite signs for negative energy (antiparticles).
For example, for a real scalar field the lagrangian is rewritten as
$$ -\ha \f ( p \sp 2 + m \sp 2 ) \f = - \f p \sb + \left( p \sb - + 
{p \sb i \sp 2 + m \sp 2 \over 2 p \sb +} \right) \f = - \f p \sb + ( 
p \sb - + H ) \f \quad ,\eqno(2.1.2)$$
where the momentum $p \sb a \equiv i \pa \sb a$, $p \sb - = 
i \pa / \pa x \sb +$ with respect to the ``time'' $x \sb +$, 
and $p \sb +$ appears like a mass in the ``hamiltonian'' $H$. 
(In the light-cone formalism, $p \sb +$ is assumed to be invertible.)
Thus, the field equations are first-order in these time derivatives, 
and the field satisfies a nonrelativistic-style Schr\"odinger equation. 
The field equation can then be solved explicitly:  In the free theory,
$$ \f ( x\sb + ) = e \sp{ix\sb + H} \f (0) \quad . \eqno(2.1.3)$$
$p\sb -$ can then be effectively replaced with $-H$.  Note that, unlike
the nonrelativistic case, the hamiltonian $H$, although hermitian, is
imaginary (in coordinate space), due to the $i$ in $p\sb + = i\pa\sb +$.
Thus, (2.1.3) is consistent with a (coordinate-space) reality condition
on the field.

For a spinor, half the components are auxiliary (nonpropagating, since 
the field
equation is only first-order in momenta), and all auxiliary components are 
eliminated in the light-cone formalism by their equations of motion
(which, by definition, don't involve inverting time derivatives $p\sb -$):
$$\li{-\ha\bar \j ( \sl p + im ) \j =& -\ha2\sp{1/4}
\pmatrix{ \j \sb + \sp \dag & \j \sb - \sp \dag \cr} 
\pmatrix{\sqrt 2 p \sb - & \s\sb i p\sb i +i m \cr
\s\sb i p\sb i -i m & -\sqrt 2 p \sb + \cr } 
2\sp{1/4}\pmatrix{ \j \sb + \cr \j \sb - \cr } \cr
=& -\j \sb + \sp \dag p \sb - \j \sb + + \j \sb - \sp 
\dag p \sb + \j \sb - \cr
&- {1\over\sqrt 2}\j \sb - \sp \dag ( \s\sb i p\sb i -i m ) \j 
\sb + - {1\over\sqrt 2}\j \sb + \sp \dag ( \s\sb i p\sb i +i m ) \j \sb - \cr
\to& \; - \j \sb + \sp \dag ( p \sb - + H ) \j \sb + \quad 
,&(2.1.4)\cr}$$
where $H$ is the {\it same} hamiltonian as in (2.1.2).  (There is an
extra overall factor of 2 in (2.1.4) for complex spinors.  We have
assumed real (Majorana) spinors.)

For the case of Yang-Mills, the covariant action is
$$ S = {1 \over g\sp 2} \int d \sp D x \; tr \; \cl \quad , \quad 
\cl = \frac14 F \sb {ab} \sp 2 \quad , \eqno(2.1.5a)$$
$$ F \sb {ab} \equiv [ \de \sb a , \de \sb b ] \quad , \quad 
\de \sb a \equiv p \sb a + A \sb a \quad , \quad 
\de \sb a ' = e \sp {i\l} \de \sb a e \sp {-i\l}\quad .\eqno(2.1.5b)$$
(Contraction with a matrix representation of the group generators is implicit.)
The light-cone gauge is then defined as
$$ A \sb + = 0 \quad . \eqno(2.1.6)$$
Since the gauge transformation of the gauge condition doesn't involve
the time derivative $\pa \sb -$, the Faddeev-Popov ghosts are
nonpropagating, and can be ignored.  The field
equation of $A \sb -$ contains no time derivatives, so $A \sb -$ is an
auxiliary field.  We therefore eliminate it by its
equation of motion:
$$ 0 = [ \de \sp a , F \sb {+a} ] = p \sb + \sp 2 A \sb - + [ \de \sp i , 
p \sb + A \sb i ] \quad \to \quad A \sb - = - {1 \over p \sb + \sp 2} [
\de \sp i , p \sb + A \sb i ] \quad . \eqno(2.1.7)$$
The only remaining fields are $A \sb i$, corresponding to the physical
transverse polarizations.  The lagrangian is then 
$$ \li{ \cl = & \ha A \sb i \bo A \sb i + [ A\sb i , A\sb j ] p\sb i A\sb j +
\frac14 [ A\sb i , A\sb j ]\sp 2 \cr
& + ( p\sb j A\sb j ) {1\over p\sb +} [ A\sb i , p\sb + A\sb i ] +
\ha \left( {1\over p\sb +} [ A\sb i , p\sb + A\sb i ] \right) ^2
\quad . & (2.1.8) \cr} $$

In fact, for {\it arbitrary} spin, after gauge-fixing 
($A \sb {+ \cdots} = 0$) and eliminating auxiliary fields 
($A \sb {- \cdots} = \cdots$), we get for the free theory
$$ \cl = - \j \sp \dag ( p \sb + ) \sp k ( p \sb - + H ) \j \quad 
,\eqno(2.1.9)$$
where $k = 1$ for bosons and $0$ for fermions.

The choice of light-cone gauges in particle mechanics will be discussed
in chapt.\ 5, and for string mechanics in sect.\ 6.3 and chapt.\ 7.
Light-cone field theory for strings will be discussed in chapt.\ 10.

\sect{2.2. Conformal algebra}

Since the free kinetic operator of any light-cone field is just $\bo$ 
(up to factors of $\pa \sb +$), the only nontrivial part of any free
light-cone field theory is the representation of the Poincar\'e group 
ISO(D$-$1,1) (see, e.g., [2.2]).  In the next section we will derive 
this representation for arbitrary massless theories (and will later 
extend it to the massive case) [2.3].   
These representations are nonlinear in the coordinates, and are constructed
from all the irreducible (matrix) representations of the light-cone's SO(D$-$2)
rotation subgroup of the spin part of the SO(D$-$1,1) Lorentz group.
One simple method of derivation involves the use of the conformal group,
which is SO(D,2) for D-dimensional spacetime (for $D>2$).  We therefore
use SO(D,2) notation by writing (D+2)-dimensional vector indices
which take the values $\pm$ as well as the usual D $a$'s:
$\ca = ( \pm , a )$.  The metric is as in (2.1.1) for the $\pm$ indices.
(These $\pm$'s should not be confused with the 
light-cone indices $\pm$, which are related but are a subset of the $a$'s.)  
We then write the conformal group generators as
$$ J\sb{\ca\cb} = ( J\sb{+a} = -ip\sb a ,\quad J\sb{-a} = -iK\sb a ,\quad
J\sb{-+} = \D ,\quad J\sb{ab} ) \quad , \eqno(2.2.1)$$
where $J\sb{ab}$ are the Lorentz generators, $\D$ is the dilatation 
generator, and $K\sb a$ are the conformal boosts.  An obvious linear
coordinate representation in terms of D+2 coordinates is
$$ J\sb{\ca\cb} = x\sb{[\ca}\pa\sb{\cb ]} + M\sb{\ca\cb} \quad ,\eqno(2.2.2)$$
where $[\quad ]$ means antisymmetrization and $M\sb{\ca\cb}$ is the intrinsic
(matrix, or coordinate-independent) part (with the same commutation
relations that follow directly for the orbital part).  The usual
representation in terms of D coordinates is obtained by imposing the
SO(D,2)-covariant constraints 
$$ x\sp\ca x\sb\ca = x\sp\ca \pa\sb\ca = M\sb\ca\sp\cb x\sb\cb +
\Sc d x\sb\ca = 0 \eqno(2.2.3a)$$
for some constant $\Sc d$ (the canonical dimension, or scale weight).
Corresponding to these constraints, which can be solved for everything
with a ``$-$'' index, are the ``gauge conditions'' which determine
everything with a ``$+$'' index but no ``$-$'' index:
$$ \pa\sb + = x\sb + - 1 = M\sb{+a} = 0 \quad . \eqno(2.2.3b)$$
This gauge can be obtained by a unitary transformation.  The solution to
(2.2.3) is then
$$ J\sb{+a} = \pa\sb a \quad , \quad J\sb{-a} = -\ha x\sb b\sp 2 \pa\sb a 
+ x\sb a x\sp b \pa\sb b + M\sb a\sp b x\sb b + \Sc d x\sb a \quad , $$
$$ J\sb{-+} = x\sp a\pa\sb a + \Sc d \quad , \quad
J\sb{ab} = x\sb{[a}\pa\sb{b]} + M\sb{ab} \quad . \eqno(2.2.4)$$

This realization can also be obtained by the usual coset space methods
(see, e.g., [2.4]), for the space SO(D,2)/ISO(D-1,1)$\otimes$GL(1).  
The subgroup corresponds to all the generators {\it except} $J\sb{+a}$.
One way to perform this construction is:  First assign the coset space
generators $J\sb{+a}$ to be partial derivatives $\pa\sb a$ (since they
all commute, according to the commutation relations which follow from
(2.2.2)).  We next equate this first-quantized coordinate representation
with a second-quantized field representation:  In general,
$$ 0 = \d \VEV{x \Big| \F} = \VEV{Jx \Big| \F} + \VEV{x \Big| \hat J \F} $$
$$ \to\quad J \VEV{x \Big| \F} = \VEV{Jx \Big| \F} = -\hat J \VEV{x \Big| \F}
= -\VEV{x \Big| \hat J \F} \quad ,\eqno(2.2.5)$$
where $J$ (which acts directly on $\sbra{x}$) is expressed in terms of 
the coordinates and their derivatives (plus ``spin'' pieces), while 
$\hat J$ (which acts directly on $\sket{\F}$) is expressed in terms of the 
{\it fields} $\F$ and their {\it functional} derivatives.  The minus
sign expresses the usual relation between active and passive transformations.
The structure constants of the second-quantized
algebra have the same sign as the first-quantized ones.  We can then
solve the ``constraint'' $J\sb{+a}=-\hat J\sb{+a}$ on $\left< x |\F\right>$ as
$$ \VEV{x \Big| \F} \equiv \F (x) = U \F (0) = 
e\sp{-x\sp a \hat J\sb{+a}} \F (0) \quad . \eqno(2.2.6)$$
The other generators can then be determined by evaluating
$$ J \F (x) = -\hat J \F (x) \quad\to\quad U\sp{-1} J U \F (0) =
-U\sp{-1} \hat J U \F (0) \quad . \eqno(2.2.7)$$
On the left-hand side, the unitary transformation replaces any $\pa\sb a$ 
with a $-\hat J\sb{+a}$ (the $\pa\sb a$ itself getting killed by the $\F
(0)$).  On the right-hand side, the transformation gives terms with $x$
dependence and other $\hat J$'s (as determined by the commutator algebra).
(The calculations are performed by expressing the transformation as a
sum of multiple commutators, which in this case has a finite number of
terms.)  The net result is (2.2.4), where $\Sc d$ is $-\hat J\sb{-+}$ on
$\F (0)$, $M\sb{ab}$ is $-\hat J\sb{ab}$, and $J\sb{-a}$ can have the 
additional term $-\hat J\sb{-a}$.  However, $\hat J\sb{-a}$ on $\F (0)$ 
can be set to zero consistently in (2.2.4), and does vanish for
physically interesting representations.

From now on, we use $\pm$ as in the light-cone notation, not SO(D,2)
notation.

The conformal equations of motion are all those which can be obtained from 
$p\sb a\sp 2 = 0$ by conformal transformations (or, equivalently, the 
irreducible
tensor operator quadratic in conformal generators which includes $p\sp 2$
as a component).  Since conformal
theories are a subset of massless ones, the massless equations of motion
are a subset of the conformal ones (i.e., the massless theories satisfy
fewer constraints).  In particular, since massless theories are scale
invariant but not always invariant under conformal boosts, the equations
which contain the generators of conformal boosts must be dropped.

The complete set of equations of motion for an arbitrary massless
representation of the Poincar\'e group are thus obtained simply by
performing a conformal boost on the defining equation, $p \sp 2 = 0$ [2.5,6]:
$$ 0 = \ha [ K \sb a , p \sp 2 ] = \ha \{ J \sb a \sp b , p \sb b \} + \ha \{
\D , p \sb a \} = M \sb a \sp b p \sb b + \left( \Sc d  - {D-2
\over 2} \right) p \sb a \quad . \eqno(2.2.8) $$
$\Sc d$ is determined by the requirement
that the representation be nontrivial (for other values of $\Sc d$ this
equation implies $p = 0$).  For nonzero spin ($M \sb {ab} \ne 0$) this equation
implies $p \sp 2 = 0$ by itself.  For example, for scalars the equation
implies only ${\Sc d} = (D-2)/2$.  For a Dirac spinor, $M \sb {ab} =
\frac14 [ \g \sb a , \g \sb b ]$ implies ${\Sc d} = (D-1)/2$ and the
Dirac equation (in the form $\g \sb a \g \cdot p \j = 0$).  For a
second-rank antisymmetric tensor, we find ${\Sc d} = D/2$ and Maxwell's
equations.  In this covariant approach to solving these equations, all
the solutions are in terms of field strengths, not gauge fields (since
the latter are not unitary representations).  We can solve these
equations in light-cone {\it notation}:  Choosing a reference frame where the
only nonvanishing component of the momentum is $p\sb +$, (2.2.8) reduces
to the equations $M\sb{-i} = 0$ and $M\sb{-+} = \Sc d - (D-2)/2$.  
The equation $M\sb{-i} = 0$ says that the only nonvanishing components
are the ones with as many (lower) ``$+$'' indices as possible (and for
spinors, project with $\g\sb +$), and no ``$-$'' indices.  In terms of 
Young tableaux, this
means 1 ``$+$'' for each column.  $M\sb{-+}$ then just counts the number
of ``$+$'' 's (plus 1/2 for a $\g\sb +$-projected spinor index), so we
find that $\Sc d - (D-2)/2 =$ the number of columns (+ 1/2 for a
spinor).  We also find that the {\it on-shell} gauge field is the
representation found by subtracting one box from each column of the
Young tableau, and in the field strength those subtracted indices are
associated with factors of momentum.

These results for massless representations can be extended to massive
representations by the standard trick of adding one spatial dimension
and constraining the extra momentum component to be the mass (operator):
Writing
$$ a \; \to ( a , m ) \quad , \quad p \sb m = M \quad , \eqno(2.2.9)$$
where the index $m$ takes one value, $p\sp 2= 0$ becomes $p\sp 2 + M\sp 2 = 0$,
and (2.2.8) becomes
$$ M\sb a\sp b p\sb b + M\sb {am} M + \left(\Sc d - {D-2\over 2}\right) 
p\sb a = 0 \quad . \eqno(2.2.10) $$
The fields (or states) are now representations of an SO(D,1) spin group
generated by $M\sb{ab}$ and $M\sb{am}$ (instead of the usual SO(D-1,1)
of just $M\sb{ab}$ for the massless case).  The fields additional to
those obtained in the massless case (on-shell field strengths)
correspond to the on-shell gauge fields in the massless limit, resulting
in a first-order formalism.  For example, for spin 1 the additional
field is the usual vector.  For spin 2, the extra fields correspond to
the on-shell, and thus traceless, parts of the Lorentz connection and
metric tensor.

For field theory, we'll be interested in real representations.  For the
massive case, since (2.2.9) forces us to work in momentum space with
respect to $p\sb m$, the reality condition should include an extra
factor of the reflection operator which reverses the ``$m$'' direction.
For example, for tensor fields, those components with an odd number of
$m$ indices should be imaginary (and those with an even number real).

In chapt.\ 4 we'll show how to obtain the off-shell fields, and thus the
trace parts, by working directly in terms of the gauge fields.  The
method is based on the light-cone representation of the Poincar\'e
algebra discussed in the next section.

\secty{2.3. Poincar\'e algebra}{2.3. Poincare algebra}

In contrast to the above covariant approach to solving (2.2.8,10), we now
consider solving them in unitary gauges (such as the light-cone gauge),
since in such gauges the gauge fields are essentially field strengths 
anyway because the gauge has been fixed: e.g., for Yang-Mills 
$A \sb a = \de \sb + \sp {-1} F \sb {+a}$, since $A \sb + = 0$.
In such gauges we work in terms of only the physical degrees of freedom
(as in the case of the on-shell field strengths), which satisfy $p\sp 2
= 0$ (unlike the auxiliary degrees of freedom, which satisfy algebraic
equations, and the gauge degrees of freedom, which don't appear in any
field equations).

In the light-cone formalism, the object is to construct all the 
Poincar\'e generators from just the manifest ones of the 
$(D-2)$-dimensional Poincar\'e subgroup, $p \sb +$, and the coordinates 
conjugate to these momenta.  The light-cone gauge is imposed by the condition
$$ M \sb {+i} = 0 \quad , \eqno(2.3.1) $$
which, when acting on the independent fields (those with only $i$
indices), says that all fields with $+$ indices have been set to vanish.
The fields with $-$ indices (auxiliary fields) are then determined as
usual by the field equations: by solving (2.2.8) for $M \sb {-i}$.  The
solution to the $i$, $+$, and $-$ parts of (2.2.8) gives
$$ M \sb {-i} = {1 \over p \sb +} ( M \sb i \sp j p \sb j + k p \sb i ) 
\quad , $$
$$ M \sb {-+} = \Sc d - {D-2 \over 2} \equiv k \quad , $$
$$ k p \sp 2 = 0 \quad . \eqno(2.3.2)$$
If (2.2.8) is solved without the condition (2.3.1), then $M\sb{+i}$ can
still be removed (and (2.3.2) regained) by a unitary transformation.
(In a first-quantized formalism, this corresponds to a gauge choice:
see sect.\ 5.3 for spin 1/2.)  The appearance of $k$ is related to 
ordering ambiguities, and we can also choose $M\sb{-+}=0$ by a 
{\it non}unitary transformation (a rescaling of the field by a power 
of $p\sb +$).  Of course, we also solve $p \sp 2 = 0$ as
$$ p \sb - = - {p \sb i \sp 2 \over 2 p \sb +} \quad . \eqno(2.3.3)$$
These equations, together with the gauge condition for $M \sb {+i}$,
determine all the Poincar\'e generators in terms of $M \sb {ij}$, $p \sb
i$, $p \sb +$, $x \sb i$, and $x \sb -$.  In the orbital pieces of $J
\sb {ab}$, $x \sb +$ can be set to vanish, since $p \sb -$ is no longer
conjugate: i.e., we work at ``time'' $x \sb + = 0$ for the
``hamiltonian'' $p \sb -$, or equivalently in the Schr\"odinger picture.
(Of course, this also corresponds to removing $x\sb +$ by a unitary
transformation, i.e., a time translation via $p\sb -$.  This is also a
gauge choice in a first-quantized formalism: see sect.\ 5.1.)
The final result is
$$ p \sb i = i \pa \sb i \quad , \quad p \sb + = i \pa \sb + \quad ,
\quad p \sb - = - {p \sb i \sp 2 \over 2 p \sb +} \quad , $$
$$ J \sb {ij} = - i x \sb {[i} p \sb {j]} + M \sb {ij} \quad , \quad
J \sb {+i} = i x \sb i p \sb + \quad , \quad 
J \sb {-+} = - i x \sb - p \sb + + k \quad , $$
$$ J \sb {-i} = - i x \sb - p \sb i - i x \sb i {p \sb j \sp 2 \over 2 p \sb +}
+ {1 \over p \sb +} ( M \sb i\sp j p\sb j + k p \sb i ) \quad . \eqno(2.3.4)$$
The generators are (anti)hermitian for the choice $k
= \ha$; otherwise, the Hilbert space metric must include a factor of 
$p \sb + \sp {1-2k}$, with respect to which all the generators are
pseudo(anti)hermitian.  In this light-cone approach to Poincar\'e
representations, where we work with the fundamental fields rather than
field strengths, $k = 0$ for bosons and $\ha$ for fermions (giving the
usual dimensions ${\Sc d}= \ha (D-2)$ for bosons and $\ha (D-1)$ for
fermions), and thus the metric is $p \sb +$ for bosons and $1$ for
fermions, so the light-cone kinetic operator (metric)$\cdot 2 ( i \pa
\sb - - p \sb - ) \sim \bo$ for bosons and $\bo / p \sb +$ for fermions.

This construction of the D-dimensional Poincar\'e algebra in terms of D$-$1
coordinates is analogous to the construction in the previous section 
of the D-dimensional
conformal algebra SO(D,2) in terms of $D$ coordinates, except that in
the conformal case (1) we start with D+2 coordinates instead of D,
(2) $x$'s and $p$'s are switched, and (3) the further constraint $x
\cdot p = 0$ and gauge condition $x\sb + =1$ are used.  Thus, $J\sb{ab}$
of (2.3.4) becomes $J\sb{\ca\cb}$ of (2.2.4) if $x \sb -$ is replaced with 
$- (1/p \sb +) x \sp j p \sb j$, $p\sb +$ is set to 1, and we then
switch $p\to x , x\to -p$.  Just as the conformal representation (2.2.4)
can be obtained from the Poincar\'e representation (in 2 extra
dimensions, by $i\to a$) (2.3.4) by eliminating one coordinate ($x\sb -$),
(2.3.4) can be reobtained from (2.2.4) by reintroducing this coordinate:
First choose $\Sc d = -ix\sb -p\sb + +k$.  Then switch $x\sb i \to p\sb i$,
$p\sb i\to -x\sb i$.  Finally, make the (almost unitary) transformation
generated by $exp[-ip\sp ix\sb i(ln~p\sb +)]$, which takes $x\sb i \to
p\sb +x\sb i$, $p\sb i\to p\sb i/p\sb +$, $x\sb -\to x\sb - +p\sp ix\sb i
/p\sb +$.

To extend these results to arbitrary representations, we use the trick
(2.2.9), or directly solve (2.2.10), giving the light-cone form of the 
Poincar\'e algebra for arbitrary representations:  (2.3.4) becomes
$$ p \sb i = i \pa \sb i \quad , \quad p \sb + = i \pa \sb + \quad ,
\quad p \sb - = - {p \sb i \sp 2 + M \sp 2 \over 2 p \sb +} \quad , $$
$$ J \sb {ij} = - i x \sb {[i} p \sb {j]} + M \sb {ij} \quad , \quad
J \sb {+i} = i x \sb i p \sb + \quad , \quad 
J \sb {-+} = - i x \sb - p \sb + + k \quad , $$
$$ J \sb {-i} = - i x \sb - p \sb i - i x \sb i {p \sb j \sp 2 + M \sp 2
\over 2 p \sb +} + {1 \over p \sb +} ( M \sb i\sp j p \sb j + M \sb {im} M
+ k p \sb i ) \quad . \eqno(2.3.5)$$
Thus, massless irreducible representations of the Poincar\'e group
ISO(D$-$1,1) are irreducible representations of the spin subgroup
SO(D$-$2) (generated by $M \sb {ij}$) which also depend on the
coordinates $( x \sb i , x \sb - )$, and irreducible massive ones are
irreducible representations of the spin subgroup SO(D$-$1) (generated by
$( M \sb {ij} , M \sb {im} )$) for some nonvanishing constant $M$.
Notice that the introduction of masses has modified only $p \sb -$ and
$J \sb {-i}$.  These are also the only generators modified when
interactions are introduced, where they become nonlinear in the fields.

The light-cone representation of the Poincar\'e algebra will be used in 
sect.\ 3.4 to derive BRST algebras, used for enforcing unitarity in 
covariant formalisms, which in turn will be used extensively to derive
gauge-invariant actions for particles and strings in the following chapters.
The general light-cone analysis of this section will be applied to the
special case of the free string in chapt.\ 7.

\sect{2.4. Interactions}

For interacting theories, the derivation of the Poincar\'e algebra is
not so general, but depends on the details of the particular type of
interactions in the theory.  We again consider the case of Yang-Mills.  
Since only $p \sb -$ and $J \sb {-i}$ obtain interacting contributions, 
we consider the derivation of only those operators.  The expression for 
$p \sb - A \sb i$ is then given directly by the field equation of $A \sb i$
$$ 0 = [ \de \sp a , F \sb {ai} ] = [ \de \sp j , F \sb {ji} ] + [ \de \sb + ,
F \sb {-i} ] + [ \de \sb - , F \sb {+i} ] = [ \de \sp j , F \sb {ji} ] +
2 [ \de \sb + , F \sb {-i} ] + [ \de \sb i , F \sb {+-} ] $$
$$ \to \quad p \sb - A \sb i = [ \de \sb i , A \sb - ] - {1 \over 2 p
\sb +} \left( [ \de \sp j , F \sb {ji} ] + [ \de \sb i , p \sb + A \sb - ]
\right) \quad , \eqno(2.4.1)$$
where we have used the Bianchi identity $[ \de \sb {[+} , F \sb {-i]} ]
= 0$.  This expression for $p \sb -$ is also used in the orbital
piece of $J \sb {-i} A \sb j$.  In the spin piece $M \sb {-i}$ we start
with the covariant-formalism equation $M \sb {-i} A \sb j = - \d \sb {ij}
A \sb -$, substitute the solution to $A \sb -$'s field equation, and
then add a gauge transformation to cancel the change of gauge induced 
by the covariant-formalism transformation $M \sb {-i} A \sb + = A \sb i$.
The net result is that in the light-cone formalism
$$ J \sb {-i} A \sb j = - i ( x \sb - p \sb i - x \sb i p \sb - ) A \sb j
- \left( \d \sb {ij} A \sb - + [ \de \sb j , {1 \over p \sb +} A \sb i ]
\right) \quad , \eqno(2.4.2)$$
with $A \sb -$ given by (2.1.7) and $p \sb - A \sb j$ by (2.4.1).  In the
abelian case, these expressions agree with those obtained by a different
method in (2.3.4).  
All transformations can then be written in functional second-quantized form as
$$ \d = -\int d \sp {D-2} x \sb i d x \sb -  \; tr \; ( \d A \sb i )
{\d \over \d A \sb i} \quad\to\quad [ \d , A\sb i ] = - ( \d A\sb i )
\quad . \eqno(2.4.3)$$
The minus sign is as in (2.2.5) for relating first- and second-quantized
operators.

As an alternative, we can consider canonical second-quantization, which
has certain advantages in the light cone, and has an interesting
generalization in the covariant case (see sect.\ 3.4).
From the light-cone lagrangian
$$ L = -i\int \F\dg p_+ \dt\F - H (\F ) \quad ,\eqno(2.4.4)$$
where $\dt{\phantom m}$ is the ``time''-derivative $i\pa / \pa x_+$,
we find that the fields have equal-time commutators similar
to those in nonrelativistic field theory:
$$ [ \F\dg (1) , \F (2) ] = -{1\over 2p_{+2}} \d (2-1) \quad ,\eqno(2.4.5)$$
where the $\d$-function is over the transverse coordinates and $x_-$
(and may include a Kronecker $\d$ in indices, if $\F$ has components). 
Unlike nonrelativistic field theory, the fields satisfy a reality
condition, in coordinate space:
$$ \F\conj = \O\F \quad , \eqno(2.4.6) $$
where $\O$ is the identity or some symmetric, unitary matrix
(the ``charge conjugation'' matrix; $\conj$ here is the hermitian
conjugate, or adjoint, in the operator sense, i.e., unlike $\dg$, it
excludes matrix transposition).  As in quantum mechanics
(or the Poisson bracket approach to classical mechanics),
the generators can then be written as functions of the dynamical variables:
$$ V = \su_n{1\over n!}\int dz_1\cdots dz_{n}\;
\cv^{(n)}  (z_1,\dots,z_n)\Phi(z_1)\cdots\Phi(z_n) \quad , \eqno(2.4.7)$$
where the arguments $z$ stand for either coordinates
or momenta and the ${\cal V}$'s are the vertex functions, which are just
functions of the coordinates (not operators).  Without
loss of generality they can be chosen to be
cyclically symmetric in the fields (or totally symmetric, if
group-theory indices are also permuted).  (Any asymmetric piece can be
seen to contribute to a lower-point function by the use of (2.4.5,6).)
In light-cone theories the coordinate-space integrals are over all
coordinates except $x_+$.
The action of the second-quantized operator $V$ on fields is calculated
using (2.4.5):
$$ [ V , \Phi(z_1)\dg ] = -{1\over 2p_{+1}}\su_n{1\over (n-1)!}
\int dz_2\cdots dz_n \; \cv^{(n)} (z_1,\dots,z_n)
\Phi(z_2)\cdots\Phi(z_n) \quad .\eqno(2.4.8)$$
A particular case of the above equations is the free case, where the operator
$V$ is quadratic in $\F$.  We will then generally write the
second-quantized operator $V$ in terms of a first-quantized operator
$\cv$ with a single integration:
$$ V = \int dz \; \F\dg p_+\cv\F \quad\to\quad
[ V , \F ] = -\cv \F \quad .\eqno(2.4.9)$$
This can be checked to relate to (2.4.7) as $\cv^{(2)}(z_1,z_2)=
2\O_1 p_{+1}\cv_1 \d (2-1)$ (with the symmetry of $\cv^{(2)}$ 
imposing corresponding conditions on the operator $\cv$).
In the interacting case, the generalization of (2.4.9) is
$$ V = {1\over N} \int dz \; \F\dg 2p\sb + (\cv\F ) \quad , \eqno(2.4.10)$$
where $N$ is just the number of fields in any particular term.  (In the
free case $N=2$, giving (2.4.9).)

For example, for Yang-Mills, we find
$$ p\sb - = \int \frac14 (F\sb{ij})\sp 2 +\ha (p\sb + A\sb -)\sp 2
\quad , \eqno(2.4.11a)$$
$$ J\sb{-i} = \int ix\sb - (p\sb + A\sb j )(p\sb i A\sb j )
+ix\sb i \left[ \frac14 (F\sb{jk})\sp 2 +\ha (p\sb + A\sb - )\sp 2
\right] - A\sb i p\sb +A\sb - \quad . \eqno(2.4.11b)$$
(The other generators follow trivially from (2.4.9).)  
$p\sb -$ is minus the hamiltonian $H$ (as in the free case (2.1.2,4,9)),
as also follows from performing the usual Legendre transformation on the
lagrangian.

In general, all the explicit $x_i$-dependence of
all the Poincar\'e generators can be determined from the commutation
relations with the momenta (translation generators) $p\sb i$.
Furthermore, since only $p_-$ and $J_{-i}$ get contributions from
interactions, we need consider only those.
Let's first consider the ``hamiltonian'' $p_-$.  Since it commutes with
$p_i$, it is translation invariant.  In
terms of the vertex functions, this translates into the condition:
$$ (p_1 +\cdots +p_n ) \Tilde\cv^{(n)}(p_1,\dots,p_n)= 0\quad ,\eqno(2.4.12)$$
where the $\Tilde{\phantom M}$ indicates Fourier transformation with
respect to the coordinate-space expression, implying that most generally
$$ \Tilde\cv^{(n)}(p_1,\dots,p_n) = \tilde f(p_1,\dots,p_{n-1})
\delta(p_1+\cdots+p_n) \quad ,\eqno(2.4.13)$$
or in coordinate space
$$\li{\cv^{(n)}(x_1,\dots,x_n) &= \tilde f
\left(i{\partial\over\partial x_1},\dots,i{\partial\over\partial x_{n-1}}
\right)\delta(x_1-x_n)\cdots\delta(x_{n-1}-x_n) \cr
&= f(x_1-x_n,\dots,x_{n-1}-x_n) \quad .&(2.4.14)\cr}$$
In this coordinate representation one can see that when ${\cal V}$
is inserted back in (2.4.7)  we have the usual
expression for a translation-invariant vertex used in field theory.
Namely, fields at the same point in coordinate space, with
derivatives acting on them, are multiplied and integrated over
coordinate space.  In this form it is clear that there is no
explicit coordinate dependence in the vertex.  As can be seen
in (2.4.14), the most general
translationally invariant vertex involves an arbitrary function
of coordinate differences, denoted as $f$ above.  For the case of
bosonic coordinates, the function $\tilde f$ may
contain inverse derivatives (that is, translational invariance does
not imply locality.)  For the case of anticommuting coordinates (see
sect.\ 2.6) the situation is simpler:  There is no locality issue, since
the most general function $f$ can always be obtained from a function 
$\tilde f$ polynomial in derivatives, acting on $\d$-functions.

We now consider $J_{-i}$.  From the commutation relations we find:
$$ [ p_i , J_{-j} \} = -\eta_{ij}p_- \quad\to\quad
[ J_{-i} , \F ] = ix_i [ p_- , \F ] + [ \Delta J_{-i} , \F ] \quad , 
\eqno(2.4.15)$$
where $\Delta J_{-i}$ is translationally invariant (commutes with $p\sb
i$), and can therefore be represented without explicit $x^i$'s.
For the Yang-Mills case, this can be seen to agree with (2.4.2) or (2.4.11).

This light-cone analysis will be applied to interacting strings in chapt.\ 10.

\sect{2.5. Graphs}

Feynman graphs for any interacting light-cone field theory can be
derived as in covariant field theory, but an
alternative not available there is to use a nonrelativistic style of
perturbation (i.e., just expanding $e \sp {iHt}$ in $H \sb {INT}$), 
since the field equations are now linear in the time derivative $p 
\sb - = i \pa / \pa x \sb + = i \pa / \pa \t$.  (As in sect.\ 2.1,
but unlike sects.\ 2.3 and 2.4, we now use $p\sb -$ to refer to this
partial derivative, as in covariant formalisms, while $-H$ refers to the
corresponding light-cone Poincar\'e generator, the two being equal on shell.)
This formalism can be derived 
straightforwardly from the usual Feynman rules (after choosing the 
light-cone gauge and eliminating auxiliary fields) by simply Fourier 
transforming from $p \sb -$ to $x \sb + = \t$ (but keeping all other 
momenta):
$$ \int_{-\infty}^\infty {dp \sb - \over 2 \p} e \sp {-ip \sb - \t}
{1 \over 2 p \sb + p \sb - + p\sb i\sp 2+m\sp 2 + i \e} =
- i \Q ( p \sb + \t ) {1 \over 2 | p \sb + |} e \sp {i \t (p\sb i\sp 2+m\sp 2) / 
2 p \sb +} \quad .\eqno(2.5.1)$$
($\Q (u)=1$ for $u>1$, $0$ for $u<1$.)
We now draw all graphs to represent the $\t$ coordinate, so that 
graphs with different $\t$-orderings of the vertices must be 
considered as separate contributions.  Then we direct all the 
propagators toward increasing $\t$, so the change in $\t$ between the 
ends of the propagator (as appears in (2.5.1)) is always positive 
(i.e., the orientation of the momenta is defined to be toward 
increasing $\t$).  We next Wick rotate $\t \to i \t$.  We also 
introduce external line factors which transform $H$ back to 
$-p \sb -$ on external lines.  The resulting rules are:
\Item{(a)}Assign a $\t$ to each vertex, and order them with respect to $\t$.
\Item{(b)}Assign $( p \sb - , p \sb + , p\sb i )$ to each external 
line, but only $( p \sb + , p\sb i )$ to each internal line, all 
directed toward increasing $\t$.  Enforce conservation of $( p \sb + , 
p\sb i )$ at each vertex, and total conservation of $p \sb -$.
\Item{(c)}Give each internal line a propagator
$$\Q ( p\sb + ) {1\over 2p\sb +} e\sp{- \t (p\sb i\sp 2+m\sp 2) / 2 p \sb +}$$
for the $( p \sb + , p\sb i )$ of that line and the positive 
difference $\t$ in the proper time between the ends.
\Item{(d)}Give each external line a factor
$$ e \sp {\t p \sb -}$$
for the $p \sb -$ of that line and the $\t$ of the vertex to which it connects.
\Item{(e)}Read off the vertices from the action as usual.
\Item{(f)}Integrate
$$ \int_0^\infty d \t$$
for each $\t$ difference between consecutive (though not necessarily 
connected) vertices.  (Performing just this integration gives the 
usual old-fashioned perturbation theory in terms of energy 
denominators [2.1], except that our external-line factors differ off shell 
in order to reproduce the usual Feynman rules.)
\Item{(g)}Integrate
$$ \int_{-\infty}^\infty {dp \sb + \; d \sp {D-2} p \sb i \over ( 2 \p 
) \sp {D - 1}}$$
for each loop.

The use of such methods for strings will be discussed in chapt.\ 10.

\sect{2.6. Covariantized light cone}

There is a covariant formalism for any field theory that has the interesting 
property that it can be obtained directly and easily from the light-cone 
formalism, without any additional gauge-fixing procedure [2.7].  Although 
this covariant gauge is not as general or convenient as the usual covariant 
gauges (in particular, it sometimes has additional off-shell infrared 
divergences), it bears strong relationship to both the light-cone and 
BRST formalisms, and can be used as a conceptual bridge.  The basic 
idea of the formalism is:  Consider a covariant theory in $D$ 
dimensions.  This is equivalent to a covariant theory in $(D+2)-2$ 
dimensions, where the notation indicates the addition of 2 extra
commuting coordinates (1 space, 1 time) and 2 (real) anticommuting coordinates,
with a similar extension of Lorentz indices [2.8].  (A similar
use of OSp groups in gauge-fixed theories, but applied to only the
Lorentz indices and not the coordinates, appears in [2.9].)
This extends the Poincar\'e group
ISO(D$-$1,1) to a graded analog IOSp(D,2$|$2).  In practice, this means 
we just take the light-cone transverse indices to be graded, 
watching out for signs introduced by the corresponding change in statistics, 
and replace the Euclidean SO(D-2) metric with the corresponding graded
OSp(D-1,1$|$2) metric:
$$ i = ( a , \a ) \quad , \quad \d \sb {ij} \; \to \;
\h \sb {ij} = ( \h \sb {ab} , C \sb {\a\b} ) \quad , \eqno(2.6.1) $$ 
where $\h \sb {ab}$ is the usual Lorentz metric and
$$ C \sb {\a\b} = C \sp {\b\a} = \s \sb 2 \eqno(2.6.2) $$
is the Sp(2) metric, which  satisfies the useful identity
$$ C\sb{\a\b}C\sp{\g\d} = \d\sb{[\a}\sp\g\d\sb{\b ]}\sp\d \quad\to\quad
A\sb{[\a}B\sb{\b ]}=C\sb{\a\b}C\sp{\g\d}A\sb\g B\sb\d \quad .\eqno(2.6.3)$$
The OSp metric is used to raise and lower graded indices as:
$$ x \sp i = \h \sp {ij} x \sb j \quad , \quad x \sb i = x \sp j \h \sb
{ji} \quad ; \qquad \h\sp{ik}\h\sb{jk} = \d\sb j\sp i \quad . \eqno(2.6.4)$$
The sign conventions are that adjacent indices are contracted with the
contravariant (up) index first.  The 
equivalence follows from the fact that, for momentum-space Feynman 
graphs, the trees will be the same if we constrain the $2-2$ extra 
``ghost'' momenta to vanish on external lines (since they'll then 
vanish on internal lines by momentum conservation); and the loops are 
then the same because, when the momentum integrands are written as 
gaussians, the determinant factors coming from the 2 extra 
anticommuting dimensions exactly cancel those from the 2 extra 
commuting ones.  For example, using the proper-time form (``Schwinger 
parametrization'') of the propagators (cf.\ (2.5.1)), 
$$ {1\over p\sp 2 + m\sp 2} = \int_0^\infty d\t \; e\sp{-\t (p\sp 2 +
m\sp 2)} \quad , \eqno(2.6.5)$$
all momentum integrations take the form

\begin{large}
$$ \li{ {1 \over \p} \int d \sp {D+2} p \; d\sp 2 p\sb\a
\; e \sp {- f ( 2p\sb + p\sb - + p\sp a p\sb a + p\sp\a p\sb\a +m\sp 2)} &=
\int d \sp D p \; e \sp {-f(p \sp a p\sb a+m\sp 2)} \cr
&=\left({\p\over f}\right)^{D/2}e\sp{-fm\sp 2}\quad ,
&\mbox{\normalsize (2.6.6)}\cr}$$
\end{large}

\noindent where $f$ is a function of the proper-time parameters.

The covariant theory is thus obtained from the light-cone one by the 
substitution
$$ ( p \sb - , p \sb + ; p \sb i ) \quad \to \quad ( p\sb - , p\sb + ; p 
\sb a , p \sb\a ) \quad ,\eqno(2.6.7a)$$
where
$$ p\sb - = p \sb \a = 0 \eqno(2.6.7b)$$
on physical states.  It's not necessary to set $p\sb + = 0$, since it only 
appears in the combination $p\sb - p\sb +$ in OSp(D,2$|$2)-invariant
products.   
Thus, $p\sb +$ can be chosen arbitrarily on external lines (but should be 
nonvanishing due to the appearance of factors of $1 / p\sb +$).  
We now interpret $x\sp\pm$ and $x\sp\a$ as the unphysical
coordinates.  Vector indices on fields are treated similarly:  Having been 
reduced to transverse ones by the light-cone formalism, they now become 
covariant vector indices with 2 additional anticommuting values ((2.6.1)).
For example, in Yang-Mills the vector 
field becomes the usual vector field plus two anticommuting scalars $A 
\sb \a$, corresponding to Faddeev-Popov ghosts.

The graphical rules become:
\Item{(a)}Assign a $\t$ to each vertex, and order them with respect to $\t$.
\Item{(b)}Assign $( p\sb + , p \sb a )$ to each external 
line, but $( p\sb + , p \sb a , p \sb\a )$ to each internal 
line, all directed toward increasing $\t$.  Enforce conservation of 
$( p\sb + , p \sb a , p \sb\a )$ at each vertex (with 
$p \sb \a = 0$ on external lines).
\Item{(c)}Give each internal line a propagator

\begin{large}
$$ \Q ( p\sb + ) {1 \over 2 p\sb +} e \sp {- \t ( p \sb a \sp 2 + p
\sp\a p \sb \a + m\sp 2 ) / 2 p\sb +}$$
\end{large}

\noindent for the $( p\sb + , p \sb a , p \sb\a )$ of 
that line and the positive difference $\t$ in the proper time between the ends.
\Item{(d)}Give each external line a factor
$$ 1 \quad .$$
\Item{(e)}Read off the vertices from the action as usual.
\Item{(f)}Integrate
$$ \int_0^\infty d \t$$
for each $\t$ difference between consecutive (though not necessarily 
connected) vertices.  
\Item{(g)}Integrate
$$ \int d\sp 2 p \sb\a $$
for each loop (remembering that for any anticommuting variable $\q$, 
$\int d \q \; 1 = 0$, $\int d \q \; \q = 1$, $\q \sp 2 = 0$).
\Item{(h)}Integrate
$$ 2 \int_{-\infty}^\infty d p\sb +$$
for each loop.
\Item{(i)}Integrate
$$ \int {d \sp D p \over ( 2 \p ) \sp D }$$
for each loop.

For theories with only scalars, integrating just (f-h) gives the 
usual Feynman graphs (although it may be necessary to add several graphs 
due to the $\t$-ordering of non-adjacent vertices).  Besides the 
correspondence of the $\t$ parameters to the usual Schwinger 
parameters, after integrating out just the anticommuting parameters 
the $p\sb +$ parameters resemble Feynman parameters.

These methods can also be applied to strings (chapt.\ 10).

\sect{Exercises}

\Item{(1)} Find the light-cone formulation of QED.  Compare with the
Coulomb gauge formulation.
\Item{(2)} Derive the commutation relations of the conformal group from
(2.2.2).  Check that (2.2.4) satisfies them.  Evaluate the commutators
implicit in (2.2.7) for each generator.
\Item{(3)} Find the Lorentz transformation $M\sb{ab}$ of a vector 
(consistent with the conventions of (2.2.2)).  (Hint:  Look at the
transformations of $x$ and $p$.)  Find the explicit form of (2.2.8) for 
that case.  Solve these equations of motion.  To what simpler
representation is this equivalent?  Study this equivalence with the
light-cone analysis given below (2.2.8).  Generalize the analysis to
totally antisymmetric tensors of arbitrary rank.
\Item{(4)} Repeat problem (3) for the massive case.  Looking at the
separate SO(D-1,1) representations contained in the SO(D,1)
representations, show that first-order formalisms in terms of the usual
fields have been obtained, and find the corresponding second-order
formulations.
\Item{(5)} Check that the explicit forms of the Poincar\'e generators
given in (2.3.5) satisfy the correct algebra (see problem (2)).  Find
the explicit transformations acting on the vector representation of the
spin group SO(D-1).  Compare with (2.4.1-2).
\Item{(6)} Derive (2.4.11).  Compare that $p\sb -$ with the light-cone
hamiltonian which follows from (2.1.5).
\Item{(7)} Calculate the 4-point amplitude in $\f\sp 3$ theory with
light-cone graphs, and compare with the usual covariant Feynman graph
calculation.  Calculate the 1-loop propagator correction in the same
theory using the {\it covariantized} light-cone rules, and again compare
with ordinary Feynman graphs, paying special attention to Feynman parameters.

%
%

\chsc{3. GENERAL BRST}{3.1. Gauge invariance and constraints}7

In the previous chapter we saw that a gauge theory can be described
either in a manifestly covariant way by using gauge degrees of freedom, 
or in a manifestly unitary way (with only physical degrees of freedom)
with Poincar\'e transformations which are nonlinear (in both coordinates
and fields).  In the gauge-covariant formalism there is a $D$-dimensional
manifest Lorentz covariance, and in the light-cone formalism a 
$D-2$-dimensional
one, and in each case a corresponding number of degrees of freedom.
There is also an intermediate formalism, more familiar from
nonrelativistic theory:  The hamiltonian formalism has a
$D-1$-dimensional manifest Lorentz covariance (rotations).  
As in the light-cone formalism,
the notational separation of coordinates into time and space suggests a
particular type of gauge condition: temporal (timelike) gauges, where
time-components of gauge fields are set to vanish.  In chapt.\ 5, this
formalism will be seen to have a particular advantage for first-quantization 
of relativistic theories:  In the classical mechanics of
relativistic theories, the coordinates are treated as functions of a
``proper time'' so that the usual time coordinate can be treated on an
equal footing with the space coordinates.  Thus, canonical quantization
with respect to this unobservable (proper) ``time'' coordinate doesn't destroy
manifest Poincar\'e covariance, so use of a hamiltonian formalism can be
advantageous, particularly in deriving BRST transformations, and the
corresponding second-quantized theory, where the proper-time doesn't
appear anyway.

We'll first consider Yang-Mills, and then generalize to arbitrary gauge
theories.  In order to study the temporal gauge, instead of the 
decomposition (2.1.1) we simply separate into time and spatial components
$$ a = ( 0 , i ) \quad , \quad A \cdot B = - A \sb 0 B \sb 0 + A \sb i B
\sb i \quad . \eqno(3.1.1)$$
The lagrangian (2.1.5) is then
$$ \cl = \frac14 F \sb {ij} \sp 2 - \ha ( p \sb 0 A \sb i - [ \de \sb i ,
A \sb 0 ] ) \sp 2 \quad . \eqno(3.1.2)$$
The gauge condition
$$ A \sb 0 = 0 \eqno(3.1.3)$$
transforms under a gauge transformation with a time derivative:  Under
an infinitesimal transformation about $A \sb 0 = 0$, 
$$ \d A \sb 0 \approx \pa \sb 0 \l \quad , \eqno(3.1.4)$$
so the Faddeev-Popov ghosts are propagating.  Furthermore, the gauge
transformation (3.1.4) does not allow the gauge choice (3.1.3) everywhere:
For example, if we choose periodic boundary conditions in time (to
simplify the argument), then
$$ \d \int_{-\infty}^\infty d x \sp 0 \; A \sb 0 \approx 0 \quad .
\eqno(3.1.5)$$ 
$A \sb 0$ can then be fixed by an appropriate initial condition, e.g.,
$A \sb 0 | \sb {x \sp 0 = 0} = 0$, but then the corresponding field
equation is lost.  Therefore, we must impose
$$ 0 = {\d S \over \d A \sb 0} = - [ \de \sb i , F \sb {0i} ] = - [ \de
\sb i , p \sb 0 A \sb i ] \quad at~x \sp 0 = 0 \eqno(3.1.6)$$
as an initial condition.  Another way to understand this is to note that
gauge fixing eliminates only degrees of freedom which don't occur in the
lagrangian, and thus can eliminate only redundant equations of motion:
Since $[ \de \sb i , F \sb {0i} ] = 0$ followed from the gauge-invariant
action, the fact that it doesn't follow after setting $A \sb 0 = 0$
means some piece of $A \sb 0$ can't truly be gauged away, and so we must
compensate by imposing the equation of motion for that piece.
Due to the original gauge invariance, (3.1.6)
then holds for all time from the remaining field equations:  In
the gauge (3.1.3), the lagrangian (3.1.2) becomes
$$ \cl = \ha A \sb i \bo A \sb i - \ha ( p\sb i A\sb i )\sp 2 +
[ A \sb i , A\sb j ] p\sb i A\sb j +\frac14 [ A\sb i , A\sb j ]\sp 2
\quad , \eqno(3.1.7)$$
and the covariant divergence of the implied field equations yields the
{\it time derivative} of (3.1.6).  (This follows from the identity
$[ \de\sp b , [ \de\sp a , F\sb{ab} ]] =0$ upon applying the field equations
$[ \de\sp a , F\sb{ia} ] =0$.  In unitary gauges, the corresponding
constraint can be derived without time derivatives, and hence is implied
by the remaining field equations under suitable boundary conditions.)  
Equivalently, if we notice that (3.1.4)
does not fix the gauge completely, but leaves time-independent gauge
transformations, we need to impose a constraint on the initial states to
make them gauge invariant.  But the generator of the residual gauge 
transformations on the remaining fields $A \sb i$ is
$$ \cg ( x \sb i ) = \left[ \de \sb i , i\fder{A\sb i} \right]
\quad , \eqno(3.1.8)$$
which is the same as the constraint (3.1.6) under canonical quantization
of (3.1.7).  Thus, the same operator (1) gives the constraint which must
be imposed in addition to the field equations because too much of $A \sb
0$ was dropped, and (2) (its transpose) gives the gauge transformations 
remaining because they left the gauge-fixing function $A \sb 0$ invariant.  
The fact that these are identical is not surprising, since in Faddeev-Popov
quantization the latter corresponds to the Faddeev-Popov ghost while the former
corresponds to the antighost.

These properties appear very naturally in a hamiltonian formulation:
We start again with the gauge-invariant lagrangian (3.1.2).  Since $A\sb
0$ has no time-derivative terms, we Legendre transform with respect to
just $\dt A\sb i$.  The result is
$$ S\sb H = {1\over g\sp 2}\int d\sp D x\; tr\; \cl\sb H \quad , \quad
\cl\sb H = \dt A\sp i \P\sb i - \ch \quad , \quad
\ch = \ch\sb 0 + A\sb 0 i \cg \quad , $$
$$ \ch\sb 0 = \ha\P\sb i\sp 2 -\frac14 F\sb{ij}\sp 2 \quad , \quad
\cg = [ \de\sb i , \P\sb i ] \quad , \eqno(3.1.9)$$
where $\dt{\phantom m} = \pa\sb 0$. 
As in ordinary nonrelativistic classical mechanics, eliminating the
momentum $\P\sb i$ from the hamiltonian form of the action (first order
in time derivatives) by its equation of motion gives back the lagrangian
form (second order in time derivatives).  Note that $A\sb 0$ appears
linearly, as a Lagrange multiplier.

The gauge-invariant hamiltonian formalism of (3.1.9) can be generalized [3.1]:
Consider a lagrangian of the form
$$ \cl\sb H = \dt z \sp M e\sb M\sp A (z) \p\sb A - \ch
\quad , \quad \ch = \ch\sb 0 ( z , \p ) + \l\sp i i\cg\sb i ( z , \p ) 
\quad , \eqno(3.1.10)$$
where $z$, $\p$, and $\l$ are the variables, representing ``coordinates,''
covariant ``momenta,'' and Lagrange multipliers, respectively.  They 
depend on the time,
and also have indices (which may include continuous indices, such as
spatial coordinates).  $e$, which is a function of $z$,
has been introduced to allow for cases with a symmetry (such as
supersymmetry) under which $dz\sp M e\sb M\sp A$ (but not $dz$ itself)
is covariant, so that $\p$ will be covariant, and thus a more convenient
variable in terms of which to express the constraints $\cg$.  When $\ch\sb
0$ commutes with $\cg$ (quantum mechanically, or in terms of Poisson
brackets for a classical treatment), this action has a gauge invariance
generated by $\cg$, for which $\l$ is the gauge field:
$$ \d ( z , \p ) = [ \z\sp i \cg\sb i , ( z , \p ) ] \quad , $$
$$ \d \left( \der t - \l\sp i\cg\sb i \right) = 0 \quad\to\quad
(\d\l\sp i)\cg\sb i = \dt\z\sp i\cg\sb i + [\l\sp j\cg\sb j ,
\z\sp i\cg\sb i ] \quad , \eqno(3.1.11)$$
where the gauge transformation of $\l$ has been determined by the
invariance of the ``total'' time-derivative $d/dt = \pa /\pa t +i\ch$.
(More generally, if $[\z\sp i\cg\sb i,\ch\sb 0]=f\sp ii\cg\sb i$, then
$\d\l\sp i$ has an extra term $-f\sp i$.)
Using the chain rule ($(d/dt)$ on $f(t,q\sb k (t))$ equals $\pa /\pa t +
\dt q\sb k (\pa /\pa q\sb k )$) to evaluate the time derivative of
$\cg$, we find the lagrangian transforms as a total derivative
$$ \d \cl\sb H = {d\over dt} \left[ (\d z\sp M)e\sb M\sp A\p\sb A -
\z\sp i i\cg\sb i \right] \quad , \eqno(3.1.12)$$
which is the usual transformation law for an action with local symmetry
generated by the current $\cg$.  When $\ch\sb 0$ vanishes (as in
relativistic mechanics), the special case
$\z \sp i = \z \l \sp i$ of the transformations of (3.1.11) are $\t$ 
reparametrizations, generated by the hamiltonian $\l \sp i \cg \sb i$.
In general, after canonical quantization, the wave
function satisfies the Schr\"odinger equation $\pa /\pa t +i\ch\sb 0 =0$,
as well as the constraints $\cg =0$ (and thus $\pa /\pa t +i\ch =0$ in any
gauge choice for $\l$).  Since $[\ch\sb 0 , \cg ]=0$, $\cg =0$ at $t=0$
implies $\cg =0$ for all $t$.  

In some cases (such as Yang-Mills), 
the Lorentz covariant form of the action can be obtained by eliminating
all the $\p$'s.  A covariant first-order form can generally be obtained
by introducing additional auxiliary degrees of freedom which enlarge $\p$
to make it Lorentz covariant.  For example, for Yang-Mills we can
rewrite (3.1.9) as
$$ \cl\sb H = \ha G\sb{0i}\sp 2 - G\sb 0\sp i F\sb{0i} +\frac14
F\sb{ij}\sp 2 $$
$$ \to\quad \cl\sb 1 = -\frac14 G\sb{ab}\sp 2 +G\sp{ab}F\sb{ab}
\quad , \eqno(3.1.13)$$
where $G\sb{0i} = i\P\sb i$, and the independent (auxiliary) fields
$G\sb{ab}$ also include $G\sb{ij}$, which have been introduced to put
$\frac14 F\sb{ij}\sp 2$ into first-order form and thus make the
lagrangian manifestly Lorentz covariant.  Eliminating $G\sb{ij}$ by
their field equations gives back the hamiltonian form.

Many examples will be given in chapts.\ 5-6 for relativistic
first-quantization, where $\ch\sb 0$ vanishes, and thus the Schr\"odinger
equation implies the wave function is proper-time-independent (i.e., we
require $\ch\sb 0 =0$ because the proper time is not physically 
observable).  Here we give an
interesting example in D=2 which will also be useful for strings.
Consider a single field $A$ with canonical momentum $P$ and choose
$$ i \cg = \frac14 (P+A')\sp 2 \quad , \quad 
\ch\sb 0 = \frac14 (P-A')\sp 2 \quad , \eqno(3.1.14)$$
where $'$ is the derivative with respect to the 1 space coordinate
(which acts as the index $M$ or $i$ from above).  From the algebra of
$P\pm A'$, it's easy to check, at least at the Poisson bracket level,
that the $\cg$ algebra closes and $\ch\sb 0$ is invariant.  (This algebra,
with particular boundary conditions, will be important in string theory:
See chapt.\ 8.  Note that $P+A'$ does not form an algebra, so its square
must be used.)  The transformation laws (3.1.11) are found to be
$$ \d A = \z \ha (P+A') \quad , \quad
\d\l = \dt\z - \l\dvec{\pa\sb 1}\z \quad . \eqno(3.1.15)$$
In the gauge $\l =1$ the action becomes the usual hamiltonian one for a
massless scalar, but the constraint implies $P+A'=0$, which means that
modes propagate only to the right and not the left.  The lagrangian form
again results from eliminating $P$, and after the redefinitions
$$ \hat\l = 2 {1-\l\over 1+\l} \quad , \quad 
\hat\z = \sqrt 2 {1\over 1+\l}\z \quad , \eqno(3.1.16)$$
we find [3.2]
$$ \cl = - (\pa\sb + A)(\pa\sb - A) + \ha\hat\l (\pa\sb - A)\sp 2 \quad ;$$
$$ \d A = \hat\z \pa\sb - A \quad , \quad
\d\hat\l = 2\pa\sb + \hat\z + \hat\z\dvec\pa\sb -\hat\l \quad ;\eqno(3.1.17)$$
where $\pa\sb\pm$ are defined as in sect.\ 2.1.

The gauge fixing (including Faddeev-Popov ghosts) and initial condition
can be described in a very concise way by the BRST method.  The basic
idea is to construct a symmetry relating the Faddeev-Popov ghosts to the
unphysical modes of the gauge field.  For example, in Yang-Mills only
$D-2$ Lorentz components of the gauge field are physical, so the
Lorentz-gauge $D$-component gauge field requires 2 Faddeev-Popov ghosts
while the temporal-gauge $D-1$-component field requires only 1.  The
BRST symmetry rotates the additional gauge-field components into the FP
ghosts, and vice versa.  Since the FP ghosts are anticommuting, the
generator of this symmetry must be, also.  

\sect{3.2. IGL(1)}

We will find that the methods of Becchi, Rouet, Stora, and Tyutin [3.3] 
are the most useful way not only to perform quantization in 
Lorentz-covariant and general nonunitary gauges, but also to derive 
gauge-invariant theories.  BRST quantization is a more general way of 
quantizing gauge theories than either canonical or path-integral 
(Faddeev-Popov), because it (1) allows more general gauges, (2) gives 
the Slavnov-Taylor identities (conditions for unitarity) 
directly (they're just the Ward identities for BRST invariance), 
and (3) can separate the gauge-invariant part of a 
gauge-fixed action.  It is defined by the conditions:  (1) BRST 
transformations form a global group with a single (abelian) 
anticommuting generator $Q$.  The group property then implies $Q \sp 2 
= 0$ for closure.  (2) $Q$ acts on physical fields as a gauge 
transformation with the gauge parameter replaced by the (real) ghost.  
(3) $Q$ on the (real) antighost gives a BRST auxiliary field 
(necessary for closure of the algebra off shell).  Nilpotence of $Q$ 
then implies that the auxiliary field is BRST invariant.  Physical 
states are defined to be those which are BRST invariant (modulo null 
states, which can be expressed as $Q$ on something) and have vanishing 
ghost number (the number of ghosts minus antighosts).

There are two types of BRST formalisms: (1) first-quantized-style BRST, 
originally found in string theory [3.4] but also applicable to ordinary field
theory, which contains all the field equations as well as the gauge 
transformations; and (2) second-quantized-style BRST, the original form 
of BRST, which contains only the gauge transformations, corresponding in
a hamiltonian formalism to those field equations (constraints) found from 
varying the time components of the gauge fields.  However, we'll find
(in sect.\ 4.4)
that, after restriction to a certain subset of the fields, BRST1 is
equivalent to BRST2.  (It's the BRST variation of the additional fields
of BRST1 that leads to the field equations for the physical fields.)
The BRST2 transformations were originally found from Yang-Mills theory.
We will first derive the YM BRST2 transformations, and by a simple
generalization find BRST operators for arbitrary theories,
applicable to BRST1 or BRST2 and to lagrangian or hamiltonian formalisms.

In the general case, there are two forms for the BRST operators,
corresponding to different classes of gauges.  
The gauges commonly used in field theory fall into three classes:
(1) unitary (Coulomb, Arnowitt-Fickler/axial, light-cone) gauges, where the 
ghosts are nonpropagating, and the constraints are solved explicitly
(since they contain no time derivatives); (2) temporal/timelike gauges, 
where the ghosts have equations of motion first-order in time derivatives 
(making them canonically conjugate to the antighosts); and (3) Lorentz
(Landau, Fermi-Feynman) gauges, where the ghost equations are second-order 
(so ghosts are independent of antighosts), and the Nakanishi-Lautrup 
auxiliary fields [3.5] (Lagrange multipliers for the gauge conditions) are
canonically conjugate to the auxiliary time-components of the gauge fields.
Unitary gauges have only physical polarizations; temporal gauges
have an additional pair of unphysical polarizations of opposite
statistics for each gauge generator; Lorentz gauges have two pairs.
In unitary gauges the BRST operator vanishes identically; in
temporal gauges it is constructed from group generators, or
constraints, multiplied by the corresponding ghosts,
plus terms for nilpotence; in Lorentz gauges it has an extra ``abelian''
term consisting of the products of the second set of unphysical fields.
Temporal-gauge BRST is defined in terms of a ghost number operator in
addition to the BRST operator, which itself has ghost number 1.  We
therefore refer to this formalism by the corresponding symmetry group
with two generators,
IGL(1).  Lorentz-gauge BRST has also an antiBRST operator [3.6], and this
and BRST transform as an ``isospin'' doublet, giving the larger group
ISp(2), which can be extended further to OSp(1,1$|$2) [2.3,3.7].  Although the
BRST2 OSp operators are generally of little value (only the IGL is 
required for quantization), the BRST1 OSp gives a powerful method
for obtaining free gauge-invariant formalisms for arbitrary
(particle or string) field theories.  In particular, for arbitrary
representations of the Poincar\'e group a certain OSp(1,1$|$2) can be
extended to IOSp(D,2$|$2) [2.3], which is derived from (but does not directly 
correspond to quantization in) the light-cone gauge.

One simple way to formulate
anticommuting symmetries (such as supersymmetry) is through the use of
anticommuting coordinates [3.8].  We therefore extend spacetime to include one
extra, anticommuting coordinate, corresponding to the one anticommuting
symmetry:
$$ a \; \to \; ( a , \a ) \eqno(3.2.1)$$
for all vector indices, including those on coordinates, with Fermi
statistics for all quantities with an odd number of anticommuting
indices.  ($\a$ takes only one value.)  Covariant derivatives and 
gauge transformations are then defined by the corresponding 
generalization of (2.1.5b), and field strengths with graded commutators 
(commutators or anticommutators, according to the statistics).  
However, unlike supersymmetry, the extra coordinate
does not represent extra physical degrees of freedom, and so we
constrain all field strengths with anticommuting indices to vanish [3.9]:
For Yang-Mills, 
$$ F \sb {\a a} = F \sb {\a\b} = 0 \quad , \eqno(3.2.2a)$$
so that gauge-invariant quantities can be constructed only from the
usual $F \sb {ab}$.  When Yang-Mills is coupled to matter fields $\f$,
we similarly have the constraints
$$ \de\sb\a \f = \de\sb\a \de\sb a \f = 0 \quad , \eqno(3.2.2b)$$
and these in fact imply (3.2.2a) (consider $\{ \de\sb\a , \de\sb\b \}$
and $[ \de\sb\a , \de\sb a ]$ acting on $\f$).
These constraints can be solved easily:
$$ \li{F \sb {\a a} = 0 \quad & \to \quad p \sb \a A \sb a = [ \de \sb a , A
\sb \a ] \quad , \cr
F \sb {\a\b} = 0 \quad & \to \quad p \sb \a A \sb \b = - \ha \{ A \sb
\a , A \sb \b \} = - A \sb \a A \sb \b \quad ; \cr
\de\sb\a \f = 0 \quad & \to\quad p\sb\a \f = - A\sb\a \f \quad .&(3.2.3)\cr}$$
(In the second line we have used the fact that $\a$ takes only one
value.)  Defining ``$~|~$'' to mean $|\sb{x\sp\a =0}$, we now interpret
$A \sb a |$ as the usual gauge field, $iA \sb \a |$ as the FP ghost, and
the BRST operator $Q$ as $Q ( \j | ) = ( p \sb \a \j ) |$.  (Similarly,
$\f |$ is the usual matter field.)  Then $\pa
\sb \a \pa \sb \b = 0$ (since $\a$ takes only one value and $\pa \sb \a$
is anticommuting) implies nilpotence
$$ Q \sp 2 = 0 \quad . \eqno(3.2.4)$$
In a hamiltonian approach [3.10] these transformations are sufficient to 
perform quantization in a temporal gauge, but for the lagrangian 
approach or Lorentz gauges we also need the FP antighost and 
Nakanishi-Lautrup auxiliary field, which we define in terms of an 
unconstrained scalar field $\Tilde A$:  $\Tilde A |$ is the antighost, and
$$ B = ( p \sb \a i \Tilde A ) | \eqno(3.2.5)$$
is the auxiliary field.  

The BRST transformations (3.2.3) can be represented in operator form as
$$ Q = C \sp i \cg \sb i + \ha C \sp j C \sp i f \sb {ij} \sp k
\der{ C \sp k} -i B \sp i \der{ \Tilde C \sp i} \quad ,\eqno(3.2.6a)$$
where $i$ is a combined space(time)/internal-symmetry index, $C$ is the
FP ghost, $\Tilde C$ is the FP antighost, $B$ is the NL auxiliary field,
and the action on the physical fields is given by the
constraint/gauge-transformation $\cg$ satisfying the algebra
$$ [ \cg \sb i , \cg \sb j \} = f \sb {ij} \sp k \cg \sb k \quad , 
\eqno(3.2.6b)$$
where we have generalized to graded algebras with graded commutator 
$[\; , \; \}$ (commutator or anticommutator, as appropriate).  In this case,
$$ \cg = \left[ \de , \cdot i\fder A\right] \quad , \eqno(3.2.7)$$
where the structure constants in (3.2.6b) are the usual group structure
constants times $\d$-functions in the coordinates.
$Q$ of (3.2.6a) is antihermitian when $C$, $\Tilde C$, and $B$ are 
hermitian and $\cg$ is antihermitian, and is nilpotent (3.2.4) as a 
consequence of (3.2.6b).
Since $\Tilde C$ and $B$ appear only in the last term in (3.2.6a), these
properties also hold if that term is dropped.  (In the notation of
(3.2.1-5), the fields $A$ and $\Tilde A$ are independent.)

When $[ \cg \sb i , f \sb {jk} \sp l \} \ne 0$, (3.2.6a) still gives
$Q\sp 2=0$.  However, when the gauge invariance has a gauge
invariance of its own, i.e., $\L\sp i \cg\sb i =0$ for some nontrivial
$\L$ depending on the physical variables implicit in $\cg$, then,
although (3.2.6a) is still nilpotent, it requires extra terms in order
to allow gauge fixing this invariance of the ghosts.  In some cases (see
sect.\ 5.4) this requires an infinite number of new terms (and ghosts).
In general, the procedure of adding in the additional ghosts and
invariances can be tedious, but in sect.\ 3.4 we'll find a method which
automatically gives them all at once.

The gauge-fixed action is required to be BRST-invariant.  The
gauge-invariant part already is, since $Q$ on physical fields is a
special case of a gauge transformation.  The gauge-invariant 
lagrangian is quantized by adding terms which are $Q$ on something 
(corresponding to integration over $x \sp \a$), and thus BRST-invariant
(since $Q\sp 2 =0$):  For example, rewriting
(3.2.3,5) in the present notation,
$$ \li{ Q A\sb a &= -i [ \de\sb a , C ] \quad , \cr
QC &= i C\sp 2 \quad , \cr
Q\tilde C &= -iB \quad , \cr
QB &= 0 \quad , &(3.2.8)\cr}$$
we can choose
$$ \cl \sb {GF} = iQ \left\{ \Tilde C \left[ f ( A ) + g ( B ) \right]
\right\} = B \left[ f ( A ) + g ( B ) \right] - \Tilde C {\pa f \over
\pa A \sb a} [ \de \sb a, C] \quad , \eqno(3.2.9)$$
which gives the usual FP term for gauge condition $f ( A ) = 0$ with
gauge-averaging function $B g ( B )$.
However, gauges more general than FP can be obtained by putting more
complicated ghost-dependence into the function on which $Q$ acts, giving
terms more than quadratic in ghosts.  In the temporal gauge 
$$ f ( A ) = A \sb 0 \eqno(3.2.10)$$ 
and $g$ contains no time derivatives in (3.2.9), so upon quantization 
$B$ is eliminated (it's nonpropagating) and $\Tilde C$ is canonically
conjugate to $C$.  Thus, in the hamiltonian formalism (3.2.6a) gives the 
correct BRST transformations without the last term, where the fields are
now functions of just space and not time, the sum in (3.2.7) runs
over just the spatial values of the spacetime index as in (3.1.8), and 
the derivatives correspond to functional derivatives which give $\d$ 
functions in just spatial coordinates.  On the other hand, in Lorentz
gauges the ghost and antighost are independent even after quantization,
and the last term in $Q$ is needed in both lagrangian and hamiltonian
formalisms; but the product in (3.2.7) and the arguments of the fields 
and $\d$ functions are as in the temporal gauge.  Therefore, in the lagrangian
approach $Q$ is gauge independent, while in the hamiltonian approach the
only gauge dependence is the set of unphysical fields, and thus the last
term in $Q$.  Specifically, for Lorentz gauges we choose
$$ f ( A ) = \pa \cdot A \quad , \quad g ( B ) = \ha \z B \quad \to $$
$$ \li{ \cl \sb {GF} & = \z \ha B \sp 2 + B \pa \cdot A - \Tilde C \pa 
\cdot [ \de , C] \cr
& = - {1 \over \z} \ha ( \pa \cdot A ) \sp 2 + \z \ha \Tilde B \sp 2
- \Tilde C \pa \cdot [ \de , C] \quad , \cr} $$
$$ \Tilde B = B + {1 \over \z} \pa \cdot A \quad , \eqno(3.2.11)$$
using (3.2.9).

The main result is that (3.2.6a) gives a general BRST operator for 
arbitrary algebras (3.2.6b), for hamiltonian or lagrangian formalisms,
for arbitrary gauges (including temporal and Lorentz), where the last term 
contains arbitrary numbers (perhaps 0) of sets of ($\Tilde C$, $B$) fields.
Since $\cg = 0$ is the field equation (3.1.6), physical states must satisfy 
$Q \j = 0$.  Actually, $\cg =0$ is satisfied only as a Gupta-Bleuler
condition, but still $Q\j =0$ because in the $C\sp i\cg\sb i$ term in
(3.2.6a) positive-energy parts of $C\sp i$ multiply negative-energy
parts of $\cg\sb i$, and vice versa.  Thus, for any value of an
appropriate index $i$, either $C\sp i\sket{\j} = \sbra{\j}\cg\sb i = 0$
or $\cg\sb i\sket{\j} = \sbra{\j} C\sp i =0$, modulo contributions from
the $C\sp 2\pa /\pa C$ term.
However, since $\cg$ is also the generator of gauge
transformations (3.1.8), any state of the form $\j + Q \l$ is equivalent
to $\j$.  The physical states are therefore said to belong to the
``cohomology'' of $Q$: those satisfying $Q \j = 0$ modulo gauge
transformations $\d \j = Q \l$.  (``Physical'' has a more restrictive
meaning in BRST1 than BRST2:  In BRST2 the physical states are just the
gauge-invariant ones, while in BRST1 they must also be on shell.)
In addition, physical states must have a specified value of the
ghost number, defined by the ghost number operator
$$ J\sp 3 = C \sp i \der{C\sp i} - \Tilde C \sp i \der{\Tilde C\sp i}
\quad , \eqno(3.2.12a)$$
where
$$ [ J\sp 3 , Q ] = Q \quad , \eqno(3.2.12b)$$
and the latter term in (3.2.12a) is dropped if the last term in (3.2.6a) is.
The two operators $Q$ and $J\sp 3$ form the algebra IGL(1), which can
be interpreted as a translation and scale transformation, respectively,
with respect to the coordinate $x \sp \a$ (i.e., the conformal group in
1 anticommuting dimension).

From the gauge generators $\cg \sb i$, which act on only the physical
variables, we can define IGL(1)-invariant generalizations which
transform also $C$, as the adjoint representation:
$$ \Hat \cg \sb i = \left\{ Q , \der{C\sp i} \right\} = 
\cg \sb i + C \sp j f \sb {ji} \sp k \der{C\sp k} \quad . \eqno(3.2.13)$$
The $\Hat \cg$'s are gauge-fixed versions of the gauge generators $\cg$.

Types of gauges for first-quantized theories will be discussed in
chapt.\ 5 for particles and chapt.\ 6 and sect.\ 8.3 for strings.  
Gauge fixing for general field theories using BRST will be described in
sect.\ 4.4, and for closed string field theory in sect.\ 11.1.
IGL(1) algebras will be used for deriving general gauge-invariant free
actions in sect.\ 4.2.  The algebra will be derived from
first-quantization for the particle in sect.\ 5.2 and for the string in
sect.\ 8.1.  However, in the next section we'll find that IGL(1) can
always be derived as a subgroup of OSp(1,1$|$2), which can be derived
in a more general way than by first-quantization.

\secty{3.3. OSp(1,1$|$2)}{3.3. OSp(1,1|2)}

Although the IGL(1) algebra is sufficient for quantization in arbitrary
gauges, in the following section we will find the larger OSp(1,1$|$2)
algebra useful for the BRST1 formalism, so we give a derivation here for
BRST2 and again generalize to arbitrary BRST.  The basic idea is to
introduce a second BRST, ``antiBRST,'' corresponding to the antighost.
We therefore repeat the procedure of (3.2.1-7) with 2 anticommuting
coordinates [3.11] by simply letting the index $\a$ run over 2 values
(cf.\ sect.\ 2.6).  The solution to (3.2.2) is now
$$ \li{F \sb {\a a} = 0 \quad & \to \quad p \sb \a A \sb a = 
[ \de \sb a , A \sb \a ] \quad , \cr
F \sb {\a\b} = 0 \quad & \to \quad p \sb \a A \sb \b = - \ha \{ A \sb
\a , A \sb \b \} - i C \sb {\a\b} B \quad ; \cr
\de\sb\a \f = 0 \quad & \to\quad p\sb\a \f = - A\sb\a \f\quad ;&(3.3.1a)\cr}$$
where $A \sb \a$ now includes both ghost and antighost.  The appearance
of the NL field is due to the ambiguity in the constraint 
$F \sb {\a\b} = p \sb {( \a} A \sb {\b )} + \cdots$.  The remaining
(anti)BRST transformation then follows from further differentiation:
$$ \{ p \sb \a , p \sb \b \} A \sb \g = 0 \quad \to \quad
p \sb \a B = - \ha [ A \sb \a , B ] + i \frac1{12} \left[ A \sp \b , \{
A \sb \a , A \sb \b \} \right] \quad . \eqno(3.3.1b)$$
The generalization of (3.2.6a) is then [3.12], defining $Q\sp\a(\j |)=
(\pa\sp\a\j )|$ (and renaming $C\sp\a = A\sp\a$),
$$ Q \sp \a = C \sp {i\a} \cg \sb i + \ha C \sp {j\a} C \sp {i\b}
f \sb {ij} \sp k \der{ C \sp {k\b}} - B \sp i
\der{ C \sp i \sb \a} + \ha C \sp {j\a} B \sp i f \sb {ij} \sp k
\der{ B \sp k}$$
$$ - \frac1{12} C \sp {k\b} C \sp {j\a} C \sp i \sb \b f \sb {ij} \sp l
f \sb {lk} \sp m \der{ B \sp m} \quad , \eqno(3.3.2)$$
and of (3.2.12a) is
$$ J\sb{\a\b} = C \sp i \sb {(\a} \der{C\sp{i\b )}} \quad , \eqno(3.3.3)$$
where $(~)$ means index symmetrization.
These operators form an ISp(2) algebra consisting of the translations $Q
\sb \a$ and rotations $J\sb{\a\b}$ on the coordinates $x \sp \a$:
$$ \{ Q \sb \a , Q \sb \b \} = 0 \quad ,$$
$$ [ J\sb{\a\b} , Q \sb \g ] = - C \sb {\g ( \a } Q \sb {\b )} \quad ,
\quad [ J\sb{\a\b} , J \sb {\g\d} ] = - C \sb {(\g (\a} J \sb {\b ) \d )}
\quad . \eqno(3.3.4)$$
In order to relate to the IGL(1) formalism, we write
$$ Q \sp \a = ( Q , \Tilde Q ) \quad , \quad C \sp \a = ( C , \Tilde C )
\quad , \quad J \sp {\a\b} = \pmatrix{J \sp + & - i J\sp 3 \cr
- i J\sp 3 & J \sp - \cr} \quad , \eqno(3.3.5)$$
and make the unitary transformation
$$ l n ~ U = -\ha C \sp j \Tilde C \sp i f \sb {ij} \sp k i \der{B \sp k} 
\quad . \eqno(3.3.6)$$
Then $U Q U \sp {-1}$ is $Q$ of (3.2.6a) and $U J\sp 3
U \sp {-1} = J\sp 3$ is $J\sp 3$ of (3.2.12a).
However, whereas there is an arbitrariness in the IGL(1) algebra in
redefining $J\sp 3$ by a constant, there is no such ambiguity in the
OSp(1,1$|$2) algebra (since it is ``simple'').

Unlike the IGL case, the NL fields now are an essential part of the
algebra.  Consequently, the algebra can be enlarged to OSp(1,1$|$2) [3.7]:
$$ J \sb {-\a} = Q \sb \a \quad , \quad 
J \sb {+\a} = 2C \sp i \sb \a \der{B\sp i} \quad , $$
$$ J \sb {\a\b} = C \sp i \sb {(\a} \der{C\sp{i\b )}} \quad , \quad
J \sb {-+} = 2 B \sp i \der{B\sp i} + C \sp{i\a} \der{C\sp{i\a}} \quad ,
\eqno(3.3.7)$$
with $Q \sb \a$ as in (3.3.2), satisfy 
$$ [ J \sb {\a\b} , J \sb {\g\d} ] = - C \sb {( \g ( \a} J \sb {\b ) \d )}
	\quad , $$
$$ [ J \sb {\a\b} , J \sb {\pm\g} ] = - C \sb {\g ( \a } J \sb {\pm\b )} 
	\quad , $$
$$ \{ J \sb {-\a} , J \sb {+\b} \} =  - C \sb {\a\b} J \sb {-+} - J \sb {\a\b}
	\quad , $$
$$ [ J \sb {-+} , J \sb {\pm\a} ] = \mp J \sb {\pm\a} \quad , $$
$$ rest = 0 \quad . \eqno(3.3.8)$$
This group is the conformal group for $x \sp \a$, with the ISp(2)
subgroup being the corresponding Poincar\'e (or Euclidean) subgroup:  
$$ \li{ J \sb {-\a} & = -\pa \sb \a \quad , \quad J \sb {+\a} = 2 x \sb
\b \sp 2 \pa \sb \a + x \sp \b M \sb {\b\a} + x \sb \a {\Sc d} \quad \cr
J \sb {\a\b} & = x \sb {(\a} \pa \sb {\b )} + M \sb {\a\b} \quad ,
\quad J\sb{-+} = -x \sp \a \pa \sb \a + {\Sc d} \quad . & (3.3.9) \cr}$$
(We define the square of an Sp(2) spinor as $(x\sb\a )\sp 2 \equiv 
\ha x\sp\a x\sb\a$.)  $J \sb {-\a}$ are the
translations, $J \sb {\a\b}$ the Lorentz transformations (rotations), $J
\sb {-+}$ the dilatations, and $J \sb {+\a}$ the conformal boosts.
As a result of constraints analogous to (3.3.1a), the translations are
realized nonlinearly in (3.3.7) instead of the boosts.  This should be
compared with the usual conformal group (2.2.4).  The action of the generators
(3.3.9) have been chosen to have the opposite
sign of those of (3.3.8), since it is a coordinate representation
instead of a field representation (see sect.\ 2.2).  In later sections
we will actually be applying (3.3.7) to coordinates, and hence (3.3.9)
should be considered a ``zeroth-quantized'' formalism.

From the gauge generators $\cg \sb i$,
we can define OSp(1,1$|$2)-invariant generalizations which
transform also $C$ and $B$, as adjoint representations:
$$ \Hat \cg \sb i = \ha \left\{ J\sb -\sp\a , \left[ J\sb{-\a} , \der{B\sp i}
\right] \right\} = \cg \sb i + C \sp{j\a} f \sb {ji} \sp k
\der{C\sp{k\a}} + B \sp j f \sb {ji} \sp k \der{B\sp k} \quad . 
\eqno(3.3.10)$$
The $\Hat \cg$'s are the OSp(1,1$|$2) generalization of the operators (3.2.13).

The OSp(1,1$|$2) algebra (3.3.7) can be extended to an inhomogeneous
algebra IOSp(1,1$|$2) when one of the generators, which we denote by
$\cg\sb 0$, is
distinguished [3.13].  We then define
$$ \li{ p\sb + &= \sqrt{ -2i\der{B\sp 0} } \quad , \cr
p\sb\a &= {1\over p\sb +} i \left( \der{C\sp{0\a}} + \ha C\sp i\sb\a
f\sb{i0}\sp j \der{B\sp j} \right) \quad , \cr
p\sb - &= - {1\over p\sb +} \left( i \Hat\cg\sb 0 + p\sb\a\sp 2
\right) \quad . & (3.3.11) \cr} $$
(The $i$ indices still include the value $0$.)
$\Hat\cg\sb 0$ is then the IOSp(1,1$|$2) invariant $i\ha (2p\sb + p\sb -
+ p\sp\a p\sb\a )$.
This algebra is useful for constructing gauge field theory for closed strings.

OSp(1,1$|$2) will play a central role in the following chapters:  In
chapt.\ 4 it will be used to derive free gauge-invariant actions.  A
more general form will be derived in the following sections, but the
methods of this section will also be used in sect.\ 8.3 to describe
Lorentz-gauge quantization of the string.

\sect{3.4. From the light cone}

In this section we will derive a general OSp(1,1$|$2) algebra from the
light-cone Poincar\'e algebra of sect.\ 2.3, using concepts developed in
sect.\ 2.6.  We'll use this general OSp(1,1$|$2) to derive a general
IGL(1), and show how IGL(1) can be extended to include interactions.

The IGL(1) and OSp(1,1$|$2) algebras of the previous section can be
constructed from an arbitrary algebra $\cg$, whether first-quantized or
second-quantized, and lagrangian or hamiltonian.  That already gives 8
different types of BRST formalisms.  Furthermore, arbitrary gauges, more
general than those obtained by the FP method, and graded algebras (where
some of the $\cg$'s are anticommuting, as in supersymmetry) can be
treated.  However, there is a ninth BRST formalism, similar to the BRST1
OSp(1,1$|$2) hamiltonian formalism, which starts from an IOSp(D,2$|$2)
algebra [2.3] which contains the OSp(1,1$|$2) as a subgroup.  This approach is
unique in that, rather than starting from the gauge covariant formalism
to derive the BRST algebra, it starts from just the usual Poincar\'e
algebra and derives both the gauge covariant formalism and BRST algebra.
In this section, instead of deriving BRST1 from first-quantization, we 
will describe this special form of BRST1, and give the OSp(1,1$|$2) 
subalgebra of which special cases will be found in the following chapters.

The basic idea of the IOSp formalism is to start from the light-cone
formalism of the theory with its nonlinear realization of the usual
Poincar\'e group ISO(D-1,1) (with manifest subgroup ISO(D-2)), extend 
this group to IOSp(D,2$|$2) (with manifest IOSp(D-1,1$|$2)) by
adding 2 commuting and 2 anticommuting coordinates, and take the
ISO(D-1,1)$\otimes$OSp(1,1$|$2) subgroup, where this ISO(D-1,1) is now
manifest and the nonlinear OSp(1,1$|$2) is interpreted as BRST.
Since the BRST operators of BRST1 contain all the
field equations, the gauge-invariant action can be derived.  Thus, not
only can the light-cone formalism be derived from the gauge-invariant
formalism, but the converse is also true.  Furthermore, for general
field theories the light-cone formalism (at least for the free theory) 
is easier to derive (although more awkward to use), and the IOSp method
therefore provides a convenient method to derive the gauge covariant formalism.

We now perform dimensional continuation as in sect.\ 2.6, but set $x\sb
+ = 0$ as in sect.\ 2.3.
Our fields are now functions of $( x \sb a , x \sb\a , x \sb - )$, and 
have indices corresponding to
representations of the spin subgroup OSp(D$-$1,1$|$2) in the massless
case or OSp(D,1$|$2) in the massive.  Of the full group IOSp(D,2$|$2) 
(obtained from extending (2.3.5)) we
are now only interested in the subgroup ISO(D$-$1,1)$\otimes$OSp(1,1$|$2).  
The former factor is the usual Poincar\'e group, acting only in the
physical spacetime directions:
$$ p \sb a = i \pa \sb a \quad , \quad J \sb {ab} = 
- i x \sb {[a} p \sb {b]} + M \sb {ab} \quad . \eqno(3.4.1)$$
The latter factor is identified as the BRST group, acting in only the
unphysical directions:
$$ J \sb {\a\b} = - i x \sb {(\a} p \sb {\b )} + M \sb {\a\b} \quad ,
\quad J \sb {-+} = - i x \sb - p \sb + + k \quad , \quad 
J \sb {+\a} = i x \sb \a p \sb + \quad , $$
$$ J \sb {-\a} = - i x \sb - p \sb \a + {1\over p\sb +} \left[
- i x \sb \a \ha ( p \sp b p \sb b + M \sp 2 + p \sp \b p \sb \b ) 
+ M \sb \a \sp \b p \sb \b + k p \sb \a + \cq\sb\a \right] \quad , $$
$$ \{ \cq\sb\a , \cq\sb\b \} = - M\sb{\a\b} (p\sp a p\sb a + M\sp 2 )
\quad ; \eqno(3.4.2a)$$
$$ \cq\sb\a = M \sb \a \sp b p \sb b + M \sb{\a m} M \quad . \eqno(3.4.2b)$$
We'll generally set $k=0$.

In order to relate to the BRST1 IGL formalism obtained from ordinary
first-quantization (and discussed in the following chapters for the
particle and string), we perform an analysis similar to that of (3.3.5,6):
Making the (almost) unitary transformation [2.3]
$$ ln~U = ( ln~p\sb + ) \left( c \der c + M\sp 3 \right) \quad ,
\eqno(3.4.3a)$$
where $x\sp\a = ( c , \tilde c )$, $M\sp{\a a} = ( M\sp{+a} , M\sp{-a} )$,
and $M\sp\a\sb m = ( M\sp +\sb m , M\sp -\sb m )$,
we get
$$ Q \quad\to\quad -i c\ha (p\sb a\sp 2 +M\sp 2) + M\sp + i\der c +
(M\sp{+a}p\sb a + M\sp +\sb m M ) + x\sb - i\der{\tilde c} \quad , $$
$$ J\sp 3 =  c\der c + M\sp 3 - \tilde c\der{\tilde c} \quad . \eqno(3.4.3b)$$
(Cf. (3.2.6a,12a).)  As in sect.\ 3.2, the extra terms in $x\sb -$ and
$\tilde c$ (analogous to $B\sp i$ and $\Tilde C\sp i$) can be dropped in
the IGL(1) formalism.  After dropping such terms, $J\sp 3\dg = 1 - J\sp
3$.  (Or we can subtract $\ha$ to make it simply antihermitian.
However, we prefer not to, so that physical states will still have
vanishing ghost number.)

Since $p\sb +$ is a momentum, this redefinition
has a funny effect on reality (but not hermiticity) properties:
In particular, $ c$ is now a momentum rather than a coordinate (because
it has been scaled by $p\sb +$, maintaining its hermiticity but making
it imaginary in coordinate space).  However, we will avoid changing
notation or Fourier transforming the fields, in order to simplify
comparison to the OSp(1,1$|$2) formalism.  The effect of (3.4.3a) on a
field satisfying $\F = \O\F\conj$ is that it now satisfies
$$ \F = (-1)\sp{c\pa / \pa c+M\sp 3}\O\F\conj \eqno(3.4.4)$$
due to the $i$ in $p\sb + = i\pa\sb +$.

These results can be extended to interacting field theory, and we use
Yang-Mills as an example [2.3].  Lorentz-covariantizing the light-cone result
(2.1.7,2.4.11), we find
$$ p\sb - = \int -\frac14 F\sp{ij}F\sb{ji} + \ha (p\sb +A\sb -)\sp 2 \quad ,$$
$$ J\sb{-\a} = \int ix\sb - (p\sb\a A\sp i)(p\sb +A\sb i) + ix\sb\a
\left[ -\frac14 F\sp{ij}F\sb{ji} + \ha (p\sb +A\sb -)\sp 2 \right]
-A\sb\a p\sb + A\sb - \quad , $$
$$ A\sb - = - {1\over p\sb +\sp 2} [\de\sp i , p\sb + A\sb i\} 
\quad . \eqno(3.4.5)$$

When working in the IGL(1) formalism, it's extremely useful to 
introduce a Lorentz covariant type of second-quantized bracket [3.14].
This bracket can be postulated independently, or derived by
covariantization of the light-cone canonical commutator, 
plus truncation of the $\tilde c, x_+$, and $x_-$ coordinates.
The latter derivation will prove useful for the derivation of IGL(1)
from OSp(1,1$|$2).  Upon covariantization of the canonical light-cone
commutator (2.4.5), the arguments of the fields
and of the $\d$-function on the right-hand side are extended accordingly.  
We now have to truncate. 
The truncation of $x_+$ is automatic:  Since the
original commutator was an equal-time one, there is no $x_+$ 
$\d$-function on the right-hand side, and it therefore 
suffices to delete the $x_+$ arguments of the fields.  
At this stage, in addition to $x_-$ dependence, the fields depend on both
$ c$ and $\tilde c$ and the right-hand side contains both $\d$-functions.
(This commutator may be useful for OSp approaches to field theory.)
We now wish to eliminate the $\tilde c$ dependence. This cannot
be done by straightforward truncation, since expansion of the field
in this anticommuting coordinate shows that one cannot eliminate
consistently the fields in the $\tilde c$ sector. We therefore
proceed formally and just delete the $\tilde c$ argument from the
fields and the corresponding $\d$-function, obtaining
$$ [ \F\dg (1) , \F (2) ]_c = -{1\over 2p_{+2}} 
\d (x_{2-}-x_{1-})\d^D (x_2-x_1) 
\d ( c_2 -  c_1)\quad ,\eqno(3.4.6)$$
which is a bracket with unusual statistics because of the anticommuting
$\d$-function on the right-hand side.  The transformation (3.4.3a) is 
performed next; its nonunitarity causes the $p_+$ dependence of (3.4.6) to 
disappear, enabling one to delete the $x_-$ argument from the
fields and the corresponding $\d$-function to find
(using $(c\pa /\pa c )c =c$)
$$ [ \F\dg (1) , \F (2) ]_c = -\ha\d^D (x_2-x_1) 
\d ( c_2- c_1)\quad . \eqno(3.4.7) $$
This is the covariant bracket.  The arguments of the fields are
$(x_a,  c)$, namely, the usual $D$ bosonic coordinates of covariant
theories and the single anticommuting coordinate of the IGL(1) formalism.
The corresponding $\d$-functions appear on the right-hand side.
(3.4.6,7) are defined for commuting (scalar) fields, but generalize
straightforwardly:  For example, for Yang-Mills, where $A\sb i$ includes
both commuting ($A\sb a$) and anticommuting ($A\sb\a$) fields, 
$[A\sb i\dg , A\sb j]$ has an extra factor of $\h\sb{ij}$.
It might be possible to define the bracket by a commutator 
$[A,B]_c = A*B - B*A$.  Classically it can be defined by a Poisson bracket:
$$ [ A\dg , B ]_c = \ha\int dz\; \left( \fder{\F (z)} A\right)
^\dagger \left( \fder{\F (z)} B\right) \quad ,\eqno(3.4.8)$$
where $z$ are all the coordinates of $\F$ (in this case, $x\sb a$ and $c$).
For $A=B=\F$, the result of equation (3.4.7) is
reproduced.  The above equation implies that the bracket is a derivation:
$$[A, BC]_c = [A,B]_c C + (-1)^{(A+1)B} B [A,C]_c\quad ,\eqno(3.4.9)$$
where the $A$'s and $B$'s in the exponent of the $(-1)$ are 0 if the
corresponding quantity is bosonic and 1 if it's fermionic.
This differs from the usual graded Leibnitz rule by a $(-1)\sp B$ due to
the anticommutativity of the $dz$ in the front of (3.4.8), which also
gives the bracket the opposite of the usual statistics:  We can write
$(-1)\sp{[A,B]\sb c} = (-1)\sp{A+B+1}$ to indicate that the bracket of 2
bosonic operators is fermionic, etc., a direct consequence of the
anticommutativity of the total $\d$-function in (3.4.7).  One can also 
verify that this bracket satisfies the other properties of a 
(generalized) Lie bracket:
$$[A,B]_c = (-1)^{AB} [B,A]_c\quad , $$
$$ (-1)\sp{A(C+1)} [A,[B,C]\sb c ]\sb c + (-1)\sp{B(A+1)} [B,[C,A]\sb c ]\sb c
+ (-1)\sp{C(B+1)}[C,[A,B]\sb c ]\sb c = 0 \quad .\eqno(3.4.10)$$ 
Thus the bracket has the opposite of the usual graded symmetry, being
antisymmetric for objects of odd statistics and symmetric otherwise.  
This property follows from the hermiticity condition (3.4.4):  
$(-1)\sp{c\pa /\pa c}$ gives $(-1)\sp{-(\pa /\pa c)c}=-(-1)\sp{c\pa /\pa c}$ 
upon integration by parts, which gives the effect of using an
antisymmetric metric.  The Jacobi identity has the same extra signs as
in (3.4.9).  These properties are sufficient to perform the
manipulations analogous to those used in the light cone.

Before applying this bracket, we make some general considerations
concerning the derivation of interacting IGL(1) from OSp(1,1$|$2).
We start with the original untransformed
generators $J\sp 3$ and $J\sb -\sp c = Q$.  The 
first step is to restrict our attention to just the fields at $\tilde c =0$.
Killing all the fields at linear order in $\tilde c$ is consistent with
the transformation laws, since the transformations  of the latter fields
include no terms which involve only $\tilde c =0$ fields.  Since the
linear-in-$\tilde c$ fields are canonically conjugate to the $\tilde
c=0$ fields, the only terms in the generators which could spoil this
property would themselves have to depend on only $\tilde c=0$ fields,
which, because of the $d\tilde c$ ($=\pa /\pa\tilde c$) integration,
would require explicit $\tilde c$-dependence.  However, from 
(2.4.15), since $\h\sp{cc} = C\sp{cc} = 0$, we see that $J\sb-\sp c$ 
anticommutes with $p\sp c$, and thus has no explicit 
\hbox{$\tilde c$-dependence}
at either the free or interacting levels.  (The only explicit 
coordinate dependence in $Q$ is from a $c$ term.)

The procedure of restricting to $\tilde c=0$ fields can then be 
implemented very simply by dropping all 
$p\sp c$($ = -\pa /\pa\tilde c$)'s in the generators.
As a consequence, we also lose all explicit $x\sb -$ terms in $Q$.
(This follows from $[ J\sb -\sp c , p\sb + ] = -p\sp c$.)
Since $\tilde c$ and $\pa /\pa\tilde c$ now occur nowhere explicitly, we
can also kill all implicit dependence on $\tilde c$:  All fields are 
evaluated at
$\tilde c=0$, the $d\tilde c$ is removed from the integral in the
generators, and the $\d (\tilde c\sb 2 - \tilde c\sb 1)$ is removed from
the canonical commutator, producing (3.4.6).  In the case of Yang-Mills
fields $A\sp i = ( A\sp a , A\sp\a ) = ( A\sp a , A\sp c , A\sp{\tilde c})$,
the BRST generator at this point is given by
$$ Q = \int ic [ -\frac14 F\sp{ij}F\sb{ji} + \ha (p\sb + A\sb -)\sp 2 ]
- A\sp c p\sb + A\sb -  \quad , \eqno(3.4.11)$$
where the integrals are now over just $x\sb a$, $x\sb -$, and $c$, and
some of the field strengths simplify:
$$ F \sp{cc} = 2 ( A\sp c )\sp 2 \quad ,
\quad other~ F\sp{ic} = [ \de\sp i , A\sp c \} 
\quad . \eqno(3.4.12)$$

Before performing the transformations which eliminate $p\sb +$
dependence, it's now convenient to expand the fields over $c$ as
$$ \li{ A\sp a &= A\sp a + c\c\sp a \quad ,\cr
	A\sp c &= iC + cB \quad , \cr
	A\sp{\tilde c} &= i\Tilde C + cD \quad , &(3.4.13)\cr}$$
where the fields on the right-hand sides are $x\sp\a$-independent.  
(The $i$'s have been chosen in accordance with (3.4.4) to make the 
final fields real.)  We next perform the $dc$ integration, and then 
perform as the first transformation the first-quantized one (3.4.3a), 
$\F\to p\sb +\sp{-J\sp 3} \F$ (using the first-quantized 
$J\sp 3 = c\pa /\pa c + M\sp 3$), which gives
$$ \li{ -iQ = \int \frac14 F\sb{ab}\sp 2 
& -\ha \left( 2B + {1\over p\sb +}i[\de\sp a , p\sb + A\sb a ]
-\{ C,\tilde C\} -{2\over p\sb +}\left\{ p\sb +\sp 2 C , 
{1\over p\sb +}\tilde C\right\} \right)^2 \cr
& - \left[ 2D +\tilde C\sp 2
-2p\sb +\sp 2 \left({1\over p\sb +}\tilde C\right)^2\right] C\sp 2 \cr
& + \left( 2\c\sp a -i [\de\sp a , \tilde C ] 
-2i \left[ p\sb + A\sp a , {1\over p\sb +}\tilde C \right] \right)
[ \de\sb a , C ] \quad . &(3.4.14)\cr}$$

This transformation also replaces (3.4.6) with (3.4.7), with an extra
factor of $\h\sb{ij}$ for $[A\sb i\dg ,A\sb j ]$, but still with
the $x\sb -$ $\d$-function.  Expanding the bracket over the $c$'s,
$$ [ \c\sb a , A\sb b ]\sb c = \ha\h\sb{ab} \quad , \quad
[ D , C ]\sb c = \ha \quad , \quad [ B , \tilde C ]\sb c = -\ha
\quad ,\eqno(3.4.15)$$
where we have left off all the $\d$-function factors (now in commuting
coordinates only).  Note that, by (3.4.10), all these brackets are
{\it symmetric}.

We might also define a second-quantized
$$ J\sp 3 = \int A\sp i p\sb + \left( c\der c + M\sp 3 \right) A\sb i
\quad , \eqno(3.4.16a)$$
but this form automatically keeps just the antihermitian part of the
first-quantized operator $c\pa /\pa c = \ha [ c,\pa /\pa c ] +\ha$:
Doing the $c$ integration and transformation (3.4.3a),
$$ J\sp 3 = \int \c\sp a A\sb a - \tilde CB -3CD \quad . \eqno(3.4.16b)$$
As a result, the terms in $Q$ of different orders in the fields have
different second-quantized ghost number.  Therefore, we use only the
first-quantized ghost operator (or second-quantize it in functional form).

As can be seen in the above equations, despite the rescaling of the fields
by suitable powers of $p_+$ there remains a fairly complicated dependence
on $p_+$.  There is no explicit $x_-$ dependence anywhere but, of course,
the fields have $x_-$ as an argument.  It would seem that there should be
a simple prescription to get rid of the $p_+$'s in the transformations.
Setting $p_+ = constant$ does not work, since it
violates the Leibnitz rule for derivatives ($p_+ \phi = a\phi$ implies
that $p_+ \phi^2 = 2a\phi^2$ and not $a\phi^2$).  Even setting
$p_+\phi_i = \lambda_i \phi_i$ does not work.  An attempt that comes very
close is the following:  Give the fields some specific $x_-$ dependence
in such a way that the $p_+$ factors can be evaluated and that afterwards
such dependence can be canceled between the right-hand side and
left-hand side of the
transformations.  In the above case it seems that the only possibility is
to set every field proportional to ${(x_-)}^0$ but then it is hard to define
$1/p_+$ and $p_+$.  One then tries setting each field proportional
to ${(x_-)}^\epsilon$ and then let $\epsilon \rightarrow 0$ at the end.  In
fact this prescription gives the correct answer for the quadratic terms
of the Yang-Mills BRST transformations.  Unfortunately it does not
give the correct cubic terms.

It might be possible to eliminate $p\sb +$-dependence simply by applying
$J\sb{-+}=0$ as a constraint.  However, this would require resolving some
ambiguities in the evaluation of the nonlocal (in $x\sb -$) operator
$p\sb +$ in the interaction vertices.

We therefore remove the explicit $p\sb +$-dependence by use of an
explicit transformation.  In the Yang-Mills case, this transformation
can be completely determined by choosing it to be the one which redefines
the auxiliary field $B$ in a way which eliminates interaction terms 
in $Q$ involving it, thus making $B+i\ha p\cdot A$ BRST-invariant.  
The resulting transformation [3.14] redefines only the BRST auxiliary fields:
$$ Q \to e\sp{L\sb\D} Q \quad , \quad L\sb A B \equiv [ A , B ]\sb c
\quad , \quad L\sb\D\sp 2 = 0 \quad , $$
$$ \D = \int \tilde C {1\over p\sb +} i[A\sp a , p\sb +A\sb a]
-\tilde C\sp 2 C +2 \left({1\over p\sb +}\tilde C\right)^2 (p\sb +\sp 2C)
\quad , \eqno(3.4.17)$$
simply redefines the auxiliary fields to absorb the awkward interaction
terms in (3.4.14).  (We can also eliminate the free terms added to 
$B$ and $\c$ by adding a term $\int\tilde Cip\sp aA\sb a$ to $\D$ to 
make the first term $\tilde C(1/p\sb +)i[\de\sp a,p\sb +A\sb a]$.)
We then find for the transformed BRST operator
$$-iQ = \int\frac14 F\sb{ab}\sp2 - \ha (2B+ip\cdot A)^2 -2DC\sp 2
+ (2\c\sp a-ip\sp a\Tilde C) [ \de\sb a , C ] \quad . \eqno(3.4.18)$$
The resulting transformations are then
$$\li{QA_a =& -i[\de_a , C ]\quad ,\cr
Q \chi_a =& i\ha [\de^b , F_{ba}] + i \{ C , \chi_a -i\ha p\sb a \Tilde C \}
	+ p_a (B+i \ha p\cdot A) \quad ,\cr
Q \Tilde C =& -2i( B+i\ha p\cdot A)\quad ,\cr
Q D =&  -i \left[ \de^a ,\chi_a -i\ha p\sb a\Tilde C \right] +i[C,D]\quad ,\cr
Q C =& iC\sp 2 \quad ,\cr
Q B =&  -\ha p^a \left[ \de_a , C \right]\quad .&(3.4.19)\cr}$$
Since all the $p\sb +$'s have been eliminated, we can now drop all 
$x\sb -$ dependence from the fields, integration, and $\d$-functions.
On the fields $A\sb a$, $C$, $\Tilde C$, $B$ of the usual BRST2
formalism, this result agrees with the corresponding transformations
(3.2.8), where this $B = \ha \Tilde B$ of (3.2.11).
By working with the second-quantized operator form of $Q$ (and of the
redefinition $\D$), we have automatically obtained a form which makes $Q\F$
integrable in $\F$, or equivalently makes the vertices which follow from
this operator cyclic in all the fields (or symmetric, if one takes
group-theory indices into account).  The significance of this property
will be described in the next chapter.

This extended-light-cone form of the OSp(1,1$|$2) algebra will be used 
to derive free
gauge-invariant actions in the next chapter.  The specific form of the
generators for the case of the free open string will be given in sect.\ 8.2,
and the generalization to the free closed string in sect.\ 11.1.
A partial analysis of the interacting string along these lines will be
given in sect.\ 12.1.

\sect{3.5. Fermions}

These results can be extended to fermions [3.15].  This requires a
slight modification of the formalism, since the Sp(2) representations
resulting from the above analysis for spinors don't include singlets.
This modification is analogous to the addition of the $B \pa / \pa
\Tilde C$ terms to $Q$ in (3.2.6a).  We can think of the OSp(1,1$|$2)
generators of (3.4.2) as ``orbital'' generators, and add ``spin''
generators which themselves generate OSp(1,1$|$2).  In particular, since
we are here considering spinors, we choose the spin generators to be
those for the simplest spinor representation, the graded generalization
of a Dirac spinor, whose generators can be expressed in terms of graded
Dirac ``matrices'':
$$ \{ \tilde \g \sp A , \tilde \g \sp B ] = 2 \h \sp{AB} \quad ,
\quad S\sb{AB} = \frac14 [ \tilde \g \sb A , \tilde \g \sb B \}
\quad , \quad J\sb{AB}' = J\sb{AB} + S\sb{AB} \quad , \eqno(3.5.1) $$
where $\{ ~ , ~ ]$ is the opposite of $[ ~ , ~ \}$.
These $\tilde \g$ matrices are not to be confused with the ``ordinary''
$\g$ matrices which appear in $M\sb{ij}$ from the dimensional
continuation of the true spin operators.  The $\tilde\g\sp A$, like
$\g\sp i$, are hermitian.  (The hermiticity of $\g \sp i$ in the
light-cone formalism follows from $( \g \sp i ) \sp 2 = 1$ for each $i$
and the fact that all states in the light-cone formalism have 
nonnegative norm, since they're physical.)
The choice of whether the $\tilde\g$'s (and also the graded $\g\sp i$'s) 
commute or anticommute with other operators (which could be arbitrarily
changed by a Klein transformation) follows from the index structure as
usual (bosonic for indices $\pm$, fermionic for $\a$).  (Thus, as usual,
the ordinary $\g$ matrices $\g\sp a$ commute with other operators,
although they anticommute with each other.)

In order to put the OSp(1,1$|$2) generators in a form more similar to
(3.4.2), we need to
perform unitary transformations which eliminate the new terms in
$J\sb{-+}$ and $J\sb{+\a}$ (while not affecting $J\sb{\a\b}$, although
changing $J\sb{-\a}$).  In general, the appropriate transformations 
$J\sb{AB}' = UJ\sb{AB}U\sp{-1}$ to eliminate such terms are:
$$ln~U = - ( ln~p\sb + ) S\sb{-+} \eqno(3.5.2a) $$
to first eliminate the $S\sb{-+}$ term from $J\sb{-+}$, and then
$$ ln~U = S\sb +\sp\a p\sb\a \eqno(3.5.2b) $$
to do the same for $J\sb{+\a}$.  The general result is
$$ J\sb{-+} = -ix\sb - p\sb + +k\quad ,\quad J\sb{+\a}=ix\sb\a p\sb +\quad ,
\quad J\sb{\a\b} = -ix\sb{(\a}p\sb{\b )} + \Hat M\sb{\a\b} \quad , $$
$$ J\sb{-\a} = -ix\sb - p\sb\a + {1\over p\sb +}\left[ -ix\sb\a \ha 
( p\sp b p\sb b + M\sp 2 + p\sp\b p\sb\b ) + \Hat M\sb\a\sp\b p\sb\b + 
k p\sb\a + \Hat\cq\sb\a \right] \quad ; \eqno(3.5.3a)$$
$$ \Hat M\sb{\a\b} = M\sb{\a\b} + S\sb{\a\b} \quad ,$$
$$ \Hat\cq\sb\a = ( M\sb\a\sp b p\sb b + M\sb{\a m} M ) +
\left[ S\sb{-\a} + S\sb{+\a} \ha ( p\sp b p\sb b + M\sp 2 ) \right] \quad .
\eqno(3.5.3b)$$
(3.5.3a) is the same as (3.4.2a), but with
$M\sb{\a\b}$ and $\cq\sb\a$ replaced by $\Hat M\sb{\a\b}$ and
$\Hat\cq\sb\a$.  In this case the last term in $\Hat\cq\sb\a$ is
$$ S\sb{-\a} + S\sb{+\a} \ha ( p\sp b p\sb b + M\sp 2 ) = 
-\ha \tilde\g\sb\a \left[ \tilde\g\sb -
+\tilde\g\sb + \ha ( p\sp b p\sb b + M\sp 2 ) \right] \quad . \eqno(3.5.4) $$
We can again choose $k=0$.  

This algebra will be used to derive free gauge-invariant actions for
fermions in sect.\ 4.5.  The generalization to fermionic strings follows
from the representation of the Poincar\'e algebra given in sect.\ 7.2.

\sect{3.6. More dimensions}

In the previous section we saw that fermions could be treated in a way
similar to bosons by including an OSp(1,1$|$2) Clifford algebra.  In the
case of the Dirac spinor, there is already an OSp(D$-$1,1$|$2) Clifford
algebra (or OSp(D,1$|$2) in the massive case) obtained by adding 2+2
dimensions to the light-cone $\g$-matrices, in terms of which $M\sb{ij}$
(and therefore the OSp(1,1$|$2) algebra) is defined.  Including the
additional $\g$-matrices makes the spinor a representation of an
OSp(D,2$|$4) Clifford algebra (OSp(D+1,2$|$4) for massive), and is thus
equivalent to adding 4+4 dimensions to the original light-cone spinor
instead of 2+2, ignoring the extra spacetime coordinates.  This suggests
another way of treating fermions which allows bosons to be treated
identically, and should thus allow a straightforward generalization to
supersymmetric theories [3.16].

We proceed similarly to the 2+2 case:  Begin by adding 4+4 dimensions to
the light-cone Poincar\'e algebra (2.3.5).  Truncate the resulting
IOSp(D+1,3$|$4) algebra to ISO(D$-$1,1)$\otimes$IOSp(2,2$|$4).
IOSp(2,2$|$4) contains (in particular) 2 inequivalent truncations to
IOSp(1,1$|$2), which can be described by (defining-representation
direct-product) factorization of the OSp(2,2$|$4) metric into the
OSp(1,1$|$2) metric times the metric of either SO(2) (U(1)) or
SO(1,1) (GL(1)):
$$ \h\sb{\ca\cb} = \h\sb{AB}\h\sb{\hat a\hat b} \quad , \quad
\ca = A\hat a $$
$$ \to\quad J\sb{AB} = \h\sp{\hat b\hat a}J\sb{A\hat a,B\hat b} \quad ,
\quad \e\sb{\hat a\hat b}\D = \h\sp{BA}J\sb{A\hat a,B\hat b} \quad , 
\eqno(3.6.1) $$
where $\D$ is the generator of the $U(1)$ or $GL(1)$ and 
$\h\sb{\hat a\hat b}= I$ or $\s\sb 1$.  These 2 OSp(1,1$|$2)'s are Wick
rotations of each other.  We'll treat the 2 cases separately.

The GL(1) case corresponds to first taking the GL(2$|$2)
(=SL(2$|$2)$\otimes$GL(1)) subgroup of
OSp(2,2$|$4) (as SU(N)$\subset$SO(2N), or GL(1$|$1)$\subset$OSp(1,1$|$2)),
keeping also half of the inhomogeneous generators to get IGL(2$|$2).
Then taking the OSp(1,1$|$2) subgroup of the SL(2$|$2) (in the same way
as SO(N)$\subset$SU(N)), we get IOSp(1,1$|$2)$\otimes$GL(1), which is
like the Poincar\'e group in (1,1$|$2) dimensions plus dilatations.
(There is also an SL(1$|$2)=OSp(1,1$|$2) subgroup of SL(2$|$2), but this
turns out not to be useful.)  The advantage of breaking down to
GL(2$|$2) is that for this subgroup the coordinates of the string (sect.\ 8.3) 
can be redefined in such a way that the extra zero-modes are separated
out in a natural way while leaving the generators local in $\s$.  This
GL(2$|$2) subgroup can be described by writing the OSp(2,2$|$4) metric as
$$ \h\sb{\ca\cb} = \pmatrix{0 & \h\sb{AB'} \cr \h\sb{A'B} & 0 \cr}
\quad , \quad \h\sb{AB'} = (-1)\sp{AB}\h\sb{B'A} \quad , \quad \ca = (A,A') 
\eqno(3.6.2a) $$
$$ \to\quad J\sb{\ca\cb} = \pmatrix{\tilde J\sb{AB} & \tilde J\sb{AB'}
\cr \tilde J\sb{A'B} & \tilde J\sb{A'B'} \cr} \quad , \quad
\tilde J\sb{AB'} = -(-1)\sp{AB}\tilde J\sb{B'A} \quad , \eqno(3.6.2b) $$
where $\tilde J\sb{A'B}$ are the GL(2$|$2) generators, to which we add the
$\tilde p\sb{A'}$ half of $p\sb\ca$ to form IGL(2$|$2).  (The metric
$\h\sb{AB'}$ can be used to eliminate primed indices, leaving covariant
and contravariant unprimed indices.)  In this notation, the original
$\pm$ indices of the light cone are now $+'$ and $-$ (whereas $+$ and
$-'$ are ``transverse'').  To reduce to the
IOSp(1,1$|$2)$\otimes$GL(1) subgroup we identify primed and unprimed
indices; i.e., we choose the subgroup which transforms them in the same way:
$$ J\sb{AB} = \tilde J\sb{A'B}+\tilde J\sb{AB'} \quad , \quad
\check p\sb A = \tilde p\sb{A'} \quad , \quad \D = \h\sp{BA'}\tilde J\sb{A'B}
\quad . \eqno(3.6.3) $$
We distinguish the momenta $\check p\sb A$ and their conjugate
coordinates $\check x\sb A$, which we wish to eliminate, from 
$p\sb A = \tilde p\sb A$ and their conjugates $x\sb A$, which 
we'll keep as the usual ones of OSp(1,1$|$2) (including the 
nonlinear $p\sb -$).  At this point these generators take the explicit form
$$ \D = i\check x\sp A \check p\sb A -ix\sb - p\sb + -ix\sp\a p\sb\a + 
M\sb{-'+} + C\sp{\b\a}M\sb{\a'\b} \quad , $$
$$ J\sb{+\a} = i\check p\sb + \check x\sb\a -i\check x\sb + \check p\sb\a
+ ip\sb + x\sb\a + M\sb{+\a'} \quad , $$
$$ J\sb{-+} = -i\check x\sb - \check p\sb + +i\check x\sb + \check p\sb -
-ix\sb - p\sb + + M\sb{-'+} \quad , $$
$$ J\sb{\a\b} = -i\check x\sb{(\a}\check p\sb{\b )} -ix\sb{(\a}p\sb{\b )}
+ M\sb{\a\b'} + M\sb{\a'\b} \quad , $$
$$ J\sb{-\a} = -i\check x\sb - \check p\sb\a +i\check x\sb\a \check p\sb -
-ix\sb - p\sb\a +ix\sb\a p\sb - + M\sb{-'\a} +{1\over\check p\sb +}\cq\sb{\a'}
\quad , $$
$$ p\sb - = -{1\over 2\check p\sb +}(p\sb a\sp 2 +M\sp 2 
+2p\sb + \check p\sb - +2p\sp\a\check p\sb\a) \quad , $$
$$ \cq\sb{\a'} = M\sb{\a'}\sp a p\sb a + M\sb{\a'm}M
+M\sb{\a'+}\check p\sb - +M\sb{\a'-'}p\sb + -M\sb{\a'\b}\check p\sp\b
-M\sb{\a'\b'}p\sp\b \quad . \eqno(3.6.4) $$
(All the $\check p$'s are linear, being unconstrained so far.)
We now use $p$ and $\D$ to eliminate the extra zero-modes.  We apply the
constraints and corresponding gauge conditions
$$ \li{ \D = 0 \quad\to\quad  &\check x\sb - = {1\over\check p\sb +}
(x\sb - p\sb + +x\sp\a p\sb\a -\check x\sb +\check p\sb -
-\check x\sp\a\check p\sb\a +iM\sb{-'+} +iC\sp{\b\a}M\sb{\a'\b}) \quad , \cr
& gauge \quad \check p\sb + = 1 \quad ; \cr
\check p\sp A\check p\sb A = 0 \quad\to\quad &\check p\sb - =
-\ha\check p\sp\a\check p\sb\a \quad , \cr
& gauge \quad \check x\sb + = 0 \quad . &(3.6.5)\cr} $$
These constraints are directly analogous to (2.2.3), which were used 
to obtain the
usual coordinate representation of the conformal group SO(D,2) from
the usual coordinate representation (with 2 more coordinates) of the
same group as a Lorentz group.  In fact, after making a unitary 
transformation of the type (3.5.2b),
$$ U = e\sp{-(ip\sb + x\sb\a + M\sb{+\a'})\check p\sp\a} \quad , \eqno(3.6.6) $$
the remaining unwanted coordinates $\check x\sb\a$ completely decouple:
$$ U J\sb{AB} U\sp{-1} = \oover J\sb{AB} + J'\sb{AB} \quad , \eqno(3.6.7a) $$
$$ \oover J\sb{+\a} = i\check x\sb\a \quad , \quad \oover J\sb{-+} =
i\check x\sp\a\check p\sb\a \quad , \quad
\oover J\sb{\a\b} = -i\check x\sb{(\a}\check p\sb{\b )} \quad , \quad
\oover J\sb{-\a} = -i\check x\sb\a\check p\sp\b\check p\sb\b 
\quad ; \eqno(3.6.7b) $$
$$ J'\sb{+\a} = 2(ip\sb + x\sb\a +M\sb{+\a'}) \quad , \quad 
J'\sb{-+} = 2(-ix\sb - p\sb + +M\sb{-'+}) + (-ix\sp\a p\sb\a
+C\sp{\b\a}M\sb{\a'\b}) \quad , $$
$$ J'\sb{\a\b} = -ix\sb{(\a}p\sb{\b )} + (M\sb{\a'\b} +M\sb{\a\b'}) \quad , $$
$$ \li{ \quad J'\sb{-\a} = & (-ix\sb - p\sb\a + M\sb{-'\a}) \cr
& + \left[ -ix\sb\a \ha (p\sb a\sp 2 +M\sp 2) + (M\sb{\a'}\sp a p\sb a
+M\sb{\a'm}M + M\sb{\a'-'}p\sb + -M\sb{\a'\b'}p\sp\b) \right]
\quad , \cr
&&(3.6.7c) \cr} $$
where $\oover J$ are the generators of the conformal group in 2
anticommuting dimensions ((3.3.9), after switching coordinates and momenta), 
and $J'$ are the desired OSp(1,1$|$2) generators.

To eliminate zero-modes, it's convenient to transform these OSp(1,1$|$2)
generators to the canonical form (3.4.2a).  This is performed [3.7] by the
redefinition 
$$ p\sb + \;\to\; \ha p\sb +\sp 2 \quad , \eqno(3.6.8a) $$
followed by the unitary transformations
$$ U\sb 1 = p\sb + \sp{-\left( -i\ha [x\sp\a , p\sb\a ] 
+2M\sb{-'+} +C\sp{\b\a}M\sb{\a'\b} \right)} \quad , $$
$$ U\sb 2 = e\sp{-2M\sb{+\a'}p\sp\a} \quad . \eqno(3.6.8b) $$
Since $p\sb +$ is imaginary (though hermitian) in ($x\sb -$) coordinate
space, $U\sb 1$ changes reality conditions accordingly (an $i$ for each
$p\sb +$).  The generators are then
$$ J\sb{+\a} = ix\sb\a p\sb + \quad , \quad J\sb{-+} = -ix\sb -p\sb +
\quad , \quad J\sb{\a\b} = -ix\sb{(\a}p\sb{\b )} +\Hat M\sb{\a\b} \quad , $$
$$ J\sb{-\a} = -ix\sb -p\sb\a +{1\over p\sb +}\left[
-ix\sb\a\ha (p\sb a\sp 2 +M\sp 2 +p\sp\b p\sb\b) +\Hat M\sb\a\sp\b p\sb\b
+\Hat\cq\sb\a \right] \quad ; \eqno(3.6.9a) $$
$$ \Hat M\sb{\a\b} = M\sb{\a\b'} +M\sb{\a'\b} \quad , $$
$$ \Hat\cq\sb\a = M\sb{-'\a} -\ha M\sb{-'\a'} +(M\sb{\a'}\sp a p\sb a
+M\sb{\a'm}M) +M\sb{+\a'}(p\sb a\sp 2 +M\sp 2) \quad . \eqno(3.6.9b) $$

For the U(1) case the derivation is a little more straightforward.
It corresponds to first taking the U(1,1$|$1,1) (=SU(1,1$|$1,1)$\otimes$
U(1)) subgroup of OSp(2,2$|$4).  From (3.6.1), instead of (3.6.2a,3) we 
now have
$$ \h\sb{\ca\cb} = \pmatrix{\h\sb{AB} & 0 \cr 0 & \h\sb{A'B'} \cr}
\quad , \quad \h\sb{AB} = \h\sb{A'B'} \quad , \quad \ca = (A,A') \quad ,$$
$$ J\sb{AB} = \tilde J\sb{AB} + \tilde J\sb{A'B'} \quad , \quad
\check p\sb A = \tilde p\sb{A'} \quad , \quad 
\D = \h\sp{BA}\tilde J\sb{A'B} \quad . \eqno(3.6.10) $$
The original light-cone $\pm$ are now still $\pm$ (no primes),
so the unwanted zero-modes can be eliminated by the constraints
and gauge choices
$$ \check p\sb A = 0 \quad\to\quad gauge \quad \check x\sb A = 0 
\quad . \eqno(3.6.11) $$
Alternatively, we could include $\check p\sb A$ among the generators,
using IOSp(1,1$|$2) as the group (as for the usual closed string: see sects.\
7.1, 11.1).  (The same result can be obtained by replacing (3.6.11) with the
constraints $\D =0$ and $\e\sp{\hat b\hat a}p\sb{A\hat a}p\sb{B\hat b}
\sim p\sb{[A}\check p\sb{B)} =0$.)  The OSp(1,1$|$2) generators are now
$$ J\sb{+\a} = ix\sb\a p\sb + + M\sb{+'\a'} \quad , \quad J\sb{-+} =
-ix\sb - p\sb + + M\sb{-'+'} \quad , $$
$$ J\sb{\a\b} = -ix\sb{(\a}p\sb{\b )} + M\sb{\a\b} + M\sb{\a'\b'} \quad , $$
$$ J\sb{-\a} = -ix\sb -p\sb\a -ix\sb\a {1\over 2p\sb +}(p\sb a\sp 2 +
M\sp 2 + p\sp\b p\sb\b) + {1\over p\sb +}
( M\sb\a\sp a p\sb a + M\sb{\a m}M + M\sb\a\sp\b p\sb\b ) + M\sb{-'\a'} 
\quad , \eqno(3.6.12a) $$
or, in other words (symbols), these OSp(1,1$|$2) generators are just
the usual ones plus the spin of a second OSp(1,1$|$2), with the same
representation as the spin of the first OSp(1,1$|$2):
$$ J\sb{AB} = \tilde J\sb{AB} + M\sb{A'B'} \quad . \eqno(3.6.12b) $$
(However, for the string $M\sb{\a m}M$ will contain oscillators from both
sets of 2+2 dimensions, so these sets of oscillators won't decouple,
even though  $\tilde J\sb{AB}$ commutes with $M\sb{A'B'}$.)
To simplify the form of $J\sb{-+}$ and $J\sb{-\a}$, we make the
consecutive unitary transformations (3.5.2):
$$ U\sb 1 = p\sb +\sp{-M\sb{-'+'}} \quad , \quad
U\sb 2 = e\sp{M\sb{+'}\sp{\a'}p\sb\a} \quad , \eqno(3.6.13) $$
after which the generators again take the canonical form:
$$ J\sb{+\a} = ix\sb\a p\sb + \quad , \quad J\sb{-+} = -ix\sb -p\sb +
\quad , J\sb{\a\b} = -ix\sb{(\a}p\sb{\b )} +\Hat M\sb{\a\b} \quad , $$
$$ J\sb{-\a} = -ix\sb -p\sb\a +{1\over p\sb +}\left[
-ix\sb\a\ha (p\sb a\sp 2 +M\sp 2 +p\sp\b p\sb\b)+\Hat M\sb\a\sp\b p\sb\b
+\Hat\cq\sb\a \right] \quad ; \eqno(3.6.14a) $$
$$ \Hat M\sb{\a\b} = M\sb{\a\b}+M\sb{\a'\b'} \quad , $$
$$ \Hat\cq\sb\a = M\sb{-'\a'} +( M\sb\a\sp b p\sb b +M\sb{\a m}M ) +
M\sb{+'\a'}\ha (p\sb a\sp 2 +M\sp 2) \quad . \eqno(3.6.14b) $$
Because of $U\sb 1$, formerly real fields now satisfy $\f\dg =
(-1)\sp{M\sb{-'+'}}\f$.  

Examples and actions of this 4+4-extended OSp(1,1$|$2) will be
considered in sect.\ 4.1, its application to supersymmetry in sect.\
5.5, and its application to strings in sect.\ 8.3.

\sect{Exercises}

\Item{(1)} Derive the time derivative of (3.1.6) from (3.1.7).
\Item{(2)} Derive (3.1.12).  Compare with the usual derivation of the
Noether current in field theory.
\Item{(3)} Derive (3.1.15,17).
\Item{(4)} Show that $Q$ of (3.2.6a) is nilpotent.  Show this directly
for (3.2.8).
\Item{(5)} Derive (3.3.1b).
\Item{(6)} Use (3.3.6) to rederive (3.2.6a,12a).
\Item{(7)} Use (3.3.2,7) to derive the OSp(1,1$|$2) algebra for
Yang-Mills in terms of the explicit independent fields (in analogy to
(3.2.8)). 
\Item{(8)} Perform the transformation (3.4.3a) to obtain (3.4.3b).
Choose the Dirac spinor representation of the spin operators (in terms
of $\g$-matrices).  Compare with (3.2.6), and identify the field
equations $\cg$ and ghosts $C$.
\Item{(9)} Check that the algebra of $\Hat\cq\sb\a$ and $\Hat
M\sb{\a\b}$ closes for (3.5.3), (3.6.9), and (3.6.14), and compare with
(3.4.2). 

%
%

\chscy{4. GENERAL GAUGE THEORIES}{4.1. OSp(1,1$|$2)}{4.1. OSp(1,1|2)}6

In this chapter we will use the results of sects.\ 3.4-6 to derive free
gauge-invariant actions for arbitrary field theories, and discuss some
preliminary results for the extension to interacting theories.

The (free) gauge covariant theory for arbitrary representations 
of the Poincar\'e group (except perhaps for those satisfying self-duality 
conditions) can be constructed from the BRST1 OSp(1,1$|$2) generators [2.3].
For the fields described in sect.\ 3.4 which are representations of 
OSp(D,2$|$2), consider the gauge invariance generated by OSp(1,1$|$2) 
and the obvious (but unusual) corresponding gauge-invariant action:
$$ \d \F = \ha J \sp {BA} \L \sb {AB} \quad\to\quad
S = \int d \sp D x \sb a d x \sb - d \sp 2 x \sb \a \; \ha \F\dg
p \sb + \d ( J \sb {AB} ) \F \quad , \eqno(4.1.1)$$
where $J \sb {AB}$ for $A = ( + , - , \a )$ (graded antisymmetric in its
indices) are the generators of OSp(1,1$|$2), and
we have set $k = 0$, so that the $p \sb +$ factor is the 
Hilbert space metric.  In particular, the $J \sb {-+}$ and $J \sb {+\a}$
transformations allow all dependence on the unphysical coordinates to be
gauged away:
$$ \d \F = - i x \sb - p \sb + \L \sb {-+} + i x \sp \a p \sb + \L \sb
{-\a} \eqno(4.1.2)$$
implies that only the part of $\F$ at $x \sb - = x \sb \a = 0$ can be
gauge invariant.  A more explicit form of $\d ( J\sb{AB} )$ is given by
$$ \li{p\sb + \d ( J\sb{AB} ) & = 
p \sb + \d ( J \sb {\a\b} \sp 2 ) i \d ( J \sb {-+} ) \d \sp 2 ( J
\sb {+\a} ) \d \sp 2 ( J \sb {-\a} ) \cr
& = \d ( x \sb - ) \d\sp 2 ( x \sb \a ) \d ( M \sb {\a\b} \sp 2 ) 
p \sb + \sp 2 J \sb {-\a} \sp 2 \quad , &(4.1.3)\cr}$$
where we have used
$$ J \sb {-+} \d ( J \sb {-+} ) = \d ( J \sb {-+} ) J \sb {-+} = 0
\quad \to \quad \d ( J \sb {-+} ) = i {1 \over p \sb +} \d ( x \sb - )
\quad , \eqno(4.1.4)$$
since $p \sb + \ne 0$ in light-cone formalisms.  The gauge invariance of
the kinetic operator follows from the fact that the $\d$-functions can
be reordered fairly freely:  $\d ( J \sb {\a\b} \sp 2 )$ (which is
really a Kronecker $\d$) commutes with all the others, while
$$ \d ( J \sb {-+} + a ) \d \sp 2 ( J \sb {\pm\a} ) = \d \sp 2 ( J \sb
{\pm\a} ) \d ( J \sb {-+} + a \mp 2 ) \quad \to \quad [ \d ( J \sb {-+} ) , 
\d \sp 2 ( J \sb {+\a} ) \d \sp 2 ( J \sb {-\a} ) ] = 0 \quad , $$
$$ [ \d \sp 2 ( J \sb {-\a} ) , \d \sp 2 ( J \sb {+\b} ) ] =
2 J \sb {-+} + ( C \sp {\a\b} J \sb {-+} + J \sp {\a\b} ) 
\ha [ J \sb {-\a} , J \sb {+\b} ] \quad , \eqno(4.1.5)$$
where the $J \sb {-+}$ and $J \sp {\a\b}$ each can be freely moved to
either side of the $[ J \sb {-\a} , J \sb {+\b} ]$.
After integration of the action
over the trivial coordinate dependence on $x \sb -$
and $x \sb \a$, (4.1.1) reduces to (using (3.4.2,4.1.3))
$$ S = \int d \sp D x \sb a \; \ha\f\dg \d ( M\sb{\a\b}\sp 2 )
( \bo - M \sp 2 + \cq\sp 2)\f \quad , \quad \d \f = -i\ha\cq\sb\a\L\sp\a
+ \ha M\sb{\a\b}\L\sp{\a\b} \quad , \eqno(4.1.6)$$
where $\f$ now depends only on the usual spacetime coordinates $x \sb a$,
and for irreducible Poincar\'e representations $\f$
has indices which are the result of starting with an irreducible
representation of OSp(D$-$1,1$|$2) in the massless case, or OSp(D,1$|$2)
in the massive case, and then truncating to the Sp(2) singlets.  
(This type of action was first proposed for the string [4.1,2].)
$\L\sp\a$ is the remaining part of the $J\sb{-\a}$ transformations after
using up the transformations of (4.1.2) (and absorbing a $1/\pa\sb +$), 
and contains the usual component gauge transformations, while
$\L\sp{\a\b}$ just gauges away the Sp(2) nonsinglets.  We have thus
derived a general gauge-covariant action by adding 2+2 dimensions to the
light-cone theory.  In sect.\ 4.4 we'll show that gauge-fixing to
the light cone gives back the original light-cone theory, proving the
consistency of this method.

In the BRST formalism the field contains not only physical 
polarizations, but also auxiliary fields (nonpropagating fields 
needed to make the action local, such as the trace of the 
metric tensor for the graviton), ghosts 
(including antighosts, ghosts of ghosts, etc.), and Stueckelberg 
fields (gauge degrees of freedom, such as the gauge part of Higgs 
fields, which allow more renormalizable and less singular formalisms 
for massive fields).  All of these but the ghosts appear in the 
gauge-invariant action.
For example, for a massless vector we start with $A\sb i =
( A \sb a , A \sb \a )$, which appears in the field $\f$ as
$$ \ket{\f} = \ket{\sp i} A\sb i \quad , \quad
\VEV{\sp i \Big| \sp j} = \h\sp{ij} \quad . \eqno(4.1.7)$$
Reducing to Sp(2) singlets, we can truncate to just $A \sb a$.  
Using the relations
$$ M\sp{ij}\ket{\sp k} = \ket{\sp{[i}}\h\sp{j)k} \quad\to\quad
M \sp {\a a} \ket{\sp b} = \h \sp {ab} \ket{\sp \a} \quad , \quad
M \sp {\a a} \ket{\sp \b} = -C \sp {\a\b} \ket{\sp a} \quad , \eqno(4.1.8)$$
where $[\quad )$ is graded antisymmetrization, we find
$$ \ha M\sp{\a b}M\sb\a\sp c\ket{\sp a} = \ha M\sp{\a b}\h\sp{ac}\ket{\sb\a}
= \h\sp{ac} \ket{\sp b} \quad , \eqno(4.1.9)$$
and thus the lagrangian
$$ \cl = \ha\bra{\f}\d (M\sb{\a\b}\sp 2)[\bo -(M\sb\a\sp b\pa\sb b)\sp 2]
\ket{\f} = \ha A\sp a ( \bo A\sb a - \pa\sb a \pa\sp b A\sb b ) \quad .
\eqno(4.1.10) $$
Similarly, for the gauge transformation
$$ \ket{\L\sp\a} = \ket{\sp i}\L\sb i\sp\a \quad\to\quad
\d A\sb a = \bra{\sb a}\ha M\sb\a\sp b\pa\sb b \ket{\L\sp\a} 
= - \ha\pa\sb a \L\sb\a\sp\a \quad . \eqno(4.1.11)$$
As a result of the $\d (M\sb{\a\b}\sp 2)$ acting on 
$\cq\sb\a\L\sp\a$, the only part of $\L$ which survives is the part
which is an overall singlet in the matrix indices and explicit $\a$
index: in this case, $\L\sb i\sp\a = \d\sb i\sp\a \l$ $\to$ $\d A\sb a =
-\pa\sb a\l$.  Note that the $\f \cq\sp 2 \f$ term can be written as a 
$(\cq\f )\sp 2$ term:  This corresponds to subtracting out a
``gauge-fixing'' term from the ``gauge-fixed'' lagrangian $\f (\bo -
M\sp 2)\f$.  (See the discussion of gauge fixing in sect.\ 4.4.)  

For a massless antisymmetric tensor we start with 
$A\sb{[ij)} = ( A \sb {[ab]} , A \sb {a\a} , A \sb {(\a\b )} )$ appearing as
$$ \ket{\f} = -\ha \ket{\sp{ij}}_A A\sb{ji} \quad , \quad
\ket{\sp{ij}}_A = {1\over\sqrt 2}\ket{\sp{[i}}\otimes\ket{\sp{j)}}
\eqno(4.1.12) $$
(and similarly for $\sket{\L\sp\a}$), and truncate to just $A\sb{[ab]}$.  
Then, from (4.1.8),
$$ \li{ \ha M\sp{\a c}M\sb\a\sp d \ket{\sp a}\ket{\sp b} 
&= \ha M\sp{\a c} \left( \h\sp{da}\ket{\sb\a}\ket{\sp b} +
\h\sp{db}\ket{\sp a}\ket{\sb\a}\right) \cr
&= \left( \h\sp{da}\ket{\sp c}\ket{\sp b} + \h\sp{db}\ket{\sp a}\ket{\sp
c}\right) + \h\sp{d(a}\h\sp{b)c}\ha\ket{\sp\a}\ket{\sb\a} \quad ,
&(4.1.13)\cr}$$
and we have
$$ \cl = \frac14 A\sp{ab} ( \bo A\sb{ab} + \pa\sp c\pa\sb{[a}A\sb{b]c} )
\quad ,\quad \d A\sb{ab} = \ha\pa\sb{[a}\L\sb{b]\a}\sp\a \quad .\eqno(4.1.14)$$

For a massless traceless symmetric tensor we start with 
$h\sb{(ij]} = ( h \sb {(ab)} , h \sb {\a b} , h \sb {[\a\b ]} )$ satisfying 
$h\sp i\sb i = h \sp a \sb a + h \sp \a \sb \a = 0$, appearing as
$$ \ket{\f} = \ha \ket{\sp{ij}}_S h\sb{ji} \quad , \quad
\ket{\sp{ij}}_S = {1\over\sqrt 2}\ket{\sp{(i}}\ket{\sp{j]}}\quad ,
\eqno(4.1.15)$$
and truncate to $( h \sb {(ab)} , h \sb {[\a\b ]})$, where $h \sb {[\a\b ]}
= \ha C \sb {\a\b} \h \sp {ab} h \sb {(ab)}$, leaving just an
unconstrained symmetric tensor.  Then, using (4.1.13), as well as
$$ \ha M\sp{\g a}M\sb\g\sp b \ket{\sp\a}\ket{\sp\b} =
\ha C\sp{\a\b}\ket{\sp{(a}}\ket{\sp{b)}} - \h\sp{ab}\ket{\sp\a}\ket{\sp\b}
\quad , \eqno(4.1.16)$$
and using the condition $h\sp\a\sb\a = - h\sp a\sb a$, we find
$$ \cl = \frac14 h\sp{ab}\bo h\sb{ab} -\ha h\sp{ab}\pa\sb b\pa\sb c h\sb
a \sp c +\ha h\sp c\sb c\pa\sb a\pa\sb b h\sp{ab} -\frac14 h\sp a\sb
a\bo h\sp b\sb b \quad ,$$
$$ \d h\sb{ab} = -\ha\pa\sb{(a}\L\sb{b)\a}\sp\a \quad . \eqno(4.1.17)$$
This is the linearized Einstein-Hilbert action for gravity.

The massive cases can be obtained by the dimensional reduction
technique, as in (2.2.9), since that's how it was done for this entire
procedure, from the light-cone Poincar\'e algebra down to (4.1.6).
(For the string, the OSp generators are
represented in terms of harmonic oscillators, and $M\sb{\a m}M$ is cubic
in those oscillators instead of quadratic, so the oscillator expressions
for the generators don't follow from dimensional reduction, and (4.1.6)
must be used directly with the $M\sb{\a m}M$ terms.)  
Technically, $p\sb m = m$ makes sense only
for complex fields.  However, at least for free theories, the resulting
$i$'s that appear in the $p\sb a p\sb m$ crossterms can be removed by
appropriate redefinitions for the complex fields, after which they can be
chosen real.  (See the discussion below (2.2.10).)
For example, for the massive vector we replace $A\sb m
\to iA\sb m$ (and then take $A\sb m$ real) to obtain
$$ \ket{\f} = \ket{\sp a}A\sb a +i\ket{\sb m}A\sb m +\ket{\sp\a}A\sb\a
\quad , \quad
\bra{\f} = A\sp a\bra{\sb a} -iA\sb m\bra{\sb m} +A\sp\a\bra{\sb\a}
\quad .\eqno(4.1.18)$$
The lagrangian and invariance then become
$$ \li{ \cl &= \ha A\sp a [( \bo - m\sp 2 ) A\sb a - \pa\sb a \pa\sp b A\sb b ]
+ \ha A\sb m \bo A\sb m +mA\sb m \pa\sp a A\sb a \cr
&= \frac14 F\sb{ab}\sp 2 -\ha (mA\sb a +\pa\sb a A\sb m)\sp 2\quad , \cr}$$
$$ \d A\sb a = -\pa\sb a\l \quad , \quad \d A\sb m = m\l\quad .\eqno(4.1.19)$$
This gives a Stueckelberg formalism for a massive vector.

Other examples reproduce all the special cases of higher-spin fields proposed 
earlier [4.3] (as well as cases that hadn't been obtained previously).
For example, for totally symmetric tensors, the usual
``double-tracelessness'' condition is automatic:  Starting from the
light cone with a totally symmetric and traceless tensor (in
transverse indices), extending $i \to (a,\a )$ and restricting to Sp(2)
singlets, directly gives a totally symmetric and traceless tensor
(in D-dimensional indices) of the same rank, and one of rank 2 lower
(but no lower than that, due to the total antisymmetry in the Sp(2) indices).

The most important feature of the BRST method of deriving
gauge-invariant actions from light-cone (unitary) representations of the
Poincar\'e group is that it automatically includes exactly the right
number of auxiliary fields to make the action local.  In the case of
Yang-Mills, the auxiliary field ($A\sb -$) was obvious, since it 
results directly from adding just 2 commuting dimensions (and not 
2 anticommuting) to the light cone, i.e., from making $D-2$-dimensional 
indices $D$-dimensional.  Furthermore, the necessity of this field for
locality doesn't occur until interactions are included (see sect.\ 2.1).
A less trivial example is the graviton:  Naively, a traceless symmetric
$D$-dimensional tensor would be enough, since this would automatically
include the analog of $A\sb -$.  However, the BRST method
automatically includes the trace of this tensor.  In general, the extra
auxiliary fields with anticommuting ``ghost-valued'' Lorentz indices 
are necessary for gauge-covariant, local formulations of field theories
[4.4,5].  In order to study this
phenomenon in more detail, and because the discussion will be useful
later in the 2D case for strings, we now give a brief discussion of
general relativity.

General relativity is the gauge theory of the Poincar\'e group.  Since
local translations (i.e., general coordinate transformations) include
the orbital part of Lorentz transformations (as translation by an amount
linear in $x$), we choose as the group generators $\pa\sb m$ and the
Lorentz spin $M\sb{ab}$.  Treating $M\sb{ab}$ as second-quantized
operators, we indicate how they act by writing explicit ``spin'' vector 
indices $a,b,\dots$ (or spinor indices) on the fields, while using $m,n,\dots$
for ``orbital'' vector indices on which $M\sb{ab}$ doesn't act, as on
$\pa\sb m$.  (The action of the second-quantized $M\sb{ab}$ follows from
that of the first-quantized:  E.g., from (4.1.8), (2.2.5), and the fact
that $(M\sp{ij})\dg = - M\sp{ij}$, we have $M\sb{ab}A\sb c =
- \h\sb{c[a}A\sb{b]}$.)  The spin indices (but not the orbital ones) can be
contracted with the usual constant tensors of the Lorentz group (the
Lorentz metric and $\g$ matrices).  The (antihermitian) generators 
of gauge transformations are thus
$$ \l = \l\sp m (x) \pa\sb m + \ha \l\sp{ab} (x) M\sb{ba}
\quad , \eqno(4.1.20)$$
and the covariant derivatives are
$$ \de\sb a = e\sb a\sp m\pa\sb m + \ha\o\sb a\sp{bc} M\sb{cb}
\quad , \eqno(4.1.21)$$
where we have absorbed the usual derivative term, since derivatives are
themselves generators, and to make the covariant derivative transform
covariantly under the gauge transformations
$$ \de\sb a ' = e\sp\l \de\sb a e\sp{-\l} \quad . \eqno(4.1.22)$$
Covariant field strengths are defined, as usual, by commutators of
covariant derivatives, 
$$ [ \de\sb a , \de\sb b ] = T\sb{ab}\sp c \de\sb c +
\ha R\sb{ab}\sp{cd} M\sb{dc} \quad , \eqno(4.1.23)$$
since that automatically makes them transform covariantly (i.e., by a 
similarity transformation, as in (4.1.22)), as a consequence of the 
transformation law (4.1.22) of the covariant derivatives themselves.
Without loss of generality, we can choose
$$ T\sb{ab}\sp c = 0 \quad , \eqno(4.1.24)$$
since this just determines $\o\sb{ab}\sp c$ in terms of $e\sb a\sp m$,
and any other $\o$ can always be written as this $\o$ plus a tensor that
is a function of just $T$.  (The theory could then always be rewritten
in terms of the $T=0$ $\de$ and $T$ itself, making $T$ an arbitrary
extra tensor with no special geometric significance.)  To solve this
constraint we first define
$$ e\sb a = e\sb a\sp m\pa\sb m \quad , $$
$$ [ e\sb a , e\sb b ] = c\sb{ab}\sp c e\sb c \quad . \eqno(4.1.25)$$
$c\sb{ab}\sp c$ can then be expressed in terms of $e\sb a\sp m$, the
matrix inverse $e\sb m\sp a$,
$$ e\sb a\sp m e\sb m\sp b = \d\sb a\sp b \quad , \quad
e\sb m\sp a e\sb a\sp n = \d\sb m\sp n \quad , \eqno(4.1.26)$$
and their derivatives.  The solution to (4.1.24) is then
$$ \o\sb{abc} = \ha ( c\sb{bca} - c\sb{a[bc]} ) \quad . \eqno(4.1.27)$$

The usual global Lorentz transformations, which include orbital and spin
pieces together in a specific way, are a symmetry of the vacuum, defined
by
$$ \langle \de\sb a \rangle = \k\pa\sb a \quad\iff\quad
\langle e\sb a\sp m \rangle = \k\d\sb a\sp m \quad .\eqno(4.1.28)$$
$\k$ is an arbitrary constant, which we can choose to be a unit of
length, so that $\de$ is dimensionless.  (In $D=4$ it's just the usual
gravitational coupling constant, proportional to the square root of
Newton's gravitational constant.)  As a result of general
coordinate invariance, any covariant object (i.e., a covariant
derivative or tensor with only spin indices uncontracted) will then also
be dimensionless.  The subgroup of the original gauge group which leaves
the vacuum (4.1.28) invariant is just the usual (global) Poincar\'e
group, which treats orbital and spin indices in the same way.
We can also treat these indices in a similar way with respect to the
full gauge group by using the ``vielbein'' $e\sb a\sp m$ and its inverse
to convert between spin and orbital indices.  In particular, the orbital
indices on all fields except the vielbein itself can be converted into
spin ones.  Also, since integration measures are antisymmetric,
converting $dx\sp m$ into $\O\sp a = dx\sp me\sb m\sp a$ converts $d\sp
D x$ into $\O\sp D = d\sp D x \; {\rm e}\sp{-1}$, where e $= det(e\sb a\sp m)$.
On such covariant fields, $\de$ always acts covariantly.
On the other hand, in the absence of spinors, all indices can be
converted into orbital ones.  In particular, instead of the vielbein we
could work with the metric tensor and its inverse:
$$ g\sb{mn} = \h\sb{ab} e\sb m\sp a e\sb n\sp b \quad , \quad
g\sp{mn} = \h\sp{ab} e\sb a\sp m e\sb b\sp n \quad . \eqno(4.1.29)$$
Then, instead of $\de$, we would need a covariant derivative which knows
how to treat uncontracted orbital indices covariantly.  

The action for gravity can be written as
$$ S = -\ha\int d\sp D x \; {\rm e}\sp{-1} \; R \quad , \quad
R = \ha R \sb{ab}\sp{ba} \quad . \eqno(4.1.30)$$
This can be rewritten in terms of $c\sb{abc}$ as
$$ {\rm e}\sp{-1} R = -\pa\sb m ( {\rm e}\sp{-1} e\sp{am} c\sb a\sp b\sb b )
+ {\rm e}\sp{-1} \left[ -\ha ( c\sp{ab}\sb b )\sp 2 +\frac18 c\sp{abc}c\sb{abc}
-\frac14 c\sp{abc}c\sb{bca} \right] \eqno(4.1.31)$$
using
$$ {\rm e}\sp{-1}e\sp a f\sb a = \pa\sb m ( {\rm e}\sp{-1}e\sp{am}f\sb a ) +
{\rm e}\sp{-1}c\sp{ab}\sb b f\sb a \quad . \eqno(4.1.32)$$
Expanding about the vacuum,
$$ e\sb a\sp m = \k\d\sb a\sp m + \k\sp{D/2}h\sb a\sp m \quad , \eqno(4.1.33)$$
where we can choose $e\sp{am}$ (and thus $h\sp{am}$) to be symmetric by
the $\l\sb{ab}$ transformation, the linearized action is just (4.1.17).
As an alternative form for the action, we can consider making the field
redefinition
$$ e\sb a\sp m \to \f\sp{-2/(D-2)}e\sb a\sp m \quad , \eqno(4.1.34)$$
which introduces the new gauge invariance of (Weyl) local scale transformations
$$ e\sb a\sp m ' = e\sp\z e\sb a\sp m \quad , \quad \f ' = e\sp{(D-2)\z /2} \f 
\quad . \eqno(4.1.35)$$
(The gauge choice $\f = constant$ returns the original fields.)
Under the field redefinition (4.1.34), the action (4.1.30) becomes
$$ S \to \int d\sp D x \; {\rm e}\sp{-1} \; \left[
2\frac{D-1}{D-2}(\de\sb a\f )\sp 2 -\ha R\f\sp 2 \right] \quad .\eqno(4.1.36)$$
We have actually started from (4.1.30) without the total-derivative
term of (4.1.31), which is then a function of just $e\sb a\sp m$ and its
first derivatives, and thus correct even at boundaries.  (We also
dropped a total-derivative term $-\pa\sb m (\ha\f\sp 2 {\rm e}\sp{-1}
e\sp{am}c\sb a\sp b\sb b )$ in (4.1.36), which will be irrelevant for
the following discussion.)  If we eliminate $\f$ by its field equation, 
but keep surface terms, this becomes
$$ \li {S &\to \int d\sp D x \; {\rm e}\sp{-1}\; 2\frac{D-1}{D-2} 
\de\cdot (\f\de\f )\cr
&= \int d\sp D x \; {\rm e}\sp{-1} \; \frac{D-1}{D-2} \bo \f\sp 2 \cr
&= \int d\sp D x \; {\rm e}\sp{-1} \; \frac{D-1}{D-2} \bo 
\left[ \langle\f\rangle\sp 2 + 2\langle\f\rangle ( \f - \langle\f\rangle )
+ ( \f - \langle\f\rangle )\sp 2\right] &(4.1.37)\cr}$$
We can solve the $\f$ field equation as
$$ \f = \langle\f\rangle \left( 1 - {1\over 4{D-1\over D-2}\bo +R}R\right)
\quad . \eqno(4.1.38)$$
(We can choose $\langle\f\rangle = 1$, or take the $\k$ out of
(4.1.28) and introduce it instead through $\langle\f\rangle =
\k\sp{-(D-2)/2}$ by a global $\z$ transformation.)
Assuming $\f$ falls off to $\langle\f\rangle$ fast enough at $\infty$,
the last term in (4.1.37) can be dropped, and, using (4.1.38), the
action becomes [4.6]
$$ S \to -\ha \int d\sp D x \; {\rm e}\sp{-1} \left( R - 
R {1\over 4{D-1\over D-2}\bo +R}R\right) \quad . \eqno(4.1.39)$$
Since this action has the invariance (4.1.35), we can gauge away the
trace of $h$ or, equivalently, gauge the determinant of $e\sb a\sp m$
to 1.  In fact, the same action results from (4.1.30) if we eliminate
this determinant by its equation of motion.

Thus, we see that, although gauge-covariant, Lorentz-covariant
formulations are possible without the extra auxiliary fields, they are
nonlocal.  Furthermore, the
nonlocalities become more complicated when coupling to nonconformal
matter (such as massive fields), in a way reminiscent of Coulomb terms
or the nonlocalities in light-cone gauges.  Thus, the construction of
actions in such a formalism is not straightforward, and requires the
use of Weyl invariance in a way analogous to the use of Lorentz
invariance in light-cone gauges.  Another
alternative is to eliminate the trace of the metric from the Einstein
action by a coordinate choice, but the remaining constrained 
(volume-preserving) coordinate invariance causes difficulties in 
quantization [4.7].

We have also seen that some properties of gravity (the ones relating to
conformal transformations) become more transparent when the scale
compensator $\f$ is introduced.  (This is particularly true for
supergravity.)  Introducing such fields into the OSp formalism requires
introducing new degrees of freedom, to make the representation larger
(at least in terms of gauge degrees of freedom).  Although such
invariances are hard to recognize at the free level, the extensions of
sect.\ 3.6 show signs of performing such generalizations.  However, 
while the U(1)-type extension can be applied to arbitrary Poincar\'e 
representations, the GL(1)-type has difficulty with fermions.  We'll
first discuss this difficulty, then show how the 2 types differ for
bosons even for the vector, and finally look again at gravity.

The U(1) case of spin 1/2 reproduces the algebra of sect.\ 3.5, 
since $M\sb{A'B'}$ of (3.6.12b) is exactly the extra term of (3.5.1):
$$ M\sb{ij} = \ha\g\sb{ij} \quad\to $$
$$ \li{ \bar\f \d (\Hat M\sb{\a\b}\sp 2) (\bo +\Hat\cq\sp 2) \f & = 
\bar\f \d (~) e\sp{i\g\sb{+'}\sl p/2}i\ha\sl p\g\sb{-'}
e\sp{i\g\sb{+'}\sl p/2} \f \cr
& = i\frac14\Bar{(\g\sb{-'}\f +i\ha\g\sb{-'}\g\sb{+'}\sl p\f )} \d (~)
\g\sb{+'}\sl p (\g\sb{-'}\f +i\ha\g\sb{-'}\g\sb{+'}\sl p\f ) \cr
& = i\ha\hat{\bar\f}\sl p\g\sb{-'}\hat\f \quad , & (4.1.40) \cr} $$
where $\g\sb{ij}=\ha [\g\sb i , \g\sb j \}$, and we have used
$$ 0 = \frac18 (\g\sp{\a\b}+\g\sp{\a'\b'})(\g\sb{\a\b}+\g\sb{\a'\b'}) =
(\g\sp\a\g\sb{\a'})\sp 2 +4 \quad\to\quad \g\sp\a\g\sb{\a'} =2i
\quad . \eqno(4.1.41) $$
(We could equally well have chosen the other sign.  This choice, with
our conventions, corresponds to harmonic-oscillator boundary conditions:
See sect.\ 4.5.)  After eliminating
$\g\sb{+'}\f$ by gauge choice or, equivalently, by absorbing it into
$\g\sb{-'}\f$ by field redefinition, this becomes just 
$\bar\varphi\sl p\varphi$.  However, in the GL(1) case, the analog to
(4.1.41) is
$$ 0 = \frac18 ( \{ \g\sp\a , \g\sp{\b'} \} + \{ \g\sp{\a'} ,
\g\sp\b \} ) ( \{ \g\sb\a , \g\sb{\b'} \} + \{ \g\sb{\a'} ,
\g\sb\b \} ) = (\g\sp\a\g\sb{\a'})(\g\sp{\b'}\g\sb\b) \quad , $$
$$ \g\sp\a\g\sb{\a'} + \g\sp{\a'}\g\sb\a = -4 \quad , \eqno(4.1.42) $$
and to (4.1.40) is
$$ \bo +\Hat\cq\sp 2 = -p\sp 2 +\frac18 (\g\sp\a\g\sb{\a'}) \g\sb{-'}
(\sl p -\g\sb + p\sp 2) +\frac18 (\g\sp{\a'}\g\sb\a)
(\sl p -\g\sb +p\sp 2) \g\sb{-'} \quad . \eqno(4.1.43) $$
Unfortunately, $\f$ and $\bar\f$ must have opposite boundary conditions
$\g\sp\a\g\sb{\a'} =0$ or $\g\sp{\a'}\g\sb\a =0$ in order to contribute
in the presence of $\d (\Hat M\sb{\a\b}\sp 2)$, as is evidenced by
the asymmetric form of (4.1.43) for either choice.  Consequently, the parts
of $\f$ and $\bar\f$ that survive are not hermitian conjugates of each
other, and the action is not unitary.  (Properly speaking, if we choose
consistent boundary conditions for both $\f$ and $\bar\f$, the action
vanishes.)  Thus, the GL(1)-type OSp(1,1$|$2) is unsuitable for
spinors unless further modified.  In any case, such a modification would
not treat bosons and fermions symmetrically, which is necessary for
treating supersymmetry.  (Fermions in the usual OSp formalism will be
discussed in more detail in sect.\ 4.5.)

For the case of spin 1 (generalizing the light-cone Hilbert space, as in
(4.1.7-8)), we expand
$$ \f = \sket{\sp a}A\sb a +i\sket{\sb{-'}}A\sb + +i\sket{\sb +}A\sb -
\quad , \quad \f\dg = A\sp a\sbra{\sb a} -iA\sb +\sbra{\sb{-'}}
-iA\sb -\sbra{\sb +} \quad , \eqno(4.1.44) $$
for the GL(1) case, and the same for U(1) with $\sket{\sb +}\to
\sket{\sb{+'}}$ (Sp(2)-spinor fields again drop out of the full $\f$).
We find for GL(1) [3.7]
$$ L = -\frac14 F\sb{ab}\sp 2 - \ha (A\sb - +\pa\cdot A)\sp 2 
= -\frac14 F\sb{ab}\sp 2 -\ha\hat A\sb -\sp 2 \quad , \eqno(4.1.45a) $$
where $F\sb{ab}=\pa\sb{[a}A\sb{b]}$, and for U(1)
$$ L = -\frac14 F\sb{ab}\sp 2 + \ha (A\sb - +\ha\bo A\sb +)\sp 2 
= -\frac14 F\sb{ab}\sp 2 +\ha\hat A\sb -\sp 2 \quad . \eqno(4.1.45b) $$
In both cases $A\sb +$ can be gauged away, and $A\sb -$ is auxiliary.
However, the sign for U(1)-type OSp(1,1$|$2) is the same as for auxiliary
fields in supersymmetry (for off-shell irreducible multiplets), whereas
the sign for GL(1) is opposite.
The sign difference is not surprising, considering the U(1) and
GL(1) types are Wick rotations of each other:  This auxiliary-field
term, together with the auxiliary component of $A\sb a$ (the light-cone
$A\sb -$), appear with the metric $\h\sb{\hat a\hat b}$ of (3.6.1), and
thus with the same sign for SO(2) (U(1)).  In fact, (4.1.45b) is
just the part of the 4D N=1 super-Yang-Mills lagrangian for fields which
are R-symmetry invariant:  $A\sb -$ can be identified with the usual 
auxiliary field, and $A\sb +$ with the $\q =0$ component of the
superfield.  Similarly, $\g\sb{+'}\f$ for spin 1/2 can be identified
with the linear-in-$\q$ part of this superfield.  This close analogy
strongly suggests that the nonminimal fields of this formalism may be
necessary for treating supersymmetry.  Note also that for GL(1) the
auxiliary automatically mixes with the spin-1 ``gauge-fixing'' function,
like a Nakanishi-Lautrup field, while for U(1) there is a kind of
``parity'' symmetry of the OSp(1,1$|$2) generators,
$\sket{\sb{A'}}\to -\sket{\sb{A'}}$, which prevents such mixing, and can
be included in the usual parity transformations to strengthen the
identification with supersymmetry.

For spin 2, for U(1) we define
$$ \li{ \f = & \ha h\sp{ab} {1\over\sqrt 2}\sket{\sb{(a}}\sket{\sb{b)}}
+iA\sp a\sb +{1\over\sqrt 2}\sket{\sb{(a}}\sket{\sb{-')}}
+iA\sp a\sb -{1\over\sqrt 2}\sket{\sb{(a}}\sket{\sb{+')}} \cr
& + \varphi\sb{++}\sket{\sb{-'}}\sket{\sb{-'}}
+\varphi\sb{+-}{1\over\sqrt 2}\sket{\sb{(+'}}\sket{\sb{-')}}
+\varphi\sb{--}\sket{\sb{+'}}\sket{\sb{+'}} \cr
& + \varphi{1\over\sqrt 2}\sket{\sp\a}\sket{\sb\a}
+\varphi'\ha\left(\ket{\sp\a}\ket{\sb{\a'}}+\ket{\sp{\a'}}\ket{\sb\a}\right)
+ \varphi''{1\over\sqrt 2}\ket{\sp{\a'}}\ket{\sb{\a'}} \quad , 
& (4.1.46a) \cr}$$
subject to the tracelessness condition ($h\sp i\sb i =0$)
$$ \ha h\sp a\sb a +\varphi\sb{+-} -\varphi-\varphi'' =0 \quad , 
\eqno(4.1.46b)$$
and find the lagrangian
$$ \li{ L = & -\frac14 h\sp{ab}(p\sp 2h\sb{ab}-2p\sb ap\sp ch\sb{cb}
+2p\sb ap\sb bh\sp c\sb c -\h\sb{ab}p\sp 2h\sp c\sb c) \cr
& + \ha (A\sb{a-}-\ha p\sp 2 A\sb{a+}+p\sb a p\sp b A\sb{b+}
-i\sqrt 2 p\sb a\varphi')\sp 2 \cr
& +(\varphi\sb{+-}-\varphi'')(\sqrt 2 \varphi\sb{--} +p\sp 2\varphi\sb{+-}
-2p\sp 2\varphi'' +{1\over 2\sqrt 2}p\sp 4\varphi\sb{++}
+p\sp 2h\sp b\sb b -p\sp bp\sp ch\sb{bc}) \cr
= & \hbox{``R''} + \ha\hat A\sb{a-}\sp 2 + 
\sqrt 2\hat\varphi\sb{+-}\hat\varphi\sb{--} \quad . & (4.1.47) \cr} $$
The second term is the square of an auxiliary ``axial'' vector (which 
again appears with sign
opposite to that in GL(1) [3.7]), which resembles the axial vector
auxiliary field of supergravity (including terms which can be absorbed,
as for spin 1).  In the last term, the redefinition $\varphi\sb{--}\to
\hat\varphi\sb{--}$ involves the (linearized) Ricci scalar.
Although it's difficult to tell from the free theory,
it may also be possible to identify some of the gauge degrees of freedom
with conformal compensators: $\varphi'$ with the compensator for local
R-symmetry, and $\varphi\sb{+-}$ (or $\varphi$ or $\varphi''$; one is
eliminated by the tracelessness condition and one is auxiliary) with the
local scale compensator.

A simple expression for interacting actions in terms of just the
OSp(1,1$|$2) group generators has not yet been found.  (However, this is
not the case for IGL(1):  See the following section.)
The usual gauge-invariant interacting
field equations can be derived by imposing $J \sb
{\a\b} \f = J \sb {-\a} \f = 0$, which are required in a (anti)BRST formalism,
and finding the equations satisfied by
the $x \sb - = x \sb \a = 0$ sector.  However, this requires use of the
other sectors as auxiliary fields, whereas in the approach described
here they would be gauge degrees of freedom.

These results for gauge-invariant actions from OSp(1,1$|$2) will 
be applied to the special case of the string in chapt.\ 11.

\sect{4.2. IGL(1)}

We now derive the corresponding gauge-invariant action in the IGL(1)
formalism and compare with the OSp(1,1$|$2) results.  We begin with the
form of the generators (3.4.3b) obtained from the transformation (3.4.3a).
For the IGL(1) formalism we can then drop the zero-modes $x\sb -$ and
$\tilde c$, and the action and invariance then are (using $\d (Q) = Q$)
$$ S = -\int d\sp Dx d c \; \F\dg \; i Q \d ( J\sp 3 ) \; \F \quad , \quad
\d \F = -iQ \L + J\sp 3 \hat\L \quad . \eqno(4.2.1) $$
This is the IGL(1) analog of (4.1.1).  (This action also was first 
proposed for the string [4.8,9].)  The $\d ( J\sp 3 )$ kills the sign
factor in (3.4.4).  However, even
though some unphysical coordinates have been eliminated, the field is
still a representation of the spin group OSp(D-1,1$|$2) (or
OSp(D,1$|$2)), and thus there is still a ``hidden'' Sp(2) symmetry
broken by this action (but only by auxiliary fields: see below).

To obtain the analog of (4.1.6), we first expand the field in the single ghost
zero-mode $ c$:
$$ \F = \f + i c\j \quad . \eqno(4.2.2)$$
$\f$ is the field of the
OSp(1,1$|$2) formalism after elimination of all its gauge zero-modes,
and $\j$ is an auxiliary field (identified with the
Nakanishi-Lautrup auxiliary fields in the gauge-fixed formalism [4.4,5]).
If we expand the action (4.2.1) in $ c$, using (3.4.3b), and the 
reality condition on the field to combine crossterms, we obtain,
with $\cq\sp\a = ( \cq\sp + , \cq\sp - )$,
$$ \li{ \cl &= -\int d c \; \F\dg iQ \d (J\sp 3) \F \cr
&= \ha\f\dg (\bo -M\sp 2) \d (M\sp 3)\f -\j\dg M\sp + \d (M\sp 3+1)\j
+2\j\dg\cq\sp +\d (M\sp 3)\f \; . &(4.2.3)\cr}$$

As an example of this action, we again consider a massless vector.
In analogy to (4.1.7),
$$\F = \left| \sp i \right> \f\sb i + ic \left| \sp i \right> \j\sb i
\quad .\eqno(4.2.4a)$$
After the $\d (J\sp 3)$ projection, the only surviving fields are
$$\F = \ket{\sp a} A\sb a + ic\ket{\sp -} B \quad ,\eqno(4.2.4b)$$
where $B$ is the auxiliary field.  Then, using the relations (from (4.1.8))
$$ M\sp 3 \Big|\sp a\Big> = 0 \quad ,\quad M\sp 3\ket{\sp\pm}=
\pm\ket{\sp\pm}\quad ;$$
$$ M\sp +\Big|\sp a\Big>=M\sp +\ket{\sp +}=0\quad ,\quad 
M\sp +\ket{\sp -}=2i\ket{\sp +}\quad ;$$
$$ M\sp{+a}\ket{\sp b}=\h\sp{ab}\ket{\sp +}\quad ,\quad M\sp{+a}\ket{\sp +}=0
\quad ,\quad M\sp{+a}\ket{\sp -}=-i\Big|\sp a\Big> \quad ;\eqno(4.2.5)$$
we find the lagrangian and invariance
$$\cl = \ha A\sp a \bo A\sb a -2B\sp 2 +2B\pa\sp a A\sb a \quad ;$$
$$ \d A\sb a=\pa\sb a\l \quad , \quad \d B=\ha\bo\l \quad ;\eqno(4.2.6)$$
which yields the usual result after elimination of $B$ by its equation
of motion.  This is the same lagrangian, including signs and
auxiliary-field redefinitions, as for the GL(1)-type 4+4-extended
OSp(1,1$|$2), (4.1.45a).

Any IGL(1) action can be obtained from a corresponding OSp(1,1$|$2)
action, and vice versa [3.13].  Eliminating $\j$ from (4.2.3) by its 
equation of motion,
$$ 0 = {\d S \over \d\j\dg} \sim M\sp +\d (M\sp 3+1)\j -\cq\sp +\d (M\sp 3)\f$$
$$ \to\quad \cl ' = \ha\f\dg (\bo -M\sp 2-2\cq\sp + M\sp +\sp{-1}\cq\sp + )
\d (M\sp 3)\f \quad ,\eqno(4.2.7)$$
the OSp(1,1$|$2) action (4.1.6) is obtained:
$$ \li{ & (\bo -M\sp 2-2\cq\sp + M\sp +\sp{-1}\cq\sp + ) \left[ \d (M\sp 3) -
	\d (M\sb{\a\b}\sp 2) \right] \cr
&\qquad =(\bo -M\sp 2-2\cq\sp + M\sp +\sp{-1}\cq\sp + )M\sp +
	M\sp +\sp{-1}\d (M\sp 3) \cr
&\qquad = \left[ (\bo -M\sp 2) M\sp + -2\cq\sp + 
	M\sp +\sp{-1}M\sp +\cq\sp + \right] M\sp +\sp{-1} \d (M\sp 3) \cr
&\qquad = \left[ (\bo -M\sp 2) M\sp + -2\cq\sp +\sp 2 \right]
	M\sp +\sp{-1} \d (M\sp 3) \cr
&\qquad = 0 \cr}$$
$$ \li{ \to\quad \cl ' &= \ha\f\dg (\bo -M\sp 2-2\cq\sp + 
M\sp +\sp{-1}\cq\sp + ) \d (M\sb{\a\b}\sp 2)\f \cr
&=\ha\f\dg (\bo -M\sp 2+\cq\sb\a\sp 2)\d (M\sb{\a\b}\sp 2)\f\quad .
&(4.2.8)\cr}$$
We have also used $\cq\sp{+2}=(\bo -M\sp 2)M\sp +$, which follows from 
the OSp commutation relations, or from $Q\sp 2=0$.
$M\sp +\sp{-1}$ is an Sp(2) lowering operator normalized so that it is the
inverse of the raising operator $M\sp +$, except that it vanishes on
states where $M\sp 3$ takes its minimum value [4.1].  It isn't an 
inverse in the strict sense, since $M\sp +$ vanishes on certain states, 
but it's sufficient for it to satisfy
$$M\sp + M\sp + \sp {-1} M\sp + = M\sp + \quad .\eqno(4.2.9)$$
We can obtain an explicit expression for $M\sp{+-1}$ using familiar 
properties of SO(3)
(SU(2)).  The Sp(2) operators are related to the conventionally normalized
SO(3) operators by $(M\sp 3, M\sp\pm ) = 2 (T\sp 3, T\sp\pm )$.
However, these are really SO(2,1) operators, and so 
have unusual hermiticity conditions:  $T \sp +$ and $T \sp -$
are each hermitian, while $T \sp 3$ is antihermitian.
Since for any SU(2) algebra $\vec T$ the commutation relations
$$[ T \sp 3 , T \sp \pm ] = \pm T \sp \pm \quad , \quad [ T \sp + , T
\sp - ] = 2 T \sp 3 \eqno(4.2.10)$$
imply
$$\vec T \sp 2 = ( T \sp 3 ) \sp 2 + \ha \{ T \sp + , T \sp - \} = 
T ( T + 1 ) \quad \to \quad T \sp \mp T \sp \pm = ( T
\mp T \sp 3 ) ( T \pm T \sp 3 + 1 ) \quad ,\eqno(4.2.11)$$
we can write
$$T \sp + \sp {-1} = {1 \over {T \sp - T \sp +}} T \sp - =
{{1 - \d \sb {T\sp 3 , T}} \over {( T - T \sp 3 ) ( T + T \sp 3 + 1 )}}
T \sp - \quad .\eqno(4.2.12)$$
We can then verify (4.2.9), as well as the identities
$$T \sp + \sp {-1} T \sp + T \sp + \sp {-1} = T \sp + \sp {-1} \quad ,$$
$$T \sp + \sp {-1} T \sp + = 1 - \d \sb {T\sp 3 , T} \quad , \quad
T \sp + T \sp + \sp {-1} = 1 - \d \sb {T \sp 3 , - T} \quad 
.\eqno(4.2.13)$$

Conversely, the IGL(1) action can be obtained by partial gauge-fixing 
of the OSp(1,1$|$2) action, by writing $\cl$ of (4.2.3) as $\cl '$ of
(4.2.7) plus a pure BRST variation.  Using the covariantly
second-quantized BRST operator of sect.\ 3.4, we can write
$$ \cl = \cl ' + \left[ Q , - \f\dg M\sp +\sp{-1} [ Q , \d (M\sp 3) \f
]\sb c \right]_c \quad . \eqno(4.2.14) $$
Alternatively, we can use functional notation, defining the operator
$$ {\Sc Q} = -\int dxd c \; ( Q \F ) {\d \over \d \F} 
\quad .\eqno(4.2.15)$$
In terms of $J\sb +\sp\a = (R,\Tilde R)$, the extra terms fix the 
invariance generated by $R$,
which had allowed $ c$ to be gauged away.  This also breaks
the Sp(2) down to GL(1), and breaks the antiBRST invariance.
Another way to understand this is by reformulating the IGL(1) in terms
of a field which has all the zero-modes of the OSp(1,1$|$2) field $\F$.
Consider the action
$$ S = \int d\sp D x d\sp 2 x\sb\a dx\sb - \; \F\dg\; p\sb + 
\d (J\sp 3) i \d (J\sb{-+}) \d (\Tilde R) \d (Q)\;\F \quad .\eqno(4.2.16) $$
The gauge invariance is now given by the 4 generators appearing as
arguments of the $\d$ functions, and is reduced from the OSp(1,1$|$2)
case by the elimination of the generators ($J\sp\pm$, $R$, 
$\Tilde Q$).  This algebra is the algebra GL(1$|$1) of $N=2$
supersymmetric quantum mechanics (also appearing in the IGL(1) formalism
for the closed string [4.10]: see sect.\ 8.2):  The 2 fermionic generators
are the ``supersymmetries,'' $J\sp 3 + J\sb{-+}$ is the O(2) generator 
which scales
them, and $J\sp 3 - J\sb{-+}$ is the ``momentum.''  If the gauge
coordinates $x\sb -$ and $\tilde c$ are integrated out, the action
(4.2.1) is obtained, as can be seen with the aid of (3.4.3).

In contrast to the light cone, where the hamiltonian operator $H$
($=-p_-$) is essentially the action ((2.4.4)), we find that with the
new covariant, second-quantized bracket of (3.4.7) the covariant 
action is the BRST operator:
Because the action (2.4.8) of a generator (2.4.7) on $\F\dg$ is equivalent to
the generator's functional derivative (because of (3.4.7)), the 
gauge-invariant action now thought of as an operator satisfies
$$[ S , \Phi\dg ]_c = \ha{{\delta S}\over {\delta \Phi}}\quad .\eqno(4.2.17)$$
Furthermore, since the gauge-covariant equations of motion of the theory
are given
by the BRST transformations generated by the operator $Q$, one has
$${{\delta S}\over{\delta \Phi}} = -2i [ Q , \Phi\dg ]_c \eqno(4.2.18)$$
$$\to\quad  S = -iQ \quad .\eqno(4.2.19)$$
(Strictly speaking, $S$ and $Q$ may differ by an irrelevant 
$\Phi$-independent term.)  This statement can be applied to any
formalism with field equations that follow from the BRST operator, 
independent of whether it originates from the light-cone, and it holds
in interacting theories as well as free ones.  In particular, for the
case of interacting Yang-Mills, the action follows directly from
(3.4.18).  After restricting the fields to $J\sp 3 =0$, this gives 
the interacting generalization of the example of (4.2.6).
The action can also be written as $ S = -2i\int d\F \; Q \F$,
where $\int d\F \;\F\sp n \equiv {1\over n+1}\int \F\dg \F\sp n$.

This operator formalism is also useful for deriving the gauge
invariances of the interacting theory, in either the IGL(1) or
OSp(1,1$|$2) formalisms (although the corresponding interacting action
is known in this form only for IGL(1), (4.2.19).)  Just as the global
BRST invariances can be written as a unitary transformation (in the
notation of (3.4.17))
$$ U = e\sp{iL\sb G} \quad , \quad G = \e\co \quad , \quad \e = constant
\quad , \eqno(4.2.20)$$
where $\co$ is any IGL(1) (or OSp(1,1$|$2)) operator (in
covariant second-quantized form), the gauge transformations can be
written similarly but with
$$ G = [f,\co ]\sb c \quad , \eqno(4.2.21)$$
where $f$ is linear in $\F$ ($f = \int \L\F$) for the usual gauge
transformation (and $f$'s higher-order in $\F$ may give field-dependent
gauge transformations).  Thus, $\F ' = U\F U\sp{-1}$, and $g (\F ) ' = 
U g(\F ) U\sp{-1}$ for any functional $g$ of $\F$.  In the free case,
this reproduces (4.1.1,4.2.1).

This relation between OSp(1,1$|$2) and IGL(1) formalisms is important
for relating different first-quantizations of the string, as will be
discussed in sect.\ 8.2.

\sect{4.3. Extra modes}

As discussed in sect.\ 3.2, extra sets of unphysical modes can be added
to BRST formalisms, such as
those which Lorentz gauges have with respect to temporal gauges, or
those in the 4+4-extended formalisms of sect.\ 3.6.
We now prove the equivalence of the OSp(1,1$|$2) actions of 
formulations with and without such modes [3.13].
Given that IGL(1) actions and equations of motion can be reduced to OSp
forms, it's sufficient to show the equivalence of the IGL actions with
and without extra modes.  The BRST and ghost-number
operators with extra modes, after the redefinition of (3.3.6),
differ from those without by the addition of
abelian terms.  We'll prove that the addition of these terms changes the IGL
action (4.2.1) only by adding auxiliary and gauge degrees of freedom.
To prove this, we consider adding such terms 2 sets of modes at a time (an even
number of additional ghost modes is required to maintain the fermionic
statistics of the integration measure):
$$ Q = Q\sb 0 + ( {\bf b\dg f - f\dg b} ) \quad , $$
$$ [ {\bf b},{\bf g}\dg ] = [ {\bf g},{\bf b}\dg ] = \{ {\bf c},
{\bf f}\dg \} = \{ {\bf f},{\bf c}\dg \} =1\quad ,\eqno(4.3.1)$$
in terms of the ``old'' BRST operator $Q\sb 0$ and the 2 new sets of modes 
${\bf b}$, ${\bf g}$, ${\bf c}$, and ${\bf f}$, and their hermitian 
conjugates.  We also assume 
boundary conditions in the new coordinates implied by the harmonic-oscillator
notation.  (Otherwise, additional unphysical fields appear, and
the new action isn't equivalent to the original one: see below.)  By an 
explicit expansion of the new field over all the new oscillators, 
$$ \F = \su_{m,n=0}^\infty ( A\sb{mn} +iB\sb{mn}{\bf c}\dg
+iC\sb{mn}{\bf f}\dg +iD\sb{mn}{\bf f\dg c\dg} ) {1\over\sqrt{m!n!}}
({\bf b}\dg )\sp m ({\bf g}\dg )\sp n \left| 0 \right> \quad ,\eqno(4.3.2)$$
we find
$$ \li{ \F\dg Q\F =& ( A\dg Q\sb 0 A + 2 B\dg Q\sb 0 C - D\dg Q\sb 0 D )
+ 2 B\dg ( {\bf b}\dg D -i {\bf b} A ) \cr
=& \su_{m,n=0}^\infty \Biggl[ ( A\dg\sb{mn}Q\sb 0A\sb{nm}
+ 2B\dg\sb{mn}Q\sb 0C\sb{nm} - D\dg\sb{mn}Q\sb 0D\sb{nm} ) \cr
&\qquad\qquad +2B\dg\sb{mn}(\sqrt nD\sb{n-1,m}-i{1\over\sqrt{m+1}}A\sb{n,m+1})
\Biggl] \quad .&(4.3.3)\cr}$$
(The ground state in (4.3.2) and the matrix elements evaluated in (4.3.3)
are with respect to only the new oscillators.)
We can therefore shift $A\sb{mn}$ by a $Q\sb 0 C\sb{m,n-1}$ term to
cancel the $B\dg Q\sb 0C$ term (using $Q\sb 0\sp 2=0$), and then $B\sb{mn}$
by a $Q\sb 0 D\sb{m,n-1}$ term to cancel the $D\dg Q\sb 0D$ term.
We can then shift $A\sb{mn}$ by $D\sb{m-1,n-1}$ to cancel the $B\dg D$
term, leaving only the $A\dg Q\sb 0A$ and $B\dg A$ terms.
(These redefinitions are equivalent to the gauge choices $C=D=0$ using the
usual invariance $\d\F =Q\L$.)  Finally, we can eliminate the Lagrange
multipliers $B$ by their equations of motion, which eliminate all of
$A$ except $A\sb{m0}$, and from the form of the remaining $A\dg Q\sb 0A$
term we find that all the remaining components of $A$ except $A\sb{00}$
drop out (i.e., are pure gauge).  This leaves only the term $A\dg\sb{00}
Q\sb 0 A\sb{00}$.  Thus, all the components except the ground
state with respect to the new oscillators can be eliminated as auxiliary
or gauge degrees of freedom.  The net result is that all the new
oscillators are eliminated from the fields and operators in the action
(4.2.1), with $Q$ thus replaced by $Q\sb 0$ (and similarly for $J\sp 3$).
(A similar analysis can be performed directly on the equations of motion
$Q\F =0$, giving this general result for the cohomology of $Q$ even in
cases when the action is not given by (4.2.1).)  This elimination of new
modes required that the creation operators in (4.3.3) be left-invertible:
$$ a\dg\sp{-1} a\dg = 1 \quad\to\quad a\dg\sp{-1} =
{1\over a\dg a+1} a = {1\over N+1}a \quad\to\quad N \ge 0 \quad ,\eqno(4.3.4)$$
implying that all states must be expressible as creation operators
acting on a ground state, as in (4.3.2) (the usual boundary conditions on
harmonic oscillator wave functions, except that here {\bf b} and {\bf g}
correspond to a space of indefinite metric).
This proves the equivalence of the IGL(1) actions, and thus, by the
previous argument, also the OSp(1,1$|$2) actions, with and without 
extra modes, and that the extra modes simply introduce more gauge and 
auxiliary degrees of freedom.

Such extra modes, although redundant in free theories, may be useful in
formulating larger gauge invariances which simplify the form of
interacting theories (as, e.g., in nonlinear $\s$ models).  The use in
string theories of such extra modes corresponding to the world-sheet
metric will be discussed in sect.\ 8.3.

\sect{4.4. Gauge fixing}

We now consider gauge fixing of these gauge-invariant actions using the
BRST algebra from the light cone, and relate this method to the standard
second-quantized BRST methods described in sects.\ 3.1-3 [4.1].  
We will find that the first-quantized BRST transformations of the fields
in the usual gauge-fixed action are {\it identical} to the
second-quantized BRST transformations, but the first-quantized BRST
formalism has a larger set of fields, some of which drop out of the
usual gauge-fixed action.  (E.g., see (3.4.19).
However, gauges exist where these fields
also appear.)  Even in the IGL(1) formalism, although all the
``propagating'' fields appear, only a subset of the BRST ``auxiliary''
fields appear, since the two sets are equal in number in the
first-quantized  IGL(1) but the BRST auxiliaries are fewer in the 
usual second-quantized formalism.  We will also consider gauge fixing to
the light-cone gauge, and reobtain the original light-cone theories to
which 2+2 dimensions were added.

For covariant gauge fixing
we will work primarily within the IGL(1) formalism, but similar methods apply 
to OSp(1,1$|$2).  Since the entire ``hamiltonian'' $\bo - M\sp 2$ vanishes 
under the constraint $Q = 0$ (acting on the field), the free gauge-fixed
action of the field theory consists of only a ``gauge-fixing'' term:
$$ \li{ S &= i \left[ Q , \int dxd c \; \ha \F \sp \dag \co \F \right]_c \cr
&= \int dxd c \; \ha \F \sp \dag [ \co , iQ \} \F \cr
&= \int dxd c \; \ha \F \sp \dag K \F \quad ,&(4.4.1)\cr}$$
for some operator ${\cal O}$, where the first $Q$, appearing in the
covariant bracket, is understood to be the second-quantized one.  
In order to get $\bo -M\sp 2$ as the 
kinetic operator for part of $\F$, we choose
$$ \co = - \left[  c , {\pa \over {\pa  c}} \right] \quad\to\quad K = 
	 c ( \bo - M\sp 2 ) - 2 \der c M \sp + \; .\eqno(4.4.2) $$
When expanding the field in $ c$, $\bo - M\sp 2$ is 
the kinetic operator for the piece containing all physical and ghost fields.
Explicitly, (3.4.3b),
when substituted into the lagrangian $L = \ha \F \sp \dag K \F$ and integrated
over $ c$, gives
$${\pa \over {\pa  c}} L = \ha \f \sp \dag ( \bo - M\sp 2 ) \f + \j 
	\sp \dag M \sp + \j  \; ,\eqno(4.4.3) $$
and in the BRST transformations $\d \F = i\e Q \F$ gives
$$\d \f = i\e ( \cq \sp + \f - M \sp + \j ) \quad ,
	\quad \d \j = i\e \left[ \cq \sp + \j - \ha (\bo -M\sp 2) \f )
	\right] \quad .\eqno(4.4.4) $$
$\f$ contains propagating fields and $\j$ contains BRST auxiliary fields.
Although the propagating fields are completely gauge-fixed, the BRST 
auxiliary fields have the gauge invariance
$$ \d \j = \l \quad , \quad M \sp + \l = 0 \quad .\eqno(4.4.5)$$

The simplest case is the scalar $\F = \varphi ( x )$.  In this case, 
all of $\j$ can be gauged away by (4.4.5), since $M\sp + = 0$.  The 
lagrangian is just $\ha \varphi ( \bo - m \sp 2 ) \varphi$.  For the
massless vector (cf.\ (4.2.4b)),
$$ \F = \ket{\sp i}A\sb i + ic\ket{\sp -}B \quad ,\eqno(4.4.6)$$
where we have again used (4.4.5).  By comparing (4.4.3) with (4.2.3), we
see that the $\f\sp 2$ term is extended to all $J\sp 3$, the $\j\sp 2$
term has the opposite sign, and the crossterm is dropped.  We thus find
$$ \cl = \ha (A\sp i)\dg\bo A\sb i +2B\sp 2 =
\ha A\sp a\bo A\sb a -i\Tilde C\bo C +2B\sp 2 \quad ,\eqno(4.4.7)$$
where we have written $A\sp\a = i( C , \Tilde C )$, due to (3.4.4,13).  
This agrees with the result (3.2.11) in the gauge $\z =1$, where this 
$B=\ha\Tilde B$.  The BRST transformations (4.4.4), using (4.2.5), are
$$ \li{ \d A\sb a &= i\e\pa\sb a C \quad ,\cr
\d C &= 0 \quad , \cr
\d\Tilde C &= \e (2B-\pa\cdot A) \quad , \cr
\d B &= i\ha\e\bo C \quad , &(4.4.8)\cr}$$
which agrees with the linearized case of (3.2.8).

We next prove the equivalence of this form of gauge fixing with the
usual approach, described in sect.\ 3.2 [4.1] (as we have just proven for
the case of the massless vector).  The steps are:  
(1) Add terms to the original BRST auxiliary fields, which vanish on 
shell, to make them BRST invariant, as they are in the usual BRST 
formulation of field theory.  (In Yang-Mills, this is the redefinition 
$B \to \Tilde B$ in (3.2.11).)  
(2) Use the BRST transformations to identify the physical fields 
(which may include auxiliary components).  We can then reobtain the 
gauge-invariant action by dropping all other fields from the lagrangian,
with the gauge transformations given by replacing the ghosts in the BRST 
transformations by gauge parameters. 

In the lagrangian (4.4.3) only the part of the BRST auxiliary field $\j$
which appears in $M\sp + \j$ occurs in the action; the rest of $\j$ is
pure gauge and drops out of the action.  Thus, we only require that
the shifted $M\sp + \j$ be BRST-invariant:
$$\j = \Tilde \j + A \f \quad ,\quad \d M\sp + \Tilde \j = 0\quad 
.\eqno(4.4.9)$$
Using the BRST transformations (4.4.4) and the identities (from (3.4.3b))
$$\li{ Q \sp 2 = 0 \quad \to \quad & [ \bo - M\sp 2 , M\sp + ] = 
[ \bo - M\sp 2 , \cq \sp + ] = [ M\sp + , \cq \sp + ] = 0 \quad , \cr
& \quad \cq \sp + \sp 2 = \ha ( \bo - M\sp 2 ) M\sp + \quad ,&(4.4.10)\cr}$$
we obtain the conditions on $A$:
$$( \cq \sp + - M\sp + A ) \cq \sp + = ( \cq \sp + - M\sp + A ) M\sp + = 0
\quad .\eqno(4.4.11)$$
The solution to these equations is
$$A = M\sp + \sp {-1} \cq \sp + \quad .\eqno(4.4.12)$$

Performing the shift (4.4.9), the gauge-fixed lagrangian takes the form
$${\pa \over {\pa c}} L = \ha \f \sp \dag \hat K \f + 2 \Tilde \j \sp 
\dag ( 1 - \d \sb {M \sp 3 , - 2T}) \cq \sp + \f + \Tilde \j \sp 
\dag M\sp + \Tilde \j \quad ,\eqno(4.4.13a)$$
$$ \hat K = \bo - M\sp 2 - 2 \cq \sp + M\sp + \sp {-1} \cq \sp + \quad 
,\eqno(4.4.13b)$$
where $T$ is the ``isospin,'' as in (4.2.11), for $M\sb{\a\b}$.
The BRST transformations can now be written as
$$\d \d \sb {M\sp 3 , - 2T} \f = \e \d \sb {M\sp 3 , - 2T} \cq \sp +
\f \quad , \quad \d ( 1 - \d \sb {M\sp 3 , - 2T}) \f = - \e M\sp +
\Tilde \j \quad ,$$
$$\d \Tilde \j = \e \d \sb {M\sp 3, 2T} ( \cq \sp + \Tilde \j -\ha \hat K
\f ) \quad .\eqno(4.4.14)$$
The BRST transformation of $\Tilde \j$ is pure gauge, and can be
dropped.  (In some of the manipulations we have used the fact that
$Q$, $\cq \sp +$, and $\bo - M\sp 2$ are symmetric, i.e., even under 
integration by parts, while $M\sp +$ is antisymmetric, and $Q$ and $\cq$
are antihermitian while $M\sp +$ and $\bo - M\sp 2$ are hermitian.
In a coordinate representation, particularly for $c$, all symmetry
generators, such as $Q$, $\cq\sp +$, and $M\sp +$, would be
antisymmetric, since the fields would be real.)

We can now throw away the BRST-invariant BRST auxiliary fields $\Tilde
\j$, but we must also separate the ghost fields in $\f$ from the
physical ones.  According to the usual BRST procedure, the physical
modes of a theory are those which are both BRST-invariant and have
vanishing ghost number (as well as satisfy the field equations).  
In particular, physical fields may transform into ghosts (corresponding
to gauge transformations, since the gauge pieces are unphysical), but
never transform into BRST auxiliary fields.  Therefore, from (4.4.14) we
must require that the physical fields have $M\sp 3 = - 2T$ to avoid
transforming into BRST auxiliary fields, but we also require vanishing
ghost number $M\sp 3 = 0$.  Hence, the physical fields (located in
$\f$) are selected by requiring the simple condition of vanishing
isospin $T = 0$.  If we project out the ghosts with the projection
operator $\d \sb {T 0}$ and use the identity (4.2.8), we
obtain a lagrangian containing only physical fields:
$$\cl\sb 1 = \ha\f\sp\dag ( \bo -M\sp 2 + \cq \sp 2 ) \d \sb {T 0} \f \quad .
\eqno(4.4.15)$$
Its gauge invariance is obtained from the BRST transformations by
replacing the ghosts (the part of $\f$ appearing on the right-hand
side of the transformation law) with the gauge parameter (the reverse
of the usual BRST quantization procedure), and we add gauge
transformations to gauge away the part of $\f$ with $T \ne 0$:
$$\d\f = -i\ha\cq\sb\a\L\sp\a + \ha M\sb{\a\b}\L\sp{\a\b}\quad ,\eqno(4.4.16)$$
where we have obtained the $\cq\sp -$ term from closure of $\cq\sp +$
with $M\sb{\a\b}$.
The invariance of (4.4.15) under (4.4.16) can be verified using the above
identities.  This action and invariance are just the original ones of
the OSp(1,1$|$2) gauge-invariant formalism (or the IGL(1) one, after
eliminating the NL auxiliary fields).
The gauge-fixing functions for the $\L$ transformations 
are also given by the BRST transformations:  They are the 
transformations of ghosts into physical fields:
$$F \sb {GF} = \cq \sp + \d \sb {T 0} \f 
= p\sp a ( M\sp +\sb a \d\sb{T0}\f ) + M ( M\sp +\sb m \d\sb{T0}\f )
\quad .\eqno(4.4.17)$$
(The first term is the usual Lorentz-gauge gauge-fixing function for
massless fields, the second term the usual addition for
Stueckelberg/Higgs fields.)
The gauge-fixed lagrangian (physical fields only) is thus
$$\cl \sb {GF} = \cl \sb 1 - \frac14 F \sb {GF} \sp \dag M \sp - F \sb {GF} = 
\ha \f \sp \dag ( \bo - M\sp 2 ) \d \sb {T 0} \f \quad ,\eqno(4.4.18)$$
in agreement with (4.4.3).

In summary, we see that this first-quantized gauge-fixing procedure is
identical to the second-quantized one with regard to (1) the
physical gauge fields, their gauge transformations, and the
gauge-invariant action, (2) the BRST transformations of the physical
fields, (3) the closure of the BRST algebra, (4) the BRST invariance
of the gauge-fixed action, and (5) the invertibility of the kinetic
operator after elimination of the NL fields.  (1) implies that the two
theories are physically the same, (2) and (3) imply that the BRST
operators are the same, up to additional modes as in sect.\ 4.3, (4)
implies that both are correctly gauge fixed (but perhaps in different
gauges), and (5) implies that all gauge invariances have been fixed,
including those of ghosts.  Concerning the extra modes, from the $\bo -
M\sp 2$ form of the gauge-fixed kinetic operator we see that they are
exactly the ones necessary to give good high-energy behavior of the
propagator, and that we have chosen a generalized Fermi-Feynman gauge.
Also, note the fact that the $c=0$ (or $x\sp\a =0$) part of the field
contains exactly the right set of ghost fields, as was manifest by the
arguments of sect.\ 2.6, whereas in the usual second-quantized formalism
one begins with just the physical fields manifest, and the ghosts and
their ghosts, etc., must be found by a step-by-step procedure.  Thus we
see that the OSp from the light cone not only gives a straightforward
way for deriving general free gauge-invariant actions, but also gives a
method for gauge fixing which is equivalent to, but more direct than, the
usual methods.

We now consider gauge fixing to the light cone.  In this gauge the gauge
theory reduces back to the original light-cone theory from which it was
heuristically obtained by adding 2+2 dimensions in sects.\ 3.4, 4.1.
This proves a general ``no-ghost theorem,'' that the OSp(1,1$|$2) (and
IGL(1)) gauge theory is equivalent on-shell to the corresponding
light-cone theory, for any Poincar\'e representation (including strings
as a special case).

Consider an arbitrary bosonic gauge field theory, with action (4.1.6).
(Fermions will be considered in the following section.)  Without loss of
generality, we can choose $M\sp 2=0$, since the massive action can be
obtained by dimensional reduction.  The light-cone gauge is then
described by the gauge-fixed field equations
$$ p\sp 2 \f = 0 \eqno(4.4.19a) $$
subject to the gauge conditions, in the Lorentz frame $p\sb a = \d\sb
a\sp + p\sb +$,
$$ M\sb{\a\b}\f = 0 \quad , \quad
\cq\sb\a\f = M\sb{\a -}p\sb +\f = 0 \quad , \eqno(4.4.19b) $$
with the residual part of the gauge invariance
$$ \d\f = -i\ha\cq\sb\a\L\sp\a \sim M\sb{\a -}\L\sp\a \quad , \eqno(4.4.19c)$$
where $\pm$ now refer to the usual ``longitudinal'' Lorentz indices.
(The light-cone gauge is thus a further gauge-fixing of the Landau
gauge, which uses only (4.4.19ab).)  (4.4.19bc) imply that the only
surviving fields are singlets of the new OSp(1,1$|$2) algebra generated
by $M\sb{\a\b}$, $M\sb{\a\pm}$, $M\sb{-+}$ (with longitudinal Lorentz
indices $\pm$): i.e., those which satisfy $M\sb{AB}\f =0$ and can't be
gauged away by $\d\f =M\sp{BA}\L\sb{AB}$.

We therefore need to consider the subgroup SO(D$-$2)$\otimes$OSp(1,1$|$2)
of OSp(D$-$1,1$|$2) (the spin group obtained by adding 2+2 dimensions to
the original SO(D$-$2)), and determine which parts of an irreducible
OSp(D$-$1,12) representation are OSp(1,1$|$2) singlets.  This is done
most simply by considering the corresponding Young tableaux (which is
also the most convenient method for adding 2+2 dimensions to the
original representation of the light-cone SO(D$-$2)).  This means
considering tensor products of the vector (defining) representation with
various graded symmetrizations and antisymmetrizations, and
(graded) tracelessness conditions on all the indices.  The obvious
OSp(1,1$|$2) singlet is given by allowing all the vector indices to take
only SO(D$-$2) values.  However, the resulting SO(D$-$2) representation
is reducible, since it is not SO(D$-$2)-traceless.  The
OSp(D$-$1,1$|$2)-tracelessness condition equates its SO(D$-$2)-traces to
OSp(1,1$|$2)-traces of representations which differ only by replacing
the traced SO(D$-$2) indices with traced OSp(1,1$|$2) ones.  However,
OSp(1,1$|$2) (or OSp(2N$|$2N) more generally) has the unusual property
that its traces are not true singlets.  The simplest example [4.11] (and
one which we'll show is sufficient to treat the general case) is the
graviton of (4.1.15).  Considering just the OSp(1,1$|$2) values of the
indices, there are 2 states which are singlets under the bosonic
subgroup GL(2) generated by $M\sb{\a\b}$, $M\sb{-+}$, namely
$\sket{\sb{(+}}\sket{\sb{-)}}$, $\sket{\sp\a}\sket{\sb\a}$.  However, 
of these two
states, one linear combination is pure gauge and one is pure auxiliary:
$$ \li{ \ket{\f\sb 1} = \ket{\sp A}\ket{\sb A} \quad\to\quad
& M\sb{AB}\ket{\f\sb 1}=0 \quad , \quad but \cr
& \ket{\f\sb 1} = -\ha M\sb\pm\sp\a \ket{\sb{(\mp}}\ket{\sb{\a )}} \quad , \cr
\ket{\f\sb 2} = \ket{\sb{(+}}\ket{\sb{-)}}-\ket{\sp\a}\ket{\sb\a} \quad\to\quad
& \ket{\f\sb 2} \ne M\sp{AB}\ket{\L\sb{BA}} \quad , \quad but \cr
& M\sb{\pm\a}\ket{\f\sb 2} = -2\ket{\sb{(\pm}}\ket{\sb{\a )}} \ne 0
\quad . &(4.4.20)\cr}$$
This result is due basically to the fact that the graded trace can't be
separated out in the usual way with the metric because of the identity
$$ \h\sp{AB}\h\sb{BA} = \h\sp A\sb A = (-1)\sp A\d\sb A\sp A =
2-2 = 0 \quad . \eqno(4.4.21) $$
Similarly, the reducible OSp(1,1$|$2) representation which consists of
the unsymmetrized direct product of an arbitrary number of vector
representations will contain no singlets, since any one trace reduces to
the case just considered, and thus the representations which result from
graded symmetrizations and antisymmetrizations will also contain none.
Thus, no SO(D$-$2)-traces of the original OSp(D$-$1,1$|$2)
representation need be considered, since they are equated to
OSp(1,1$|$2)-nonsinglet traces by the OSp(D$-$1,1$|$2)-tracelessness
condition.  Hence, the only surviving SO(D$-$2) representation is the
original irreducible light-cone one, obtained by restricting all vector
indices to their SO(D$-$2) values and imposing SO(D$-$2)-tracelessness.

These methods apply directly to open strings.  The modification for
closed strings will be discussed in sect.\ 11.1.

\sect{4.5. Fermions}

The results of this section are based on the OSp(1,1$|$2) algebra of
sect.\ 3.5.

The action and invariances are again given by (4.1.1), with the modified
$J\sb{AB}$ of sect.\ 3.5, and (4.1.3) is unchanged except for
$M\sb{\a\b} \to \Hat M\sb{\a\b}$.  We also allow for the inclusion of a
matrix factor in the Hilbert-space metric to maintain the
(pseudo)hermiticity of the spin operators (e.g., $\g\sp 0$ for a Dirac
spinor, so $\F\dg \to \bar\F = \F\dg\g\sp 0$).  Under the action of the $\d$ 
functions, we can make the replacement
$$ \li{ p \sb + \sp 2 J \sb {-\a} \sp 2 \to & - \frac34 ( p \sp 2 + M
\sp 2 ) + ( M \sb \a \sp b p \sb b + M \sb {\a m} M ) \sp 2 \cr
& - \ha \tilde \g \sp \a ( M \sb \a \sp b p \sb b + M \sb {\a m} M ) 
\left[ \tilde\g\sb - +\tilde\g\sb +\ha ( p \sp 2 + M \sp 2 ) \right] 
\quad , &(4.5.1)\cr}$$
where $p \sp 2 \equiv p \sp a p \sb a$.  Under the action of the same
$\d$ functions, the gauge transformation generated by $J \sb {-\a}$ is
replaced with
$$ \d \F = J \sb - \sp \a \L \sb \a \quad \to \quad
\left\{ ( M \sp {\a b} p \sb b + M \sp \a \sb m M ) -
\ha \tilde \g \sp \a \left[ \tilde\g\sb - +\tilde\g\sb +\ha 
( p \sp 2 + M \sp 2 ) \right] \right\} \L \sb \a \quad . \eqno(4.5.2) $$
Choosing $\L \sb \a = \tilde \g \sb \a \L$, the $\tilde\g\sb -$ part of
this gauge transformation can be used to choose the partial gauge 
$$ \tilde\g\sb + \F = 0 \quad . \eqno(4.5.3)$$
The action then becomes
$$ S = \int d \sp D x \; \bar \f \; \d ( \Hat M \sb {\a\b} \sp 2 )
i \tilde \g \sp \a ( M \sb \a \sp b p \sb b + M \sb {\a m} M ) \; \f
\quad , \eqno(4.5.4) $$
where we have reduced $\F$ to the half represented by $\f$ by using
(4.5.3).  (The $\tilde\g\sb\pm$ can be represented as 2$\times$2 matrices.)
The remaining part of (4.5.2), together with the $J \sb {\a\b}$ 
transformation, can be written in terms of $\tilde\g\sb\pm$-independent 
parameters as
$$ \li{ \d \f = & ( M \sp {\a b} p \sb b + M \sp \a \sb m M )
( A \sb \a + \ha \tilde \g \sb \a \tilde \g \sp \b A \sb \b )
+ \left[ - \frac14 ( p \sp 2 + M \sp 2 ) + ( M \sb \a \sp b p \sb b +
M \sb {\a m} M ) \sp 2 \right] B \cr
& + \ha \Hat M \sp {\a\b} \L \sb {\a\b} \quad . & (4.5.5) \cr}$$

One way to get general irreducible spinor representations of orthogonal groups
(except for chirality conditions) is to take the direct product of a
Dirac spinor with an irreducible tensor representation, and then
constrain it by setting to zero the result of
multiplying by a $\g$ matrix and contracting vector
indices.  Since the OSp representations used here are obtained by dimensional 
continuation, this means we use the same constraints, with the vector 
indices $i$ running over all commuting and anticommuting values
(including $m$, if we choose to define $M \sb {im}$ by dimensional
reduction from one extra dimension).  The OSp spin operators can then be
written as
$$ M \sb {ij} = \check M \sb {ij} + \frac14 [ \g \sb i , \g \sb j \}
\quad , \eqno(4.5.6) $$
where $\check M$ are the spin operators for some tensor representation
and $\g \sb i$ are the OSp Dirac matrices, satisfying the OSp Clifford algebra
$$ \{ \g \sb i , \g \sb j ] = 2 \h \sb {ij} \quad . \eqno(4.5.7) $$
We choose similar relations between $\g$'s and $\tilde\g$'s:
$$ \{ \g \sb i , \tilde\g\sb B ] = 0 \quad . \eqno(4.5.8) $$
Then, noting that $\ha ( \g \sp \a +i
\tilde \g \sp \a )$ and $\ha ( \g \sb \a -i \tilde \g \sb \a )$ satisfy
the same commutation relations as creation and annihilation operators,
respectively, we define
$$ \g \sb \a = a \sb \a + a \sp\dag \sb \a \quad , \quad
\tilde \g \sb \a = i\left( a \sb \a - a \sp\dag \sb \a \right) \quad ; \quad
\left[ a \sb \a , a \sp{\dag\b} \right] = \d \sb \a \sp\b\quad .\eqno(4.5.9)$$
We also find
$$ \Hat M \sb {\a\b} = \check M \sb {\a\b} + a \sp\dag \sb {(\a} a \sb{\b )}
\quad . \eqno(4.5.10) $$
This means that an arbitrary representation $\j \sb {(\a \cdots \b )}$ 
of the part of the Sp(2) generated by $\check M \sb {\a\b}$ that is also a 
singlet under the full Sp(2) generated by $\Hat M \sb {\a\b}$ can be written as
$$ \j \sb {(\a \cdots \b )} = a \sp \dag \sb \a \cdots a \sp\dag \sb \b
\j \quad , \quad a \sb \a \j = 0 \quad . \eqno(4.5.11) $$

In particular, for a Dirac spinor $\check M = 0$, so the action (4.5.4) 
becomes simply (see also (4.1.40))
$$ S = \int d \sp D x \; \bar \f ( \sl p + \SL M ) \f \quad , \eqno(4.5.12)$$
where $\sl p \equiv \g \sp a p \sb a$, $\SL M \equiv \g \sb m M$ ($\g
\sb m$ is like $\g \sb 5$), all dependence on $\g \sp \a$ and $\tilde \g
\sp \a$ has been eliminated, and the gauge transformation (4.5.5) vanishes.
(The transformation $\f\to e\sp{i\g\sb m\p /4}\f$ takes
$\sl p + \SL M \to \sl p +iM$.)

In the case of the gravitino, we start with $\f = \sket{\sp i}\f\sb i$,
where $\sket{\sp i}$ is a basis for the representation of $\check M$ (only).
$\f$ must satisfy not
only $\Hat M \sb {\a\b} \f = 0$ but also the irreducibility condition
$$ \g \sp i \f \sb i = 0 \quad \to \quad \f \sb \a = \ha \g \sb \a \g
\sp a \f \sb a \quad , \quad a \sb \a \f \sp a = 0 \quad . \eqno(4.5.13) $$
($a \sp\dag \sb \a$ or $i \tilde \g \sb \a$ could be used in place of
$\g \sb \a$ in the solution for $\f \sb \a$.)  Then using (4.1.8) for
$\check M$ on $\f$,
straightforward algebra gives the action
$$ \li{ S & = \int d \sp D x \; \bar \f \sb a ( \h \sp {ab} \sl p - 
p \sp {(a} \g \sp {b)} + \g \sp a \sl p \g \sp b ) \f \sb b \cr
& = \int d \sp D x \; \bar \f \sb a \g \sp {abc} p \sb b \f \sb c \quad ,
&(4.5.14)\cr}$$
where $\g \sp {abc} = (1/3!) \g \sp {[a} \g \sp b \g \sp {c]}$, giving the
usual gravitino action for arbitrary $D$.  The gauge invariance
remaining in (4.5.5) after using (4.5.13-14), and making suitable
redefinitions, reduces to the usual
$$ \d \f \sb a = \pa \sb a \l \quad . \eqno(4.5.15) $$

We now derive an alternative form of the fermionic action which
corresponds to actions given in the literature for fermionic strings.
Instead of explicitly
solving the constraint $\Hat M \sb {\a\b} = 0$ as in (4.5.11), we use the
$\Hat M \sb {\a\b}$ gauge invariance of (4.5.5)
to ``rotate'' the $a \sp{\a\dag}$'s.  For example, writing $a\sp\a =
( a\sp + , a\sp - )$, we can rotate them
so they all point in the ``$+$'' direction:  Then we need consider only
$\f$'s of the form $\f ( a \sp{+\dag} ) \left| 0 \right>$.  (The $+$
value of $\a$ should not be confused with the $+$ index on $p \sb +$.)
The $\d ( \Hat M \sb {\a\b} \sp 2 )$ then picks out the piece of the
form (4.5.11).  (It ``smears'' over directions in Sp(2).  This use of $a
\sp \a$ is similar to the ``harmonic coordinates'' of harmonic
superspace [4.12].)  We can also pick
$$ \f = e \sp {-i a \sp {+\dag} a \sp{-\dag}} \f ( a \sp{+\dag}
) \left| 0 \right> \quad , \eqno(4.5.16a) $$
since the exponential (after $\d ( \Hat M )$ projection) just redefines 
some components by shifting by components of lower $\check M$-spin.  In
this gauge, writing $\g\sp\a = (s,u)$, $\tilde\g\sp\a = (\tilde s ,
\tilde u )$, we can rewrite $\f$ as
$$ \left| \tilde 0 \right> = e \sp {-i a \sp {+\dag} a \sp{-\dag}} 
\left| 0 \right> \quad \iff \quad \tilde s \left| \tilde 0 \right> 
= u \left| \tilde 0 \right> = 0 \quad , $$
$$ \f = \f ( s ) \left| \tilde 0 \right> \quad \iff \quad
\tilde s \f = 0 \quad . \eqno(4.5.16b) $$
By using an appropriate (indefinite-metric) coherent-state formalism, we
can choose $s$ to be a coordinate (and $u = -2 i \pa / \pa s$).  
We next make the replacement
$$ \d ( \Hat M \sb {\a\b} \sp 2 ) \quad \to \quad \int ds \;
\m (s) \; \d ( 2\check M \sp + + s \sp 2 )
\d \left( \check M \sp 3 + s \der s \right) \eqno(4.5.17)$$
after pushing it to the right in (4.5.4) so it hits the $\f$, where we
have just replaced projection onto $\Hat M \sb {\a\b} \sp 2 = 0$ with
$\Hat M \sp + = \Hat M \sp 3 = 0$ (which implies $\Hat M \sp -
= 0$).  The first $\d$ function factor is a Dirac $\d$ function, while
the second is a Kronecker $\d$.  
$\m (s)$ is an appropriate measure factor; instead of
determining it by an explicit use of coherent states, we fix it by
comparison with the simplest case of the Dirac spinor:  Using
$$ \int ds \; \e (s) \d ( s\sp 2 + r\sp 2 ) (a + bs) = b\quad ,\eqno(4.5.18a)$$
we find
$$ \m ( s ) \sim \e ( s ) \quad . \eqno(4.5.18b) $$
Then only the $\tilde u$ term of (4.5.4) contributes in this gauge,
and we obtain (the $\tilde u$ itself having already been absorbed
into the measure (4.5.18b))
$$ S = -2 \int d \sp D x \; d s\; \e ( s ) \; \bar \f \; \cq\sp +
\d ( 2\check M\sp + + s\sp 2 ) \d\left( \check M\sp 3 + s
\der s\right) \; \f \quad . \eqno(4.5.19)$$
(All dependence on $\g \sp \a$ and $\tilde \g \sp \a$ has been reduced
to $\f$ being a function of just $s$.  This action was first
proposed for the string [4.13].)    
Since the first $\d$ function can be used to replace any $s \sp
2$ with a $-2\check M \sp +$, we can perform all such replacements, or
equivalently choose the gauge
$$ \f ( s ) = \f \sb 0 + s \f \sb 1 \quad . \eqno(4.5.20) $$
(An equivalent procedure was performed for the string in [4.14].)
For the Dirac spinor, after integration over $s$ (including that in
$\cq\sp + = \ha s ( \sl p + \SL M )$), 
the Dirac action is easily found ($\f\sb 1$ drops out).  For general
spinor representations, $\cq\sp +$ has an additional $\check\cq\sp +$
term, and $s$ integration gives the lagrangian
$$\cl = \bar\f\sb 0 ( \sl p + \SL M ) \d ( \check M\sp 3 ) \f\sb 0
+ 2\left[ \bar\f\sb 1 \check\cq\sp + \d ( \check M\sp 3 ) \f\sb 0
-\bar\f\sb 0\check\cq\sp +\d (\check M\sp 3+1)\f\sb 1\right]\quad .
\eqno(4.5.21)$$
However, $\g$-matrix trace constraints (such as (4.5.13)) must still be 
solved to relate the components.

The explicit form of the OSp(1,1$|$2) operators for the fermionic string
to use with these results follows from the light-cone Poincar\'e
generators which will be derived in sect.\ 7.2.  The $s$
dependence of $\cq\sp +$ is then slightly more complicated (it also has
a $\pa / \pa s$ term).  (The resulting action first appeared in [4.14,15].)

The proof of equivalence to the light cone is similar to that for bosons
in the previous section.  Again considering the massless case, the basic
difference is that we now have to use, from (3.5.3b),
$$ \Hat M\sb{\a\b} = M\sb{\a\b} + S\sb{\a\b} \quad , \quad
\Hat\cq\sb\a = M\sb{\a -}p\sb + + S\sb{-\a} \quad , \eqno(4.5.22) $$
and other corresponding generators, as generating the new OSp(1,1$|$2).
This is just the diagonal OSp(1,1$|$2) obtained from $S\sb{AB}$ and the
one used in the bosonic case.  In analogy to the bosonic case, we
consider reducible OSp(D$-$1,1$|$2) representations corresponding to
direct products of arbitrary numbers of vector representations with one
spinor representation (represented by graded $\g$-matrices).  We then
take the direct product of this with the $S\sb{AB}$ representation,
which is an OSp(1,1$|$2) spinor but an SO(D$-$2) scalar.  Since the
direct product of 2 OSp(1,1$|$2) spinors gives the direct sum of all
(graded) antisymmetric tensor representations, each once (by the usual
$\g$-matrix decomposition), from the bosonic result we see that the only
way to get an OSp(1,1$|$2) singlet is if all vector indices again take
only their SO(D$-$2) values.  The OSp(D$-$1,1$|$2) spinor is the direct
product of an SO(D$-$2) spinor with an OSp(1,1$|$2) spinor, so the net
result is the original light-cone one.  In the bosonic case traces in
OSp(1,1$|$2) vector indices did not give singlets because of (4.4.21);
a similar result holds for $\g$-matrix traces because of
$$ \g\sp A \g\sb A = \h\sp A\sb A = 0 \quad . \eqno(4.5.23) $$

More general representations for $S\sb{AB}$ could be considered,
e.g., as in sect.\ 3.6.  The action can then be rewritten as (4.1.6), 
but with $M\sb{\a\b}$ and $\cq\sb\a$ replaced by
$\Hat M\sb{\a\b}$ and $\Hat\cq\sb\a$ of (3.5.3b).  In
analogy to (4.5.2,3), the $S\sb{-\a}$ part of the $J\sb{-\a}$
transformation can be used to choose the gauge $S\sb{+\a}\F =0$.  Then,
depending on whether the representation allows application of the
``lowering'' operators $S\sb{-\a}$ 0,1, or 2 times, only the terms of
zeroth, first, or second order in $S\sb{-\a}$, respectively, can contribute in 
the kinetic operator.  Since these terms are respectively second, first, and
zeroth order in derivatives, they can be used to describe bosons,
fermions, and auxiliary fields.  

The argument for equivalence to the light cone directly generalizes to
the U(1)-type 4+4-extended OSp(1,1$|$2) of sect.\ 3.6.  Then $\Hat
M\sb{AB} = M\sb{AB} +S\sb{AB}$ has a singlet only when $M\sb{AB}$ and
$S\sb{AB}$ are both singlets (for bosons) or both Dirac spinors (for fermions).

\sect{Exercises}

\Item{(1)} Derive (4.1.5).
\Item{(2)} Derive (4.1.6).
\Item{(3)} Find the gauge-invariant theory resulting from the light-cone
theory of a totally symmetric, traceless tensor of arbitrary rank.
\Item{(4)} Find the explicit infinitesimal gauge transformations of
$e\sb a\sp m$, $e\sb m\sp a$, e$\sp{-1}$, $g\sp{mn}$, $g\sb{mn}$,
and $\o\sb{abc}$ from (4.1.20-22).  Linearize, and show
the gauge $e\sp{[am]} =0$ can be obtained with $\l\sb{ab}$.
Find the transformation for a covariant vector $A\sb a$ (from a
similarity transformation, like (4.1.22)).
\Item{(5)} Write $c\sb{abc}$ explicitly in terms of $e\sb a\sp m$.
Find $T\sb{abc}$ and $R\sb{abcd}$ in terms of $c\sb{abc}$ and $\o\sb{abc}$.
Derive (4.1.31).  Linearize to get (4.1.17).
\Item{(6)} Find an expression for $\o\sb{abc}$ when (4.1.24) is not
imposed, in terms of $T\sb{abc}$ and the $\o$ of (4.1.27).
\Item{(7)} Derive global Poincar\'e transformations by finding the
subgroup of (4.1.20) which leaves (4.1.28) invariant.
\Item{(8)} Find the field equation for $\f$ from (4.1.36), and show that
(4.1.38) satisfies it.
\Item{(9)} Derive the gauge-covariant action for gravity in the
GL(1)-type 4+4-extension of OSp(1,1$|$2), and compare with the U(1)
result, (4.1.47).
\Item{(10)} Find the BRST transformations for the IGL(1) formalism of
sect.\ 4.2 (BRST1, derived from the light cone) for free gravity.  Find 
those for the usual IGL(1) formalism of sect.\ 3.2 (BRST2, derived from
second-quantizing the gauge-invariant field theory).  After suitable 
redefinitions of the BRST1 fields (including auxiliaries and ghosts), 
show that a subset of these fields that corresponds to the complete set 
of fields in the BRST2 formalism has identical BRST transformations.
\Item{(11)} Formulate $\f\sp 3$ theory as in (4.2.19), using the bracket
of (3.4.7).
\Item{(12)} Derive the gauge transformations for interacting Yang-Mills
by the covariant second-quantized operator method of (4.2.21), in both
the IGL(1) and OSp(1,1$|$2) formalisms.
\Item{(13)} Find the free gauge-invariant action for gravity in the
IGL(1) formalism, and compare with the OSp(1,1$|$2) result (4.1.17).
Find the gauge-fixed action by (4.4.1-5).
\Item{(14)} Perform IGL(1) gauge-fixing, as in sect.\ 4.4, for a
second-rank antisymmetric tensor gauge field.  Perform the analogous
gauge fixing by the method of sect.\ 3.2, and compare.  Note that there
are scalar commuting ghosts which can be interpreted as the ghosts for
the gauge invariances of the vector ghosts (``ghosts for ghosts'').
\Item{(15)} Derive (4.5.14,15).

%
%

\chsc{5. PARTICLE}{5.1. Bosonic}6

If coordinates are considered as fields, and their arguments as the 
coordinates of small spacetimes, then the mechanics of particles and 
strings can be considered as 1- and 2-dimensional field theories, 
respectively (see sect.\ 1.1).  (However, to avoid confusion, we will
avoid referring to mechanics theories as field theories.)
Thus, the particle is a useful analog 
of the string in one lower ``dimension'', and we here review its 
properties that will be found useful below for the string.

As described in sect.\ 3.1, the mechanics action for any relativistic
particle is completely determined by the constraints it satisfies, which
are equivalent to the free equations of motion of the corresponding
field theory.  The first-order (hamiltonian) form ((3.1.10)) is more convenient
than the second-order one because (1) it makes canonical conjugates explicit, 
(2) the inverse propagator (and, in more general cases, all other 
operator equations of motion) can be directly read off as the 
hamiltonian, (3) path-integral quantization is easier, and (4) 
treatment of the supersymmetric case is clearer.
The simplest example is a massless, spinless particle,
whose only constraint is the Klein-Gordon equation $p\sp 2 =0$.  From (3.1.10),
the action [5.1] can thus be written in first-order form as
$$ S = \int d \t \; \cl \quad , \quad 
\cl = \dt x \cdot p - g \ha p \sp 2 \quad,\eqno(5.1.1)$$
where $\t$ is the proper time, of which the position $x$, momentum 
$p$, and 1-dimensional metric $g$ are functions, and $\dt{\phantom m} 
= \pa / \pa \t$.  The action is invariant under Poincar\'e transformations 
in the higher-dimensional spacetime described by $x$, as well as 1D 
general coordinate transformations ($\t$-reparametrizations).  
The latter can be obtained from (3.1.11):
$$ \d x = \z p \quad , \quad \d p = 0 \quad , \quad \d g = \dt \z
\quad . \eqno(5.1.2)$$
These differ from the usual transformations by terms which vanish on
shell:  In general, any action with more than one field is invariant 
under $\d \f \sb i = \l \sb {ij} \d S / \d \f \sb j$, where $\l \sb {ij}$ is 
antisymmetric.  Such invariances may be necessary for off-shell
closure of the above algebra, but are irrelevant for obtaining the field
theory from the classical mechanics.  (In fact, in the component
formalism for supergravity, gauge invariance is more easily proven using
a first-order formalism with the type of transformations in (5.1.2)
rather than the usual transformations which follow from the second-order
formalism [5.2].)  In this case, if we add the transformations
$$ \d ' x = \e{\d S\over \d p} \quad , \quad
\d ' p = - \e{\d S\over \d x} \eqno(5.1.3)$$
to (5.1.2), and choose $\e = g\sp{-1}\z$,
we obtain the usual general coordinate transformations (see sect.\ 4.1)
$$ \d '' x = \e \dt x \quad , \quad \d '' p = \e \dt p \quad , \quad
\d '' g = \dt{(\e g )} \quad . \eqno(5.1.4)$$

The second-order form is obtained by eliminating $p$:
$$ \cl = g \sp {-1} \ha \dt x \sp 2 \quad .\eqno(5.1.5)$$
  The transformations (5.1.4) for $x$ and $g$ also follow directly 
from (5.1.2) upon eliminating $p$ by its equation of motion.
The massive case is obtained by replacing $p \sp 2$ with $p \sp 2 + m \sp 2$
in (5.1.1).  When the additional term is carried over to (5.1.5), we get
$$ \cl = \ha g \sp {-1} \dt x \sp 2 - \ha g m \sp 2 \quad . \eqno(5.1.6)$$
$g$ can now also be eliminated by its equation of motion, producing
$$ S = - m \int d \t \; \sqrt{-\dt x \sp 2} = - m \int \sqrt{-dx\sp 2}
\quad ,\eqno(5.1.7)$$
which is the length of the world line.

Besides the 1D invariance of (5.1.1) under reparametrization of $\t$, 
it also has the discrete invariance of $\t$ reversal.  If we choose $x 
( \t ) \to x ( - \t )$ under this reversal, then $p ( \t ) \to - p ( - 
\t )$, and thus this proper-time reversal can be identified as the 
classical mechanical analog of charge (or complex) conjugation in 
field theory [5.3], where $\f ( x ) \to \f \conj ( x )$ implies $\f ( p ) 
\to \f \conj ( - p )$ for the fourier transform.  (Also, the 
electromagnetic coupling $q \int d \t \; \dt x \cdot A ( x )$ when 
added to (5.1.1) requires the charge $q \to - q$.)

There are two standard gauges for quantizing (5.1.1).  In the 
light-cone formalism the gauge is completely fixed (for $p \sb + \ne 
0$, up to global transformations, which are eliminated by boundary 
conditions) by
$$ x \sb + = p \sb + \t \quad .\eqno(5.1.8)$$
We then eliminate $p \sb -$ as a lagrange multiplier with 
field equation $g = 1$.  The lagrangian then simplifies to
$$ \cl =  \dt x \sb - p \sb + + \dt x \sb i p \sb i
- \ha ( p \sb i \sp 2 + m\sp 2 ) \quad , \eqno(5.1.9)$$
with (retarded) propagator
$$ - i \Q ( \t ) e \sp {i \t \ha ( p \sb i \sp 2 + m\sp 2 )}\eqno(5.1.10a)$$
(where $\Q ( \t ) = 1$ for $\t > 0$ and $0$ otherwise) or, fourier 
transforming with respect to $\t$,
$$ {1 \over {i {\pa \over {\pa \t}} + \ha ( p \sb i \sp 2 + m\sp 2 ) 
+ i \e }} = {1 \over {p \sb + p \sb - + \ha ( p \sb i \sp 2 + m\sp 2 ) 
+ i \e}} = {1 \over {\ha ( p \sp 2 + m\sp 2 ) + i \e }} \quad .\eqno(5.1.10b)$$
For interacting particles, it's preferable to choose
$$ x \sb + = \t \quad , \eqno(5.1.11)$$
so that the same $\t$ coordinate can be used for all particles.  Then 
$g = 1 / p \sb +$, so the hamiltonian $\ha ( p \sb i \sp 2 + m\sp 2 )$ gets 
replaced with $( p \sb i \sp 2 + m\sp 2 ) / 2 p \sb +$, which more 
closely resembles the nonrelativistic case.  If we also use the
remaining (global) invariance of $\t$ reparametrizations (generated by
$p\sp 2$), we can choose the gauge $x\sb + =0$, which is the same as
choosing the Schr\"odinger picture.

Alternatively, in the covariant formalism one chooses the gauge
$$ g = constant \quad ,\eqno(5.1.12)$$
where $g$ can't be completely gauge-fixed because of the invariance of 
the 1D volume $\ct = \int d \t \; g$.  The functional integral over 
$g$ is thus replaced by an ordinary integral over $\ct$ [5.4], and the 
propagator is [5.3,5]
$$- i \int_0^\infty d \ct \; \Q ( \ct ) e \sp {i \ct \ha ( p \sp 2 + m\sp 2 )} 
= {1 \over {\ha ( p \sp 2 + m\sp 2 ) + i \e}} \quad .\eqno(5.1.13)$$

The use of the mechanics approach to the particle is somewhat pointless
for the free theory, since it contains no information except the
constraints (from which it was derived), and it requires treatment of
the irrelevant ``off-shell'' behavior in the ``coordinate'' $\t$.
However, the proper-time is useful in interacting theories for studying
certain classical limits and various properties of
perturbation theory.  In particular, the form of the propagator given in
(5.1.13) (with Wick-rotated $\t$: see sect.\ 2.5) is the most convenient 
for doing loop integrals using dimensional regularization:  The momentum 
integrations become simple Gaussian integrals, which can be trivially 
evaluated in arbitrary dimensions by analytic continuation from integer ones:
$$ \int d\sp D p \; e\sp{-ap\sp 2} = \left( \int d\sp 2p\; e\sp{-ap\sp 2}
\right)^{D/2} = \left( {\p\over a} \right)^{D/2} \quad . \eqno(5.1.14)$$
(The former integral factors into 1-dimensional ones; the latter is
easily performed in polar coordinates.)
The Schwinger parameters $\t$ are then converted into the usual Feynman
parameters $\a$ by inserting unity as $\int_0^\infty d\l\; \d (\l -\su\t )$,
rescaling $\t\sb i = \l\a\sb i$, and integrating out $\l$, which now
appears in standard $\G$-function integrals, to get the usual Feynman 
denominators.  An identical procedure is applied in string theory, but
writing the parameters as $x=e\sp{-\t}$, $w=e\sp{-\l}$.  (See (9.1.10).)
By not converting the $\t$'s into $\a$'s, the high-energy behavior of
scattering amplitudes can be analyzed more easily [5.6].  Also, the
singularities in an amplitude correspond to classical paths of the
particles, and this identification can be seen to be simply the
identification of the $\t$ parameters with the classical proper-time
[5.7].  1-loop calculations can be performed by introducing external
fields (see also sect.\ 9.1) and treating the path of the particle in
spacetime as closed [5.5,8].  Such calculations can treat arbitrary
numbers of external lines (or nonperturbative external fields) for
certain external field configurations (such as constant gauge-field
strengths).  Finally, the introduction of such expressions for
propagators in external fields allows the study of classical limits of
quantum field theories in which some quantum fields (represented by the
external field) become classical fields, as in the usual classical
limit, while other fields (represented by the particles described by the
mechanics action) become classical particles [5.9].

This classical mechanics analysis will be applied to the string in chapt.\ 6.

\sect{5.2. BRST}

In this section we'll apply the methods of sect.\ 3.2-3 to study BRST
quantization of particle mechanics, and find results equivalent to those
obtained by more general methods in sect.\ 3.4.

In the case of particle mechanics (according to sect.\ 3.1), for the
action of the previous section we have $\cg = -i \ha ( p \sp 2 + m\sp 2 )$,
and thus [4.4], for the ``temporal'' gauge $g=1$, from (3.2.6)
$$ Q = -i  c \ha ( p \sp 2 + m\sp 2 ) \quad , \eqno(5.2.1)$$
which agrees with the general result (3.4.3b).
We could write $ c = \pa / \pa {\bf C}$ so that in the classical field 
theory which follows from the quantum mechanics the field $\f ( x , {\bf C} )$
could be real (see sect.\ 3.4).  This also follows from the fact that the 
($\t$-reparametrization) gauge-parameter corresponding to $ c$ carries 
a (proper-)time index (it's a 1D vector), and thus changes sign under 
$\t$-reversal (mechanics' equivalent of field theory's complex 
conjugation), and so $ c$ is a momentum ($\f ( x , {\bf C} ) =\f\conj
( x , {\bf C} )$, $\f ( p ,  c ) = \f \conj ( - p , -  c )$).  

We now consider extending IGL(1) to OSp(1,1$|$2) [3.7].  By (3.3.2),
$$ Q \sp \a = -ix\sp \a \ha ( p \sp 2 + m \sp 2 ) - b\pa \sp \a \quad .
\eqno(5.2.2)$$
In order to compare with sect.\ 3.4, we make the redefinitions (see (3.6.8))
$$ b = i \der g \quad , \quad g = \ha p\sb + \sp 2 \quad , \eqno(5.2.3a)$$
(where $g$ is the world-line metric) and the unitary
transformation
$$ l n ~ U = - ( l n ~ p\sb + ) \ha \left[ x \sp \a , \pa \sb \a\right] 
\quad , \eqno(5.2.3b)$$
finding
$$ U Q \sp \a U \sp {-1} = -ix\sp \a {1 \over 2p\sb +} \left( 
p \sp 2 + m \sp 2 + p\sp\a p\sb\a \right) -ix\sb - p\sb\a\quad , \eqno(5.2.4)$$
which agrees with the expression given in (3.4.2) for the generators 
$J \sb {-\a}$
for the case of the spinless particle, as does the rest of the OSp(1,1$|$2) 
obtained from (3.3.7).  

In a lagrangian formalism, for the action (5.1.6) with invariance (5.1.4),
(3.3.2) gives the BRST transformation laws
$$ \li{ Q \sp \a x\sp a & = x \sp \a \dt x\sp a \quad , \cr
Q \sp \a g & = \dt {( x \sp \a g )} \quad , \cr
Q \sp \a x \sp \b & = \ha x \sp {(\a} \dt x \sp {\b )} - C \sp
{\a\b} b \quad , \cr
Q \sp \a b & = \ha ( x \sp \a \dt b - b \dt x \sp \a )
- \frac14 ( \dt x\sb\b \sp 2 x \sp \a + x\sb\b \sp 2 \ddt x \sp \a ) \quad . 
& (5.2.5) \cr}$$
We first make the redefinition
$$ \tilde b = b - \ha \dt {(x\sb\a \sp 2 )} \eqno(5.2.6)$$
to simplify the transformation law of $x \sp \a$ and thus $b$:
$$ \li{ Q \sp \a x \sp \b & = x \sp \a \dt x \sp \b - C \sp {\a\b} \tilde
b \quad ,\cr
Q \sp \a \tilde b & = x \sp \a \dt {\tilde b} \quad . & (5.2.7) \cr}$$
We then make further redefinitions
$$ x \sp \a \; \to \; g \sp {-1} x \sp \a \quad , \quad \tilde b \; \to \;
g \sp {-1} \left[ b - 2 \dt {( g \sp {-1} x\sb\a \sp 2 )}\right] \quad ,
\eqno(5.2.8)$$ 
which simplify the $g$ transformation, allowing a further simplification
for $b$:
$$ \li{ Q \sp \a x\sp a & = x \sp \a g \sp {-1} \dt x\sp a = 
x \sp \a p\sp a \quad ,\cr
Q \sp \a g & = \dt x \sp \a \quad ,\cr
Q \sp \a x \sp \b & = - C \sp {\a\b} b \quad ,\cr
Q \sp \a b & = 0 \quad . & (5.2.9) \cr}$$
To get just a BRST operator (as for the IGL(1) formalism), we can
restrict the Sp(2) indices in (5.2.9) to just one value.  Then $x\sp\a$
for the other value of $\a$ (the antighost) and $b$ can be dropped.
(They form an independent IGL(1) multiplet, as described in sect.\ 3.2.)

To get the OSp(1,1$|$2) formalism, we choose a ``Lorentz'' gauge.
We then quantize with the ISp(2)-invariant gauge-fixing term
$$ \cl \sb 1 = -Q\sb\a \sp 2 f ( g ) = f '' ( g ) ( \dt g b -
\dt x\sb\a \sp 2 ) \eqno(5.2.10)$$
for some arbitrary function $f$ such that $f '' \ne 0$.  Canonically
quantizing (where $f'(g)$ is conjugate to $b$), and using the equations 
of motion, we find $Q \sp \a$ from its Noether current (which in $D=1$ 
is also the charge) to be given by (5.2.2).  For an IGL(1) formalism, we
can use the temporal gauge (writing $x\sp\a = ( c , \tilde c )$)
$$ \cl\sb 1 = iQ[\tilde c f(g)] = bf(g) - if'(g)\tilde c \dt c \quad .
\eqno(5.2.11)$$
(Compare the discussions of gauge choices in sect.\ 3.2-3.)

Although Lorentz-gauge quantization gave a result equivalent to that
obtained from the light cone in sect.\ 3.4, we'll find in sect.\ 8.3
for the string a
result equivalent to that obtained from the light cone in sect.\ 3.6.

\sect{5.3. Spinning}

The mechanics of a relativistic spin-1/2 particle [5.1] is obtained by 
symmetrizing the particle action for a spinless particle with respect 
to {\it one-dimensional} (local) supersymmetry.  We thus generalize
$x(\t )\to X(\t ,\q )$, etc., where $\q$ is a single, real,
anticommuting coordinate.  We first define global supersymmetry by the 
generators
$$ q = {\pa \over \pa \q} - \q i \pa \quad , \quad 
i \pa \equiv i {\pa \over \pa \t} = -q \sp 2 \quad ,\eqno(5.3.1a)$$
which leave invariant the derivatives
$$ {\bf d} = {\pa \over \pa \q} + \q i \pa \quad , \quad \pa = - i 
{\bf d} \sp 2 \quad . \eqno(5.3.1b)$$
The local invariances are then generated by (expanding covariantly)
$$ K = \k i {\bf d} + k i \pa \quad ,\eqno(5.3.2a)$$
which act covariantly (i.e., as $( \; ) ' = e \sp {iK} ( \; ) e \sp 
{-iK}$) on the derivatives
$$\cd = G{\bf d} + \Sc G \pa \quad , \quad - i \cd \sp 2 \quad .\eqno(5.3.2b)$$
This gives the infinitesimal transformation
$$ \d\cd = i[K,\cd ] = i(KG-i\cd\k ){\bf d} +i(K\Sc G -i\cd k)\pa
+2i\k G\pa \quad . \eqno(5.3.2c)$$
We next use $\k$ by the last part of this transformation to choose 
the gauge
$$ \Sc G = 0 \to \k = i\ha {\bf d} k \quad .\eqno(5.3.3)$$
The action (5.1.1) becomes
$$ \li{S &= \int d \t \; d \q \; G \sp {-1} \left[ ( - i \cd \sp 2 X ) 
	\cdot P - \ha P \cdot \cd P \right] \cr
&= \int d \t \; d \q \; \left[ - i G ( {\bf d} X ) \cdot ( {\bf d} P ) 
	- \ha P \cdot {\bf d} P \right] \quad .&(5.3.4)\cr}$$

When expanded in components by $\int d \q \to {\bf d}$, and defining
$$ \li{ X = x \quad ,&\quad \cd X = i \g \quad ;\cr
	P = \z \quad ,&\quad \cd P = p \quad ;\cr
	G = g \sp {-1/2} \quad ,&\quad {\bf d} G = i g \sp {-1} \j \quad ;
&(5.3.5)\cr}$$
when evaluated at $\q = 0$ (in analogy to sect.\ 3.2), we find
$$ S = \int d \t \; \left( \dt x \cdot p + i \j \g \cdot p - i \g 
\cdot \dt \z - g \ha p \sp 2 + \ha i \z \cdot \dt \z \right) \quad 
.\eqno(5.3.6)$$
The $( g , x , p )$ sector works as for the bosonic case.  In the $( 
\j , \g , \z )$ sector we see that the quantity $i ( \g - \ha \z )$ is 
canonically conjugate to $\z$, and thus
$$ \g = {\pa \over \pa \z} + \ha \z \quad ,\eqno(5.3.7a)$$
which has $\g$-matrix type commutation relations.  It anticommutes with
$$ \hat \g = \g - \z = {\pa \over \pa \z} - \ha \z \quad ,\eqno(5.3.7b)$$
which is an independent set of $\g$-matrices.  However, it is $\g$ 
which appears in the Dirac equation, obtained by varying $\j$.

In a light-cone formalism, we again eliminate all auxiliary ``$-$'' components 
by their equations of motion, and use the gauge invariance (5.3.2-3) to 
fix the ``$+$'' components
$$ X \sb + = p \sb + \t \quad \to \quad x \sb + = p \sb + \t \quad , 
\quad \g \sb + = 0 \quad .\eqno(5.3.8)$$
We then find $G = 1$, and (5.3.6) reduces to
$$ \cl = \dt x \sb - p \sb + + \dt x \sb i p \sb i - \ha p \sb i \sp 2 
- i ( \g \sb - - \z \sb - ) \dt \z \sb + - i \g \sb i
\dt \z \sb i + \ha i \z \sb i \dt \z \sb i \quad 
.\eqno(5.3.9)$$

In order to obtain the usual spinor field, it's necessary to add a 
lagrange multiplier term to the action constraining $\hat \g = 0$.  
This constraint can either be solved classically (but only for even 
spacetime dimension $D$) by determining half of the $\z$'s to be the 
canonical conjugates of the other half (consider $\z \sb 1 + i \z \sb 
2$ vs.\ $\z \sb 1 - i \z \sb 2$, etc.), or by imposing it quantum 
mechanically on the field Gupta-Bleuler style.  The former approach 
sacrifices manifest Lorentz invariance in the coordinate approach; 
however, if the $\g$'s are considered simply as operators (without 
reference to their coordinate representation), then the field is the 
usual spinor representation, and both can be represented in the usual 
matrix notation.  This constrained action is equivalent to the 
second-order action
$$ \li{ S &= \int d \t \; d \q \; \ha G \sp {-1} ( \cd \sp 2 X ) \cdot ( 
	\cd X ) \cr
&= \int d \t \; d \q \; \ha ( G {\bf d} X ) {\bf d} ( G {\bf d} X ) \quad ,\cr
&= \int d\t\; ( \ha g\sp{-1} \dt x\sp 2 + ig\sp{-1}\j\g\cdot\dt x
-\ha i \g\cdot\dt\g ) \quad ,&(5.3.10a)\cr}$$
or, in first-order form for $x$ only,
$$ S = \int d\t\; ( \dt x\cdot p - g\ha p\sp 2 + \ha i\dt\g\cdot\g
+i\j\g\cdot p ) \quad . \eqno(5.3.10b)$$
The constraint $\g\cdot p =0$ (the Dirac equation) is just a factorized
form of the constraint (2.2.8) for this particular representation of the
Lorentz group.

A further constraint is necessary to get an irreducible Poincar\'e 
representation in even $D$.  Since any function of an anticommuting coordinate 
contains bosonic and fermionic terms as the coefficients of even and 
odd powers of that coordinate, we need the constraint $\g \sb D = \pm 
1$ on the field (where $\g \sb D$ means just the product of all the 
$\g$'s) to pick out a field of just one statistics (in this case, a 
Weyl spinor: notice that $D$ is even in order for the previous 
constraint to be applied).  In the OSp approach
this Weyl chirality condition can also be obtained by an
extension of the algebra [4.10]:  OSp(1,1$|$2)$\otimes$U(1),
where the U(1) is chiral transformations, results in an extra Kronecker
$\d$ which is just the usual chirality projector.  This U(1) generator
(for at least the special case of a Dirac spinor or Ramond string)
can also be derived as a constraint from first-quantization:  The classical
mechanics action for a Dirac spinor, under the global transformation 
$\d\g\sb a \sim \e\sb{abc\cdots d}\g\sp b \g\sp c\cdots \g\sp d$, varies
by a boundary term $\sim \int d\t \der\t \g\sb D$, where as usual 
$\g\sb D \sim \e\sb {abc\cdots d}\g\sp a \g\sp b \g\sp c\cdots \g\sp d$.
By adding a lagrange multiplier term
for $\g\sb D \pm 1$, this symmetry becomes a local one, gauged by the
lagrange multiplier (as for the other equations of motion).  
By 1D supersymmetrization, there is also a lagrange multiplier for 
$\e\sb {abc\cdots d} p\sp a \g\sp b\g\sp c\cdots\g\sp d$.
The action then describes a Weyl spinor.

Many supersymmetric gauges are possible for $g$ and $\j$.  The
simplest sets both to constants (``temporal'' gauge $G=1$),
but this gauge doesn't allow an OSp(1,1$|$2) algebra.  The next simplest 
gauge, ${\bf d}G=0$, does the same to $\j$ but sets the $\t$
derivative of $g$ to vanish, making it an extra coordinate in the field
theory (related to $x\sb -$, or $p\sb +$), giving the generators of
(3.4.2).  However, the gauge which also keeps $\j$ as a coordinate 
(and as a partner to $g$) is $\dt G =0$.  In order to get the maximal 
coordinates (or at least zero-modes, for the string) we choose an 
OSp(1,1$|$2) which keeps $\j$ (related to $\tilde\g\sb\pm$, and the
corresponding extra ghost, related to $\tilde\g\sb\a$).  This gives 
the modified BRST algebra of (3.5.1).

An {\it ``isospinning'' particle} [5.10] can be described similarly.  By 
dropping the $\j$ term in (5.3.6,10b) it's possible to have a different 
symmetry on the indices of $( \g , \z )$ than on those of $( x , p )$. 
In fact, even the range of the indices and the metric can be 
different.  Thus, spin separates from orbital angular momentum and 
becomes isospin.  There is no longer an anticommuting gauge 
invariance, but with a positive definite metric on the isospinor 
indices it's no longer necessary to have one to maintain unitarity.  
If we use the constraint $\hat \g = 0$ we get an isospinor, but if we 
don't we get a matrix, with the $\g$'s acting on one side and the 
$\hat \g$'s on the other.  Noting that $\t$ reversal switches $\g$ 
with $- \hat \g$, we see that the matrix gets transposed.  Therefore, 
the complex conjugation that is the quantum-mechanical analog of $\t$ 
reversal is actually {\it hermitian} conjugation, particularly on a 
field which is a matrix representation of some group.  (When $\hat \g$ 
is constrained to vanish, $\t$ reversal is not an invariance.)  By 
combining these anticommuting variables with the previous ones we get 
an isospinning spinning particle.

At this point we take a slight diversion to discuss properties of
spinors in arbitrary dimensions with arbitrary spacetime signature.
This will complete our discussion of spinors in this section, and will
be useful in the following section, where representations of
supersymmetry, which is itself described by a spinor generator, will be
found to depend qualitatively on the dimension.  The analysis of spinors
in Euclidean space (i.e, the usual spinor representations of SO(D)) can
be obtained by the usual group theoretical methods (see, e.g., [5.11]),
using either Dynkin diagrams or an explicit representation of the
$\g$-matrices.  The properties of spinors in SO($D\sb +$,$D\sb -$) can
then be obtained by Wick rotation of $D\sb -$ space directions into time
ones.  (Of particular interest are the D-dimensional Lorentz group
SO(D$-$1,1) and the D-dimensional conformal group SO(D,2).)
This affects the spinors with respect to only complex conjugation
properties.  A useful notation to classify spinors and their properties
is:  Denote a fundamental spinor (``spin 1/2'') as $\j\sb\a$, and its
hermitian conjugate as $-\bar\j\sb{\dt\a}$.  Denote another spinor
$\j\sp\a$ such that the contraction $\j\sp\a\j\sb\a$ is invariant under
the group, and its hermitian conjugate $\bar\j\sp{\dt\a}$.  The
representation of the group on these various spinors is then related by
taking complex conjugates and inverses of the matrices representing the
group on the fundamental one.  For SO($D\sb +$,$D\sb -$) there are 
always some of these representations that are equivalent, since SO(2N)
has only 2 inequivalent spinor representations and SO(2N+1) just 1.
(In $\g$-matrix language, the Dirac spinor can be reduced to 2
inequivalent Weyl spinors by projection with $\ha (1\pm\g\sb D)$ in even
dimensions.)  In cases where there is another fundamental spinor 
representation not included in this set, we also introduce a 
$\j\sb{\a'}$ and the corresponding 3 other spinors.  (However, in that 
case all 4 in each set will be equivalent, since there are at most 2
inequivalent altogether.)  Many properties of the spinor representations
can be described by classifying the index structure of: (1) the
inequivalent spinors, (2) the bispinor invariant tensors, or
``metrics,'' which are just
the matrices relating the equivalent spinors in the sets of 4, and (3)
the $\s$-matrices ($\g$-matrices for $D$ odd, but in even $D$ the
matrices half as big which remain after Weyl projection), which are
simply the Clebsch-Gordan coefficients for relating spinor$\otimes$spinor
to vector.  In the latter 2 cases, we also classify the symmetry in the
2 spinor indices, where appropriate.

The metrics are of 3 types (along with their complex conjugates and 
inverses): (1) $M\sb\a\sp{\dt\b}$,
which gives charge conjugation for (pseudo)real representations, and is
related to complex conjugation properties of $\g$-matrices,
(2) $M\sb{\a\dt\b}$, which is the matrix which relates the Dirac spinors
$\J$ and $\bar\J$, if it commutes with Weyl projection, and is related
to hermitian conjugation properties of $\g$-matrices, and
(3) $M\sb{\a\b}$, which is the Clebsch-Gordan coefficients
for spinor$\otimes$same-representation spinor to scalar, and is related
to transposition properties of $\g$-matrices.  For all of
these it's important to know whether the metric is symmetric or
antisymmetric; in particular, for the first type we get either real or
pseudoreal representations, respectively.  In $\g$-matrix language, this
charge conjugation matrix is straightforwardly constructed in the
representation where the $\g$-matrices are expressed in terms of direct
products of the Pauli matrices for the 2-dimensional subspaces.  Upon
Wick rotation of 1 direction each in any number of pairs corresponding
to these 2-dimensional subspaces, the corresponding Pauli matrix factor
in the charge conjugation matrix must be dropped (with perhaps some
change in the choice of Pauli matrix factors for the other subspaces).
It then follows that (pseudo)reality is the same in 
SO($D\sb +$+1,$D\sb -$+1) as in SO($D\sb +$,$D\sb -$), so all cases
follow from the Euclidean case.  For the second type of metric, 
$\bar\J = \J\dg$
in the Euclidean case, so $M\sb{\a\dt\b}$ is just the identity matrix
(i.e., the spinor representations are unitary).  After Wick rotation,
this matrix becomes the product of all the $\g$-matrices in the Wick
rotated directions, since those $\g$-matrices got factors of $i$ in the
Wick rotation, and thus need this extra factor to preserve the reality
of the tensors $\bar\J\g\cdots\g\J$.  The symmetry properties of this 
metric then follow from those of the $\g$-matrices.  Also, because of
the signature of the $\g$-matrices, it follows that this metric, except
in the Euclidean case, has half its eigenvalues +1 and half $-1$.
The last type of metric has only undotted indices and thus has nothing
to do with complex conjugation, so its properties are unchanged by Wick
rotation.  It's identical to the first type in Euclidean space (since
the second type is the identity there; in general, if 2 of the metrics
exist, the third is just their product), which thus determines it in the
general case.  Various types of groups are defined by these metrics
alone (real, unitary, orthogonal, symplectic, etc.), 
with the SO($D\sb +$,$D\sb -$) group as a subgroup.  (In fact,
these metrics {\it completely} determine the SO($D\sb +$,$D\sb -$)
group, up to abelian factors, in $D \equiv D\sb + + D\sb - \le 6$,
and allow all vector indices to be replaced by pairs of spinor indices.
They also determine the group in $D=8$ for $D\sb -$ even, due to ``triality,''
the discrete symmetry which permutes the vector representation with the
2 spinors.)  We also classify the $\s$-matrices by their symmetry
properties only when its 2 spinor indices are for equivalent
representations, so they are unrelated to complex conjugation (both
indices undotted), and thus their symmetry is determined by the
Euclidean case.

We now summarize the results obtained by the methods sketched above for
spinors $\j$, metrics $\h$ (symmetric) and $\O$ (antisymmetric), and
$\s$-matrices, in terms of $D$ mod 8 and $D\sb -$ mod 4:

\vskip.2in

\begin{tabular}{|cc||c|c|c|c|} \hline
& $D\sb -$ & 0 & 1 & 2 & 3 \\
$D$ & & Euclidean & Lorentz & conformal & \\ \hline\hline
& & $\j\sb\a$ $\j\sb{\a'}$ & $\j\sb\a$ $\j\sb{\dt\a}$ &
$\j\sb\a$ $\j\sb{\a'}$ & $\j\sb\a$ $\j\sb{\dt\a}$ \\
0 & & $\h\sb{\a\b}$ $\h\sb\a\sp{\dt\b}$ $\h\sb{\a\dt\b}$ & $\h\sb{\a\b}$ &
$\h\sb{\a\b}$ $\O\sb\a\sp{\dt\b}$ $\O\sb{\a\dt\b}$ & $\h\sb{\a\b}$ \\
& & $\s\sb{\a\b'}$ & $\s\sb{\a\dt\b}$ & $\s\sb{\a\b'}$ & $\s\sb{\a\dt\b}$ 
\\ \hline
& & $\j\sb\a$ & $\j\sb\a$ & $\j\sb\a$ & $\j\sb\a$ \\
1 & & $\h\sb{\a\b}$ $\h\sb\a\sp{\dt\b}$ $\h\sb{\a\dt\b}$ & 
$\h\sb{\a\b}$ $\h\sb\a\sp{\dt\b}$ $\h\sb{\a\dt\b}$ & 
$\h\sb{\a\b}$ $\O\sb\a\sp{\dt\b}$ $\O\sb{\a\dt\b}$ & 
$\h\sb{\a\b}$ $\O\sb\a\sp{\dt\b}$ $\O\sb{\a\dt\b}$ \\
& & $\s\sb{(\a\b )}$ & $\s\sb{(\a\b )}$ & $\s\sb{(\a\b )}$ & $\s\sb{(\a\b )}$ 
\\ \hline
& & $\j\sb\a$ $\j\sp\a$ & $\j\sb\a$ $\j\sp\a$ & $\j\sb\a$ $\j\sp\a$ &
$\j\sb\a$ $\j\sp\a$ \\
2 & & $\h\sb{\a\dt\b}$ & $\h\sb\a\sp{\dt\b}$ & 
$\O\sb{\a\dt\b}$ & $\O\sb\a\sp{\dt\b}$ \\
& & $\s\sb{(\a\b )}$ $\s\sp{(\a\b )}$ & $\s\sb{(\a\b )}$ $\s\sp{(\a\b )}$ &
$\s\sb{(\a\b )}$ $\s\sp{(\a\b )}$ & $\s\sb{(\a\b )}$ $\s\sp{(\a\b )}$ 
\\ \hline
& & $\j\sb\a$ & $\j\sb\a$ & $\j\sb\a$ & $\j\sb\a$ \\
3 & & $\O\sb{\a\b}$ $\O\sb\a\sp{\dt\b}$ $\h\sb{\a\dt\b}$ & 
$\O\sb{\a\b}$ $\h\sb\a\sp{\dt\b}$ $\O\sb{\a\dt\b}$ & 
$\O\sb{\a\b}$ $\h\sb\a\sp{\dt\b}$ $\O\sb{\a\dt\b}$ & 
$\O\sb{\a\b}$ $\O\sb\a\sp{\dt\b}$ $\h\sb{\a\dt\b}$ \\
& & $\s\sb{(\a\b )}$ & $\s\sb{(\a\b )}$ & $\s\sb{(\a\b )}$ & $\s\sb{(\a\b )}$ 
\\ \hline
& & $\j\sb\a$ $\j\sb{\a'}$ & $\j\sb\a$ $\j\sb{\dt\a}$ &
$\j\sb\a$ $\j\sb{\a'}$ & $\j\sb\a$ $\j\sb{\dt\a}$ \\
4 & & $\O\sb{\a\b}$ $\O\sb\a\sp{\dt\b}$ $\h\sb{\a\dt\b}$ & $\O\sb{\a\b}$ &
$\O\sb{\a\b}$ $\h\sb\a\sp{\dt\b}$ $\O\sb{\a\dt\b}$ & $\O\sb{\a\b}$ \\
& & $\s\sb{\a\b'}$ & $\s\sb{\a\dt\b}$ & $\s\sb{\a\b'}$ & $\s\sb{\a\dt\b}$ 
\\ \hline
& & $\j\sb\a$ & $\j\sb\a$ & $\j\sb\a$ & $\j\sb\a$ \\
5 & & $\O\sb{\a\b}$ $\O\sb\a\sp{\dt\b}$ $\h\sb{\a\dt\b}$ & 
$\O\sb{\a\b}$ $\O\sb\a\sp{\dt\b}$ $\h\sb{\a\dt\b}$ & 
$\O\sb{\a\b}$ $\h\sb\a\sp{\dt\b}$ $\O\sb{\a\dt\b}$ & 
$\O\sb{\a\b}$ $\h\sb\a\sp{\dt\b}$ $\O\sb{\a\dt\b}$ \\
& & $\s\sb{[\a\b ]}$ & $\s\sb{[\a\b ]}$ & $\s\sb{[\a\b ]}$ & $\s\sb{[\a\b ]}$ 
\\ \hline
& & $\j\sb\a$ $\j\sp\a$ & $\j\sb\a$ $\j\sp\a$ & $\j\sb\a$ $\j\sp\a$ &
$\j\sb\a$ $\j\sp\a$ \\
6 & & $\h\sb{\a\dt\b}$ & $\O\sb\a\sp{\dt\b}$ & 
$\O\sb{\a\dt\b}$ & $\h\sb\a\sp{\dt\b}$ \\
& & $\s\sb{[\a\b ]}$ $\s\sp{[\a\b ]}$ & $\s\sb{[\a\b ]}$ $\s\sp{[\a\b ]}$ &
$\s\sb{[\a\b ]}$ $\s\sp{[\a\b ]}$ & $\s\sb{[\a\b ]}$ $\s\sp{[\a\b ]}$ 
\\ \hline
& & $\j\sb\a$ & $\j\sb\a$ & $\j\sb\a$ & $\j\sb\a$ \\
7 & & $\h\sb{\a\b}$ $\h\sb\a\sp{\dt\b}$ $\h\sb{\a\dt\b}$ & 
$\h\sb{\a\b}$ $\O\sb\a\sp{\dt\b}$ $\O\sb{\a\dt\b}$ & 
$\h\sb{\a\b}$ $\O\sb\a\sp{\dt\b}$ $\O\sb{\a\dt\b}$ & 
$\h\sb{\a\b}$ $\h\sb\a\sp{\dt\b}$ $\h\sb{\a\dt\b}$ \\
& & $\s\sb{[\a\b ]}$ & $\s\sb{[\a\b ]}$ & $\s\sb{[\a\b ]}$ & $\s\sb{[\a\b ]}$ 
\\ \hline
\end{tabular}

\vskip.2in

\noindent (We have omitted the vector indices on the $\s$-matrices.
We have also omitted metrics which are complex conjugates or inverses of
those shown, or are the same but with all indices primed, where relevant.)
Also, not indicated in the table is the fact that
$\h\sb{\a\dt\b}$ is positive definite for the Euclidean case and 
half-positive, half-negative otherwise.  Finally, the dimension of the
spinors is $2\sp{(D-1)/2}$ for $D$ odd and $2\sp{(D-2)/2}$ (Weyl spinor)
for $D$ even.  These $N\times N$ metrics define classical groups as
subgroups of $GL(N,C)$:
$$ \li{ \h\sb{\a\b} & \to SO(N,C) \cr
\O\sb{\a\b} & \to Sp(N,C) \cr
\h\sb\a\sp{\dt\b} & \to GL(N,R) \cr
\O\sb\a\sp{\dt\b} & \to GL\conj (N) \quad (\equiv U\conj (N)) \cr
\h\sb{\a\dt\b} & \to U(N) \quad (or~U(\frac N2 ,\frac N2 )) \cr
\O\sb{\a\dt\b} & \to U(\frac N2 ,\frac N2 ) \cr
\h\sb{\a\b} ~ \h\sb\a\sp{\dt\b} ~ \h\sb{\a\dt\b} & \to
	SO(N) \quad (or~SO(\frac N2 ,\frac N2 )) \cr
\h\sb{\a\b} ~ \O\sb\a\sp{\dt\b} ~ \O\sb{\a\dt\b} & \to SO\conj (N) \cr
\O\sb{\a\b} ~ \h\sb\a\sp{\dt\b} ~ \O\sb{\a\dt\b} & \to Sp(N) \cr
\O\sb{\a\b} ~ \O\sb\a\sp{\dt\b} ~ \h\sb{\a\dt\b} & \to 
	USp(N) \quad (or~USp(\frac N2 ,\frac N2 )) \cr}$$
(When the matrix has a trace, the group can be factored into the
corresponding ``$S$''-group times an abelian factor $U(1)$ or $GL(1,R)$.)

The $\s$-matrices satisfy the obvious relation analogous to the
$\g$-matrix anticommutation relations:  Contract a pair of spinors on 2
$\s$-matrices and symmetrize in the vector indices and you get (twice) the
metric for the vector representation (the SO($D\sb +$,$D\sb -$) metric)
times a Kronecker $\d$ in the remaining spinor indices:
$$ \s\sp{(a}\sb{\a\dt\a}\s\sp{b)\b\dt\a} = \s\sp{(a}\sb{\a\a'}\s\sp{b)\b\a'} =
\s\sp{(a}\sb{[\a\g ]}\s\sp{b)[\b\g ]} = \s\sp{(a}\sb{(\a\g )}\s\sp{b)(\b\g )} =
2\h\sp{ab}\d\sb\a\sp\b \quad , \eqno(5.3.11)$$
and similarly for expressions with dotted and undotted (or primed and
unprimed) indices switched.  (We have raised indices with spinor metrics
when necessary.)

Although (irreducible) spinors thus have many differences in different
dimensions, there are some properties which are dimension-independent,
and it will prove useful to change notation to emphasize those
similarities.  We therefore define spinors which are real in all
dimensions (or would be real after a complex similarity transformation,
and therefore satisfy a generalized Majorana condition).  For those
kinds of spinors in the above table which are complex or pseudoreal,
this means making a bigger spinor which contains the real and imaginary
components of the previous one as independent components.  If the
original spinor was complex ($D\sb + - D\sb -$ twice odd), the new
spinor is reducible to an irreducible spinor and its inequivalent 
complex conjugate representation, which transform oppositely with 
respect to an internal U(1) generator (``$\g\sb 5$'').  If the 
original spinor was pseudoreal ($D\sb + - D\sb - = 3,4,5~mod~8$),
the new spinor reduces to 2 equivalent irreducible spinor
representations, which transform as a doublet with respect to an 
internal SU(2).

The net result for these real spinors is that we have the following
analog of the above table for those properties which hold for all values
of $D\sb +$:

\vskip.1in

\begin{center}
\begin{tabular}{|c||c|c|c|c|} \hline
$D\sb -$ & 0 & 1 & 2 & 3 \\
& Euclidean & Lorentz & conformal & \\ \hline\hline
& $\j\sb\a$ $\j\sb{\a'}$ & $\j\sb\a$ $\j\sp\a$ &
$\j\sb\a$ $\j\sb{\a'}$ & $\j\sb\a$ $\j\sp\a$ \\
& $\h\sb{\a\b}$ & & $\O\sb{\a\b}$ & \\
& $\g\sb{\a\b'}$ & $\g\sb{(\a\b )}$ $\g\sp{(\a\b )}$ &
$\g\sb{\a\b'}$ & $\g\sb{[\a\b ]}$ $\g\sp{[\a\b ]}$ \\ \hline
\end{tabular}
\end{center}

\vskip.1in

\noindent These $\g$-matrices satisfy the same relations as the
$\s$-matrices in (5.3.11).  (In fact, they are identical for $D\sb + =
D\sb -~mod~8$.)  Their additional, $D\sb +$-dependent properties can be
described by additional metrics: (1) the internal symmetry generators
mentioned above; and (2) for $D$ odd, a metric $M\sb{\a\b}$ or
$M\sb{\a\b'}$ which relates the 2 types of spinors (since there are 2
independent irreducible spinor representations only for $D$ even).

Similar methods of first-quantization will be applied in sect.\ 7.2
to the spinning string, which has spin-0 and spin-1/2 ground states.
Classical mechanics actions for particles with other spins (or strings
with ground states with other spins), i.e., gauge fields, are not known.
(For the superstring, however, a nonmanifestly supersymmetric formalism
can be obtained by a truncation of the spinning string, eliminating some
of the ground states.)  On the other hand, the BRST approach of chap.\ 3 allows
the treatment of the quantum mechanics of arbitrary gauge fields.
Furthermore, the superparticle, described in the following section, is
described classical-mechanically by a spin-0 or spin-1/2 superfield,
which includes component gauge fields, just as the string has component
gauge fields in its excited modes.

\sect{5.4. Supersymmetric}

The superparticle is obtained from the spinless particle by 
symmetrizing with respect to the supersymmetry of the 
higher-dimensional space in which the one-dimensional world line of 
the particle is imbedded.  (For reviews of supersymmetry, see [1.17].)
As for the spinless particle, a full understanding of this action 
consists of just understanding the algebras of the covariant 
derivatives and equations of motion.  In order to describe arbitrary
$D$, we work with the general real spinors of the previous section.
The covariant
derivatives are $p\sb a$ (momentum) and $d\sb\a$ (anticommuting spinor), with
$$\{ d\sb\a , d\sb\b \} = 2 \g\sp a\sb{\a\b}p\sb a \eqno(5.4.1)$$
(the other graded commutators vanish), where the $\g$ matrices are 
symmetric in their spinor indices and satisfy
$$ \g \sp {(a} \sb {\a \g} \g \sp {b) \b \g} = 2 \h \sp {ab} \d \sb \a 
\sp \b \quad ,\eqno(5.4.2)$$
as described in the previous section.
This algebra is represented in terms of coordinates $x\sp a$ (spacetime)
and $\q\sp\a$ (anticommuting), and their partial derivatives $\pa\sb a$
and $\pa\sb\a$, as
$$ p\sb a = i\pa\sb a \quad , \quad d\sb\a = \pa\sb\a +i\g\sp
a\sb{\a\b}\q\sp\b\pa\sb a \quad . \eqno(5.4.3)$$
These covariant derivatives are invariant under supersymmetry
transformations generated by $p\sb a$ and $q\sb\a$, which form the algebra
$$ \{ q\sb\a , q\sb\b \} = -2\g\sp a\sb{\a\b}p\sb a \quad . \eqno(5.4.4)$$
$p\sb a$ is given above, and $q\sb\a$ is represented in terms of the
same coordinates as
$$ q\sb\a = \pa\sb\a -i\g\sp a\sb{\a\b}\q\sp\b\pa\sb a \quad . \eqno(5.4.5)$$
(See (5.3.1) for $D=1$.)  All these objects also transform covariantly
under Lorentz transformations generated by
$$ J\sb{ab} = -ix\sb{[a}p\sb{b]} + \frac14 \q\sp\a
\g\sb{[a\a\g}\g\sb{b]}\sp{\g\b} \pa\sb\b + M\sb{ab} \quad , \eqno(5.4.6)$$
where we have included the (coordinate-independent) spin term $M\sb{ab}$.
(In comparison with (2.2.4), the spin operator here gives just the spin of
the superfield, which is a function of $x$ and $\q$, whereas the spin
operator there includes the $\q\pa$ term, and thus gives the spin of the
component fields resulting as the coefficients of a Taylor expansion of
the superfield in powers of $\q$.)

As described in the previous section, even for ``simple'' supersymmetry (the
smallest supersymmetry for that dimension), these spinors are
reducible if the irreducible spinor representation isn't real,
and reduce to the direct sum of an irreducible spinor and its
complex conjugate.  However, we can further generalize by letting the
spinor represent more than one of such real spinors (and some of each of
the 2 types that are independent when $D$ is twice odd), and still use
the same notation, with a single index representing all spinor
components.  (5.4.1-6) are then unchanged (except for the range of the
spinor index).  However, the nature of the supersymmetry representations
will depend on $D$, and on the number of minimal supersymmetries.  In
the remainder of this section we'll stick to this notation to manifest
those properties which are independent of dimension, and include such
things as internal-symmetry generators when required for
dimension-dependent properties.

For a massless, real, scalar field, $p \sp 2 = 0$ is the only equation
of motion, but for a massless, real, scalar superfield, the additional 
equation $\sl p d = 0$ (where $d$ is the spinor derivative) is necessary 
to impose that the superfield is a unitary representation of
(on-shell) supersymmetry [5.12]:  Since the hermitian supersymmetry 
generators $q$ satisfy $\{ q,q\}\sim p$, we have that $\{\sl pq,\sl pq\}
\sim p\sp 2\sl p =0$, but on unitary representations any hermitian
operator whose square vanishes must also vanish, so $0=\sl pq =\sl pd$
up to a term proportional to $p\sp 2 =0$.  This means that only half the
$q$'s are nonvanishing.  We can further divide these remaining $q$'s in
(complex) halves as creation and annihilation operators.  A massless,
irreducible representation of supersymmetry is then specified in this
nonmanifestly Lorentz covariant approach by fixing the ``Clifford''
vacuum of these creation and annihilation operators to be an irreducible
representation of the Poincar\'e group.

Unfortunately, the $p\sp 2$ and $\sl pd$ equations are not sufficient to
determine an irreducible representation of supersymmetry, even for a
scalar superfield (with certain exceptions in $D\le 4$) since, although
they kill the unphysical half of the $q$'s, they don't restrict the
Clifford vacuum.  The latter restriction requires extra constraints in a
manifestly Lorentz covariant formalism.  There are several ways to find 
these additional constraints:  One is to consider coupling 
to external fields.  The simplest case is external super-Yang-Mills
(which will be particularly relevant for strings).  
The generalization of the covariant derivatives is
$$\li{d \sb \a &\to \de \sb \a = d \sb \a + \G \sb \a \quad ,\cr
p \sb a &\to \de \sb a = p \sb a + \G \sb a \quad .&(5.4.7)\cr}$$
We thus have a graded covariant derivative $\de\sb A$, $A=(a,\a )$.
Without loss of generality, we consider cases where the 
only physical fields in the super-Yang-Mills multiplet are a vector 
and a spinor.  The other cases (containing scalars) can be obtained 
easily by dimensional reduction.  Then the commutation relations of 
the covariant derivatives become [5.13]
$$\li{\{ \de \sb \a \; , \; \de \sb \b \} &= 2 \g \sp a \sb {\a\b} \de \sb a
	\quad ,\cr
[ \de \sb \a \; , \; \de \sb a ] &= 2 \g \sb {a\a\b} W \sp \b \quad,\cr
[ \de \sb a \; , \; \de \sb b ] &= F \sb {ab} \quad,&(5.4.8)\cr}$$
where $W \sp \a$ is the super-Yang-Mills field strength (at $\q = 0$, 
the physical spinor field), and consistency of the Jacobi (Bianchi) 
identities requires
$$ \g \sb {a(\a\b} \g \sp a \sb {\g )\d} = 0 \quad.\eqno(5.4.9)$$
This condition (when maximal 
Lorentz invariance is assumed, i.e., SO(D-1,1) for $a$ taking $D$ 
values) implies spacetime dimensions $D = 3,4,6,10$ , and ``antispacetime'' 
dimensions (the number of values of the index $\a$) $D ' = 2 ( D - 2 )$.
(The latter identity follows from multiplying (5.4.9) by $\g\sp{b\a\b}$
and using (5.4.2).)
The generalization of the equations of motion is [5.14]
$$\li{\sl p d &\to \g \sp {a\a\b} \de \sb a \de \sb\b \quad,\cr
\ha p \sp 2 &\to \ha \de \sp a \de \sb a + W \sp \a \de \sb \a 
\quad,&(5.4.10)\cr}$$
but closure of this algebra also requires new equations of motion 
which are certain Lorentz pieces of $\de \sb {[\a} \de \sb {\b ]}$.
Specifically, in $D=3$ 
there is only one Lorentz piece (a scalar), and it gives the usual 
field equations for a scalar multiplet [5.15]; in $D=4$ the scalar piece 
again gives the usual equations for a chiral scalar multiplet, but 
the (axial) vector piece gives the chirality condition (after 
appropriate normal ordering); in $D=6$ only the self-dual third-rank 
antisymmetric tensor piece appears in the algebra, and it gives 
equations satisfied by scalar multiplets (but not by tensor 
multiplets, which are also described by scalar superfield strengths, 
but can't couple minimally to Yang-Mills) [5.16]; but in $D=10$ no multiplet 
is described because the only one possible would be Yang-Mills itself, 
but its field strength $W \sp \a$ carries a Lorentz index, and the 
equations described above (which apply only to scalars) need extra 
terms containing Lorentz generators.  

Another way to derive the modifications is to use superconformal 
transformations.  The superconformal groups [5.17] are actually easier to
derive than the supersymmetry groups because they are just graded
versions of classical groups.  Specifically, the classical supergroups
(see [5.18] for a review)
have defining representations defined in terms of a metric 
$M\sb{A\dt B}$, which makes them unitary (or pseudounitary, if the
metric isn't positive definite), and sometimes also a graded-symmetric metric
$M\sb{AB}$, and thus $M\sb A\sp{\dt B}$ by combining them (and their
inverses).  The generators which have bosonic-bosonic or
fermionic-fermionic indices are bosonic, and those with
bosonic-fermionic are fermionic.  (The choice of which parts of the $A$
index are bosonic and which are fermionic can be reversed, but this
doesn't affect the statistics of the group generators.)  Since the
bosonic subgroup of the supergroups with just the $M\sb{A\dt B}$ metric
is the direct product of 2 unitary groups, those supergroups are called
(S)SU(M$|$N) for M values of the index of one statistics and N of the 
other, where
the S is because a (graded) trace condition is imposed, and there is a
second S for M=N because then a second trace can be removed (so each of
the 2 unitary subgroups becomes SU(N)).  The supergroups which also have
the graded-symmetric metric $M\sb{AB}$ have a bosonic subgroup which is
orthogonal in the sector where the metric is symmetric and symplectic in
the sector where it is antisymmetric.  In this case we choose the metric
$M\sb{A\dt B}$ also to have graded symmetry, in such a way that the
metric $M\sb A\sp{\dt B}$ obtained from their product is totally
symmetric, so the defining representation is real, or totally
antisymmetric, so the representation is pseudoreal.  The former is
generally called OSp(M$|$2N), and we call the latter OSp$\conj$(M$|$2N).

We next assume that the anticommuting generators of these supergroups
are to be identified with the conformal generalization of the
supersymmetry generators.  Thus, one index is to be identified with an
internal symmetry, and the other with a conformal spinor index.  The
conformal spinor reduces to 2 Lorentz spinors, one of which is the usual
supersymmetry, the ``square root'' of translations, and the other of
which is ``S-supersymmetry,'' the square root of conformal boosts.
The choice of supergroup then follows immediately from the graded
generalization of the conformal spinor metrics appearing in the table of
the previous section [5.19]:

\vskip.2in
\begin{center}
\begin{tabular}{|c||c|c|c|}\hline
D mod 8 & superconformal & bosonic subgroup & dim-0/dilatations \\ \hline\hline
0,4 & (S)SU(N$|\n$,$\n$) & 
	SU($\n$,$\n$)$\otimes$SU(N)($\otimes$U(1)) &
	SL($\n$,C)$\otimes$(S)U(N) \\ \hline
1,3 & OSp(N$|$2$\n$) & Sp(2$\n$)$\otimes$SO(N) &
	SL($\n$,R)$\otimes$SO(N) \\ \hline
2 & OSp(N$|$2$\n$) & " & " \\
& {\it or} (")$\sb + \otimes$(")$\sb -$ & (")$\sb + \otimes$(")$\sb -$ &
	(")$\sb + \otimes$(")$\sb -$ \\ \hline
5,7 & OSp$\conj$(2$\n |$2N) & SO$\conj$(2$\n$)$\otimes$USp(2N) &
	SL$\conj$($\n$)$\otimes$USp(2N) \\ \hline
6 & OSp$\conj$(2$\n |$2N) & " & " \\
& {\it or} (")$\sb + \otimes$(")$\sb -$ & (")$\sb + \otimes$(")$\sb -$ &
	(")$\sb + \otimes$(")$\sb -$ \\ \hline
\end{tabular}
\end{center}
\vskip.2in

\noindent where ``dim-0'' are the generators which commute with
dilatations (see sect.\ 2.2), so the last column gives
Lorentz$\otimes$internal symmetry (at least).
$\n$ is the dimension of the (irreducible) Lorentz spinor (1/2
that of the conformal spinor), N is the number of minimal
supersymmetries, and $D$($>$2) is the dimension of (Minkowski) spacetime
(with conformal group SO(D,2), so 2 less than the $D$ in the previous 
table).  The 2 choices for twice-odd $D$ depend on whether we choose to
represent the superconformal group on both primed and unprimed spinors.
If so, there can be a separate N and N$'$.  Again, we have used $\conj$ to
indicate groups which are Wick rotations such that the defining
representation is pseudoreal instead of real.  (SL$\conj$ is sometimes
denoted SU$\conj$.)

Unfortunately, as discussed in the previous section, the bosonic
subgroup acting on the conformal spinor part of the defining
representation, as defined by the spinor metrics (plus the trace
condition, when relevant) gives a group bigger than the conformal group
unless $D\le 6$.  However, we can still use $D\le 6$, and perhaps some
of the qualitative features of $D>6$, for our analysis of massless field
equations.  We then generalize our analysis of sect.\ 2.2 from conformal
to superconformal.  It's sufficient to apply just the S-supersymmetry 
generators to just the Klein-Gordon operator.  We then find [2.6,5.19]:
$$ \ha p \sp 2 \to \sl pq = \sl p d \to \left\{ \matrix{
\ha \{ p \sp b \; , \; J \sb {ab} \} + \ha \{ p \sb a \; , \; \D \} =
	p \sp b M \sb {ab} + p \sb a \left( {\Sc d} - {D-2 \over 
	2} \right) \cr
q\sb{[\a}q\sb{\b ]} +\cdots =
d \sb {[\a} d \sb {\b ]} + \cdots \hfill \cr}\right.\quad,\eqno(5.4.11)$$
where the last expression 
means certain Lorentz pieces of $dd$ plus certain terms containing 
Lorentz and internal symmetry generators.  In particular,
$\frac1{16}\g\sb{abc}\sp{\a\b}d\sb\a d\sb\b +\ha p\sb{[a}M\sb{bc]}$
is the supersymmetric analog of the Pauli-Lubansky vector [5.20].
The vector equation is (2.2.8) again, derived in essentially the same way.

For the constraints we therefore choose [5.21]
$$ \li{ \ca &= \ha p\sp 2 \quad ,\cr
	\cb\sp\a &= \g\sp{a\a\b}p\sb a d\sb\b \quad ,\cr
	\cc\sb{abc} &= \frac1{16}\g\sb{abc}\sp{\a\b}d\sb\a d\sb\b 
		+\ha p\sb{[a}M\sb{bc]} \quad , \cr
	\cd\sb a &= M\sb a\sp b p\sb b + kp\sb a \quad ; &(5.4.12a)\cr}$$
or, in matrix notation,
$$ \li{ \ca &= \ha p\sp 2 \quad ,\cr
	\cb &= \sl pd \quad ,\cr
	\cc\sb{abc} &= \frac1{16}d\g\sb{abc}d +\ha p\sb{[a}M\sb{bc]}\quad ,\cr
	\cd\sb a &= M\sb a\sp b p\sb b + kp\sb a \quad ; &(5.4.12b)\cr}$$
where out of (5.4.11) we have chosen $\ca$ and $\cd$ as for
nonsupersymmetric theories (sect.\ 2.2), $\cb$ for unitarity (as
explained above), and just the Pauli-Lubansky part of the rest (which 
is all of it for D=10), the significance of which will be explained below.

These constraints satisfy the algebra
$$\li{ \{ \cb , \cb \} &= 4\sl p \ca \quad , \cr
[ \cd\sb a , \cd\sb b ] &= -2M\sb{ab}\ca -p\sb{[a}\cd\sb{b]} \quad ,\cr
[ \cc\sb{abc} , \cb ] &= -8\g\sb{abc}d\ca \quad , \cr
[ \cc\sb{abc} , \cd\sb d ] &= \h\sb{d[a}p\sb b \cd\sb{c]} \quad , \cr
[ \cc\sp{abc} , \cc\sb{def} ] &= -\frac14 [ \d\sb{[d}\sp{[a}p\sb e
	\cc\sp{bc]}\sb{f]}-(abc\iff def)] -\frac1{128}d
(4\g\sp{abc}\sb{def}-2\d\sb{[d}\sp{[a}\d\sb e\sp b \g\sp{c]}\sb{f]})\cb
	\quad , \cr
rest &=0 \quad , &(5.4.13)\cr}$$
with some ambiguity in how the right-hand side is expressed due to the
relations
$$ \li{ p\cdot\cd &= 2k\ca \quad , \cr
\sl p\cb &= 2d\ca \quad , \cr
d\cb &= 2(tr~I)\ca \quad , \cr
\frac16p\sb{[a}\cc\sb{bcd]} &= -\frac1{16}d\g\sb{abcd}\cb \quad , \cr
p\sp c\cc\sb{cab} &= 2M\sb{ab}\ca + p\sb{[a}\cd\sb{b]} +
	\frac1{16}d\g\sb{ab}\cb \quad . &(5.4.14)\cr}$$
($\g\sb{ab}=\ha\g\sb{[a}\g\sb{b]}$, etc.)

In the case of supersymmetry with an internal symmetry group
(extended supersymmetry, or even simple supersymmetry in D=5,6,7), there is an
additional constraint analogous to $\cc\sb{abc}$ for superisospin:
$$ \cc\sb{a,int} = \frac18d\g\sb a\s\sb{int}d + p\sb a M\sb{int} \quad .
\eqno(5.4.15)$$
$\s\sb{int}$ are the matrix generators of the internal symmetry group,
in the representation to which $d$ belongs, and $M\sb{int}$ are those
which act on the external indices of the superfield.

Unfortunately, there are few superspin-0 multiplets that are contained
within spin-0, isospin-0 superfields (i.e., that themselves contain spin-0,
isospin-0 component fields).  In fact, the only such multiplets of
physical interest in D$>$4 are N=1 Yang-Mills in D=9 and N=2 nonchiral
supergravity in D=10.  (For a convenient listing of multiplets, see
[5.22].)  However, by the method described in the previous section,
spinor representations for the Lorentz group can be introduced.
By including ``$\g$-matrices'' for internal symmetry, we can also introduce
defining representations for the internal symmetry groups for which they are
equivalent to the spinor representations of orthogonal groups (i.e., 
SU(2)=USp(2)=SO(3), USp(4)=SO(5), SU(4)=SO(6), 
SO(4)-vector=SO(3)-spinor$\otimes$SO(3)$'$-spinor, 
SO(8)-vector=SO(8)$'$-spinor). 
Furthermore, arbitrary U(1) representations can be described by adding
extra terms without introducing additional coordinates.  This allows the
description of most superspin-0 multiplets, but with some notable
exceptions (e.g., 11D supergravity).  However, these equations are not
easily generalized to nonzero superspins, since, although the superspin
operator is easy to identify in the light-cone formalism (see below),
the corresponding operator would be nonlocal in a covariant description
(or appear always with an additional factor of momentum).

We next consider the construction of mechanics actions.  These equations
describe only multiplets of superspin 0, i.e., the smallest
representations of a given supersymmetry algebra, for reasons to be
described below.  (This is no restriction in D=3,
where superspin doesn't exist, and in D=4 arbitrary superspin can be
treated by a minor modification, since there superspin is abelian.)
As described in the previous section, only spin-0 and spin-1/2
superfields can be described by classical mechanics, and we begin with
spin-0, dropping spin terms in (5.4.12), and the generator $\cd$.  The
action is then given by (3.1.10), where [5.23]
$$ z \sp M = ( x \sp m , \q \sp \m ) \quad , \quad \p \sb A = ( p \sb 
a , i d \sb \a ) \quad ,$$
$$ \dt z \sp M e \sb M \sp A ( z ) = ( \dt x - i \dt \q \g \q , 
\dt \q ) \quad ,$$
$$ i \cg \sb i ( \p ) = ( \ca , \cb\sp\a , \cc\sb{abc} ) \quad .\eqno(5.4.16)$$
Upon quantization, the covariant derivatives become
$$ \p \sb A = i e \sb A \sp M \pa \sb M = i ( \pa \sb a , \pa \sb \a + 
i \g \sp a \sb {\a \b} \q \sp \b \pa \sb a ) \quad ,\eqno(5.4.17)$$
which are invariant under the supersymmetry transformations
$$ \d x = \x - i \e \g \q \quad , \quad \d \q = \e \quad .\eqno(5.4.18)$$
The transformation laws then follow directly from (3.1.11), with the aid of
(5.4.13) for the $\l$ transformations.

The classical mechanics action can be quantized covariantly by BRST
methods.  In particular, the transformations generated by $\cb$ [5.24]
(with parameter $\k$) close on those generated by $\ca$ (with parameter $\x$):
$$ \d x = \x p + i \k ( \g d + \sl p \g \q ) \quad ,\quad \d \q = \sl p
\k \quad ,$$
$$ \d p = 0 \quad ,\quad \d d = 2 p \sp 2 \k \quad ,$$
$$ \d g = \dt \x + 4 i \k \sl p \j \quad ,\quad \d \j = \dt \k 
\quad .\eqno(5.4.19) $$
Because of the second line of (5.4.14), the ghosts have a gauge 
invariance similar to the
original $\k$ invariance, and then the ghosts of those ghosts again have
such an invariance, etc., ad infinitum.  This is a consequence of the
fact that only half of $\q$ can be gauged away, but there is generally
no Lorentz representation with half the components of a spinor, so the
spinor gauge parameter must itself be half gauge, etc.  Although
somewhat awkward, the infinite set of ghosts is straightforward to find.
Furthermore, if derived from the light cone, the OSp(1,1$|$2) generators
automatically contain this infinite number of spinors:  There, $\q$ is
first-quantized in the same way as the Dirac spinor was second-quantized
in sect.\ 3.5, and $\q$ obtains an infinite number of components (as an
infinite number of ordinary spinors) as a result of being a representation
of a graded Clifford algebra (specifically, the Heisenberg algebras of
$\g\sp\a$ and $\tilde\g\sp\a$).  This analysis will be made in the next
section. 

On the other hand,
the analysis of the constraints is simplest in the light-cone formalism.
The $\ca$, $\cb$, and $\cd$ equations can be solved directly, because
they are all of the form $ p\cdot f = p\sb + f\sb - + \cdots$ :
$$ \li{ \ca = 0 \quad&\to\quad p\sb - = -{1\over 2p\sb +}p\sb i\sp 2\quad ,\cr
\cb = 0 \quad&\to\quad \g\sb - d = 
	-{1\over 2p\sb +}\g\sb i p\sb i \g\sb -\g\sb + d \quad ,\cr
\cd = 0 \quad&\to\quad M\sb{-i} = {1\over p\sb +}(M\sb i\sp j p\sb j +
	kp\sb i ) \quad , \quad M\sb{-+} = k \quad ,&(5.4.20a)\cr}$$
where we have chosen the corresponding gauges
$$ \li{ x\sb + &= 0 \quad ,\cr
	\g\sb +\q &= 0 \quad ,\cr
	M\sb{+i} &= 0\quad .&(5.4.20b)\cr}$$
These solutions restrict the $x$'s, $\q$'s, and Lorentz indices,
respectively, to those of the light cone.  (Effectively, D is reduced by
2, except that $p\sb +$ remains.)  

However, a superfield which is a function of a light-cone $\q$ is not 
an irreducible representation of supersymmetry (except sometimes in 
D$\le$4), although it is a unitary one.  In fact, $\cc$ is just the 
superspin operator which separates the representations:  Due to the 
other constraints, all its components are linearly related to
$$ \cc\sb{+ij} = p\sb + M\sb{ij} + \frac1{16}d\g\sb +\g\sb{ij}d 
\quad .\eqno(5.4.21)$$
Up to a factor of $p\sb +$, this is the light-cone superspin:  On an
irreducible representation of supersymmetry, it acts as an irreducible
representation of SO(D-2).  In D=4 this can be seen easily by
noting that the irreducible representations can be represented in terms
of chiral superfields ($\bar d =0$) with different numbers of $d$'s
acting on them, and the $d\bar d$ in $\cc$ just counts the numbers of
$d$'s.  In general, if we note that the full light-cone Lorentz
generator can be written as
$$ \li{ J\sb{ij} &= -ix\sb{[i}p\sb{j]}+\ha\q\g\sb{ij}\der\q +M\sb{ij}\cr
	&= -ix\sb{[i}p\sb{j]} -{1\over 16p\sb +}q\g\sb +\g\sb{ij}q
		+{1\over 16p\sb +}d\g\sb +\g\sb{ij}d +M\sb{ij}\cr
	&= \hat J\sb{ij} + {1\over p\sb +}\cc\sb{+ij} \quad ,&(5.4.22)\cr}$$
then, by expressing any state in terms of $q$'s acting on the Clifford
vacuum, we see that $\hat J$ gives the correct transformation for those
$q$'s and the $x$-dependence of the Clifford vacuum, so $\cc /p\sb +$
gives the spin of the Clifford vacuum less the contribution of the $qq$
term on it, i.e., the superspin.
Unfortunately, the mechanics action can't handle spin operators for
irreducible representations (either for $M\sb{ij}$ or the superspin), so
we must restrict ourselves not only to spin 0 (referring to the external
indices on the superfield), but also superspin 0 (at least at the
classical mechanics level).  Thus, the remaining constraint $\cc =0$ is
the only possible (first-class) constraint which can make the
supersymmetry representation irreducible.  The constraints (5.4.12) are
therefore necessary and sufficient for deriving the mechanics action.
However, if we allow the trivial kind of second-class constraints that
can be solved in terms of matrices, we can generalize to spin-1/2.  In
principle, we could also do superspin-1/2, but this leads to covariant
fields which are just those for superspin-0 with an extra spinor index
tagged on, which differs by factors of momentum (with appropriate index
contractions) from the desired expressions.  (Thus, the superspin
operator would be nonlocal on the latter.)  Isospin-1/2 can be treated
similarly (and superisospin-1/2, but again as a tagged-on index).

The simplest nontrivial example is D=4.  (In D=3, $\cc\sb{+ij}$
vanishes, since the transverse index $i$ takes only one value.)
There light-cone spinors have only 1 (complex) component, and so does 
$\cc\sb{+ij}$.  For this case, we can (and must, for an odd number N of
supersymmetries) modify $\cc\sb{abc}$:
$$ \cc\sb{abc} = \frac1{16}d\g\sb{abc}d +\ha p\sb{[a}M\sb{bc]} +
	iH \e\sb{abcd}p\sp d \quad , \eqno(5.4.23)$$
where $H$ is the ``superhelicity.''  We then find
$$ \cc\sb{+ij} = p\sb + \e\sb{ij} \left( \cm +iH - i{1\over 4p\sb +}
[ d\sb{\bf a} , \bar d\sp{\bf a} ] \right) \quad , \eqno(5.4.24)$$
where $M\sb{ij} = \e\sb{ij} \cm$ and $\{ d\sb{\bf a} , \bar d\sp{\bf b} \}
= p\sb + \d\sb{\bf a}\sp{\bf b}$, and $\bf a$ is an SU(N) index.
The ``helicity'' $h$ is given by $\cm =-ih$, and we then find by expanding
the field over chiral fields $\f$ [5.25-27] ($\bar d\f =0$)
$$\j\sim (d)\sp n\f \quad\to\quad H = h + \frac14 (2n-N) \quad .\eqno(5.4.25)$$
Specifying both the spin and superhelicity of the original superfield
fixes both $h$ and $H$, and thus determines $n$.  Note that this
requires $H$ to be quarter-(odd-)integral for odd N.  In general, the
SU(N) representation of $\f$ also needs to be specified, and the
relevant part of (5.4.15) is
$$ \cc\sb{+\bf a}\sp{\bf b} = p\sb + \left[ M\sb{\bf a}\sp{\bf b} 
-i{1\over 4p\sb +}\left( [ d\sb{\bf a} , \bar d\sp{\bf b} ] - \frac14
\d\sb{\bf a}\sp{\bf b} [ d\sb{\bf c} , \bar d\sp{\bf c} ] \right) \right]
\quad , \eqno(5.4.26)$$
and the vanishing of this quantity forces $\f$ to be an SU(N)-singlet.
(More general cases can be obtained simply by tacking extra indices onto
the original superfield, and thus onto $\f$.)

We next consider 10D super Yang-Mills.  The appropriate superfield is a
Weyl or Majorana spinor, so we include terms as in the previous section 
in the mechanics action.
To solve the remaining constraint, we first decompose SO(9,1)
covariant spinors and $\g$-matrices to SO(8) light-cone ones as
$$ d \sb \a = 2\sp{1/4}\pmatrix{d \sb + \cr d \sb - \cr} \quad , $$
$$ \sl p \sb {\a\b} = \pmatrix{\sqrt 2 p \sb + & \sl p \sb T \cr \sl p \sb T 
	\sp \dag & - \sqrt 2 p \sb - \cr} \pmatrix{\S & 0 \cr 0 & \S \cr} 
\quad , \quad \sl p \sp {\a\b} = \pmatrix{\S & 0 \cr 0 & \S \cr} 
	\pmatrix{\sqrt 2 p \sb - & \sl p \sb T \cr \sl p \sb T \sp \dag & - 
	\sqrt 2 p \sb + \cr} \quad ,$$
$$ \sl a \sb T \sl b \sb T \sp \dag + \sl b \sb T \sl a \sb T \sp \dag 
= \sl a \sb T \sp \dag \sl b \sb T + \sl b \sb T \sp \dag \sl a \sb T 
= 2 a \sb T \cdot b \sb T \quad ,$$
$$\sl p \sb T \conj = \S \sl p \sb T \S \quad , \quad \S = \S \sp 
\dag = \S \conj \quad , \quad \S \sp 2 = I \quad .\eqno(5.4.27)$$
(We could choose the Majorana representation $\S = I$, but other 
representations can be more convenient.)  The independent
supersymmetry-covariant derivatives are then
$$ d \sb + = {\pa \over \pa \q \sp +} + p \sb + \S \q \sp + \quad , 
\quad p \sb T \quad , \quad p \sb + \quad .\eqno(5.4.28)$$
In order to introduce chiral light-cone superfields, we further reduce
SO(8) to \\
SO(6)$\otimes$SO(2)=U(4) notation:
$$ d \sb + = \sqrt 2\pmatrix{d \sb {\bf a} \cr \bar d \sp {\bf a} \cr} \quad 
, \quad \sl p \sb T = \pmatrix{p \sb {\bf ab} & \d \sb {\bf a} \sp 
{\bf b} p \sb L \cr \d \sb {\bf b} \sp {\bf a} \bar p \sb L & \bar p 
\sp{\bf ab} \cr} \quad , \quad \S = \pmatrix{0 & I \cr I & 0 \cr }$$
$$ ( \bar p \sp {\bf ab} = \ha \e \sp {\bf abcd} p \sb {\bf cd} ) 
\quad .\eqno(5.4.29)$$
In terms of this ``euphoric'' notation, the constraints $\cc\sb{+ij}$ are
written on the SO(6)-spinor superfield as 
$$\li{ {1\over p\sb +} \cc\sb{+\bf a}\sp{\bf b} 
\pmatrix{\j\sb {\bf c} \cr \bar\j\sp{\bf c} \cr}
&= i\ha \pmatrix{\d\sb{\bf c}\sp{\bf b}\j\sb{\bf a}
-\frac14\d\sb{\bf a}\sp{\bf b}\j\sb{\bf c} \cr
-\d\sb{\bf a}\sp{\bf c}\bar\j\sp{\bf b}
+\frac14\d\sb{\bf a}\sp{\bf b}\bar\j\sp{\bf c} \cr} -i{1\over 4p\sb +}
\left( [ d\sb{\bf a} , \bar d\sp{\bf b} ] -\frac14 \d\sb{\bf a}\sp{\bf b}
[ d\sb{\bf d} , \bar d\sp{\bf d} ] \right) 
\pmatrix{\j\sb {\bf c} \cr \bar\j\sp{\bf c} \cr} \; , \cr
{1\over p\sb +} \cc\sb{+} \pmatrix{\j\sb {\bf c} \cr \bar\j\sp{\bf c} \cr} 
&= i\ha \pmatrix{-\j\sb {\bf c} \cr \bar\j\sp{\bf c} \cr} 
-i{1\over 4p\sb +} [ d\sb{\bf a} , \bar d\sp{\bf a} ]
\pmatrix{\j\sb {\bf c} \cr \bar\j\sp{\bf c} \cr}  \quad , \cr
{1\over p\sb +}\cc\sb{+\bf ab}\pmatrix{\j\sb {\bf c} \cr \bar\j\sp{\bf c} \cr} 
&= i\ha\pmatrix{\e\sb{\bf abcd}\bar\j\sp{\bf d} \cr 0 \cr}
-i{1\over 2p\sb +}d\sb{\bf a}d\sb{\bf b}
\pmatrix{\j\sb {\bf c} \cr \bar\j\sp{\bf c} \cr} \quad ,&(5.4.30)\cr}$$
and the complex conjugate equation for $\cc\sb +\sp{\bf ab}$.  (Note
that it is crucial that the original SO(10) superfield $\j\sp\a$ was a spinor 
of chirality opposite to that of $d\sb\a$ in order to obtain soluble 
equations.)  The solution to the first 2 equations gives $\j$ in terms
of a chiral superfield $\f$,
$$ \j\sb {\bf a} = d\sb{\bf a}\f \quad , \eqno(5.4.31a) $$
and that to the third equation imposes the self-duality condition [5.25-27]
$$ \frac1{24}\e\sp{\bf abcd}d\sb{\bf a}d\sb{\bf b}d\sb{\bf c}d\sb{\bf d}
	\f = p\sb +\sp 2\bar\f \quad . \eqno(5.4.31b) $$
This can also be written as
$$ \left\{ \pr \left[ \left( \ha p \sb + \right) \sp {-1/2} d 
\right] \right\} \f = \bar \f \quad \to 
\quad \int d \q \; e \sp {\q \sp {\bf a} \p \sb {\bf a} p \sb +/2 } \f 
( \q \sp {\bf a} ) = \left[ \f ( \bar \p \sp {\bf a} ) \right] \conj 
\quad .\eqno(5.4.32)$$
This corresponds to the fact that in the mechanics action $\t$ 
reversal on $\q \sp \a$ includes multiplication by the charge 
conjugation matrix, which switches $\q \sp {\bf a}$ with
$\bar \q \sb {\bf a}$, which equals $- ( 2/ p \sb +) \pa / \pa \q 
\sp {\bf a}$ by the chirality condition $\bar d\sp{\bf a} =0$.

These results are equivalent to those obtained from first-quantization
of a mechanics action with $d\sb\a =0$ as a second-class constraint [5.28].
(This is the analog of the 
constraint $\hat \g = 0$ of the previous section.)  This is 
effectively the same as dropping the $d$ terms from the action, which 
can then be written in second-order form by eliminating $p$ by its 
equation of motion.
This can be solved either by using a chiral superfield [5.27] as a solution to
this constraint in a Gupta-Bleuler formalism,
$$ \bar d \sp {\bf a} \f = 0 \quad \to \quad d \sb {\bf a} \bar \f = 0 
\quad \to \quad \int \bar \f d \sb + \f = 0 \quad ,\eqno(5.4.33)$$
or by using a superfield with a {\it real} 4-component $\q$ [5.26] as a
solution to this constraint before quantization (but after going to a
light-cone gauge), determining half the remaining components of $\g\sb -\q$
to be the canonical conjugates of the other half.  However, whereas
either of these methods with second-class constraints requires the
breaking of manifest Lorentz covariance just for the formulation of the
(field) theory, the method we have described above has constraints on
the fields which are manifestly Lorentz covariant ((5.4.12)).
Furthermore, this second-class approach requires that (5.4.32) be
imposed in addition, whereas in the first-class approach it and the
chirality condition automatically followed together from (5.4.21) (and
the ordinary reality of the original SO(9,1) spinor superfield).

On the other hand, the formalism with second-class constraints can be
derived from the first-class formalism {\it without} $M\sb{ab}$ terms
(and thus without $\cd$ in (5.4.12)) [5.29]:  Just as $\ca$ and $\cb$ were
solved at the classical level to obtain (5.4.20), $\cc =0$ can also be
solved classically.  To be specific, we again consider $D=10$.
Then $\cc\sb{+ij}=0$ is equivalent to $d\sb{[\m}d\sb{\n ]}=0$ (where
$\m$ is an 8-valued light-cone spinor index).
(They are just different linear combinations of the same 28
antisymmetric quadratics in $d$, which are the only nonvanishing $d$
products classically.)  This constraint implies the components of $d$
are all proportional to the same anticommuting scalar, times different
commuting factors:
$$ d\sb\m d\sb\n = 0 \quad\to\quad d\sb\m = c\z\sb\m \quad , \eqno(5.4.34a)$$
where $c$ is anticommuting and $\z$ commuting.  Furthermore,
$\cc\sb{+ij}$ are just SO(8) generators on $d$, and thus their gauge
transformation can be used to rotate it in any direction, thus
eliminating all but 1 component [5.30].  (This is clear from triality, since
the spinor 8 representation is like the vector 8.)  Specifically, we
choose the gauge parameters of the $\cc$ transformations to depend on
$\z$ in such a way as to rotate $\z$ in any one direction, and then
redefine $c$ to absorb the remaining $\z$ factor:
$$ \cc~gauge:\quad d\sb\m = \d\sb m\sp 1 c \quad . \eqno(5.4.34b)$$
In this gauge, the
$\cc$ constraint itself is trivial, since it is antisymmetric in $d$'s.
Finally, we quantize this one remaining component $c$ of $d$ to find 
$$ quantization: \quad c\sp 2 = p\sb + \quad\to\quad c = \pm\sqrt{p\sb +}
\quad . \eqno(5.4.34c)$$
$c$ has been determined only up to a
sign, but there is a residual $\cc$ gauge invariance, since the $\cc$
rotation can also be used to rotate $\z$ in the opposite direction,
changing its sign.  After using the gauge invariance to make all but one
component of $\z$ vanish, this sign change is the only part of the gauge
transformation which survives.  It can then be used to choose the sign
in (5.4.34c).  Thus, all the 
$d$'s are determined (although 1 component is nonvanishing), and we 
obtain the same set of coordinates ($x$ and $q$, no $d$) as in the 
second-class formalism.  The $\cc$ constraint can also be solved
completely at the quantum mechanical level by Gupta-Bleuler methods
[5.29].  The SO(D-2) generators represented by $\cc$ are then divided up
into the Cartan subalgebra, raising operators, and lowering operators.
The raising operators are imposed as constraints (on the ket, and the
lowering operators on the bra), implying only the highest-weight state
survives, and the generators of the Cartan subalgebra are imposed only
up to ``normal-ordering'' constants, which are just the weights of that state.

The components of this chiral superfield can be identified with the
usual vector + spinor [5.26,27]:
$$ \f ( x , \q \sp {\bf a} ) = $$
$$( p \sb + ) \sp {-1} A \sb L ( x ) + 
\q \sp {\bf a} ( p \sb + ) \sp {-1} \c \sb 
{\bf a} ( x ) + \q \sp {2 {\bf ab}} A \sb {\bf ab} ( x ) + \q \sp 3 
\sb {\bf a} \bar \c \sp {\bf a} ( x ) + \q \sp 4 ( p \sb + ) 
\bar A \sb L ( x ) \quad ,\eqno(5.4.35)$$
and the $\q =0$ components of $\g\sb +\j = ( \j\sb{\bf a} ,
\bar\j\sp{\bf a} )$ can be identified with the spinor.
Alternatively, the vector + spinor content can be obtained directly from
the vanishing of $\cc\sb{+ij}$ of (5.4.21), without using euphoric
notation:  We first note that the $\G$-matrices of $M\sb{ij}$ are 
represented by 2 spinors, corresponding to the 2 different chiralities
of spinors in SO(8).  SO(8) has the property of ``triality,'' which is
the permutation symmetry of these 2 spinors with the vector
representation.  (All are 8-component representations.)  Since the
anticommutation relations of $d$ are just a triality transformation of
those of the $\G$'s (modulo $p\sb +$'s), they are represented by the
other 2 representations: a spinor of the other chirality and a vector.
The same holds for the representation of $q$.  Thus, the direct product
of the representations of $\G$ and $d$ includes a singlet (superspin 0),
picked out by $\cc\sb{+ij}=0$, so the total SO(8) representation
(generated by $J\sb{ij}$ of (5.4.22)) is just that of $q$, a spinor (of
opposite chirality) and a vector.

There is another on-shell method of analysis of (super)conformal
theories that is manifestly covariant and makes essential use of
spinors.  This method expresses the fields in terms of the ``spinor''
representation of the superconformal group.  (The ordinary conformal 
group is the case N=0.)  The spinor is defined in terms of generalized
$\g$-matrices (``twistors'' [5.31] or ``supertwistors'' [5.32]): 
$$ \{ \bar \g\sb A , \g\sp B ] = \d\sb A\sp B \quad , \eqno(5.4.36)$$
where the index has been lowered by $M\sb{A\dt B}$, and the grading is
such that the conformal spinor part has been chosen {\it commuting} and
the internal part {\it anticommuting}, just as ordinary $\g$-matrices
have bosonic (vector) indices but are anticommuting.  The anticommuting
$\g$'s are then closely analogous to the light-cone supersymmetry
generators $\g\sb - q$.  The generators are then represented as
$$ G\sp A\sb B \sim \g\sp A\bar \g\sb B \quad , \eqno(5.4.37)$$
with graded (anti)symmetrization or traces subtracted, as appropriate.
The case of OSp(1,1$|$2) has been treated in (3.5.1).
A representation in terms of the usual superspace coordinates can then
be generated by coset-space methods, as described in sect.\ 2.2.
We begin by identifying the subgroup of the supergroup which corresponds
to supersymmetry (by picking 1 of the 2 Lorentz spinors in the conformal
spinor generator) and translations (by closure of supersymmetry).
We then equate their representation in (5.4.37) (analogous to the 
$\hat J$'s of sect.\ 2.2) with their representation in (5.4.3,5)
(analogous to the $J$'s of sect.\ 2.2).  (The constant $\g$-matrices of
(5.4.3,5) should not be confused with the operators of (5.4.36).)
This results in an expression analogous to (2.2.6), where $\F (0)$ 
is a function of half of (linear combinations of) the $\g$'s
of (5.4.36) (the other half being their canonical conjugates).  
(For example, the bosonic part of $\g\sp A$ is a conformal spinor which 
is expressed as a Lorentz spinor $\z\sb\a$ and its canonical conjugate.
(5.4.37) then gives $p\sb a = \g\sb a\sp{\a\b}\z\sb\a\z\sb\b$, which
implies $p\sp 2 =0$ in $D=3,4,6,10$ due to (5.4.9).)  We then
integrate over these $\g$'s to obtain a function of just the usual
superspace coordinates $x$ and $\q$.  Due to the quadratic form of the
momentum generator in terms of the $\g$'s, it describes only positive
energy.  Negative energies, for antiparticles, can be introduced by
adding to the field a term for the complex conjugate representation.
At least in $D$=3,4,6,10 this
superfield satisfies $p\sp 2 =0$ as a consequence of the explicit form
of the generators (5.4.37), and as a consequence all the equations which
follow from superconformal transformations.  These equations form a
superconformal tensor which can be written covariantly as an expression
quadratic in $G\sp A\sb B$.  In $D$=4 an additional U(1) acting on the twistor
space can be identified as the (little group) helicity, and in $D$=6 a
similar SU(2)($\otimes$SU(2)$'$ if the primed supergroup is also
introduced) appears.  ((5.4.34a) is also a supertwistor type of relation.)

The cases $D$=3,4,6,10 [5.33] are especially
interesting not only for the above reason and (5.4.9) but also because
their various spacetime groups form an interesting pattern if we
consider these groups to be the same for these different dimensions
except that they are over different generalized number systems {\bf A}
called ``division algebras.''  These are generalizations of complex
numbers which can be written as $z=z\sb 0+\su_1^n z\sb ie\sb i$,
$\{ e\sb i , e\sb j\} =-2\d\sb{ij}$, where $n$=0,1,3, or 7.
Choosing for the different dimensions the division algebras

\begin{center}
\begin{tabular}{r|l}
D & {\bf A} \\ \hline
3 & real \\
4 & complex \\
6 & quaternion \\
10 & octonion \\
\end{tabular}
\end{center}
\vskip.2in

\noindent we have the correspondence
$$ \left. \matrix{ SL\sb 1 (1,{\bf A}) \hfill &= SO(D-2) \hfill\cr
SU(2,{\bf A}) \hfill &= SO(D-1) \hfill\cr
SU(1,1,{\bf A}) \hfill &= SO(D-2,1) \hfill\cr
SL(2,{\bf A}) \hfill &= SO(D-1,1) \hfill\cr
SU'(4,{\bf A}) \hfill &= SO(D,2) \hfill\cr
SU(N|4,{\bf A}) \hfill &= superconformal \hfill\cr} \right\} / SO(D-3) $$
where $SL\sb 1$ means only the real part ($z\sb 0$) of the trace of the
defining representation vanishes, by $SU'$ we mean traceless and having
the metric $\O\sb{\a\dt\b}$ (vs.\ $\h\sb{\a\dt\b}$
for $SU$), and the graded $SU$ has metric $M\sb{A\dt B} = ( \h\sb{a\dt b} ,
\O\sb{\a\dt\b} )$ (in that order).  The $\dt{\phantom m}$ refers to
generalized conjugation $e\sb i\to -e\sb i$ (and the $e\sb i$ are invariant
under transposition, although their ordering inside the matrices changes).
The ``$/SO(D-3)$'' refers to the fact that to get the desired groups we
must include rotations of the $D-3$ $e\sb i$'s among themselves.  The
only possible exception is for the $D=10$ superconformal groups, which
don't correspond to the $OSp(N|32)$ above, and haven't been shown to
exist [5.34].  The light-cone form of the identity (5.4.9) ((7.3.17)) is
equivalent to the division algebra identity $|xy|=|x||y|$
($|x|\sp 2 = xx\conj = x\sb 0\sp 2 +\su x\sb i\sp 2$), where both the
vector and spinor indices on the light-cone $\g$-matrices correspond to
the index for $(z\sb 0 , z\sb i)$ (all ranging over $D-2$ values) [5.35].

A similar first-quantization analysis will be made for the superstring
in sect.\ 7.3.

\sect{5.5. SuperBRST}

Instead of using the covariant quantization which would follow directly
from the constraint analysis of (5.4.12), we will derive here the BRST
algebra which follows from the light-cone by the method of sect.\ 3.6,
which treats bosons and fermions symmetrically [3.16].  We begin with 
any (reducible)
light-cone Poincar\'e representation which is also a supersymmetry
representation, and extend also the light-cone supersymmetry
generators to 4+4 extra dimensions.  The resulting
OSp(D+1,3$|$4) spinor does not commute with the BRST OSp(1,1$|$2)
generators, and thus mixes physical and unphysical states.  Fortunately,
this extended supersymmetry operator $q$ can easily be projected down to
its OSp(1,1$|$2) singlet piece $q\sb 0$.  We begin with the fact that
the light-cone supersymmetry generator is a tensor operator in a spinor
representation of the Lorentz group:
$$ [ J\sb{ab} , q ] = -\ha\g\sb{ab}q \quad , \eqno(5.5.1) $$
where $\g\sb{ab} = \ha\g\sb{[a}\g\sb{b]}$.  (All $\g$'s are now Dirac
$\g$-matrices, not the generalized $\g$'s of (5.4.1,2).)  
As a result, its extension to 4+4 extra dimensions transforms with 
respect to the U(1)-type OSp(1,1$|$2) as
$$ [ J\sb{AB} , q \} = -\ha (\g\sb{AB}+\g\sb{A'B'}) q \quad . \eqno(5.5.2) $$
It will be useful to combine $\g\sb A$ and $\g\sb{A'}$ into creation
and annihilation operators as in (4.5.9):
$$ \g\sb A = a\sb A + a\dg\sb A \quad , \quad
\g\sb{A'} = i (a\sb A - a\dg\sb A) \quad ; \quad
\{ a\sb A , a\dg\sb B ] = \h\sb{AB} \eqno(5.5.3) $$
$$ \to\quad \ha (\g\sb{AB} + \g\sb{A'B'}) = a\dg\sb{[A}a\sb{B)}
\quad . \eqno(5.5.4) $$
($a\dg\sb A a\sb B$, without symmetrization, are a representation of
$U(1,1|1,1)$.) 
We choose boundary conditions such that all ``states'' can be created
by the creation operators $a\dg$ from a ``vacuum'' annihilated by the
annihilation operators $a$.  (This choice, eliminating states obtained
from a second vacuum annihilated by $a\dg$, is a type of Weyl
projection.)  This vacuum is a fermionic spinor (acted on by $\g\sb a$)
whose statistics are changed by $a\dg\sp\a$ (but not by $a\dg\sb\pm$).
If $q$ is a real spinor, we can preserve
this reality by choosing a representation where $\g\sp A$ is real and
$\g\sp{A'}$ is imaginary.  (In the same way, for the ordinary harmonic
oscillator the ground state can be chosen to be a real function of $x$,
and the creation operator $\sim x-\pa /\pa x$ preserves the reality.)
The corresponding charge-conjugation matrix is $C=i\g\sb{5'}$, where
$$ \g\sb{5'} = \ha [\g\sb{+'} , \g\sb{-'}]
e\sp{\p\ha\{\g\sp{c'} , \g\sp{\tilde c'} \} } \quad , \eqno(5.5.5) $$
with $\g\sp{\a'} = (\g\sp{c'} , \g\sp{\tilde c'})$.
($e\sp{i\g\sb{5'}\p /4}$ converts to the representation where both
$\g\sp A$ and $\g\sp{A'}$ are real.)

We now project to the OSp(1,1$|$2) singlet
$$ [ J\sb{AB} , q\sb 0 \} = 0 \quad\to\quad q\sb 0 = \d (a\dg\sp{A}a\sb A) 
q \quad , \eqno(5.5.6) $$
where the Kronecker $\d$ projects down to ground states with respect to
these creation operators.  It satisfies
$$ \d (a\dg a) a\dg = a \d (a\dg a) = 0 \quad . \eqno(5.5.7) $$
This projector can be rewritten in various forms:
$$ \d (a\dg\sp A a\sb A) = \d (a\dg\sp\a a\sb\a)a\sb + a\sb - a\dg\sb +
a\dg\sb - = \int_{-\p}^\p{du\over 2\p}\; e\sp{iua\dg\sp A a\sb A}
\quad . \eqno(5.5.8) $$

We next check that this symmetry of the physical states is the usual
supersymmetry.  We start with the light-cone commutation relations
$$ \{ q , \bar q \} = 2\cp\sl p \quad , \eqno(5.5.9) $$
where $\cp$ is a Weyl projector, when necessary, and, as usual,
$\bar q = q\dg\h$, with $\h$ the hermitian spinor metric satisfying
$\g\dg\h =\h\g$.  ($\h$'s explicit form will change upon adding
dimensions because of the change in signature of the Lorentz metric.)
We then find
$$ \{ q\sb 0 , \bar q\sb 0 \} = \d (a\dg a) 2\cp\sl p \d (a\dg a) =
2\cp\d (a\dg a) \g\sp a p\sb a \quad , \eqno(5.5.10) $$
where the $\g\sb A$ and $\g\sb{A'}$ terms have been killed by the 
$\d (a\dg a)$'s on the left and right.  The factors other than
$2\g\sp a p\sb a$ project to the physical subspace (i.e., restrict the
range of the extended spinor index to that of an ordinary Lorentz spinor).
The analogous construction for the GL(1)-type OSp(1,1$|$2) fails,
since in that case the corresponding projector
$\d (\h\sp{BA}\g\sb A\g\sb{B'})=\d (\h\sp{BA}\g\sb{A'}\g\sb B)\dg$
(whereas $\d (a\dg a)$ is hermitian), and the 2 $\d$'s then kill all
terms in $\sl p$ except $\h\sp{BA}\g\sb{A'}p\sb B$.

As a special case, we consider arbitrary massless representations of
supersymmetry.  The light-cone representation of the supersymmetry
generators is (cf.\ (5.4.20a))
$$ q = q\sb + - {1\over 2p\sb +}\g\sb +\g\sp i p\sb i q\sb +
\quad , \eqno(5.5.11a) $$
where $q\sb +$ is a self-conjugate light-cone spinor:
$$ \g\sb - q\sb + = 0 \quad , \quad \{ q\sb + , \bar q\sb + \}
= 2\cp \g\sb - p\sb + \quad . \eqno(5.5.11b) $$
Thus, $q\sb +$ has only half as many nonvanishing components as a
Lorentz spinor, and only half of those are independent, the other half
being their conjugates.  The Poincar\'e algebra is then specified by
$$ M\sb{ij} = {1\over 16p\sb +}\bar q \g\sb +\g\sb{ij}q + \check M\sb{ij}
\quad , \eqno(5.5.12) $$
where $\check M$ is an irreducible representation of SO(D-2), the
superspin, specifying the spin of the Clifford vacuum of $q\sb +$.
(Cf.\ (5.4.21,22).  We have normalized the $\bar qq$ term for Majorana $q$.)

After adding 4+4 dimensions, $q\sb +$ can be Lorentz-covariantly further
divided using $\g\sb{\pm'}$:
$$ q\sb + = \sqrt{p\sb +}\left(\der{\bar\q} +2\g\sb - \q\right) \quad , $$
$$ \g\sb - \der{\bar\q} = \g\sb + \q = \g\sb{+'}\der{\bar\q}
= \g\sb{-'}\q = 0 \quad , $$
$$ \left\{ \der{\bar\q} , \bar\q \right] = \cp (\ha\g\sb -\g\sb +)
(\ha\g\sb{+'}\g\sb{-'}) \quad . \eqno(5.5.13) $$
After substitution of (5.5.11,12) into (5.5.6), we find
$$ q\sb 0 = \d (a\dg a) \left[ \sqrt{p\sb +}\der{\bar\q} +
{2\over\sqrt{p\sb +}}(\g\sp a p\sb a +\g\sp\a p\sb\a)\q \right] . 
\eqno(5.5.14) $$
From (5.5.12) we obtain the corresponding spin operators (for Majorana
$\q$)
$$ M\sb{ab} = \ha\bar\q\g\sb{ab}\der{\bar\q} +\check M\sb{ab} \quad , \quad
M\sb{\a\b} + M\sb{\a'\b'} = \bar\q a\dg\sb{(\a}a\sb{\b )}\der{\bar\q}
+\check M\sb{\a\b} +\check M\sb{\a'\b'} \quad , $$
$$ M\sb{-'+'} = -\ha\bar\q\der{\bar\q} +\check M\sb{-'+'} \quad , \quad
M\sb{+'\a'} = \ha\bar\q\g\sb -\g\sb{+'}\g\sb{\a'}\q 
+\check M\sb{+'\a'} \quad , $$
$$ M\sb{\a a} = \ha\bar\q\g\sb\a\g\sb a\der{\bar\q} +\check M\sb{\a a}
\quad , \quad M\sb{-'\a'} = 
\frac1{16}\der\q\g\sb +\g\sb{-'}\g\sb{\a'}\der{\bar\q}
+\check M\sb{-'\a'} \quad . \eqno(5.5.15) $$
Finally, we perform the unitary transformations (3.6.13) to find
$$ q\sb 0 = \d (a\dg a) \left( \der{\bar\q}+2\g\sp a p\sb a\q \right)
\quad . \eqno(5.5.16) $$
The $\d$ now projects out just the OSp(1,1$|$2)-singlet part of $\q$
(i.e., the usual Lorentz spinor):
$$ q\sb 0 = \der{\bar\q\sb 0} +\g\sp a p\sb a\q\sb 0 \quad , $$
$$ \der{\bar\q\sb 0} = \d (a\dg a)\der{\bar\q} \quad , \quad
\q\sb 0 = 2\d (a\dg a)\q \quad , \quad
\left\{ \der{\bar\q\sb 0} , \bar\q\sb 0 \right\} = \cp\d (a\dg a)
\quad . \eqno(5.5.17) $$

(5.5.15) can be substituted into (3.6.14).  We then find the OSp(1,1$|$2)
generators
$$ J\sb{+\a} = ix\sb\a p\sb + \quad , \quad J\sb{-+} = -ix\sb - p\sb +
\quad , \quad
J\sb{\a\b} = -ix\sb{(\a}p\sb{\b )} + \Hat M\sb{\a\b} \quad , $$
$$ J\sb{-\a} = -ix\sb -p\sb\a +{1\over p\sb +}\left[ -ix\sb\a \ha (p\sb a\sp 2
+p\sp\b p\sb\b) +\Hat M\sb\a\sp\b p\sb\b +\Hat\cq\sb\a\right] \quad ; 
\eqno(5.5.18a) $$
$$ \Hat M\sb{\a\b} = \bar\q a\dg\sb{(\a}a\sb{\b )}\der{\bar\q}
+\check M\sb{\a\b} +\check M\sb{\a'\b'} \quad , $$
$$ \li{ \Hat\cq\sb\a = & -i\frac18\Bar{\left(\der{\bar\q}+i\g\sb
-\g\sb{+'}\g\sp a p\sb a\q\right)}\g\sb +\g\sb{-'}a\dg\sb\a
\left(\der{\bar\q}-i\g\sb -\g\sb{+'}\g\sp a p\sb a\q\right) \cr
& +\left( \check M\sb{-'\a'} +\check M\sb\a\sp a p\sb a
+\ha\check M\sb{+'\a'}p\sb a\sp 2 \right) \quad , & (5.5.18b) \cr}$$
and $J\sb{ab} = -ix\sb{[a}p\sb{b]}+\ha\bar\q\g\sb{ab}\pa /\pa\bar\q
+\check M\sb{ab}$ for the Lorentz generators.
Finally, we can remove all dependence on $\g\sb\pm$ and $\g\sb{\pm'}$ by
extracting the corresponding $\g\sb 0$ factors contributing to the
spinor metric $\h$:
$$ \g\sb 0 = -i\ha ( \g\sb + -\g\sb - ) ( \g\sb{+'} -\g\sb{-'} ) 
= \g\sb 0\dg \quad , \quad \d (a\dg a)\g\sb 0 = \g\sb 0\d (a\dg a) =
\d (a\dg a) \quad , $$
$$ \der{\bar\q} \;\to\; \g\sb 0\der{\bar\q} = i\ha\g\sb -\g\sb{+'}
\der{\bar\q} \quad , \quad \bar\q \;\to\; \bar\q\g\sb 0 =
i\ha\bar\q\g\sb +\g\sb{-'} \quad , $$
$$ (\g\sb +,\g\sb{-'})\der{\bar\q} = (\g\sb +,\g\sb{-'})\q = 
\bar\q (\g\sb -,\g\sb{+'}) = \der\q (\g\sb -,\g\sb{+'}) = 0 \quad , $$
$$ \left\{ \der{\bar\q} , \bar\q \right] = \cp (\ha\g\sb +\g\sb -)
(\ha\g\sb{-'}\g\sb{+'}) \quad ; \eqno(5.5.19a) $$
and then convert to the harmonic oscillator basis with respect to these $\g$'s:
$$ \der{\bar\q} \to e\sp{a\dg\sb + a\dg\sb -}\der{\bar\q} \quad , \quad
\q \to \ha e\sp{a\dg\sb + a\dg\sb -}\q \quad , $$
$$ \bar\q \to \ha \bar\q e\sp{a\sb + a\sb -} \quad , \quad
\der\q \to \der\q e\sp{a\sb + a\sb -} \quad ; $$
$$ a\sb\pm \der{\bar\q} = a\sb\pm \q = \bar\q a\dg\sb\pm = \der\q
a\dg\sb\pm = 0 \quad ; $$
$$ \left\{ \der{\bar\q} , \bar\q \right] = \cp (a\sb +a\sb -a\dg\sb +
a\dg\sb -) \quad . \eqno(5.5.19b) $$
All $a\sb\pm$'s and $a\dg\sb\pm$'s can then be eliminated, and the
corresponding projection operators (the factor multiplying $\cp$ in
(5.5.19b)) dropped. 
The only part of (5.5.18) (or the Lorentz generators) which gets modified is
$$ \Hat\cq\sb\a = -\ha\rlap{\hbox{$\bar{\phantom q}$}}\Sc q a\dg\sb\a\Sc d
+\left( \check M\sb{-'\a'} +\check M\sb\a\sp a p\sb a
+\ha\check M\sb{+'\a'}p\sb a\sp 2 \right) \quad , $$
$$ \Sc q = \der{\bar\q}+\g\sp a p\sb a\q \quad , \quad
\Sc d = \der{\bar\q}-\g\sp a p\sb a\q\ \quad , $$
$$ q\sb 0 = \d (a\dg\sp\a a\sb\a) \Sc q \quad , \quad
d\sb 0 = \d (a\dg\sp\a a\sb\a) \Sc d \quad . \eqno(5.5.20)$$
If we expand the first term in $\Hat\cq\sb\a$ level by level in 
$a\dg\sb\a$'s, we find a
$q$ at each level multiplying a $d$ of the previous level.  In
particular, the first-level ghost $q\sb{1\a}$ multiplies the physical $d\sb 0$.
This means that $d\sb 0=0$ is effectively imposed for only half of its spinor
components, since the components of $q$ are not all independent.

An interesting characteristic of this type of BRST (as well as more
conventional BRST obtained by first-quantization) is that spinors obtain
infinite towers of ghosts.  In fact, this is necessary to allow the most
general possible gauges.  The simplest explicit example is BRST 
quantization of the action of sect.\ 4.5 for the Dirac spinor quantized 
in a gauge where the gauge-fixed kinetic operators are all $p\sp 2$ 
instead of $\sl p$.  However, these ghosts are not all necessary
for the gauge invariant theory, or for certain types of gauges.  For
example, for the type of gauge invariant actions for spinors 
described in sect.\ 4.5, the only parts of the infinite-dimensional 
OSp(D-1,1$|$2) spinors which are not pure gauge are the usual Lorentz
spinors.  (E.g., the OSp(D-1,1$|$2) Dirac spinor reduces to an ordinary
SO(D-1,1) Dirac spinor.)  For gauge-fixed, 4D N=1 supersymmetric 
theories, supergraphs use chains of ghost superfields which always
terminate with chiral superfields.  Chiral superfields can be
irreducible off-shell representations of supersymmetry since they
effectively depend on only half of the components of $\q$.  (An analog
also exists in 6D, with or without the use of harmonic superspace
coordinates [4.12].)  However, no chiral division of $\q$ exists in 10D ($\q$
is a real representation of SO(9,1)), so an infinite tower of ghost
superfields is necessary for covariant background-field gauges.
(For covariant non-background-field gauges, all but the usual finite
Faddeev-Popov ghosts decouple.)  Thus, the infinite tower is not just a
property of the type of first-quantization used, but is an inherent
property of the second-quantized theory.  However, even in
background-field gauges the infinite tower (except for the
Faddeev-Popovs) contribute only at one loop to the effective action, so
their evaluation is straightforward, and the only expected problem would
be their summation.  

The basic reason for the tower of $\q$'s is
the fact that only 1/4 (or, in the massive case, 1/2) of them appear 
in the gauge-invariant theory on-shell, but if $\q$ is an irreducible 
Lorentz representation it's impossible to cancel 3/4 (or 1/2) of it 
covariantly.  We thus effectively obtain the sums
$$ 1-1+1-1+\cdots = \ha \quad , \eqno(5.5.21a) $$
$$ 1-2+3-4+\cdots = \frac14 \quad . \eqno(5.5.21b) $$
(The latter series is the ``square'' of the former.)
The positive contributions represent the physical spinor (or $\q$) 
and fermionic ghosts at even levels, the negative contributions 
represent bosonic ghosts at odd levels (contributing in loops with 
the opposite sign), and the $\ha$ or $\frac14$ represents the desired
contribution (as obtained directly in light-cone gauges).  Adding
consecutive terms in the sum gives a nonconvergent (but nondivergent in
case (5.5.21a)) result which oscillates about the desired result.  
However, there are unambiguous ways to regularize these sums.  For 
example, if we represent the levels in terms of harmonic oscillators
(one creation operator for (5.5.21a), and the 2 $a\dg\sb\a$'s for (5.5.21b)),
these sums can be represented as integrals over coherent states (see
(9.1.12)).  For (5.5.21a), we have:
$$ \li{ str(1) & = tr\left[(-1)\sp N\right] = \int{d\sp 2z\over\p}\;
e\sp{-|z|\sp 2}\bra{z}(-1)\sp{a\dg a}\ket{z} \cr
& = \int{d\sp 2z\over\p}\; e\sp{-|z|\sp 2}\left<z|-z\right>
= \int{d\sp 2z\over\p}\; e\sp{-2|z|\sp 2}
= \ha\int{d\sp 2z\over\p}\; e\sp{-|z|\sp 2} \cr
& = \ha \quad , &(5.5.22) \cr} $$
where $str$ is the supertrace; for (5.5.21b), the supertrace over the
direct product corresponding to 2 sets of oscillators factors into the
square of (5.5.22).  (The corresponding partition function is
$str(x\sp N)=1/(1+x)=1-x+x\sp2-\cdots$, and for 2 sets of oscillators
$1/(1+x)\sp 2=1-2x+3x\sp 2-\cdots$.)  

An interesting consequence of (5.5.21) is the preservation of the identity
$$ D' = 2\sp{(D-k)/2} \quad ; \qquad D' = str\sb S (1) \quad , \quad
D = str\sb V (1) \quad ; \eqno(5.5.23) $$
upon adding $(2,2|4)$ dimensions, where $D'$ and $D$ are the
``superdimensions'' of a spinor and vector, defined in terms of
supertraces of the identity for that representation, and $k$ is an
integer which depends on whether the dimension is even or odd and
whether the spinor is Weyl and/or Majorana (see sect.\ 5.3).  $k$ 
is unchanged by adding
$(2,2|4)$ dimensions, $D$ changes by addition of $4-4=0$, and, because 
of (5.5.21), $D'$ changes by a factor of $2\sp 2\cdot (\ha )\sp 2 =1$.
This identity is important for super-Yang-Mills and superstrings.

Before considering the action for arbitrary supersymmetric theories,
we'll first study the equations of motion, since the naive kinetic
operators may require extra factors to write a suitable lagrangian.
Within the OSp(1,1$|$2) formalism, the gauge-fixed field equations are 
(cf.\ (4.4.19))
$$ ( p\sp 2 +M\sp 2 ) \f = 0 \eqno(5.5.24a) $$
when subject to the (Landau) gauge conditions [2.3]
$$ \Hat\cq\sb\a \f = \Hat M\sb{\a\b} \f = 0 \quad . \eqno(5.5.24b)$$
Applying these gauge conditions to the gauge transformations, we find
the residual gauge invariance 
$$ \d\f = -i\ha\Hat\cq\sb\a\L\sp\a \quad , \quad 
\left[ -\frac32 ( p\sp 2 +M\sp 2 ) + \Hat\cq\sp 2 \right] \L\sp\a =
\Hat M\sb{\a\b}\L\sb\g + C\sb{\g (\a}\L\sb{\b )} = 0 \quad . \eqno(5.5.24c) $$
(In the IGL(1) formalism, sect.\ 4.2, the corresponding equations involve
just the $\Hat\cq\sp +$ component of $\cq\sp\a$ and the $\Hat M\sp +$
and $\Hat M\sp 3$ components of $\Hat M\sp{\a\b}$, but are
equivalent, since $\Hat M\sp +=\Hat M\sp 3 =0 \to
\Hat M\sp -=0$, and $\Hat\cq\sp + = \Hat M\sp -=0\to \Hat\cq\sp -=0$.)

For simplicity, we consider the massless case, and $\check
M\sb{ij} =0$.  We can then choose the reference frame where 
$p\sb a = \d\sb a\sp + p\sb +$, and solve these equations in 
light-cone notation.  (The $+$'s and $-$'s now refer to the usual 
Lorentz components; the unphysical $x\sb -$, $p\sb +$, and $\g\sb{\pm'}$
have already been eliminated.)  The gauge conditions (5.5.24b)
eliminate auxiliary degrees of freedom (as $\pa\cdot A = p\sb +A\sb - =0$
eliminates $A\sb -$ in Yang-Mills), and (5.5.24c) eliminates remaining
gauge degrees of freedom (as $A\sb +$ in light-cone-gauge Yang-Mills).
We divide the spinors $\Sc d$, $\Sc q$, 
$\pa / \pa\q$, and $\q$ into halves using $\g\sb\pm$, and then further divide
those into complex conjugate halves as creation and annihilation
operators, as in (5.4.27,29):
$$ \Sc d \; \to \; \g\sb +\Sc d , \g\sb -\Sc d \; \to \;
d\sb{\bf a} , \bar d\sp{\bf a} , \pa\sb{\bf a} , \bar\pa\sp{\bf a} \quad , $$
$$ \Sc q \; \to \; \g\sb +\Sc q , \g\sb -\Sc q \; \to \;
q\sb{\bf a} , \bar q\sp{\bf a} , \pa\sb{\bf a} , \bar\pa\sp{\bf a} \quad , 
\eqno(5.5.25) $$
where the ``$-$'' parts of $\Sc d$ and $\Sc q$ are both just partial
derivatives because the momentum dependence drops out in this frame, and
$d,\bar d$ and $q,\bar q$ have graded harmonic oscillator commutators
(up to factors of $p\sb +$).  (5.5.20) then becomes
$$ \Hat\cq\sb\a \sim \bar q\sp{\bf a}a\dg\sb\a\pa\sb{\bf a} +
q\sb{\bf a}a\dg\sb\a\bar\pa\sp{\bf a} +\bar\pa\sp{\bf a}a\dg\sb\a d\sb{\bf a} +
\pa\sb{\bf a}a\dg\sb\a\bar d\sp{\bf a} \quad . \eqno(5.5.26) $$
Since $\Hat\cq\sb\a$ consists of terms of the form $AB$, either $A$ or
$B$ can be chosen as the constraint in (5.5.24b), and the other will
generate gauge transformations in (5.5.24c).  We can thus choose either
$\pa\sb{\bf a}$, or $\bar d\sp{\bf a}$ and $\bar q\sp{\bf a}$, and
similarly for the complex conjugates, except for the Sp(2) singlets,
where the choice is between $\pa\sb{\bf a}$ and just $\bar d\sp{\bf a}$
(and similarly for the complex conjugates).  However, choosing both $d$
and $\bar d$ (or both $q$ and $\bar q$) for constraints causes the field
to vanish, and choosing them both for gauge generators allows the field
to be completely gauged away.  As a result, the only consistent
constraints and gauge transformations are
$$ \bar d\sp{\bf a}\f = (a\sb\a\bar q\sp{\bf a})\f = \bar\pa\sp{\bf a}\f = 0 
\quad , \quad \d\f = \pa\sb{\bf a}\bar\l\sp{\bf a} \quad , \eqno(5.5.27a) $$
subject to the restriction that the residual gauge transformations
preserve the gauge choice (explicitly, (5.5.24c), although it's more
convenient to re-solve for the residual invariance in light-cone notation).  
(There is also a complex conjugate term in $\f$ if it satisfies a reality 
condition.  For each value of the index {\bf a}, the choice of which 
oscillator is creation and which is annihilation is arbitrary, and 
corresponding components of $d$ and $\bar d$ or $q$ and $\bar q$ can be 
switched by changing gauges.)  Choosing the gauge
$$ \q\sp{-\bf a}\f = 0 \quad , \eqno(5.5.27b) $$
for the residual gauge transformation generated by $\pa\sb{\bf a}=
\pa /\pa\q\sp{-\bf a}$, the field becomes
$$ \f ( \q\sp + , \bar\q\sp + , \q\sp - , \bar\q\sp - ) = 
\d ( a\sb\a\q\sp + ) \d (\q\sp -) \varphi (\q\sb 0\sp + ,\bar\q\sb
0\sp +) \quad , \quad \bar d\sb 0 \varphi = 0 \quad . \eqno(5.5.28) $$
$\varphi$ is the usual chiral light-cone superfield (as in sect.\ 5.4), 
a function of only 1/4 of the usual Lorentz spinor $\q\sb 0$.
This agrees with the general result of equivalence to the light cone
for U(1)-type 4+4-extended BRST given at the end of sect.\ 4.5.

Since the physical states again appear in the middle of the $\q$
expansion (including ghost $\q$'s), we can again use
(4.1.1) as the action:  In the light-cone gauge, from (5.5.28), integrating 
over the $\d$-functions,
$$ S = \int dx\; d\q\sb 0\sp +\; d\bar\q\sb 0\sp + \; \bar\varphi\bo\varphi
\quad , \eqno(5.5.29) $$
which is the standard light-cone superspace action [5.25].  As usual for the 
expansion of superfields into light-cone superfields,
the physical light-cone superfield appears in the middle of the
non-light-cone-$\q$ expansion of the gauge superfield, with auxiliary
light-cone superfields appearing at higher orders and pure gauge ones at
lower orders.  Because some of the ghost $\q$'s are commuting,
we therefore expect an infinite number of auxiliary fields in the
gauge-covariant action, as in the
harmonic superspace formalism [4.12].  This may be necessary in general,
because this treatment includes self-dual multiplets, such as 10D
super-Yang-Mills.  (This multiplet is superspin 0, and thus does not
require the superspin $\check M\sb{ij}$ to be self-dual, so it can be
treated in the OSp(1,1$|$2) formalism.  However, an additional
self-conjugacy condition on the light-cone superfield is required, (5.4.31b),
and a covariant OSp(1,1$|$2) statement of this condition would be necessary.) 
However, in some cases (such as
4D N=1 supersymmetry) it should be possible to truncate out all but a
finite number of these auxiliary superfields.  This would require an
(infinite) extension of the group OSp(1,1$|$2) (perhaps involving part
of the unphysical supersymmetries $a\sb\a q$), in the same way that
extending IGL(1) to OSp(1,1$|$2) eliminates Nakanishi-Lautrup auxiliary fields.

The unusual form of the $OSp(1,1|2)$ operators for supersymmetric
particles may require new mechanics actions for them.
It may be possible to derive these actions by inverting the quantization
procedure, first using the BRST algebra to derive the hamiltonian and
then finding the gauge-invariant classical mechanics lagrangian.

\sect{Exercises}

\Item{(1)} Derive (5.1.2) from (5.1.1) and (3.1.11).
\Item{(2)} Show that, under the usual gauge transformation $A\to A
+\pa\l$, $exp [ -iq \int_{\t\sb i}^{\t\sb f} d\t \; \dt x \cdot A(x)]$
transforms with a factor $exp\{-iq[\l (x(\t\sb f)) -\l (x(\t\sb i))]\}$.
(In a Feynman path integral, this corresponds to a gauge transformation
of the ends of the propagator.)
\Item{(3)} Fourier transform (5.1.13), using (5.1.14).  Explicitly
evaluate the proper-time integral in the massless case to find the
coordinate-space Green function satisfying $\bo G(x,x') = \d\sp D
(x-x')$ for arbitrary $D>2$.  Do $D=2$ by differentiating with respect
to $x\sp 2$, then doing the proper-time integral, and finally
integrating back with respect to $x\sp 2$.  (There is an infinite
constant of integration which must be renormalized.)  For comparison, do
$D=2$ by taking the limit from $D>2$.
\Item{(4)} Use the method described in (5.1.13,14) to evaluate the
1-loop propagator correction in $\f\sp 3$ theory.  Compare the
corresponding calculation with the covariantized light-cone method of
sect.\ 2.6.  (See exercise (7) of that chapter.)
\Item{(5)} Derive (5.2.1), and find $J\sp 3$.  Derive (5.2.2) and the
rest of the OSp(1,1$|$2) algebra.  Show these results agree with those
of sect.\ 3.4.
\Item{(6)} Quantize (5.3.10) in the 3 supersymmetric gauges described in
that section, and find the corresponding IGL(1) (and OSp(1,1$|$2), when
possible) algebras in each case, using the methods of sects.\ 3.2-3.
Note that the methods of sect.\ 3.3 require some generalization, since
commuting antighosts can be conjugate to the corresponding ghosts and
still preserve Sp(2):  $[ C\sb\a , C\sb\b ] \sim C\sb{\a\b}$.  Show
equivalence to the appropriate algebras of sects.\ 3.4-5.
\Item{(7)} Derive the tables in sect.\ 5.3. (Review the group theory of
SO(N) spinors, if necessary.)  Use the tables to derive the groups,
equivalent to SO($D\sb +$,$D\sb -$) for $D\le 6$, for which these
spinors are the defining representation.
\Item{(8)} Express the real-spinor $\g$-matrices of sect.\ 5.3 in terms
of the $\s$-matrices there for arbitrary $D\sb +$ and $D\sb -$.
Use the Majorana representation where the spinor is not necessarily 
explicitly real, but equivalent to a real one, such that: (1) for
complex representations, the bottom half of the spinor is the complex
conjugate of the top half (each being irreducible); (2) for pseudoreal
representations, the bottom half is the complex conjugate again but with
the index converted with a metric to make it explicitly the same
representation as the top; (3) for real representations, the spinor is
just the real, irreducible one.  Find the matrices representing the
internal symmetry (U(1) for complex and SU(2) for pseudoreal).
\Item{(9)} Check the Jacobi identities for the covariant derivatives
whose algebra is given in (5.4.8).  Check closure of the algebra (5.4.10)
in $D=10$, including the extra generator described in the text.
\Item{(10)} Derive (5.4.13).
\Item{(11)} Write the explicit action and transformation laws for
(5.4.16).
\Item{(12)} Write the explicit equations of motion (5.4.12), modified by
(5.4.23), for a scalar superfield for N=1 supersymmetry in D=4.  Show
that this gives the usual covariant constraints and field equations (up to
constants of integration) for the chiral scalar superfield (scalar
multiplet).  Do the same for a spinor superfield, and obtain the
equations for the vector-multiplet field strength.
\Item{(13)} Derive the explicit form of the twistor fields for $D$=3,4,6.
Find an explicit expression for the supersymmetrized Pauli-Lubansky
vector in $D$=4 in terms of supertwistors, and show that it
automatically gives an explicit expression for the superhelicity $H$ of
(5.4.23) as an operator in supertwistor space.  Show the supertwistor
Pauli-Lubansky vector automatically vanishes in $D$=3, and derive an
expression in $D$=6.

%
%

\chsc{6. CLASSICAL MECHANICS}{6.1. Gauge covariant}4

In this chapter we'll consider the mechanics action for the string and
its gauge fixing, as a direct generalization of the treatment of the
particle in the previous chapter.

The first-order action for string mechanics is obtained by 
generalizing the 1-dimensional particle mechanics world-line of 
(5.1.1) to a 2-dimensional world sheet [6.1]:
$$ S = {1 \over \a '} \int {d \sp 2 \s \over 2 \p} \; \left[ ( \pa \sb {\bf m} 
X ) \cdot P \sp {\bf m} + {\Sc g} \sb {\bf mn} \ha P \sp {\bf m} \cdot P
\sp{\bf n} \right] \quad ,\eqno(6.1.1)$$
where $X ( \s \sp {\bf m} )$ is the position in the higher-dimensional 
space in which the world sheet is imbedded of the point whose location 
in the world sheet itself is given by $\s \sp {\bf m} = ( \s \sp{\bf 0} , 
\s \sp{\bf 1} ) = ( \t , \s )$, 
$d \sp 2 \s = d \s \sp{\bf 0} d \s \sp{\bf 1} = d \t d \s$, $\pa \sb 
{\bf m} = ( \pa \sb{\bf 0} , \pa \sb{\bf 1} ) = ( \pa / \pa \t , \pa / \pa \s 
)$, and ${\Sc g}\sb {\bf mn} = ( - g ) \sp {-1/2} g \sb {\bf mn}$ is the 
unit-determinant part of the 2D metric.  (Actually, it has determinant $-1$.)
$1 / 2 \p \a '$ is both the 
string tension and the rest-mass per unit length.  (Their ratio, the 
square of the velocity of wave propagation in the string, is
unity in units of the speed of light:  The string is relativistic.)
This action is invariant under 2D general coordinate transformations
(generalizing (5.1.4); see sect.\ 4.1):
$$ \li{ \d X & = \e \sp{\bf m} \pa \sb{\bf m} X \quad , \cr
\d P \sp {\bf m} & = \pa \sb{\bf n} ( \e \sp{\bf n} P \sp{\bf m} ) -
P \sp {\bf n} \pa \sb{\bf n} \e \sp {\bf m} \quad , \cr
\d \Sc g \sp {\bf mn} & = \pa \sb{\bf p} ( \e \sp{\bf p} \Sc g \sp{\bf mn} ) -
\Sc g \sp {\bf p(m} \pa \sb{\bf p} \e \sp {\bf n)} \quad . & (6.1.2) \cr}$$

Other forms of this action which 
result from eliminating various combinations of the auxiliary fields 
$P \sp {\bf m}$ and ${\Sc g}\sb {\bf mn}$ are
$$\li{S &= {1\over\a '}\int{d\sp 2\s\over 2 \p} \; \left[ \dt X \cdot P 
	\sp{\bf 0} - {1\over\Sc g\sb{\bf 11}}\ha( P\sp{{\bf 0}2} + X ' \sp 2 )
	- {{\Sc g}\sb{\bf 01}\over{\Sc g}\sb{\bf 11}}
	P\sp{\bf 0} \cdot X ' \right] &(6.1.3a)\cr
&= {1\over\a '}\int{d\sp 2\s\over 2 \p} \; \left[ \dt X \cdot P 
	\sp{\bf 0} - \l\sb +\frac14 ( P\sp{\bf 0} + X ') \sp 2
	- \l\sb - \frac14 ( P\sp{\bf 0} - X ' ) \sp 2 \right] &(6.1.3b)\cr
&= - {1 \over \a '} \int {d \sp 2 \s \over 2 \p} \; {\Sc g} \sp {\bf mn}
	\ha ( \pa \sb {\bf m} X ) \cdot ( \pa \sb {\bf n} X ) &(6.1.3c)\cr
&= {1 \over \a '} \int {d \sp 2 \s \over 2 \p} \; \left[ ( \pa \sb {\bf m} X ) 
	\cdot P \sp {\bf m} + \sqrt {- det \; P \sp {\bf m} \cdot P \sp 
	{\bf n}} \right] &(6.1.3d)\cr
&= - {1 \over \a '} \int {d \sp 2 \s \over 2 \p}\; \sqrt {- det \; ( \pa \sb 
	{\bf m} X ) \cdot ( \pa \sb {\bf n} X )} = - {1 \over \a '} \int \sqrt
	{- ( d X \sp a \wedge d X \sp b ) \sp 2} \quad ,\cr
&&(6.1.3e)\cr}$$
where ${1 \over \a '} P \sp{\bf 0}$ is the momentum ($\s$-)density 
(the momentum $p = {1 \over \a '} \int {d \s \over 2 \p } \; P \sp{\bf 0}$),
\hbox{$' = \pa / \pa \s$,} and to obtain (6.1.3d) we have used the 
determinant of the $\Sc g$ equations of motion
$$ P \sp {\bf m} \cdot P \sp {\bf n} - \ha {\Sc g} \sp {\bf mn} 
{\Sc g} \sb {\bf pq} P \sp {\bf p} \cdot P \sp {\bf q} = 0 
\quad .\eqno(6.1.4)$$
As a consequence of this equation and the equation of motion for $P$,
the 2D metric is proportional to the ``induced'' metric $\pa\sb{\bf
m}X\cdot\pa\sb{\bf n}X$ (as appears in (6.1.3e)), which results from 
measuring distances in the usual Minkowski way in the D-dimensional 
space in which the 2D surface is imbedded (using 
$dX= d\s\sp{\bf m}\pa\sb{\bf m}X$).  (The equations of motion don't 
determine the proportionality factor, since only the unit-determinant 
part of the metric appears in the action.)
In analogy to the particle, (6.1.4) also represents the 
generators of 2D general-coordinate transformations.
(6.1.3a) is the hamiltonian form, (6.1.3b) is a rewriting of the
hamiltonian form to resemble the example (3.1.14) (but with the {\it
indefinite-metric sum} of squares of both left- {\it and} right-handed 
modes constrained),
(6.1.3c) is the second-order form, and (6.1.3e) is the area swept out by
the world sheet [6.2].  If the theory is derived from the form (6.1.3b), there
are 2 sets of transformation laws of the form (3.1.15) (with appropriate
sign differences), and (6.1.2,3c) can then be obtained as (3.1.16,17).

For the open string, the $X$ equations of motion also imply certain 
boundary conditions in $\s$.  (By definition, the closed string has no 
boundary in $\s$.)  Varying the $(\pa X ) \cdot P$ term and 
integrating by parts to pull out the $\d X$ factor, besides the 
equation of motion term $\pa P$ we also get a surface term $n \sb 
{\bf m} P \sp {\bf m}$, where $n \sb {\bf m}$ is the normal to the 
boundary.  If we assume 2D coordinates such that the position of the 
$\s$ boundaries are constant in $\t$, then we have the boundary condition
$$ P \sp{\bf 1} = 0 \quad .\eqno(6.1.5)$$

It's convenient to define the quantities
$$ \Hat P\sp{(\pm)} = {1\over\sqrt{2\a '}}(P\sp{\bf 0}\pm X') \eqno(6.1.6a)$$
because the hamiltonian constraints appearing in (6.1.3b) (equivalent
to (6.1.4)) can be expressed very simply in terms of them as
$$ \Hat P\sp{(\pm )2} =0 \quad , \eqno(6.1.6b)$$
and because they have simple Poisson brackets with each other.
For the open string, it's further useful to extend $\s$:  If we choose
coordinates such that $\s =0$ for all $\t$ at one end of the string,
and such that (6.1.5) implies $X'=0$ at that end, then we can define
$$ X(\s ) = X(-\s ) \quad , \quad \Hat P (\s ) = {1\over\sqrt{2\a '}}
(P\sp{\bf 0}+X') = \Hat P\sp{(\pm )}(\pm\s )~for~\pm\s >0\quad ,\eqno(6.1.7a)$$
so the constraint (6.1.6b) simplifies to
$$ \Hat P\sp 2 = 0 \quad . \eqno(6.1.7b)$$

\sect{6.2. Conformal gauge}

The conformal  gauge is given by the gauge conditions (on (6.1.1,3ac))
$$ {\Sc g} \sb {\bf mn} = \h \sb {\bf mn} \quad ,\eqno(6.2.1)$$
where $\h$ is the 2D flat (Minkowski) space metric.  (Since $\Sc g$ is
unit-determinant, it has only 2 independent components, so the 2
gauge parameters of (6.1.2) are sufficient to determine it completely.)
As for the 
particle, this gauge can't be obtained everywhere, so it's imposed 
everywhere except the boundary in $\t$.  Then variation of $\Sc g$ 
at initial or final $\t$ implies (6.1.4) there, and the remaining 
field equations then imply it everywhere.  In this gauge those 
equations are
$$ P \sb {\bf m} = - \pa \sb {\bf m} X \quad , \quad \pa \sb {\bf m} P 
\sp {\bf m} = 0 \quad \to \quad \pa \sp 2 X = 0 \quad .\eqno(6.2.2)$$
It is now easy to see that the endpoints of the string travel at the
speed of light.  (With slight generalization, this can be shown in
arbitrary gauges.)  From (6.1.4,5) and (6.2.1,2), we find that $dX =
d\t\dt X$, and thus $dX\cdot dX = d\t\sp 2 \dt X\sp 2 = d\t\sp 2 ( \dt
X\sp 2 + X'\sp 2 ) = 0$.
(6.2.2) is most easily solved by the use of 2D light-cone coordinates
$$ \s \sp{\bf\pm} = {1 \over \sqrt 2} ( \s \sp{\bf 1} \mp \s \sp{\bf 0} ) \to
\h \sb {\bf mn} = \pmatrix{0&1\cr 1&0\cr} = \h \sp {\bf mn} \quad ,
\eqno(6.2.3)$$
where 2D indices now take the values ${\bf\pm}$.  We then have
$$ \pa\sb{\bf +}\pa\sb{\bf -}X = 0 \to X = \ha [ \hat X \sp{(+)} ( \t + \s ) + 
\hat X \sp{(-)} ( \t - \s ) ] \quad .\eqno(6.2.4a)$$
For the open string, the boundary condition at one boundary, chosen to 
be $\s = 0$, is
$$ ( \pa\sb{\bf +}+\pa\sb{\bf -}) X = 0 \to\dt{\hat X} \sp{(+)} ( \t ) = \dt 
{\hat X} \sp{(-)} ( \t ) \to \hat X \sp{(+)}( \t ) =\hat X\sp{(-)}( \t )\quad ,
\eqno(6.2.4b)$$
without loss of generality, since the constant parts of $\hat X \sp{(\pm )}$
appear in $X$ only as their sum.  Thus, the modes of the open string 
correspond to the modes of one handedness of the closed string.  The 
boundary condition at the other boundary of the open string, taken as 
$\s = \p$, and the ``boundary'' condition of the closed string, which is 
simply that the ``ends'' at $\s = \pm \p$ are the same point (and thus 
the closed string $X$ is periodic in $\s$ with period $2 \p$, or
equivalently $X$ and $X'$ have the same values at $\s = \p$ as 
at $\s = -\p$) both take the form
$$ \dt{\hat X} \sp{(\pm )} ( \t + 2 \p ) = \dt{\hat X} \sp{(\pm )} ( \t
) \quad\to\quad \hat X\sp{(\pm )} (\t +2\p ) = \hat X\sp{(\pm )} (\t ) +
4\p\a ' p\sp{(\pm )} \quad , \eqno(6.2.4c)$$
$$ p\sp{(+)} = p\sp{(-)} \quad . \eqno(6.2.4d)$$
The constraints (6.1.4) also simplify:
$$ P \sp {{\bf\pm} 2}( \t , \s ) = \ha \hat X \sp{(\pm )} ' \sp 2 (\t \pm \s ) 
= 0 \quad .\eqno(6.2.5)$$
These constraints will be used to build the BRST algebra in chapt.\ 8.

The fact that the modes of the open string correspond to half the 
modes of the closed string (except that both have 1 zero-mode) means 
that the open string can be formulated as a closed string with modes of 
one handedness (clockwise or counterclockwise).  This is accomplished 
by adding to the action (6.1.1) for the closed string the term
$$ S \sb 1 = \int {d \sp 2 \s \over 2 \p} \; \ha u\sb a u\sb b\l\sb{\bf mn}
\ha ( \Sc g \sp {\bf mp} - \e \sp {\bf mp} )
\ha ( \Sc g \sp {\bf nq} - \e \sp {\bf nq} )
( \pa \sb {\bf p} X \sp a ) ( \pa \sb {\bf q} X \sp b ) \quad 
,\eqno(6.2.6)$$
where $u$ is a constant, timelike or lightlike (but not spacelike)
vector ($u\sp 2 \le 0$), and $\e\sp{+-}=-1$.
$\l$ is a lagrange multiplier which constrains $( u\cdot\pa\sb{\bf -} X 
)\sp 2 = 0$, and thus $u\cdot \pa\sb{\bf -} X = 0$, in the 
gauge (6.2.1).  Together with (6.2.5), this implies the Lorentz
covariant constraint $\pa\sb - X\sp a = 0$, so $X$ depends only on 
$\t - \s$, as in (3.1.14-17).  Thus, the formulation using (6.2.6) is
Lorentz covariant even though $S\sb 1$ is not manifestly so (because of
the constant vector $u$).
We can then identify the new $X$ with the $\hat X$ of (6.2.4).
Since $\l$ itself appears multiplied by $\pa\sb{\bf -} X$ 
in the equations of motion, it thus drops out, implying that it's a 
gauge degree of freedom which, like $\Sc g$, can be gauged away except 
at infinity.  

Similar methods can be applied to the one-handed modes of the heterotic
string [1.13].  (Then in (6.2.6) only spacelike $X$'s appear, so instead
of $u\sb a u\sb b$ any positive-definite metric can be used, effectively
summing over the one-handed $X$'s.)
Various properties of the actions (3.1.17, 6.2.6) have been discussed 
in the literature [6.3], particularly in relation to anomalies in 
the gauge symmetry of the lagrange multipliers upon naive lagrangian 
quantization.  One simple way to avoid these anomalies while keeping a 
manifestly covariant 2D lagrangian is to add scalars $\f$ with the 
squares of both $\pa\sb -\f$ and $\pa\sb +\f$ appearing in 
lagrange-multiplier terms [6.4].
Alternatively (or additionally), one can add Weyl-Majorana 2D spinors 
(i.e., real, 1-component, 1-handed spinors) whose nonvanishing 
energy-momentum tensor component couples to the appropriate lagrange 
multiplier.  (E.g., a spinor with kinetic term $\j\pa\sb +\j$ appears 
also in the term $\l [ (\pa\sb -\f)\sp 2 + \j\pa\sb -\j ]$.)  
These nonpropagating fields appear together with scalars with only one 
or the other handedness or neither constrained, and unconstrained 
fermions which are Weyl and/or Majorana or neither.  There are (at most)
2 lagrange multipliers, one for each handedness.

In the conformal gauge there is still a residue of the gauge 
invariance, which originally included not only 2D general coordinate 
transformations but also local rescalings of the 2D metric (since only 
its unit-determinant part appeared in the action).  By definition, the 
subset of these transformations which leave (6.2.1) invariant is
the conformal group.  Unlike in higher dimensions, the 2D 
conformal group has an infinite number of generators.  It can easily 
be shown that these transformations consist of the coordinate 
transformations (restricted by appropriate boundary conditions)
$$ \s \sp{\bf\pm} ' = \z \sp{\bf\pm} ( \s \sp{\bf\pm} ) \quad ,\eqno(6.2.7a)$$
with $\pm$'s not mixing (corresponding to 2 1D general coordinate 
transformations), since these coordinate transformations have an effect 
on the metric which can be canceled by a local scaling:
$$ d\s\sp 2 ' = 2d\s\sp{\bf +}' d\s\sp{\bf -} ' = \z\sp{\bf +}' \z\sp{\bf -}' 
2 d \s \sp{\bf +} d \s \sp{\bf -} \quad .\eqno(6.2.7b)$$
On shell, these transformations are sufficient to gauge away one 
Lorentz component of $X$, another being killed by the constraint (6.2.5).
These 2 Lorentz components can be eliminated more directly by 
originally choosing stronger gauge conditions, as in the light-cone gauge.

The conformal gauge is a temporal gauge, since it is equivalent to
setting the time components of the gauge field to constants:
$ \Sc g\sb{\bf m0} = \h\sb{\bf m0}$.  When generalized to $D>2$, it is
the choice of Gaussian normal coordinates.  We can instead choose a
Lorentz gauge, $\pa\sb{\bf m}\Sc g\sp{\bf mn} = 0$.  This is the De
Donder gauge, or harmonic coordinates, which is standardly used in $D>2$.
We'll discuss this gauge in more detail in sect.\ 8.3.

\sect{6.3. Light cone}

In a light-cone formalism [6.5] not only are more gauge degrees of 
freedom eliminated than in covariant gauges, but 
also more (Lorentz) auxiliary fields.  We do the latter first by 
varying the action (6.1.1) with respect to all fields carrying a 
``$-$'' Lorentz index ($X \sb -$, $P \sp {\bf m} \sb -$):
$$ \li{{\d \over \d X \sb -} &\to \pa \sb {\bf m} P \sp {\bf m} \sb + = 0
	\quad ;&(6.3.1a)\cr
{\d \over \d P \sp {\bf m} \sb -} &\to \Sc g \sb {\bf mn} = ( A \sp 
	{\bf r} B \sb {\bf r} ) \sp {-1} ( \e \sb {\bf pm} A \sp {\bf p} \e 
	\sb {\bf qn} A \sp {\bf q} - B \sb { \bf m} B \sb { \bf n} ) 
	\quad , \cr}$$
$$ A \sp {\bf m} = P \sp {\bf m} \sb + \quad , \quad B \sb {\bf m} = 
\pa \sb {\bf m} X \sb + \quad .\eqno(6.3.1b)$$

We next eliminate all fields with a ``$+$'' index by gauge conditions:
$$ \li{\t : \quad& X \sb + = k \t \cr
\s : \quad& P \sp{\bf 0} \sb + = k \quad ,&(6.3.2a)\cr}$$
where $k$ is an arbitrary constant.  (The same procedure is applied 
in light-cone Yang-Mills, where $A \sb -$ is eliminated as an 
auxiliary field and $A \sb +$ as a gauge degree of freedom: see sect.\ 
2.1.)  The 
latter condition determines $\s$ to be proportional to the amount of 
$+$-momentum between $\s = 0$ and the point at that value of $\s$ (so 
the string length is proportional to $\int d \s P \sp{\bf 0} \sb +$, which 
is a constant, 
since $\pa \sb {\bf m} P \sp {\bf m} \sb + = 0$).  Thus, $\s$ is 
determined up to a function of $\t$ (corresponding to the choice of 
where $\s = 0$).  However, $P \sp{\bf 1} \sb +$ is also 
determined up to a function of $\t$ (since now $\pa \sb{\bf 1} P \sp{\bf 1} \sb
+ = 0$), so $\s$ is completely determined, up to global 
translations $\s \to \s + constant$, by the further condition 
$$P \sp{\bf 1} \sb + = 0 \quad .\eqno(6.3.2b)$$
(6.3.2) implies (6.2.1).
For the open string, by the converse of the argument leading to the 
boundary condition (6.1.5), this determines the values of $\s$ at the 
boundaries up to constants, so the remaining global invariance is used 
to choose $\s = 0$ at one boundary.  For the closed string, the global 
invariance remains, and is customarily dealt with in the quantum 
theory by imposing a constraint of invariance under this transformation
on the field or first-quantized wave function.  

The length of the string is then given by integrating 
(6.3.2a):  $p \sb + = {1 \over \a '} \int {d \s \over 2 \p} P \sp{\bf 0} \sb +=
(1 / 2 \p \a ' ) k \cdot length$.  The two most convenient choices are
$$\li{k = 1 \quad &\to \quad length = 2 \p \a ' p \sb + \cr
length = \p \; ( 2 \p ) \quad &\to \quad k = 2 \a ' p \sb + \; 
( \a ' p \sb + ) ~ for ~ open ~ (closed) \quad .&(6.3.3)\cr}$$
For the free string the latter choice is more convenient for the 
purpose of mode expansions.  In the case of interactions $k$ must be 
constant even through interactions, and therefore can't be identified 
with the value of $p \sb +$ of each string, so the former choice is 
made.  Note that for $p \sb + < 0$ the string then has negative length.  It 
is then interpreted as an {\it anti}string (or outgoing string, as 
opposed to incoming string).  The use of negative 
lengths is particularly useful for interactions, since then the 
vertices are (cyclically) symmetric in all strings: e.g., a string of length 
$1$ breaking into 2 strings of length $\ha$ is equivalent to a string 
of length $\ha$ breaking into strings of length $1$ and $- \ha$.

The action now becomes
$$ S = \int d \t \left\{ \dt x \sb - p \sb + + {1 \over \a '} \int {d \s \over
2 \p} \left[ ( \pa \sb {\bf m} X \sb i ) P \sp {\bf m} \sb i + \h \sb 
{\bf mn} \ha P \sp {\bf m} \sb i P \sp{\bf n} \sb i \right]\right\}
\quad ,\eqno(6.3.4a)$$
or, in hamiltonian form,
$$ S = \int d \t \left\{ \dt x \sb - p \sb + + {1 \over \a '} \int {d \s \over
2 \p} \left[ \dt X \sb i P \sp{\bf 0} \sb i - \ha ( P \sp{\bf 0} 
\sb i \sp 2 + X ' \sb i \sp 2 ) \right]\right\} \quad .\eqno(6.3.4b)$$
$X \sb i$ is found as in (6.2.4), but $X \sb +$ is given by (6.3.2), and $X 
\sb -$ is given by varying the original action with respect to the 
auxiliary fields ${\Sc g} \sb {\bf mn}$ and $P \sp {\bf m} \sb +$,
conjugate to those varied in (6.3.1):
$$\li{{\d \over \d {\Sc g} \sb {\bf mn}} \to& P \sp {\bf m} \sb - = - P 
	\sp {\bf m} \sb i P \sp{\bf 0} \sb i - \ha \d \sb{\bf 0} \sp {\bf m} \h
	\sb {\bf np} P \sp {\bf n} \sb i P \sp {\bf p} \sb i \cr
{\d \over \d P \sp {\bf m} \sb +} \to& X \sb - = x \sb - - \int {d \s 
	\over 2 \p} \; P \sp 	1 \sb - + constant \quad ,&(6.3.5)\cr}$$
where the constant is chosen to cancel the integral when integrated 
over $\s$, so that ${1 \over \a '} \int {d \s \over 2 \p} \dt X \sb - 
P \sp{\bf 0} \sb + = \dt x \sb - p \sb +$ in (6.3.4).  (This term has 
been restored in (6.3.4) in order to avoid using equations of motion to 
eliminate any coordinates
whose equations of motion involve time derivatives, and are thus not 
auxiliary.  In (6.3.1a) all but the zero-mode part of the $X \sb -$ 
equation can be used to solve for all but a $\s$-independent part of 
$P \sp{\bf 1} \sb +$, without inverting time derivatives.)

After the continuous 2D symmetries have been eliminated by coordinate 
choices, certain discrete symmetries remain: $\s$ and $\t$ reversal.
As in the particle case, $\t$ reversal corresponds to a form of charge 
conjugation.  However, the open string has a group theory factor associated 
with each end which is the complex-conjugate representation of that at 
the other end (so for unitary representations they can cancel for 
splitting or joining strings), so for charge conjugation the 2 ends 
should switch, which requires $\s$ reversal.  Furthermore, the closed 
string has clockwise and counterclockwise modes which are 
distinguishable (especially for the heterotic string), so again we 
require $\s$ reversal to accompany $\t$ reversal to keep $\s \sp{\bf\pm}$ 
from mixing.  We therefore define charge conjugation to be the 
simultaneous reversal of $\t$ and $\s$ (or $\s \to 2\p\a'p\sb + - \s$ to 
preserve the positions of the boundaries of the open string).  On the 
other hand, some strings are {\it nonoriented} (as opposed to the {\it 
oriented} ones above) in that solutions with $\s$ reversed are not 
distinguished (corresponding to open strings with real representations 
for the group theory factors, or closed strings with clockwise modes 
not separated from counterclockwise).  For such strings we also need 
to define a $\s$ reversal which, because of its action on the 2D 
surface, is called a ``twist''.   In the quantum theory these 
invariances are imposed as constraints on the fields or first-quantized 
wave functions (see chapt.\ 10).

\sect{Exercises}

\Item{(1)} Derive (6.1.3) from (6.1.1).  Derive (6.1.5).
\Item{(2)} Derive (6.3.1).

%
%

\Chsc{7. }{LIGHT-CONE QUANTUM }{MECHANICS}{7.1. Bosonic}4

In this section we will quantize the light-cone gauge bosonic string
described in sect.\ 6.3 and derive the Poincar\'e algebra, which is a
special case of that described in sect.\ 2.3.

The quantum mechanics of the free bosonic string is described in the 
light-cone formalism [6.5] in terms of the independent coordinates $X \sb i 
( \s )$ and $x \sb -$, and their canonical conjugates $P \sp{\bf 0} \sb i ( 
\s )$ and $p \sb +$, with ``time'' ($x \sb + = 2\a' p\sb + \t$)
dependence given by the hamiltonian (see (6.3.4b))
$$ H = {1 \over \a '} \int {d \s \over 2 \p} \; \ha ( P \sp{\bf 0} \sb i
\sp 2 + X \sb i ' \sp 2 ) \quad .\eqno(7.1.1)$$
Functionally,
$$ {1 \over \a '} P \sp{\bf 0} \sb i ( \s ) = i {\d \over \d X \sb i ( \s )} 
\quad , \quad \left[ {\d \over \d X \sb i ( \s \sb 1 )} , X \sb j ( \s \sb 2 ) 
\right] = \d \sb {ij} 2 \p \d ( \s \sb 2 - \s \sb 1 ) \quad .\eqno(7.1.2)$$
(Note our unconventional normalization for the functional derivative.)

For the open string, it's convenient to extend $\s$ from $[ 0 , \p ]$ 
to $[ - \p , \p ]$ as in (6.1.7) by defining
$$ X ( - \s ) = X ( \s ) \quad , \quad \Hat P ( \s ) = {1 \over \sqrt 
{2\a '}} \left( \a ' i {\d \over \d X} + X ' \right) 
= {1\over\sqrt{\a '}}P\sp\pm (\pm\s )~for~\pm\s >0 \quad ,\eqno(7.1.3a)$$
so that $\Hat P$ is periodic (and $\sim P \sp +$ or $P \sp -$, as in 
(6.2.5), for $\s >$ or $< 0$), {\it or} to express the open string as a 
closed string with modes which propagate only clockwise (or only 
counterclockwise) in terms of $\hat X$ of sect.\ 6.2:
$$\Hat P( \s ) = {1 \over \sqrt {2 \a '}} \hat X ' ( \s )\quad ,\eqno(7.1.3b)$$
which results in the same $\Hat P$ (same boundary conditions and 
commutation relations).  The latter interpretation will prove useful 
for graph calculations.  (However, the form of the interactions will 
still look different from a true closed string.)  Either way, the new 
boundary conditions (periodicity) allow (Fourier) expansion of all 
operators in terms of exponentials instead of cosines or sines.  
Furthermore, $\Hat P$ contains all of $X$ and $\d / \d X$ (except $x$, 
which is conjugate to $p \sim \int d \s \Hat P$; i.e., all the 
translationally invariant part).  In particular,
$$ H = \int_{- \p}^\p {d \s \over 2 \p} \; \ha \Hat P \sb i \sp 2 \quad . 
\eqno(7.1.4)$$
The commutation relations are
$$ [ \Hat P \sb i ( \s \sb 1 ) , \Hat P \sb j ( \s \sb 2 ) ] =
2 \p i \d ' ( \s \sb 2 - \s \sb 1 ) \d \sb {ij} \quad .\eqno(7.1.5)$$

For the closed string, we define 2 $\Hat P$'s by
$$ \Hat P \sp{(\pm )} ( \s ) = {1 \over \sqrt{2\a '}} \left[ \a ' i 
\fder{X ( \pm \s ) } \pm X ' ( \pm \s ) \right] = 
{1 \over \sqrt{2 \a '}} \hat X \sp{(\pm )} ' ( \s ) \quad . \eqno(7.1.6)$$
Then operators which are expressed in terms of integrals over $\s$ 
also include sums over $\pm$: e.g., $H = H \sp{(+)} + H \sp{(-)} $, with $H 
\sp{(\pm )}$ given in terms of $\Hat P \sp{(\pm )}$ by (7.1.4).

In order to compare with particles, we'll need to expand all operators 
in Fourier modes.  In practical calculations, functional techniques 
are easier, and mode expansions should be used only as the final step: external
line factors in graphs, or expansion of the effective action.  (Similar 
remarks apply in general field theories to expansion of superfields 
in components and expansion of fields about vacuum expectation 
values.)  For $\Hat P$ for the open string (suppressing Lorentz indices),
$$ \Hat P ( \s ) = \su _{n=-\infty}^\infty \a 
\sb n e \sp {-in\s} \quad,$$
$$ \a \sb 0 = \sqrt{2 \a '} p \quad , \quad \a \sb n = ( \a \sb {-n} ) 
\sp \dag = - i \sqrt n a \sb n \sp \dag \quad ,$$
$$ [ p , x ] = i \quad , \quad [ \a \sb m , \a \sb n ] = n \d \sb 
{m+n,0} \quad , \quad [ a \sb m , a \sb n \sp \dag ] = \d \sb {mn} 
\quad .\eqno(7.1.7a)$$
We also have
$$ \hat X ( \s ) = \hat X \sp \dag ( \s ) = \hat X \conj ( - \s ) = 
( x + 2 \a ' p \s ) + \sqrt {2 \a '} \su_1^\infty {1 \over 
\sqrt n} ( a \sb n \sp \dag e \sp {-in\s} + a \sb n e \sp {in\s} ) \quad ,$$
$$ \quad X ( \s ) = \ha \left[ \hat X ( \s ) + \hat X ( - \s ) \right] \quad , 
\quad P \sp{\bf 0} ( \s ) = \a ' i {\d \over \d X ( \s )} = \ha 
\left[ \hat X ' ( \s ) + \hat X ' ( - \s ) \right] \quad .\eqno(7.1.7b)$$
(The $\sp \dag$ and $\conj$ have the usual matrix interpretation if 
the operators are considered as matrices acting on the Hilbert space:
The $\sp \dag$ is the usual operatorial hermitian conjugate, whereas 
the $\conj$ is the usual complex conjugate as for functions, which 
changes the sign of momenta.  Combined they give the operatorial 
transpose, which corresponds to integration by parts, and thus also 
changes the sign of momenta, which are derivatives.)
$H$ is now defined with normal ordering:
$$ \li { H &= \ha : \su_{-\infty}^\infty \a \sb n \cdot \a \sb {-n} : 
+~ constant \cr
&= \a ' p \sb i \sp 2 + \su_1^\infty n a \sb n \sp \dag \cdot a \sb n + 
constant = \a' p \sb i \sp 2 + N + constant \quad .&(7.1.8a)\cr}$$
In analogy to ordinary field theory, $i \pa / \pa \t + H = \a' p \sp 2 
+ N + constant$.

The constant in $H$ has been introduced as a finite renormalization 
after the infinite renormalization done by the normal ordering:  As 
in ordinary field theory, wherever infinite renormalization is necessary 
to remove infinities, finite renormalization should also be considered 
to allow for the ambiguities in renormalization prescriptions.  
However, also as in ordinary field theories, the renormalization is 
required to respect all symmetries of the classical theory possible (otherwise 
the symmetry is anomalous, i.e., not a symmetry of the quantum 
theory).  In the case of any light-cone theory, the one symmetry which 
is never manifest (i.e., an automatic consequence of the notation)
is Lorentz invariance.  (This 
sacrifice was made in order that unitarity would be manifest by 
choosing a ghost-free gauge with only physical, propagating degrees of 
freedom.)  Thus, in order to prove Lorentz invariance isn't violated, 
the commutators of the Lorentz generators $J \sb {ab} = {1 \over \a '} \int {d 
\s\over 2\p}X\sb{[ a} P \sp{\bf 0} \sb {b]}$ must be checked.  All are trivial 
except $[ J\sb{-i} , J\sb{-j} ] = 0$ because $X\sb -$ and $P \sp{\bf 0} \sb -$ 
are quadratic in $X \sb i$ and $P \sp{\bf 0} \sb i$ by (6.3.5).  The proof 
is left as an (important) exercise for the reader that the desired 
result is obtained only if $D = 26$ and the renormalization constant in $H$ and
$P\sp{\bf 0}\sb -$ ($H = - {1\over\a '}\int{d\s\over 2\p}P\sp{\bf 0} \sb -$) is
given by
$$ H = \a ' p \sb i \sp 2 + N - 1 \equiv \a ' ( p \sb i \sp 2 + M \sp 2 ) 
\quad .\eqno(7.1.8b)$$
The constant in $H$ also follows from noting that the first excited level 
($a\sb{1i}\dg\sket{0}$) is just a transverse vector, which must be massless.
The mass spectrum of the open string, given by the operator $M \sp 2$, is 
then a harmonic oscillator spectrum:  The possible values of the mass squared 
are $-$1, 0, 1, 2, ... times $1 / \a '$, with spins at each mass level running 
as high as $N = \a ' M \sp 2 + 1$.  (The highest-spin state is created by 
the symmetric traceless part of $N$ $a \sb 1 \sp \dag$'s.)

Group theory indices are associated with the ends of the open string, so 
the string acts like a matrix in that space.  If the group is 
orthogonal or symplectic, the usual (anti)symmetry of the adjoint 
representation (as a matrix acting on the vector representation) is 
imposed not just by switching the indices, but by flipping the whole 
string (switching the indices and $\s \iff \p - \s$).  As a result, 
$N$ odd gives matrices with the symmetry of the adjoint representation 
(by including an extra ``$-$'' sign in the condition to put the massless 
vector in that representation), while $N$ even gives matrices with the 
opposite symmetry.  (A particular curiosity is the ``SO(1)'' open string, 
which has {\it only} $N$ even, and no massless particles.)
Such strings, because of their symmetry, are thus 
``nonoriented''.  On the other hand, if the group is just unitary, 
there is no symmetry condition, and the string is ``oriented''.
(The ends can be labeled with arrows pointing in opposite directions, 
since the hermitian conjugate state can be thought of as the 
corresponding ``antistring''.)

The closed string is treated analogously.  
The mode expansions of the coordinates and operators are given in 
terms of 2 sets of hatted operators $\hat X \sp{(\pm )}$ or $\Hat P \sp
{(\pm )}$ expanded over $\a \sp {(\pm )}\sb n$:
$$ X ( \s ) = \ha \left[ \hat X \sp{(+)} ( \s ) + \hat X \sp{(-)} ( - \s ) 
\right] \quad , \quad P \sp{\bf 0} ( \s ) = \ha \left[ \hat X \sp{(+)} ' ( \s )
+ \hat X \sp{(-)} ' ( - \s ) \right] \quad ;$$
$$ \hat X \sp{(\pm )} ( \s ) = \hat X \sp{(\pm )} \sp \dag ( \s ) = \hat X \sp
{(\pm )} \conj ( - \s ) \quad \to \quad \a \sp{(\pm )}\sb n = \a \sp{(\pm )}\sb
{-n} \sp \dag = - \a \sp{(\pm )}\sb n \conj \quad . \eqno(7.1.9)$$
However, because of the periodicity condition (6.2.4c), and
since the zero-modes (from (7.1.7)) appear only as their sum, 
they are not independent:
$$ x \sp{(\pm )} = x \quad , \quad p \sp{(\pm )} = \ha p \quad .\eqno(7.1.10)$$
Operators which have been integrated over $\s$ (as in (7.1.4)) can be expressed
as sums over two sets ($\pm$) of open-string operators: e.g., 
$$ H = H \sp{(+)} + H \sp{(-)} = \ha \a ' p \sb i \sp 2 + N - 2 \quad , 
\quad N = N \sp{(+)} + N \sp{(-)} \quad .\eqno(7.1.11)$$
Remembering the constraint under global $\s$ translations
$$ i {d \over d \s} \equiv \int {d \s \over 2 \p} \; X ' \cdot i {\d \over 
\d X} = \D N \equiv N \sp{(+)} - N \sp{(-)} = 0 \quad ,\eqno(7.1.12)$$
the spectrum is then given by the direct product of 2 open strings, 
but with the states constrained so that the 2 factors (from the 2 open 
strings) give equal contributions to the mass.  The masses squared are thus 
$-$4, 0, 4, 8, ... times $1 / \a '$, and the highest spin at any mass level 
is $N = \ha \a ' M \sp 2 + 2$ (with the corresponding state given by the 
symmetric traceless product of $N$ $a \sb 1 \sp \dag$'s, half of which 
are from one open-string set and half from the other).  Compared with 
$\a ' M \sp 2 + 1$ for the open string, this means that the ``leading Regge 
trajectory'' $N ( M \sp 2 )$ (see sec.\ 9.1) has half the slope and twice the 
intercept for the closed string as for the open string.  
If we apply the additional constraint of symmetry of the state under 
interchange of the 2 sets of string operators, the state is symmetric under 
interchange of $\s \iff - \s$, and is therefore ``nonoriented'';
otherwise, the string is ``oriented'', the clockwise and 
counterclockwise modes being distinguishable (so the string can carry 
an arrow to distinguish it from a string that's flipped over).

From now on we choose units 
$$ \a ' = \ha \quad . \eqno (7.1.13)$$

In the case of the open bosonic string, the free light-cone Poincar\'e 
generators can be obtained from the covariant expressions (the obvious
generalization of the particle expressions, because of (7.1.2), since
$X\sb a$ and $P\sp{\bf 0}\sb a$ are defined to Lorentz transform as
vectors and $X$ to translate by a constant),
$$ J \sb {ab} = - i \int_{-\p}^\p {d \s \over 2 \p} X \sb {[a} ( \s ) 
P\sp{\bf 0}\sb {b]} ( \s ) \quad , \quad p \sb a = \int {d \s \over 2 \p}
P\sp{\bf 0}\sb a ( \s ) \quad ,\eqno(7.1.14a)$$
by substituting the gauge condition (6.3.2) and free field equations
(6.1.7) (using (7.1.7))
$$ \hat X \sb + ( \s ) = p \sb + \s \quad , \quad \Hat P \sp 2 -2 = 0 
\quad \to \quad \Hat P \sb - = - {\Hat P \sb i \sp 2 -2 \over 2 p \sb +}$$
$$ \to \quad J \sb {ij} = - i x \sb {[i} p \sb {j]} + \su a \sp \dag \sb
{n[i} a \sb {nj]} \quad , $$
$$ J \sb {+i} = i x \sb i p \sb + \quad , \quad J \sb {-+} = - i x \sb -
p \sb + \quad , $$
$$ J \sb {-i} = - i ( x \sb - p \sb i - x \sb i p \sb - ) + 
\su a \sp \dag \sb {n[-} a \sb {ni]} \quad , $$
$$ p \sb - = - {1 \over 2 p \sb +} \left[ p \sb i \sp 2 + 2 \left( \su n
a \sp \dag \sb n \sp i a \sb {ni} - 1 \right) \right] \quad , \quad
a \sb {n-} = - {1 \over p \sb +} ( p \sp i a \sb {ni} + \D \sb n )
\quad , $$
$$ \D \sb n = i {1 \over \sqrt n} \left( \ha \su_{m=1}^{n-1} \sqrt{m(n-m)} 
a \sb m \sp i a \sb {n-m,i} - \su_{m=1}^\infty \sqrt{m(n+m)} 
a \sp \dag \sb m \sp i a \sb {n+m,i} \right) \quad . \eqno(7.1.14b)$$
(We have added the normal-ordering constant of (7.1.8) to the constraint
(6.1.7).)
As for arbitrary light-cone representations, the Poincar\'e generators
can be expressed completely in terms of the independent coordinates
$x\sb i$ and $x\sb -$, the corresponding momenta $p\sb i$ and $p\sb +$,
the mass operator $M$, and the generators of a spin SO(D$-$1), $M\sb{ij}$ 
and $M\sb{im}$.  For the open bosonic string, we have
the usual oscillator representation of the SO(D$-$2) spin generators
$$ M \sb {ij} = -i \int^\prime \hat X\sb i \Hat P\sb j =
\su_n a \sp \dag \sb {n[i} a \sb {nj]} \quad ,
\eqno(7.1.15a)$$
where $\int^\prime$ means the zero-modes are dropped.  The mass operator
$M$ and the remaining SO(D$-$1) operators $M \sb {im}$ are given by
$$ M \sp 2 = -2p\sb +\int^\prime\Hat P\sb - = \int^\prime (\Hat P\sb i\sp 2 -2)
= 2 \left( \su n a\dg\sb n\sp i a\sb{ni} -1 \right) = 2(N-1) \quad , $$
$$ M \sb {im} M = ip\sb + \int^\prime \hat X\sb i \Hat P\sb - =
\su \left( a\dg\sb{ni} \D\sb n - \D\dg\sb n a\sb{ni} \right) 
\quad . \eqno(7.1.15b) $$

The usual light-cone formalism for the closed string is not a true
light-cone formalism, in the sense that not all constraints have been
solved explicitly by eliminating variables:  The one constraint that
remains is that the contribution to the ``energy'' $p\sb -$ from the
clockwise modes is equal to that from the counterclockwise ones.
As a result, the naive Poincar\'e algebra does not close [4.10]:  Using the
expressions 
$$ J \sb {-i} = - i ( x \sb - p \sb i - x \sb i p \sb - ) + 
\su_{n,\pm} a \sp \dag \sp{(\pm )} \sb {n[-} a \sp{(\pm )} \sb {ni]} \quad , $$
$$ p \sb - = - {1 \over 2 p \sb +} \left[ p \sb i \sp 2 + 4 \left(
N\sp{(+)} + N\sp{(-)} - 2 \right) \right] \quad , \eqno(7.1.16a) $$
we find
$$ [ J \sb{-i} , J\sb{-j} ] = - {4\over p\sb + \sp 2} \D N \D J\sb{ij} 
\quad , \eqno(7.1.16b)$$
where $\D J\sb{ij}$ is the difference between the $(+)$ and $(-)$ parts
of $J\sb{ij}$.

We instead define 2 sets of open-string light-cone Poincar\'e 
generators $J\sp{(\pm )}\sb{ab}$ and $p\sp{(\pm )}\sb a$, built out of
independent zero- and nonzero-modes.  The
closed-string Poincar\'e generators are then [4.10]
$$ J\sb{ab} = J\sp{(+)}\sb{ab} + J\sp{(-)}\sb{ab} \quad , \quad
p\sb a = p\sp{(+)}\sb a + p\sp{(-)}\sb a \quad . \eqno(7.1.17a)$$
Since the operators
$$ \D p\sb a = p\sp{(+)}\sb a - p\sp{(-)}\sb a \eqno(7.1.17b)$$
commute with themselves and transform as a vector under the Lorentz
algebra, we can construct a Poincar\'e algebra from just $J\sb{ab}$ 
and $\D p\sb a$.  This is the Poincar\'e algebra whose extension is
relevant for string field theory; it closes off shell.  (This holds in
either light-cone or covariant quantization.)
However, as described above, this results in an unphysical doubling of
zero-modes.  This can be fixed by applying the constraints (see (7.1.10))
$$ \D p\sb a = 0 \quad . \eqno(7.1.18) $$
In the light-cone formalism, 
$$ \D p\sb + = \D p\sb i = 0 \eqno(7.1.19a) $$ 
eliminates independent zero-modes, while
$$ 0 = \D p\sb - = - {1\over p\sb +}( M\sp{2(+)} - M\sp{2(-)} ) 
= - {2\over p\sb +} \D N \eqno(7.1.19b) $$
is then the usual remaining light-cone constraint equating the
numbers of left-handed and right-handed modes.

The generators of the Lorentz subgroup again take the form (2.3.5), as
for the open string, and the operators appearing in $J\sb{ab}$ are 
expressed in terms of the open-string ones appearing in (7.1.15) as
$$ M\sb{ij} = \su M\sp{(\pm )}\sb{ij} \quad , $$
$$ M\sp 2 = 2\su M\sp{2(\pm )} = 4(N-2) \quad , \quad
N = \su N\sp{(\pm )} \quad , \quad \D N = N\sp{(+)} - N\sp{(-)} \quad ,$$
$$ M\sb{im}M = 2\su (M\sb{im}M)\sp{(\pm )} \quad . \eqno(7.1.20) $$
(Since $J\sb{ab}$ and $p\sb a$ are expressed as sums, so are $M\sb{ab}$
and $M$.  This causes objects quadratic in these operators to be
expressed as twice the sums in the presence of the constraint $\D p\sb a
= 0 \to p\sp{(\pm )}\sb a = \ha p\sb a$.)

These Poincar\'e algebras will be used to derive the OSp(1,1$|$2)
algebras used in finding gauge-invariant actions in sects.\ 8.2 and 11.2.

\sect{7.2. Spinning}

In this section we'll describe a string model with fermions obtained 
by introducing a 2D supersymmetry into the world sheet [7.1,5.1],
in analogy to sect.\ 5.3, and derive the corresponding Poincar\'e algebra.
The description of the superstring obtained by this method isn't
manifestly spacetime supersymmetric, so we'll only give a brief
discussion of this formalism before giving the present status of the 
manifestly supersymmetric formulation.  In both cases, (free) quantum
consistency requires $D=10$.

For simplicity, we go directly to the \hbox{(first-)}quantized 
formalism, since as far as the field theory is concerned the only 
relevant information from the free theory is how to construct the 
covariant derivatives from the coordinates, and then the equations of 
motion from the covariant derivatives.  
After quantization, when (in the Schr\"odinger picture) the coordinates 
depend only on $\s$, there still remains a 1D supersymmetry in
$\s$-space [7.2].  The covariant derivatives are 
simply the 1D supersymmetrization of the $\Hat P$ operators of sect.\ 7.1 
(or those of sect.\ 6.1 at the classical level, using Poisson brackets):
$$ \{ \hat D \sb a ( \s \sb 1 , \q \sb 1 ) , \hat D \sb b ( \s \sb 2 , 
\q \sb 2 ) \} = \h \sb {ab} {\bf d} \sb 2 2 \p \d ( \s \sb 2 - \s \sb 1 ) 
\d ( \q \sb 2 - \q \sb 1 ) \quad ,\eqno(7.2.1a)$$
$$ \hat D ( \s , \q ) = \hat \J ( \s ) + \q \Hat P ( \s ) $$
$$ \to [ \Hat P , \Hat P ] = as ~ before \quad , \quad
\{ \hat \J \sb a ( \s \sb 1 ) , \hat \J \sb b ( \s \sb 2 ) \} =
\h \sb {ab} 2 \p \d ( \s \sb 2 - \s \sb 1 ) \quad .\eqno(7.2.1b)$$
The 2D superconformal generators (the supersymmetrization of the generators
(6.1.7b) of 2D general coordinate transformations, or of just the
residual conformal transformations after the lagrange multipliers have
been gauged away) are then
$$ \ha \hat D {\bf d} \hat D = ( \ha \hat \J \cdot \Hat P ) + \q ( \ha \Hat 
P \sp 2 + \ha i \hat \J ' \cdot \hat \J ) \quad .\eqno(7.2.2)$$
There are 2 choices of boundary conditions:
$$ \hat D ( \s , \q ) = \pm \hat D ( \s + 2 \p , \pm \q ) \to
\Hat P ( \s ) = \Hat P ( \s + 2 \p ) \quad , \quad \hat \J ( \s ) = 
\pm \hat \J ( \s + 2 \p ) \quad .\eqno(7.2.3)$$
The $+$ choice gives fermions (the Ramond model), while the $-$ gives 
bosons (Neveu-Schwarz).

Expanding in modes, we now have, in addition to (7.1.7),
$$ \hat \J ( \s ) = \su_{-\infty}^\infty \g \sb n 
e \sp {-in\s} \quad \to \quad \{ \g \sp a \sb m , \g \sp b \sb n \} = 
\h \sp {ab} \d \sb {m+n,0} \quad , \quad \g \sb {-n} = \g \sb n \sp 
\dag \quad ,\eqno(7.2.4)$$
where $m,n$ are integral indices for the fermion case and 
half-(odd)integral for the bosonic.  The assignment of statistics 
follows from the fact that, while the $\g \sb n$'s are creation 
operators $\g\sb n = d \sb n \sp \dag$ for $n > 0$, they are $\g$ matrices 
$\g\sb 0 = \g / \sqrt 2$ for $n = 0$ (as in the particle case, but 
in relation to the usual 
$\g$ matrices now have Klein transformation factors for both $d \sb 
n$'s and, in the BRST case, the ghost $\Hat C$, or else the $d \sb n$'s and 
$\Hat C$ are related to the usual by factors of $\g \sb {11}$).  
However, as in the particle case, a functional 
analysis shows that this assignment can only be maintained if the 
number of anticommuting modes is even; in other words,
$$\g \sb {11} ( - 1 ) \sp {\su d \sp \dag d} = 1 \quad .\eqno(7.2.5)$$
(In terms of the usual $\g$-matrices, the $\g$-matrices here also
contain the Klein
transformation factor $( - 1 ) \sp {\su d \sp \dag d}$.  However, in
practice it's more convenient to use the $\g$-matrices of \hbox{(7.2.4-5),}
which anticommute with all fermionic operators, and equate them directly
to the usual matrices after all fermionic oscillators have been eliminated.)
As usual, if $\g\sb 0$ is represented explicitly as matrices, the
hermiticity in (7.2.4) means pseudohermiticity with respect to the time
component $\g\sb 0\sp 0$ (the indefinite metric of the Hilbert space of
a Dirac spinor).  However, if $\g\sb 0$ is instead represented as operators
(as, e.g., creation and annihilation operators, as for the usual
operator representation of SU(N)$\subset$SO(2N)), no explicit metric is
necessary (being automatically included in the definition of hermitian
conjugation for the operators).

The rest is similar to the bosonic formalism, and is straightforward 
in the 1D superfield formalism.  For example,
$$ H = \int {d \s \over 2 \p} \; d \q \; \ha \hat D {\bf d} \hat D = 
\int {d \s \over 2 \p} \; ( \ha \Hat P \sp 2 + \ha i \hat \J ' \hat \J ) 
= \ha (p \sp 2 + M \sp 2) \quad ,$$
$$ M \sp 2 = 2\su_{n>0} n ( a \sb n \sp \dag \cdot a \sb n + d \sb n 
\sp \dag \cdot d \sb n ) \quad .\eqno(7.2.6a)$$
For the fermionic sector we also have (from (7.2.2))
$$ \int {d \s \over 2 \p} \; \hat \J \cdot \Hat P = {1\over\sqrt 2}
( \sl p + \Tilde M ) \quad , \quad \Tilde M\sp 2 = M\sp 2 \quad .
\eqno(7.2.6b)$$
(As described above, the $d$'s anticommute with $\g$, and thus
effectively include an implicit factor of $\g\sb{11}$.  $\Tilde M$ is
thus analogous to the $\SL M$ of (4.5.12).)
The fermionic ground state is massless (especially due to the above 
chirality condition), but the bosonic ground state is a tachyon.
(The latter can most easily be seen, as for the bosonic string, by
noting that the first excited level consists of only a massless vector.)  
However, consistent quantum interactions require truncation to the 
spectrum of the superstring described in the next section.  This means,
in addition to the chirality condition (7.2.5) in the fermionic sector,
the restriction in the bosonic sector to even $M\sp 2$ [7.3].  (Unlike the
fermionic sector, odd $M\sp 2$ is possible because of the half-integral
mode numbers.)  As for the bosonic string, besides determining the
ground-state masses Lorentz invariance also fixes the dimension, now $D=10$.

In the light-cone formulation of the spinning string we have instead of
(7.1.14) [7.4]
$$ J \sb {ab} = \int {d\s \over 2\p} \left( - i X \sb {[a} P\sp{\bf 0}\sb {b]}
+ \ha \hat\J \sb {[a} \hat\J \sb {b]} \right) \quad , $$
$$ \hat X \sb + = p \sb + \s \quad , \quad \hat\J \sb + = 0 \quad ; \quad
\Hat P \sp 2 + i \hat\J ' \cdot \hat\J = \hat\J \cdot \Hat P = 0 $$
$$ \to \quad \Hat P \sb - = - {1 \over 2 p \sb +} \left(
\Hat P \sb i \sp 2 + i \hat\J ' \sp i \hat\J \sb i \right) \quad , \quad
\hat\J\sb - = - {1\over p\sb +} \hat\J\sp i \Hat P\sb i \quad .\eqno(7.2.7)$$
The resulting component expansion for the Neveu-Schwarz string is
similar to the non-spinning bosonic string, with extra contributions
from the new oscillators.  For the case of the Ramond string, comparing to
(2.3.5), we find in place of (7.1.15)
$$ M \sp 2 = 2 \su n \left( a \sp\dag \sb n \sp i a \sb {ni} + 
d \sp\dag \sb n \sp i d \sb {ni} \right) \quad , $$
$$ M \sb {ij} = \frac14 \g \sb {[i} \g \sb {j)} + 
\su \left( a \sp\dag \sb {n[i} a \sb {nj)} + d \sp\dag \sb {n[i} d \sb {nj)} 
\right) \quad , $$
$$ M \sb {im} M = \ha \left( \g \sb i \Tilde M + \ha \g \sp j \S \sb
{ji} + \S \sb i \sp j \g \sb j \right) + \left[ \left(
a \sp\dag \sb {ni} \D \sb n - \D \sp\dag \sb n a \sb {ni} \right) +
\left( d \sp\dag \sb {ni} \X \sb n - \X \sp\dag \sb n d \sb {ni}
\right) \right] \; , $$
$$ \Tilde M = \S \sp i \sb i \quad , \quad \S \sb {ij} = i \su
\sqrt{2n} \left( d \sp\dag \sb {ni} a \sb {nj} - d \sb {ni} a \sp\dag
\sb {nj} \right) \quad , $$
$$ p \sb - = - {1 \over 2 p \sb +} ( p \sb i \sp 2 + M \sp 2 ) \quad ,
\quad \g \sb - = - {1 \over p \sb +} ( \g \sp i p \sb i + \Tilde M ) \quad , $$
$$ a \sb {n-} = - {1 \over p \sb +} \left( p \sp i a \sb {ni} - i \ha
\sqrt{\frac n2} \g \sp i d \sb {ni} + \D \sb n \right) \quad , \quad
d \sb {n-} = - {1 \over p \sb +} \left( p \sp i d \sb {ni} + i
\sqrt{\frac n2} \g \sp i a \sb {ni} + \X \sb n \right) \quad , $$
$$ \li{ \D \sb n = & i {1 \over \sqrt n} \Bigg\{ \ha \su_{m=1}^{n-1} \left[
\sqrt{m(n-m)} \; a \sb m \sp i a \sb {n-m,i} + ( m - \frac n2 ) d \sb m \sp
i d \sb {m-n,i} \right] \cr
& - \su_{m=1}^\infty \left[ \sqrt{m(n+m)} \; a
\sp\dag \sb m \sp i a \sb {n+m,i} + ( m + \frac n2 ) d \sp\dag \sb m \sp i
d \sb {n+m,i} \right] \Bigg\} \quad , \cr} $$
$$ \X \sb n = i \left[ \su_{m=1}^{n-1} \sqrt m \; a \sb m \sp i d \sb
{n-m,i} + \su_{m=1}^\infty \left( \sqrt{n+m} \; d \sp\dag \sb m \sp i a \sb
{n+m,i} - \sqrt m \; a \sp\dag \sb m \sp i d \sb {n+m,i} \right) \right]
\quad . \eqno(7.2.8) $$

This algebra can be applied directly to obtain gauge-invariant actions,
as was described in sect.\ 4.5.

\sect{7.3. Supersymmetric}

We now obtain the superstring [7.3] as a combined generalization of the 
bosonic string and the superparticle, which was described in sect.\ 5.4.   

Although the superstring can be formulated as a 
truncation of the spinning string, a manifestly supersymmetric 
formulation is expected to have the usual advantages that superfields 
have over components in ordinary field theories: simpler constructions 
of actions, use of supersymmetric gauges, easier quantum calculations, 
no-renormalization theorems which follow directly from analyzing 
counterterms, etc.  As usual, the free theory can be 
obtained completely from the covariant derivatives and equations of 
motion [7.5].  The covariant derivatives are defined by their affine Lie, or
Ka\v c-Moody
(or, as applied to strings, ``Ka\v c-Kradle''), algebra of the form 
$$ {1 \over 2 \p} [ \cg\sb i (\s\sb 1) , \cg\sb j (\s\sb 2) \} = 
\d (\s\sb 2 -\s\sb 1) f\sb{ij}\sp k \cg\sb k (\s\sb 1) + 
i \d ' (\s\sb 2 -\s\sb 1) g\sb{ij} \quad , \eqno(7.3.1)$$
where $f$ are the algebra's structure constants and $g$ its
(not necessarily Cartan) metric (both
constants).  The zero-modes of these generators give an ordinary
(graded) Lie algebra with structure constants $f$.
The Jacobi identities are satisfied if and only if
$$ f\sb{[ij|}\sp l f\sb{l|k)}\sp m = 0 \quad , \eqno(7.3.2a)$$
$$ f\sb{i(j|}\sp l g\sb{l|k]} = 0 \quad , \eqno(7.3.2b)$$
where the first equation is the usual Jacobi identity of a Lie algebra
and the second states the total (graded) antisymmetry of the structure
constants with index lowered by the metric $g$.  In this case, we wish to
generalize $\{ d \sb \a , d \sb \b \} = 2 \g \sp a \sb {\a\b} p \sb a$ for the 
superparticle and $[ \Hat P \sb a ( \s \sb 1 ) , \Hat P \sb b ( \s \sb 
2 ) ] = 2 \p i \d ' ( \s \sb 2 - \s \sb 1 ) \h \sb {ab}$ for the bosonic 
string.  The simplest generalization consistent with the Jacobi 
identities is:
$$\li{ \{ D \sb \a ( \s \sb 1 ) , D \sb \b ( \s \sb 2 ) \} &= 
2 \p \d ( \s \sb 2 - \s \sb 1 ) 2 \g \sp a \sb {\a\b} P \sb a ( \s \sb 1 )
\quad ,\cr
[ D \sb \a ( \s \sb 1 ) , P \sb a ( \s \sb 2 ) ] &= 
2 \p \d ( \s \sb 2 - \s \sb 1 ) 2 \g \sb {a\a\b} \O \sp \b ( \s \sb 1 )
\quad ,\cr
\{ D \sb \a ( \s \sb 1 ) , \O \sp \b ( \s \sb 2 ) \} &=
2 \p i \d ' ( \s \sb 2 - \s \sb 1 ) \d \sb \a \sp \b \quad ,\cr
[ P \sb a ( \s \sb 1 ) , P \sb b ( \s \sb 2 ) ] &= 
2 \p i \d ' ( \s \sb 2 - \s \sb 1 ) \h \sb {ab} \quad ,\cr
[ P , \O ] = \{ \O , \O \} &= 0 \quad ,&(7.3.3a)\cr}$$
$$ \g \sb {a(\a\b} \g \sp a \sb {\g )\d} = 0 \quad .\eqno(7.3.3b)$$
(7.3.2b) requires the introduction of the operator $\O$, and (7.3.2a)
then implies (7.3.3b).
This supersymmetric set of modes (as $\Hat P$ for the bosonic string) 
describes a complete open string or half a closed string, so two such 
sets are needed for the closed superstring, while the heterotic string 
needs one of these plus a purely bosonic set.  

Note the analogy with the super-Yang-Mills algebra (5.4.8):
$$ ( D \sb \a , P \sb a , \O \sp \a ) \iff ( \de \sb \a , \de \sb a , W
\sp \a ) \quad , \eqno(7.3.4) $$
and also that the constraint (7.3.3b) occurs on the $\g$-matrices, which
implies $D = 3, 4, 6,$ or 10 [7.6] when the maximal Lorentz invariance 
is assumed (i.e., all of \hbox{SO(D$-$1,1)} for the $D$-vector $P \sb a$).

This algebra can be solved in terms of $\Hat P \sb a$, a spinor
coordinate $\Q \sp \a ( \s ) $, and its derivative $\d / \d \Q \sp \a$:
$$ \li { D \sb \a  &= {\d \over {\d \Q \sp \a}} + \g \sp a \sb {\a \b} \Hat
P \sb a \Q \sp \b + \ha i \g \sp a \sb {\a \b} \g \sb {a \g \d} \Q \sp
\b \Q \sp \g \Q ' \sp \d \quad ,\cr
P \sb a &= \Hat P \sb a + i \g \sb {a \a \b} \Q \sp \a \Q ' \sp \b \quad ,\cr
\O \sp \a &= i \Q ' \sp \a \quad . &(7.3.5) \cr}$$
These are invariant under supersymmetry generated by
$$ \li { q \sb \a &= \int {d \s \over 2 \p} \; \left( {\d \over {\d \Q \sp 
\a}} - \g \sp a
\sb {\a \b} \Hat P \sb a \Q \sp \b - \frac16 i \g \sp a \sb {\a \b} \g \sb 
{a \g \d} \Q \sp \b \Q \sp \g \Q ' \sp \d \right) \quad ,\cr
p \sb a &= \int {d \s \over 2 \p} \; \Hat P \sb a \quad , &(7.3.6) \cr}$$
where $\{ q \sb \a , q \sb \b \} = - 2 \g \sp a \sb {\a \b} p \sb a$.

The smallest (generalized Virasoro) algebra which includes generalizations 
of the operators $\ha \Hat P \sp 2$ of the bosonic string and $\ha p \sp 2$ 
and $\sl p d$ of the superparticle is generated by
$$ \li { \ca &= \ha P \sp 2 + \O \sp \a D \sb \a = \ha \Hat P \sp 2 + i
\Q ' \sp \a { \d \over {\d \Q \sp \a}} \quad ,\cr
\cb \sp \a &= \g \sp {a \a \b} P \sb a D \sb \b \quad ,\cr
\cc \sb {\a \b} &= \ha D \sb {[ \a} D \sb {\b ]} \quad ,\cr
\cd \sb a &= i \g \sb a \sp {\a \b} D \sb \a D ' \sb \b \quad . &(7.3.7) \cr}$$
Note the similarity of ${\cal A }$ to (5.4.10), (7.2.6a), and (8.1.10,12).  
The algebra generated by these operators is (classically)
$$ \li { {1 \over 2 \p} [ \ca ( 1 ) , \ca ( 2 ) ] = 
&i \d ' ( 2 - 1 ) [ \ca ( 1 ) + \ca ( 2 ) ] \quad ,\cr
{1\over 2 \p} [ \ca ( 1 ) , \cb \sp \a ( 2 ) ] = &
i \d ' ( 2 - 1 ) [ \cb \sp \a ( 1 ) + \cb \sp \a ( 2 ) ] \quad ,\cr
{1 \over 2 \p} [ \ca ( 1 ) , \cc \sb {\a \b} ( 2 ) ] = &i \d ' 
( 2 - 1 ) [ \cc \sb {\a \b} ( 1 ) + \cc \sb {\a \b} 
( 2 ) ] \quad ,\cr
{1 \over 2 \p} [ \ca ( 1 ) , \cd \sb a ( 2 ) ] = &
i \d ' ( 2 - 1 ) [ \cd \sb a ( 1 ) + 2 \cd \sb a ( 2 ) ] 
\quad ,\cr
{1\over 2 \p} \{ \cb \sp \a ( 1 ) , \cb \sp \b ( 2 ) \} = &i \d '
(2 - 1 ) \ha \g \sp {a \a \g} \g \sb a \sp {\b \d} [ \cc \sb {\g 
\d} ( 1 ) + \cc \sb {\g \d} ( 2 ) ]
+ 4 \d ( 2 - 1 ) \cdot \cr
&\cdot [ \g \sp {a \a \b} ( P \sb a \ca + \frac18
\cd \sb a ) + ( \d \sb \g \sp {( \a} \d \sb \d \sp {\b )} - \ha \g \sb a \sp
{\a \b} \g \sp a \sb {\g \d} ) \O \sp \g \cb \sp \d ] \quad ,\cr
{1 \over 2 \p} [ \cb \sp \a ( 1 ) , \cc \sb {\b \g} ( 2 ) ] = 
&4\d ( 2 - 1 ) [ \d \sb {[ \b} \sp \a D \sb {\g ]} \ca + ( \d \sb
\d \sp \e \d \sb {[ \b} \sp \a - \ha \g \sb a \sp {\a \e} \g \sp a \sb 
{\d [ \b} ) \O \sp \d \cc \sb {\g ] \e} ] \quad ,\cr
{1 \over 2 \p} [ \cb \sp \a ( 1 ) , \cd \sb a ( 2 ) ] = &- 2 i 
\d ' ( 2 - 1 ) [ 2 \g \sb a \sp {\a \b} D \sb \b \ca + ( 3 \d 
\sb \g \sp \e \g \sb a \sp {\a \d} - \g \sb {ab \g} \sp \e \g \sp {b \a \d} ) 
\cdot\cr
&\cdot \O \sp \g \cc \sb {\d \e} ] ( 1 ) + 2 i \d ( 2
- 1 ) [ 4 \g \sb a \sp {\a \b} D ' \sb \b \ca \cr
&+ ( 3 \d \sb \g \sp \e \g \sb a \sp {\a \d} - \g \sb {ab \g} \sp \e \g \sp {b
\a \d} ) \O \sp \g \cc ' \sb {\d \e} - i \g \sb {a \b \g} \g \sp {b \a \g} \O 
\sp \b \cd \sb b ] \quad ,\cr
{1\over 2 \p} [ \cc \sb {\a \b} ( 1 ) , \cc \sb {\g \d} ( 2 ) ] =
&- 2 \d ( 2 - 1 ) P \sb a \g \sp a \sb {[ \g [ \a} \cc \sb {\b ] 
\d ]}\quad ,\cr
{1\over 2 \p} [ \cc \sb {\a \b} ( 1 ) , \cd \sb a ( 2 ) ] = &2 i 
\d ' (2 - 1 ) \g \sb a \sp {\g \d} \g \sb {b \g [ \a} P \sp b \cc
\sb {\b ] \d} ( 1 )\cr
&- 4 i \d ( 2 - 1 ) ( \g \sb {a \g [ \a} D ' \sb {\b ]} \cb \sp
\g + P \sp b \g \sb {b a [ \a} \sp \d \cc ' \sb {\b ] \d} ) \quad ,\cr
{1 \over 2 \p} [ \cd \sb a ( 1 ) , \cd \sb b ( 2 ) ] = 
&- \d ' ' (2 - 1 ) \g \sb {ab \a} \sp \b D \sb \b \cb \sp \a ( 1 ) + ( 2 )\cr
&+ \d ' ( 2 - 1 ) \left[ -4i P \sb {( a} \cd \sb {b )} 
+ \h \sb {ab} ( 3 D ' \sb \a \cb \sp \a - D\sb\a\cb\sp\a' ) \right]
( 1 ) - ( 2 )\cr
&+ \d (2 - 1 ) \left[ -2i ( 3 P ' \sb {[ a} \cd \sb {b ]} 
- P\sb{[a}\cd'\sb{b]}) \right. \cr
&\left. +2 \g \sb {ab \a}\sp \b ( 3D ' \sb \b \cb ' \sp \a -D''\sb\b\cb\sp\a ) 
+\g\sb{abc}\sp{\a\b} ( 3 P ' \sp c \cc ' \sb {\a \b}
- P''\sp c \cc\sb{\a\b} ) \right] \quad . \cr
&&(7.3.8) \cr}$$
(Due to identities like $\Sl P D D \sim \cb D \sim \Sl P \cc$,
there are other forms of some of these relations.)

BRST quantization can again be performed, and there are an infinite
number of ghosts, as in the particle case.  However, a remaining problem
is to find the appropriate ground state (and corresponding string field).  
Considering the results of sect.\ 5.4, this may
require modification of the generators (7.3.7) and BRST operator, perhaps to
include Lorentz generators (acting on the ends of the string?) or separate
contributions from the BRST transformations of Yang-Mills field theory
(the ground state of the open superstring, or of a set of modes of one
handedness of the corresponding closed strings).  On the
other hand, the ground state, rather than being Yang-Mills, might be purely
gauge degrees of freedom, with Yang-Mills appearing at some excited level,
so modification would be unnecessary.  The condition $Q \sp 2 = 0$ 
should reproduce the conditions $D = 10$, $\a \sb 0 = 1$.

The covariant derivatives and constraints can also be derived from a 2D 
lagrangian of the general form (3.1.10), as for the superparticle [7.5].
This classical mechanics Lagrangian imposes weaker constraints than the 
Green-Schwarz one [7.6] (which sets $D\sb \a = 0$ via Gupta-Bleuler), 
and thus should not impose stronger conditions.  

On the other hand, quantization in the light-cone formalism is 
understood.  Spinors are 
separated into halves, with the corresponding separation of the $\g$ 
matrices giving the splitting of vectors into transverse and 
longitudinal parts, as in (5.4.27).  The light-cone gauge is then chosen as
$$ P \sb + ( \s ) = p \sb + \quad , \quad \Q \sp - = 0 \quad .\eqno(7.3.9)$$
Other operators are then eliminated by auxiliary field equations:
$$ \li {\ca &= 0 \quad \to \quad P \sb - = -{1\over p\sb +}
\left( \ha \Hat P\sb i\sp 2 +i\Q\sp + ' D\sb + \right) \quad , \cr
\cb &= 0 \quad \to \quad D \sb - = {1\over \sqrt 2 p\sb +}
\Hat{\Sl P}\sb T\dg D\sb + \quad .&(7.3.10)\cr}$$
The remaining coordinates are $x \sb \pm$, $X \sb i ( \s )$, and $\Q 
\sp + ( \s )$, and the remaining operators are
$$ D \sb + = {\d \over \d \Q \sp +} + p \sb + \S \Q \sp + \quad , \quad
P \sb i = \Hat P \sb i  \quad , \quad \O \sp + = i \Q \sp + ' \quad .
\eqno(7.3.11)$$
However, instead of imposing $\cc = \cd = 0$ quantum mechanically, we
can solve them classically, in analogy to the particle case.  The
$\cc\sb{+ij} (\s )$ are now {\it local} (in $\s$) SO(8) generators, and
can be used to gauge away all but 1 Lorentz component of $D\sb +$,
by the same method as (5.4.34ab) [5.30,29].  After this, $\cd\sb +$ is
just the product of this one component times its $\s$-derivative.
Furthermore, $\cd\sb +$ is a Virasoro algebra for $D\sb +$,
and can thus be used to gauge away all but the zero-mode [5.30,29] (as the
usual one $\ca$ did for $P\sb +$ in (7.3.9)), after this constraint
implies $D\sb +$ factors in a way analogous to (5.4.34a):
$$ \cd\sb + = 0 \quad\to\quad D\sb + = c \x (\s ) \quad . \eqno(7.3.12)$$  
(The proof is identical, since $\cd\sb +=0$ is equivalent to
$D\sb + (\s\sb 1) D\sb + (\s\sb 2) =0$.)  We are thus back to the
particle case for $D$, with a single mode remaining, satisfying the
commutation relation $c\sp 2 = p\sb + \to c = \pm\sqrt{p\sb +}$, so $D$
is completely determined.  Alternatively, as for the particle, we could
consider $D=0$ as a second-class constraint [7.6], or impose the 
condition $D \sb + = 0$ (which eliminates all auxiliary string 
fields), as a Gupta-Bleuler constraint.  This requires a further 
splitting of the spinors, as in (5.4.29), and 
the Gupta-Bleuler constraint is again a chirality condition, as in 
(5.4.33).  $\cc = \cd = 0$ are then also satisfied ala Gupta-Bleuler (with 
appropriate ``normal ordering'').  Thus, in a ``chiral'' representation 
(as in ordinary supersymmetry) we have a chiral, ``on-shell'' 
string superfield, or wave function, 
$\F [ x \sb \pm , X \sb i ( \s ) , \Q \sp {\bf a} ( \s ) ]$, 
which satisfies a light-cone field equation
$$ \left( i {\pa \over \pa x \sb +} + H \right) \F = 0 \quad , $$
$$ H = -p\sb - = - {1 \over \a '} \int {d \s \over 2 \p} \; \Hat P \sb -
= \int {d \s \over 2 \p} \; \left( \ha \Hat P \sb i \sp 2 + i \Q \sp {\bf a} ' 
{\d \over \d \Q \sp {\bf a}} \right) \quad .\eqno(7.3.13)$$
The dimension of spacetime $D=10$ and the constant in $H$ 
(zero) are determined by considerations similar to those of the 
bosonic case (Lorentz invariance in the light-cone formalism, or BRST 
invariance in the yet-to-be-constructed covariant formalism).

Similarly, light-cone expressions for $q \sb \a$ can be obtained from 
the covariant ones:
$$ \li{ q \sb + &= \int {d \s \over 2 \p} \; \left( {\d \over \d \Q \sp +} - 
p \sb + \S \Q \sp + \right) \equiv \int {d\s\over 2\p}\; Q\sb +\quad ,\cr
q\sb - &= \int {d\s\over 2\p}\; {1\over\sqrt 2 p\sb +}\Hat{\Sl P}\sb
T\dg Q\sb + \quad .&(7.3.14)\cr}$$
If the superstring is formulated directly in the light cone, (7.3.14)
can be used as the starting point.  $Q\sb +$ and $D\sb +$ can be
considered as independent variables (instead of $\Q\sp +$ and $\d
/\d\Q\sp +$), defined by their self-conjugate commutation relations
(analogous to those of $\Hat P$):
$$ \li{ {1\over 2\p} \{ Q\sb + (\s\sb 1) , Q\sb + (\s\sb 2) \} & = 
- \d (\s\sb 2 -\s\sb 1) 2p\sb + \S \quad , \cr
{1\over 2\p} \{ D\sb + (\s\sb 1) , D\sb + (\s\sb 2) \} & = 
\d (\s\sb 2 -\s\sb 1) 2p\sb + \S \quad .&(7.3.15)\cr}$$
However, as described above and for the particle, $D\sb +$ is unnecessary
for describing physical polarizations, so we need not introduce it.
In order to more closely study the closure of the algebra (7.3.14), we
introduce more light-cone spinor notation (see sect.\ 5.3):  
Working in the Majorana representation $\S =I$, we introduce 
$(D-2)$-dimensional Euclidean $\g$-matrices as
$$ \sl p\sb T \to \g\sp i\sb{\m\n '}p\sb i \quad , $$
$$ \g\sp{(i}\sb{\m\m'}\g\sp{j)}\sb{\n\m'} = 2\d\sp{ij}\d\sb{\m\n} \quad ,
\quad \g\sp{(i}\sb{\m\m'}\g\sp{j)}\sb{\m\n'} = 2\d\sp{ij}\d\sb{\m'\n'} \quad ,
\eqno(7.3.16)$$
where not only vector indices $i$ but also spinor indices $\m$ and $\m'$
can be raised and lowered by Kronecker $\d$'s, and primed and unprimed
spinor indices are not necessarily related.  (However, as for the
covariant indices, there may be additional relations satisfied by the
spinors, irrelevant for the present considerations, that differ in
different dimensions.)  Closure of the supersymmetry algebra (on the
momentum in the usual way (5.4.4), but in light-cone notation, and with
the light-cone expressions for $p\sb i$, $p\sb +$, and $p\sb -$) then
requires the identity (related to (7.3.16) by ``triality'')
$$ \g\sb{i(\m |\m'}\g\sp i\sb{|\n )\n'} = 2 \d\sb{\m\n}\d\sb{\m'\n'}
\quad .\eqno(7.3.17)$$
This identity is actually (7.3.3b) in light-cone notation, and the
equality of the dimensions of the spinor and vector
can be derived by tracing (7.3.17) with $\d\sb{\m\n}$.

Returning to deriving the light-cone formalism from the covariant one, 
we can also obtain the light-cone expressions for the
Poincar\'e generators, which should prove important for covariant
quantization, via the OSp(1,1$|$2) method.  As in general, they are
completely specified by $M\sb{ij}$, $M\sb{im}M$, and $M\sp 2$:
$$ \li{ M\sb{ij} &= \int^\prime -i\hat X\sb i \Hat P\sb j + \cm\sb{ij} 
\quad , \cr
M\sb{im}M &= \int^\prime i\hat X\sb i p\sb +\Hat P\sb - + 
\cm\sb i\sp j\Hat P\sb j \quad , \cr
M\sp 2 &= \int^\prime -2p\sb + \Hat P\sb - \quad ,&(7.3.18)\cr}$$
where
$$ -p\sb +\Hat P\sb - = \ha\Hat P\sb i\sp 2 + i{1\over 8p\sb +} Q\g\sb +Q'
-i{1\over 8p\sb +}\cd\sb + \quad , $$
$$ \cm\sb{ij} = -{1\over 16p\sb +}Q\g\sb +\g\sb{ij}Q
+{1\over 16p\sb +}\cc\sb{+ij} \quad , \eqno(7.3.19)$$
contain all $D$ dependence (as opposed to $X$ and $Q$ dependence) only
in the form of $\cc$ and $\cd$, which can therefore be dropped.
(Cf.\ (5.4.22).  $\g\sb +$ picks out $Q\sb +$ from $Q$, as in (5.4.27).)

We can now consider deriving the BRST algebra by the method of adding
4+4 dimensions to the light-cone (sects.\ 3.6, 5.5).  Unfortunately,
adding 4+4 dimensions doesn't preserve (7.3.17).  In fact, from the
analysis of sect.\ 5.3, we see that to preserve the symmetries of the
$\s$-matrices requires increasing the number of commuting dimensions by
a multiple of 8, and the number of time dimensions by a multiple of 4.
This suggests that this formalism may need to be generalized to 
adding 8+8 dimensions to the light-cone (4 space, 4 time, 8 fermionic).
Coincidentally, the light-cone superstring has 8+8 physical
($\s$-dependent) coordinates, so this would just double the number of
oscillators.  Performing the reduction from OSp(4,4$|$8) to OSp(2,2$|$4)
to OSp(1,1$|$2), if one step is chosen to be U(1)-type and the other
GL(1)-type, it may be possible to obtain an algebra which has the
benefits of both formalisms.

As for the bosonic string, the closed superstring 
is constructed as the direct product of 2 open strings (1 for the
clockwise modes and 1 for the counterclockwise):  The
hamiltonian is the sum of 2 open-string ones, and the closed-string 
ground state is the product of 2 
open-string ones.  In the case of type I or IIB closed strings, the 2 
sets of modes are the same kind, and the former (the bound-state of 
type I open strings) is nonoriented (in order to be consistent with the
$N=1$ supersymmetry of the open string, rather than the $N=2$
supersymmetry generated by the 2 sets of modes of oriented, type II
strings).  Type IIA closed 
strings have $\Q$'s with the opposite chirality between the two sets 
of modes (i.e., one set has a $\Q \sp \a$ while the other has a $\Q \sb 
\a$).  The ground states of these closed strings are supergravity 
($N=1$ supergravity for type I and $N=2$ for type II).
The heterotic string is a closed string for which one set of 
modes is bosonic (with the usual tachyonic scalar ground state)
while the other is supersymmetric (with the usual supersymmetric 
Yang-Mills ground state).  The lowest-mass physical states, due to the 
$\D N = 0$ restriction, are the product of the massless sector of each 
set (since now $\D N = H \sp{(+)} - H\sp{(-)} = ( N\sp{(+)} - 1 ) -N\sp{(-)}$).
The dimension of spacetime for the 2 sets of modes is made consistent by 
compactification of some of the 26 dimensions of one and some (or 
none) of the 10 of the other onto a torus, leaving the same 
number of noncompactified dimensions (at least the physical 4) for 
both sets of modes.  These compactified bosonic modes can also be 
fermionized (see the next section), giving an equivalent formulation 
in which the extra dimensions don't explicitly appear:  For example,
fermionization of 16 of the dimensions produces 32 (real) fermionic
coordinates, giving an SO(32) internal symmetry (when the fermions are
given the same boundary conditions, all periodic or all antiperiodic).
The resulting spectrum for the massless sector of heterotic strings 
consists of supergravity coupled to supersymmetric 
Yang-Mills with $N=1$ (in 10D counting) supersymmetry.  The vectors 
gauging the Cartan subalgebra of the full Yang-Mills group are the 
obvious ones coming from the toroidal compactification (i.e., those 
that would be obtained from the noncompactified theory by just dropping 
dependence on the compactified coordinates), while the rest correspond 
to ``soliton'' modes of the compactified coordinates for which the 
string winds around the torus.  As for the dimension and ground-state 
mass, quantum consistency restricts the allowed compactifications, and 
in particular the toroidal compactifications are restricted to those 
which, in the case of compactification to $D=10$, give Yang-Mills 
group SO(32) or E$\sb 8 \otimes$E$\sb 8$.  (These groups give 
anomaly-free 10D theories in their massless sectors.  There is also an
SO(16)$\times$SO(16) 10D-compactification which can be considered to
have broken N=1 supersymmetry.  There are other 10D-compactifications which
have tachyons.)

Some aspects of the interacting theory will be described in chapts.\ 9
and 10.

\newpage

\sect{Exercises}

\Item{(1)} Use (7.1.7) as the {\it classical} solution for $\hat X\sp i$, 
and set $a\sb{ni} =0$ for $n\ne 1$ and $i\ne 1$.  Find $X\sp a (\s ,\t )$. 
In the center-of-mass frame, find the energy and spin, and relate them.
\Item{(2)} Do (1) in the conformal gauge by using (7.1.7) for $\hat X\sp a$ 
for $a=0,1$ (same $n$), and applying the constraint (6.2.5).  Compare results.
\Item{(3)} Prove the light-cone Poincar\'e algebra closes only for $D=26$,
and determines the constant in (7.1.8a).
\Item{(4)} Find explicit expressions for all the states at the 4 lowest
mass levels of the open bosonic string.  For the massive levels, combine
SO(D$-$2) representations into SO(D$-$1) ones.  Do the same for the 4 lowest
(nontrivial) mass levels of the closed string.
\Item{(5)} Derive (7.1.15), including the expressions in terms of
$\s$-integrals.  What happens to the part of this integral symmetric in
$ij$ for $M\sb{ij}$?
\Item{(6)} Derive (7.2.8).
\Item{(7)} Derive (7.3.17), both from (7.3.3b) and closure of (7.3.14).
Show that it implies $D-2=1,2,4,8$.
\Item{(8)} Show that (7.3.5) satisfies (7.3.3).  Show that (7.3.6) gives
a supersymmetry algebra, and that the operators of (7.3.5) are
invariant.  Check (7.3.8) till you drop.
\Item{(9)} Find all the states in the spinning string at the tachyonic,
massless, and first massive levels.  Show that, using the truncation of
sect.\ 7.2, there are equal numbers of bosons and fermions at each level.
Construct the same states using the $X$ and $Q$ oscillators of sect.\ 7.3.

%
%

\def\alf{\propto\sb 0}

\chsc{8. BRST QUANTUM MECHANICS}{8.1. IGL(1)}4

We first describe the form of the BRST algebra obtained by first-quantization 
of the bosonic string by the method of sect.\ 3.2, using the constraints
found in the conformal (temporal) gauge in sect.\ 6.2.

The residual gauge invariance in the 
covariant gauge is conformal transformations (modified by the 
constraint that they preserve the position of the boundaries).  After 
quantization in the Schr\"odinger picture (where the coordinates have 
no $\t$ dependence), the Virasoro operators [8.1]
$$\cg ( \s ) = -i\left[\ha \Hat P \sp 2 ( \s ) -1\right]\quad ,\eqno(8.1.1)$$
with $\Hat P$ as in (7.1.3) but for all Lorentz components,
generate only these transformations (instead of the complete set of 2D 
general coordinate transformations they generated when left as 
arbitrary off-shell functions of $\s$ and $\t$ in the classical 
mechanics).  Using the hamiltonian form of BRST quantization, we first 
find the classical commutation relations (Poisson brackets, neglecting
the normal-ordering constant in (8.1.1))
$$ \li{ [ \cg ( \s \sb 1 ) , \cg ( \s \sb 2 ) ] 
&= 2 \p \d ' ( \s \sb 2 - \s \sb 1 ) [ \cg ( \s \sb 1 ) + 
\cg ( \s \sb 2 ) ] \cr
&= 2\p \left[ \d (\s\sb 2 - \s\sb 1 ) \cg ' (\s\sb 2) +
2 \d ' (\s\sb 2 - \s\sb 1) \cg (\s\sb 2) \right] \quad , &(8.1.2a)\cr}$$
or in mode form
$$ i \cg ( \s ) = \su {\bf L} \sb n e \sp {-in\s} \quad \to $$
$$ [ {\bf L} \sb m , {\bf L} \sb n ] = ( n - m ) {\bf L} \sb {m+n} 
\quad .\eqno(8.1.2b)$$
These commutation relations, rewritten as
$$ \left[ \int {d\s\sb 1\over 2\p}\; \l\sb 1 (\s\sb 1) \cg (\s\sb 1) ,
\int {d\s\sb 2\over 2\p}\; \l\sb 2 (\s\sb 2) \cg (\s\sb 2) \right] =
\int {d\s \over 2\p}\; \l\sb{[2} (\s ) \l\sb{1]}' (\s ) \cg (\s )
\quad , \eqno(8.1.3a)$$
correspond to the 1D general coordinate transformations (in
``zeroth-quantized'' notation)
$$ \left[ \l \sb 2 ( \s ) \der{ \s} , \l \sb 1 ( \s ) \der{ \s} \right] 
= \l \sb {[2} \l \sb {1]} ' \der{ \s} \quad , \eqno(8.1.3b)$$
giving the structure constants
$$ f ( \s \sb 1 , \s \sb 2 ; \s \sb 3 ) = 2 \p \d ' ( \s \sb 2 - \s \sb 1 )
\left[ \d ( \s \sb 1 - \s \sb 3 ) + \d ( \s \sb 2 - \s \sb 3 ) \right]
\quad , \eqno(8.1.4a)$$
or
$$ \l \sb 1 \sp j \l \sb 2 \sp i f \sb {ij} \sp k \quad \iff \quad \l \sb
1 \dvec{\der\s} \l \sb 2 \quad . \eqno(8.1.4b)$$
More generally, we have operators whose commutation relations
$$ [ \cg (\s\sb 1 ) , \co (\s\sb 2 ) ] = 2\p\left[ \d (\s\sb 2 -\s\sb 1)
\co ' (\s\sb 2) + w \d ' (\s\sb 2 -\s\sb 1 )\co (\s\sb 2) \right]
\eqno(8.1.5a)$$
represent the transformation properties of a 1D tensor of (covariant)
rank $w$, or a scalar density of weight $w$:
$$ \left[ \int \l\cg , \co \right] = \l\co ' + w\l ' \co \quad .\eqno(8.1.5b)$$
Equivalently, in terms of 2D conformal transformations, it has scale
weight $w$.  (Remember that conformal transformations in $D=2$ are
equivalent to 1D general coordinate transformations on $\s\sp{\pm}$: See
(6.2.7).)  In particular, we see from (8.1.2a) that $\cg$ itself is a
2nd-rank-covariant (as opposed to contravariant) tensor:  It is the
energy-momentum tensor of the mechanics action.  (It was derived by
varying that action with respect to the metric.)  The finite form of
these transformations follows from exponentiating the Lie algebra
represented in (8.1.3):  (8.1.5) can then also be rewritten as the usual
coordinate transformations
$$ \left( \pa\s '\over\pa\s \right)^w \co ' ( \s ' ) = \co (\s )
\quad , \eqno(8.1.6a)$$
where the primes here stand for the transformed quantities (not
$\s$-derivatives) or as
$$ ( d\s ' )\sp w \co ' (\s ' ) = (d\s )\sp w \co (\s ) \quad ,\eqno(8.1.6b)$$
indicating their tensor structure.  In particular, a covariant vector
($w=1$) can be integrated to give an invariant.  $\Hat P$ is such a
vector (and the momentum $p$ the corresponding conformal invariant),
which is why $\cg$ has twice its weight (by (8.1.1)).

Before performing the BRST quantization of this algebra, we relate it to
the light-cone quantization of the previous chapter.  The constraints
(8.1.1) can be solved in a Gupta-Bleuler fashion in light-cone notation.
The difference between that and actual light-cone quantization is that
in the light-cone quantization $\Hat P\sb -$ is totally eliminated at
the classical level, whereas in the light-cone notation for the
covariant-gauge quantization the constraint is used to determine the
dependence on the $\pm$ oscillators in terms of the transverse
oscillators.  One way to do this would be to start with a state
constructed from just transverse oscillators (as in light-cone
quantization) and add in terms involving longitudinal oscillators until
the constraints are satisfied (or actually half of them, ala
Gupta-Bleuler).  A simpler way is to start at the classical level in an
arbitrary conformal gauge with transverse oscillators, and then conformally
transform them to the light-cone gauge to see what a transverse oscillator
(in the physical sense, not the light-cone-index sense) looks like.  We
thus wish to consider
$$ d\s ' \Hat P\sb i ' (\s ') = d\s\Hat P\sb i (\s ) \quad , $$
$$ \s ' = {1\over p\sb +} \hat X\sb + (\s ) = \s + oscillator\hbox{-}terms
\quad . \eqno(8.1.7)$$
(Without loss of generality, we can work at $x\sb + =0$.  Equivalently,
we can explicitly subtract $x\sb +$ from $\hat X\sb +$ everywhere the
latter appears in this derivation.)  If we consider the same
transformation on $\Hat P\sb +$ (using $\pa\hat X /\pa\s \equiv\Hat P$),
we find $\Hat P\sb + ' = p\sb +$, the light-cone gauge.  (8.1.7) can be
rewritten as
$$ \Hat P ' (\s\sb 1 ) = \int d\s\sb 2 \;\d \left( \s\sb 1 - {1\over p\sb +}
\hat X\sb + ( \s\sb 2 )\right) \Hat P (\s\sb 2 ) \quad . \eqno(8.1.8)$$
(8.1.7) follows upon replacing $\s\sb 1$ with $\s\sb 1 '$ and
integrating out the $\d$-function (with the Jacobian giving the
conformal weight factor).  A more convenient form for quantization comes
from the mode expansion:  Multiplying by $e\sp{in\s\sb 1}$ and integrating,
$$ \a\sb n ' = \int {d\s\over 2\p}\; e\sp{in\hat X\sb +(\s )/p\sb +}
\Hat P (\s ) \quad . \eqno(8.1.9)$$
These (``DDF'') operators [8.2] (with normal ordering, as usual, upon
quantization, and with transverse Lorentz index $i$, and $n>0$) can be 
used to create all physical states.  Due to their definition in terms of
a conformal transformation from an arbitrary conformal gauge to a
completely fixed (light-cone) gauge, they are automatically conformally
invariant: i.e., they commute with $\cg$.  (This can be verified to
remain true after quantization.)  Consequently, states constructed from
them satisfy the Gupta-Bleuler constraints, since the conformal
generators push past these operators to hit the vacuum.
Thus, these operators allow the
construction of the physical Hilbert space within the formalism of
covariant-gauge quantization, and allow a direct comparison with
light-cone quantization.

On the other hand, for most purposes it is more convenient to solve the
constraints as covariantly as possible (which is we why we are working
with covariant-gauge quantization in the first place).  The next step is
the IGL(1) algebra [3.4] 
$$ Q = \int {d \s \over 2 \p} \; \Hat C
\left( -i\ha \Hat P \sp 2 + \Hat C ' \fder{\Hat C} +i \alf \right) \equiv
-i \int {d \s \over 2 \p} \; \Hat C \ca \quad ,$$
$$ J\sp 3 = \int{d\s\over 2\p}\;\Hat C\fder{\Hat C} \quad .\eqno(8.1.10)$$
Expanding in the ghost zero-mode
$$ c = \int {d\s\over 2\p}\; \Hat C  \eqno(8.1.11) $$
we also find (see (3.4.3b))
$$ p\sp 2 +M\sp 2 = 2\int\ca \quad , \quad
M\sp + = -i\int \Hat C\Hat C ' \quad . \eqno(8.1.12)$$
$\alf$ (the intercept of the leading Regge 
trajectory) is a constant introduced, as in the light-cone 
formalism, because of implicit normal ordering.  The only ambiguous
constant in $J\sp 3$ is an overall one, which we choose to absorb into
the zero-mode term so that it appears as $c \pa / \pa c$, so that
physical fields have vanishing ghost number.  (This also makes
$J\sp{3\dag}=1-J\sp 3$.)  In analogy to the 
particle, $\Hat C$ is a momentum (defined to be periodic on 
$\s \in [ - \p , \p ]$), as follows from consideration of $\t$ reversal 
in the classical mechanics action, but here $\t$ reversal is accompanied 
by $\s$ reversal in order to avoid switching $+$ and $-$ modes.  (In the 
classical action the ghost is odd under such a transformation, since it 
carries a 2D vector index, as does the gauge parameter, while the 
antighost is even, carrying 2 indices, as does the gauge-fixing function 
${\Sc g}\sb{\bf mn}$.)  $\Hat C$ can also be separated into odd and even 
parts, which is useful when similarly separating $\Hat P$ as in (7.1.3a):
$$ \Hat C = C + \Tilde C \quad , \quad C ( - \s 
) = C ( \s ) \quad , \quad \Tilde C ( - \s ) = - \Tilde C ( \s ) \quad .
\eqno(8.1.13)$$

We now pay attention to the quantum effects.  Rather than examining the
BRST algebra, we look at the IGL(1)-invariant Virasoro operators (from
(3.2.13))
$$ \hat\cg = \cg + \Hat C ' \fder{\Hat C} + \left( \Hat C \fder{\Hat C}
\right)^\prime +i(\alf -1) \quad .\eqno(8.1.14)$$
(The $\alf -1$ just replaces the 1 in (8.1.1) with $\alf$.)
Corresponding to (8.1.2b), we now have the exact quantum mechanical
commutation relations (after normal ordering)
$$ [ \Lhat\sb m , \Lhat\sb n ] = (n-m)\Lhat\sb{m+n} +
\left[ {D-26\over 12}(m\sp 3-m) +2(\alf -1)m \right] \d\sb{m,-n}
\quad .\eqno(8.1.15)$$
The terms linear in $m$ in the anomalous terms (those not appearing in 
the classical result (8.1.2)) are trivial, and can be arbitrarily
modified by adding a constant to $\Lhat\sb 0$.  That the remaining term
is $\sim m\sp 3$ follows from (1D) dimensional analysis:  $[\hat\cg ,
\hat\cg ] \sim \d ' \hat\cg + \d '''$, since the first term implies
$\hat\cg \sim 1/\s\sp 2$ dimensionally, so only $\d ''' \sim 1/\s\sp 4$
can be used.  The values of the coefficients in these terms can also be
determined by evaluating just the vacuum matrix elements
$\sbra{0}[\Lhat\sb{-n} , \Lhat\sb n ]\sket{0}$ for $n=1,2$.  Further examining
these terms, we see that the ghost contributions are necessary to
cancel those from the physical coordinates (which have coefficient $D$),
and do so only for $D=26$.  The remaining anomaly cancels for $\alf =1$.
Under the same conditions one can show that $Q \sp 2 = 0$.
Thus, in the covariant formalism, where
Lorentz covariance is manifest and not unitarity (the opposite of the 
light-cone formalism), $Q \sp 2 = 0$ is the analog of the light cone's 
$[ J \sb {-i} , J \sb {-j} ] = 0$ (and the calculation is almost 
identical, so the reader will have little trouble modifying his 
previous calculation).  $\Hat C$ and $\d / \d\Hat C$ can be expanded in
zero-modes and creation and annihilation operators, as $\Hat P$
((7.1.7a)), but the creation operators in $\Hat C$ are canonically
conjugate to the annihilation operators in $\d / \d\Hat C$, and vice
versa:
$$ \li{ \Hat C &= c + \su_1^\infty {1\over\sqrt n}
( c\sb n\dg e\sp{-in\s} + c\sb n e\sp{in\s} ) \quad , \cr
\fder{\Hat C} &= \der c + \su_1^\infty \sqrt n
( -i\tilde c\sb n\dg e\sp{-in\s} +i\tilde c\sb n e\sp{in\s} ) \quad ;\cr}$$
$$ \{ c\sb m , \tilde c\sb n\dg \} = i\d\sb{mn} \quad , \quad
\{ \tilde c\sb m , c\sb n\dg \} = -i\d\sb{mn} \quad . \eqno(8.1.16)$$
(Since the IGL(1) formalism is directly related to the
OSp(1,1$|$2), as in sect.\ 4.2, we have normalized the oscillators in a
way that will make the Sp(2) symmetry manifest in the next section.)
The physical states are obtained by hitting $\left| 0 \right>$ with $a 
\sp \dag$'s but also requiring $Q \left| \j \right> = 0$; states 
$\left| \j \right> = Q \left| \c \right>$ are null states (pure 
gauge).  The condition of being annihilated by $Q$ is equivalent to 
being annihilated by ${\bf L}\sb n$ for $n \le 0$ (i.e., the 
``nonpositive energy'' part of $\cg ( \s )$, which is now normal ordered 
and includes the $- \alf$ term of 
(8.1.10,14)), which is just the constraint in Gupta-Bleuler quantization.
${\bf L} \sb 0$ is simply the Lorentz-covariantization of $H$ of 
(7.1.8) (i.e., all transverse indices replaced with Lorentz indices).

An interesting fact about the Virasoro algebra (8.1.15) (and its
generalizations, see below) is that, after an appropriate shift in 
{\bf L}$\sb 0$ (namely, the choice of $\alf =1$ in this case), the
anomaly does not appear in the Sp(2) (=SL(2)=SU(1,1)=SO(2,1)=
projective group) subalgebra given by $n=0,\pm 1$ [8.3],
independent of the representation (in this case, $D$).  Furthermore,
unlike the whole Virasoro algebra (even when the anomaly cancels), we
can define a state which is left invariant by this Sp(2).  Expanding
(8.1.14) in modes (as in (7.1.7a, 8.1.2b)), the only term in $\Lhat\sb 1$
containing no annihilation operators is $\sim p\cdot a\sb 1\dg$, so we
choose $p=0$.  Then $\Lhat\sb 0 =0$ requires the state be on-shell, which
means it's in the usual massless sector (Yang-Mills).  Further
examination then shows that this state is uniquely determined to be the
state corresponding to a constant ($p=0$) Yang-Mills ghost field $C$.
It can also be shown that this is the only gauge-invariant,
BRST-invariant state (i.e., in the ``cohomology'' of $Q$) of that
ghost-number $J\sp 3$ [8.4].  Since it has the same ghost number as the
gauge parameter $\L$ (see (4.2.1)), this means that it can be identified
as the only gauge invariance of the theory which has no inhomogeneous term:
Any gauge parameter of the form $\L =Q\e$ not only leaves the free
action invariant, but also the interacting one, since upon gauge fixing
it's a gauge invariance of the ghosts (which means the ghosts themselves
require ghosts), which must be maintained at the interacting level for 
consistent quantization.  However, any parameter satisfying $Q\L =0$
won't contribute to the free gauge transformation of the physical
fields, but may contribute at the interacting level.  In fact, gauge
transformations in the cohomology of $Q$ are just the global invariances
of the theory, or at least those which preserve the second-quantized 
vacuum about which the decomposition into free and interacting has 
been defined.  Since
the BRST transformation $\d\F =Q\F$ is just the gauge transformation
with the gauge parameter replaced by the ghost, this transformation
parameter appears in the field in the same position as would the
corresponding ghost.  For the bosonic string, the only massless 
physical field is Yang-Mills, and thus the only global invariance 
is the usual global nonabelian symmetry.  Thus, the state invariant
under this Sp(2) directly corresponds to the global invariance of the
string theory, and to its ghost.  This Sp(2) symmetry can be maintained
at the interacting level in tree graph calculations (see sect.\ 9.2), 
especially for vertices, basically due to the fact that tree graphs have the
same global topology as free strings.  In such calculations it's
therefore somewhat more convenient to expand states about this
Sp(2)-invariant ``vacuum'' instead of the usual one.  (We now refer to
the first-quantized vacuum with respect to which free fields are defined.
It's redefinition is unrelated to the usual vacuum redefinitions of
field theory, which are inhomogeneous in the fields.)  This effectively
switches the role of the corresponding pair of ghost oscillators (just
the $n=1$ mode) between creation and annihilation operators.

The closed string [4.5] is quantized similarly, but with 2 sets of modes
($\pm$; except that there are still just one $x$ and $p$), and we can separate 
$$ \Hat C \sb \pm ( \s ) = ( C \pm \Tilde C ) ( \pm \s ) \eqno(8.1.17)$$
corresponding to (7.1.6).

Since $\ca$ commutes with both $M\sp 3$ and $M\sp +$, it is
Sp(2)-invariant.  Thus, the modified Virasoro operators $\Lcheck
\sb n$ it gives (in
analogy to (8.1.2b), or, more specifically, the nonnegative-energy ones), 
and in particular their fermionic parts, can be used to generate (BRST)
Sp(2)-invariant states, with the exception of the zeroth and first 
fermionic Virasoro operators (the projective subgroup), which vanish on 
the vacuum.  We will now show that these operators, together with the bosonic
oscillators, are sufficient to generate {\it all} such states, i.e., the
complete set of physical fields [4.1].  (By physical fields we mean all
fields appearing in the gauge-invariant action, including Stueckelberg
fields and unphysical Lorentz components.)  This is seen by
bosonizing the two fermionic coordinates into a single additional
bosonic coordinate, whose contribution to the Virasoro operators
includes a term linear in the new oscillators, but lacking the first
mode.  This corresponds to the fact that $M\sp +$ contains a term
linear in the annihilation operator of the first mode.  Thus, the
Virasoro operators generate excitations in all but the first mode of
the new coordinate, and the condition $M\sp + =0$ kills only
excitations in the first mode.  $J \sp 3$ is just the zero mode of the
new coordinate, so its vanishing (which then implies $T = 0$)
completes the derivation.

The bosonization is essentially the same as the standard procedure
[8.5], except for differences due to the indefinite metric of the
Hilbert space of the ghosts.  The fermionic coordinates can be expressed 
in terms of a bosonic coordinate $\hat \c$ (analogous to $\hat X$) as
$$\Hat C = e \sp {\hat \c} \quad , \quad 
{\d \over {\d \Hat C}} = e \sp {-\hat \c} \quad ,\eqno(8.1.18)$$
with our usual implicit normal ordering (with both terms $\hat q +\hat
p\s$ of the zero
mode appearing in the same exponential factor).  Note the hermiticity of
these fermionic coordinates, due to the lack of $i$'s in the
exponents.  (For physical bosons and fermions, we would use
$\hat\j = e\sp{i\hat\varphi}$, $\hat\j\dg = e\sp{-i\hat\varphi}$, with
$\hat\j$ canonically conjugate to $\hat\j\dg$.)
$\hat \c$ has the mode expansion
$$\hat \c = ( \hat q + \hat p \s ) + 
\su _{n=1} ^{\infty} {1 \over {\sqrt n}} ( \hat a \sb n
e \sp {i n \s} + \hat a \sp \dag \sb n e \sp {- i n \s}) \quad ;$$
$$[ \hat p , \hat q ] = - i \quad , \quad [ \hat a \sb m , \hat a 
\sp \dag \sb n ] = - \d \sb
{mn} \quad .\eqno(8.1.19)$$
By comparison with (7.1.7), we see that this coordinate has a timelike
metric (i.e., it's a ghost).  Using
$$ : e\sp{a\hat\c (\s )} : ~ : e\sp{b\hat\c (\s ')} : ~~ = ~~
: e\sp{a\hat\c (\s )+b\hat\c (\s')} : \left[ 2i~sin
\frac{\s'-\s}2 +\e \right]^{ab} \quad , \eqno(8.1.20)$$
we can verify the fermionic anticommutation relations, as well as
$$J\sp 3 = i \hat p +\ha \quad , \quad 
M\sp + = \int {d \s \over 2 \p} \; e \sp {2 \hat \c} \quad ;\eqno(8.1.21a)$$
$$\ca =\ha \left( \Hat P \sp 2 - \hat \c ' \sp 2 - \hat \c '' - \frac94 \right)
\quad .\eqno(8.1.21b)$$
Since $J\sp 3$ is quantized in integral values, 
$\c$ is defined to exist on a circle of imaginary length with anticyclic 
boundary conditions.  (The imaginary eigenvalues of this hermitian 
operator are due to the indefiniteness induced by the ghosts into the 
Hilbert-space metric.)  Conversely, choosing such values for $i
\hat p$ makes $\Hat C$ periodic in $\s$.
The SU(2) which follows from $J\sp 3$ and $M\sp +$ is not the usual 
one constructed in bosonization [8.6] because of the extra factors and 
inverses of $\pa / \pa \s$ involved (see the next section).

Since we project onto $\hat p = i \ha$ when acting on $\F$, we find
for the parts of $M\sp +$ and $\ca$ linear in $\c$ oscillators when
acting on $\F$
$$M\sp + = e \sp {2\hat q} 2\hat a \sb 1 + \cdots \quad , \quad
\Lcheck \sb n =
\ha \sqrt n ( n - 1 ) \hat a \sp \dag \sb n + \cdots \quad .\eqno(8.1.22)$$
This shows how the constraint $T = 0$ essentially just eliminates the
zeroth and first oscillators of $\c$.

We have seen some examples above of Virasoro operators defined as
expressions quadratic in functions of $\s$ (and their functional
derivatives).  More generally, we can consider a bosonic (periodic) 
function $\hat f (\s )$ with arbitrary weight $w$.  In order to obtain
the transformation law (8.1.5), we must have
$$ \cg (\s ) = \hat f ' \fder{\hat f} -w \left(\hat f \fder{\hat f}
\right)^\prime \quad , \eqno(8.1.23a)$$
up to an overall normal-ordering constant (which we drop).  By
manipulations like those above, we find
$$ [ {\bf L}\sb m , {\bf L}\sb n ] = (n-m){\bf L}\sb{m+n} +
\left\{ \left[ (w-\ha )\sp 2 -\frac1{12} \right] m\sp 3 - \frac16 m \right\}
\d\sb{m,-n} \quad . \eqno(8.1.23b)$$
Since $\hat f '$ and $\d /\d\hat f$ (or $\hat f$ and $-\d /\d\hat f '$)
have the same commutation relations as two $\Hat
P$'s, but with off-diagonal metric $\h\sb{ab} = ({0\atop 1}{1\atop 0})$, 
for $w=0$ (or $w=1$) the algebra (8.1.23) must give just twice the
contribution to the anomaly as a single $\Hat P$.  This agrees exactly
with (8.1.15) (the $D$ term).  For fermionic $\hat f$, the anomalous terms in
(8.1.23) have the opposite overall sign.  In that case, $\hat f$ and $\d
/\d\hat 
f$ have the anticommutation relations of 2 physical fermions (see sect.\ 7.2),
again with the off-diagonal metric, and $w=\ha$ gives the Virasoro
operators for 2 physical fermions (i.e., as in the above bosonic case, 
$\cg$ can be rewritten as the sum of 2 independent $\cg$'s).  The
anomaly for a single physical fermion in thus given by half of that in
(8.1.23b), with opposite sign.  Another interesting case is $w=-1$ (or
2), which, for fermions, gives the ghost contribution of (8.1.15) (the
non-$D$ terms, for $\alf =0$; comparing (8.1.14) with (8.1.23a), we see
$\Hat C$ has $w=-1$ and thus $\d /\d\Hat C$ has $w=2$).  
Thus, (8.1.23) is sufficient to give all
the Virasoro algebras which are homogeneous of second order in 1D
functions.  By the method of bosonization (8.1.18), the fermionic case
of (8.1.23a) can be rewritten as
$$ \cg = i \left[ \ha \Hat\cp\sp 2 + ( \ha - w ) \Hat\cp ' + \frac18
\right] \quad , \eqno(8.1.23c)$$
where $\hat f = exp\;\hat\c$ in terms of a timelike coordinate $\c$
($\Hat\cp = \hat\c '$).  For $w=\ha$, this gives an independent
demonstration that 2 physical fermions give the same anomaly as 1
physical boson (modulo the normal-ordering constant), since they are
physically equivalent (up to the boundary conditions on the zero-modes).
(There are also factors of $i$ that need to be inserted in various
places to distinguish physical bosons and fermions from ghost ones, but
these don't affect the value of the anomaly.)

As before, these Virasoro operators correspond to 2D energy-momentum
tensors obtained by varying an action with respect to the 2D metric.
Using the vielbein formalism of sect.\ 4.1, we first note that the
Lorentz group has only one generator, which acts very simply on the
light-cone components of a covariant tensor:
$$ M\sb{\bf ab} = \e\sb{\bf ab}\cm \quad (\e\sb{+-}=1) \quad\to\quad
[ \cm , \j\sb{(s)} ] = s\j\sb{(s)} \quad , \eqno(8.1.24)$$
where for tensors $s$ is the number of ``$+$'' indices minus ``$-$''
indices.  However, since the 2D Lorentz group is abelian, this
generalizes to arbitrary ``spin,'' half-integral as well as irrational.
The covariant derivative can then be written as
$$ \de\sb{\bf a} = e\sb{\bf a} + \o\sb{\bf a} \cm \quad , $$
$$ \o\sb{\bf a} = \ha\e\sp{\bf cb}\o\sb{\bf abc} = \ha\e\sp{\bf
cb}c\sb{\bf bca} = -\e\sb{\bf ab} {\rm e}\pa\sb{\bf m} {\rm e}\sp{-1} 
e\sp{\bf bm} \quad . \eqno(8.1.25)$$
We also have the only nonvanishing component of the curvature
$R\sb{\bf abcd}$ given by (4.1.31):
$$ {\rm e}\sp{-1}R = \e\sp{\bf mn}\pa\sb{\bf n}\o\sb{\bf m} =
\pa\sb{\bf m} \left[ e\sp{\bf am} \pa\sb{\bf n} 
( {\rm e}\sp{-1}e\sb{\bf a}\sp{\bf n} )\right] \quad . \eqno(8.1.26)$$
The covariant action corresponding to (8.1.23a) is
$$ S \sim \int d\sp 2 x \; {\rm e}\sp{-1} \; ( \j\sp{(+)}\sb{1-w} \de\sb -
\j\sp{(+)}\sb w + \j\sp{(-)}\sb{1-w} \de\sb + \j\sp{(-)}\sb w ) \quad , $$
$$ [ \cm , \j\sp{(\pm )}\sb w ] = \pm w \j\sp{(\pm )}\sb w 
\quad , \eqno(8.1.27)$$
where $\j\sp{(\pm )}\sb w$ corresponds to $\hat f$ and 
$\j\sp{(\pm )}\sb{1-w}$ to $\d /\d\hat f$.  For open-string boundary
conditions, $\j\sp{(\pm )}\sb w$ are combined to form $\hat f$ (as, for
$w=1$, $P\sp{\pm}$ combined to form $\Hat P$ in (7.1.3a)); for closed
strings, the 2 functions can be used independently (as the usual $(\pm )$ 
modes for closed strings).  We thus see that the spin is related to the
weight (at least for these free, classical fields) as $s=\pm w$ for
$\j\sp{(\pm )}\sb w$.  The action corresponding to
(8.1.23c) (neglecting the normal-ordering constant) is [8.7]
$$ S \sim \int d\sp 2 x \; {\rm e}\sp{-1} \; \left[ \ha \c\bo\c + (w-\ha )R\c
\right] \quad . \eqno(8.1.28a)$$
(We have dropped some surface terms, as in (4.1.36).)
The fact that (8.1.28a) represents a particle with spin can be seen in
(at least) 2 ways.  One way is to perform a duality transformation [8.8]:
(8.1.28a) can be written in first-order form as
$$ S \sim \int d\sp 2 x \; {\rm e}\sp{-1} \; \left[ \ha (F\sb{\bf a})\sp 2 
+ F\sp{\bf a} \de\sb{\bf a} \c  + (w-\ha )R\c \right] \quad . \eqno(8.1.28b)$$
(Note that $\de\c$ is the field strength for $\c$ under the global
invariance $\c\to\c~+~constant$.  In that respect, the last term in
(8.1.28b) is like a ``Chern-Simons'' term, since it can only be written
as the product of 1 field with 1 field strength, in terms of the fields
$\o\sb{\bf a}$ and $\c$ and their field strengths $R$ and $\de\sb{\bf a}\c$.)
Eliminating $F$ by its equation of motion gives back (8.1.28a), while
eliminating $\c$ gives
$$ S \sim \int d\sp 2 x \; {\rm e}\sp{-1} \; \ha (G\sb{\bf a})\sp 2 \quad , $$
$$ G\sb{\bf a} = -\e\sb{\bf ab}\de\sp{\bf b} \f\quad , \quad 
[ \cm , \f ] = w-\ha \quad . \eqno(8.1.28c)$$
(Actually, since (8.1.28a) and (8.1.28c) are equivalent on shell, we
could equally well have started with (8.1.28c) and avoided this
discussion of duality transformations.  However, (8.1.28a) is a little
more conventional-looking, and the one that more commonly appears in the
literature.)  The unusual Lorentz transformation law of $\f$ follows
from the fact that it's the logarithm of a tensor:
$$ [ \cm , e\sp\f ] = (w-\ha ) e\sp\f \quad , \quad
[ \cm , e\sp{-\f} ] = (\ha -w) e\sp{-\f} \quad . \eqno(8.1.29)$$
This is analogous to (8.1.18), but  the weights there are increased by
$\ha$ by quantum effects.  (More examples of this effect will be
discussed in sect.\ 9.1.)

Another way to see that $\c$ has an effective Lorentz weight $w$ is to
look at the relationship between Lorentz weights and weights under 2D
general coordinate transformations (or 1D, or 2D conformal
transformations), as in (8.1.5).  This follows from the fact that
conformally invariant theories, when coupled to gravity, become locally
scale invariant theories (even without introducing the scale compensator
$\f$ of (4.1.34)).  (Conversely, conformal
transformations can be defined as the subgroup of general coordinate +
local scale transformations which leaves the vacuum invariant.)
This means that we can gauge-transform e to 1, or, equivalently,
redefine the nongravitational fields to cancel all dependence on e.
Then $e\sb{\bf a}\sp{\bf m}$ appears only in the unit-determinant combination
$ \Sc e\sb{\bf a}\sp{\bf m} =$ e$\sp{-1/2}e\sb{\bf a}\sp{\bf m}$.  
The weights $w$ then appear
in the scale transformation which leaves (8.1.27) invariant:
$$ \j\sb w ' = e\sp{w\z}\j\sb w \quad . \eqno(8.1.30)$$
(This has the same form as a local Lorentz transformation, but with
different relative signs for the fields in the $(\pm )$ terms of
(8.1.27).  This is related to the fact that, upon applying the equations
of motion, and in the conformal gauge, the 2 sets of fields depend
respectively on $\t\mp\s$, and therefore have independent conformal
transformations on these $\pm$ coordinates, except as related by
boundary conditions for the open string.)
Choosing $e\sp\z =$ e$\sp{-1/2}$ then replaces
the $\j$'s with fields which are scale-invariant, but transform under
general coordinate transformations as densities of weight $w$ (i.e., as
tensors times e$\sp{-w/2}$).  In the conformal gauge, these densities
satisfy the usual free (from gravity) field equations, since the
vielbein has been eliminated (the determinant by redefinition, the rest
by choice of general coordinate gauge).  Similar remarks apply to (8.1.28), but
it's not scale invariant.  To isolate the scale noninvariance of that
action, rather than make the above scale transformation, we make a
nonlocal redefinition of $\c$ in (8.1.28a) which reduces to the above
type of scale transformation in the conformal gauge 
$\Sc e\sb{\bf a}\sp{\bf m} = \d\sb{\bf a}\sp{\bf m}$:
$$ \c \to \c - (w-\ha ) {1\over\bo} R \quad . \eqno(8.1.31)$$
In the conformal gauge, $R = -\ha\bo~ln$ e.  (Remember:  $\c$ is like the
logarithm of a tensor.)  Under this redefinition, the action becomes
$$ S \to \int d\sp 2 x \; {\rm e}\sp{-1} \; \left[ \ha\c\bo\c 
-\ha (w-\ha )\sp 2 R {1\over\bo} R \right] \quad . \eqno(8.1.32)$$
(Note that $\c$ now satisfies the usual scalar field equation.)
The redefined field is now scale invariant, and the scale noninvariance
can now be attributed to the second term, which is the same kind of term
responsible for the conformal (Virasoro) anomalies at the quantum level
(i.e., the 1-loop contribution to the 2D field theory in a background
gravitational field).  In fact, the conformally invariant action
(4.1.39), with a factor proportional to $1/(D-2)$, is the dimensionally
regularized expression responsible for the anomaly:  Although (4.1.39) is
conformally invariant in arbitrary $D$, subtraction of the divergent
(i.e., with a coefficient $1/(D-2)$) $R$ term, which is conformally
invariant only in $D=2$ (as follows from considering the $D\to 2$ limit
of (4.1.39) without multiplying by $1/(D-2)$), leaves a renormalized
(finite) action which, in the limit $D\to 2$, is just the second term of
(8.1.32).  Thus, the second term in (8.1.32) contributes classically to
the anomaly of (8.1.23b), the remaining contribution being the usual
quantum contribution of the scalar.  (On the other hand, in the
fermionic theory from which (8.1.28) can be derived by quantum
mechanical bosonization, all of the anomaly is quantum mechanical.)

$D < 26$ can also be quantized 
(at least at the tree level), but there is an anomaly in 2D local 
scale invariance which causes $det ( g \sb {\bf mn} )$ to reappear at 
the quantum level [8.9] (or, in the light-cone formalism, an extra 
``longitudinal'' Lorentz component of $X$ [1.3,4]); however, there are 
complications at the one-loop level which have not yet been resolved.

Presently the covariant formulation of string interactions is understood
only within the IGL(1) formalism (although in principle it's
straightforward to obtain the OSp(1,1$|$2) formalism by eliminating the
auxiliary fields, as in sect.\ 4.2).  These interactions will be
discussed in sect.\ 12.2.

\secty{8.2. OSp(1,1$|$2)}{8.2. OSp(1,1|2)}

We next use the light-cone Poincar\'e algebra of the string, obtained in
sect.\ 7.1,  to derive the OSp(1,1$|$2) formulation as in sect.\ 3.4,
which can be used to find the gauge-invariant action.  We then relate
this to the first-quantized IGL(1) of the previous section by the
methods of sect.\ 4.2.  This OSp(1,1$|$2) formalism can also be derived 
from first quantization simply by treating the zero-mode of $\Sc
g$ as $g$ for the particle (sect.\ 5.1), and the other modes of the 
metric as in the conformal gauge (sect.\ 6.2). 

All operators come directly from the light-cone expressions of sect.\
7.1, using the dimensional extension of sect.\ 2.6, as described in the
general case in sect.\ 3.4.  In particular, to evaluate the action and
its gauge invariance in the form (4.1.6), we'll need the expressions for
$M\sb{\a\b}$, $M\sp 2$, and $\cq\sb\a = M\sb\a\sp a p\sb a + M\sb{\a m}M$ 
given by using $i=(a,\a )$ in (7.1.14,15).  Thus,
$$ M \sb {\a\b} = -i\int^\prime \hat X\sb\a \Hat P\sb\b
= \su_n a \sp \dag \sb {n(\a} a \sb {n\b )} \quad ,$$
$$ M \sb {\a a} = -i\int^\prime \hat X\sb\a \Hat P\sb a
= \su_n a \sp \dag \sb {n[\a} a \sb {na]} \quad ,$$
$$ M \sp 2 = \int^\prime (\Hat P\sb i\sp 2 -2)
= 2 \left( \su n a \sp \dag \sb n \sp i a \sb {ni} - 1 \right) 
= 2 ( N - 1 ) \quad , $$
$$ M \sb {\a m} M = -i\ha\int^\prime \hat X\sb\a \Hat P\sb j\sp 2
= \su \left( a \sp\dag \sb {n\a} \D \sb n - \D \sp\dag
\sb n a \sb {n\a} \right) \quad ,$$
$$ \D \sb n = i {1 \over \sqrt n} \left( \ha \su_{m=1}^{n-1} \sqrt{m(n-m)} 
a \sb m \sp i a \sb {n-m,i} - \su_{m=1}^\infty \sqrt{m(n+m)} 
a \sp \dag \sb m \sp i a \sb {n+m,i} \right) \quad , \eqno(8.2.1)$$
where again the $i$ summation is over both $a$ and $\a$, representing
modes coming from both the physical $X\sp a (\s )$ (with $x\sp a$
identified as the usual spacetime coordinate) and the ghost modes
$X\sp\a (\s )$ (with $x\sp\a$ the ghost coordinates of sect.\ 3.4), as
in the mode expansion (7.1.7).

To understand the relation of the first-quantized BRST quantization
[4.4,5] to that derived from the light cone (and from the OSp(1,1$|$2)), 
we show the Sp(2) symmetry of the ghost coordinates.  We first combine all
the ghost oscillators into an ``isospinor'' [4.1]:
$$\cc \sp \a = {1 \over {\pa / \pa \s}} \left(
\Hat C ' , \fder{\Hat C} \right) \quad .\eqno(8.2.2)$$
This isospinor directly corresponds (except for lack of zero modes) to 
$\hat X\sp\a$ of the OSp(1,1$|$2) formalism from the light cone:  We identify
$$ \hat X\sp\a = (x\sp\a + p\sp\a \s ) + \cc\sp\a \quad , \eqno(8.2.3)$$
and we can thus directly construct
objects which are manifestly covariant under the Sp(2) of $M\sb{\a\b}$.
The periodic inverse derivative 
in (8.2.2) is defined in terms of the saw-tooth function
$${1 \over {\pa / \pa \s}} f ( \s ) = \int d \s ' \; \ha \left[ \e ( \s -
\s ' ) - {1 \over \p} ( \s - \s ' ) \right] f ( \s ' ) \quad 
,\eqno(8.2.4)$$
where $\e ( \s ) = \pm 1$ for $\pm \s > 0$.
The product of the derivative with this inverse derivative, in either
order, is simply the projection operator which subtracts out the zero
mode.  (For example, $\cc \sp +$ is just $\Hat C$ minus its zero
mode.)  Along with $\Hat P\sb a$, this completes the identification of the
nonzero-modes of the two formalisms.  We can then rewrite the other 
relevant operators (8.1.10,12) in terms of $\cc\sp\a$:
$$ p\sp 2 + M\sp 2 = \int {d \s \over 2 \p} \;
( \Hat P\sb a \sp 2 + \cc\sp\a ' \cc\sb\a ' - 2 ) \quad ,$$
$$ \cq \sb \a = -i\ha \int {d \s \over 2 \p} \; \cc \sb \a 
( \Hat P\sb a \sp 2 + \cc\sp\a ' \cc\sb\a ' - 2 ) \quad ,$$
$$ M \sb {\a \b} = - i\ha \int {d \s \over 2 \p} \; \cc \sb 
{( \a} \cc \sb {\b )} ' \quad .\eqno(8.2.5)$$
Again, all definitions include normal ordering.  This first-quantized 
IGL(1) can then be seen to agree with that derived from OSp(1,1$|$2)
in (8.2.1) by expanding in zero-modes.

For the closed string, the OSp(1,1$|$2) algebra is extended to an
IOSp(1,1$|$2) algebra following the construction of (7.1.17):  As for
the open-string case, the D-dimensional indices of the light-cone
formalism are extended to (D+4)-dimensional indices, but just the values
$A = (\pm , \a )$ are kept for the BRST algebra.
To obtain the analog of the IGL(1) formalism, we perform the 
transformation (3.4.3a) for both 
left and right-handed modes.  The extension of IGL(1) to GL(1$|$1) 
analogous to that of OSp(1,1$|$2) to IOSp(1,1$|$2) uses the subalgebras
$( Q , J\sp 3 , p\sb - ,$ \hbox{$ p\sp{\tilde c}=\pa / \pa c )$}
of the IOSp(1,1$|$2)'s of each set of open-string operators.
After dropping the terms containing $\pa / \pa \tilde c$, $x\sb -$ drops
out, and we can set $p\sb + =1$ to obtain:
$$ \li{ Q \quad\to\quad & -i c \ha ( p\sb a\sp 2 + M\sp 2 ) + M\sp + i\der c
+ \cq\sp + \quad , \cr
J\sp 3 \quad\to\quad &  c\der c + M\sp 3 \quad , \cr
p\sb - \quad\to\quad & -\ha ( p\sb a \sp 2 + M\sp 2 ) \quad , \cr
p \sp {\tilde c} \quad\to\quad & \der c \quad . &(8.2.6) \cr}$$
These generators have the same algebra as N=2 supersymmetry in one
dimension, with $Q$ and $p\sp{\tilde c}$ corresponding to the two
supersymmetry generators (actually the complex combination and its
complex conjugate), $J\sp 3$ to the O(2) generator which scales them,
and $p\sb -$ the 1D momentum.  The closed-string algebra GL(1$|$1) is then
constructed in analogy to the IOSp(1,1$|$2), taking sums for the $J$'s
and differences for the $p$'s.

The application of this algebra to obtaining the gauge-invariant action
will be discussed in chapt.\ 11.

\sect{8.3. Lorentz gauge}

We will next consider the OSp(1,1$|$2) algebra which
corresponds to first-quantization in the Lorentz gauge [3.7], as obtained
by the methods of sect.\ 3.3 when applied to the Virasoro algebra of
constraints.

From (3.3.2), for the OSp(1,1$|$2) algebra we have
$$ \li{ Q \sp \a = -i\int {d\s \over 2\p} \bigg[ & C \sp \a (\ha\Hat P \sp
2 -1) + \ha C \sp {(\a} C \sp {\b )} ' i \fder{C \sp \b} -
B i \fder{C \sb \a } + \ha ( C \sp \a B ' - C \sp \a ' B ) i \fder{B} \cr
& + \frac14 C \sp \b ( C \sb \b ' C \sp \a ) ' i \fder{B} \bigg] 
\quad . & (8.3.1) \cr}$$
\vskip.1in\noindent
$B$ is conjugate to the time-components of the gauge field, which in
this case means the components $\Sc g \sp {00}$ and $\Sc g \sp {01}$  of the
unit-determinant part of the world-sheet metric (see chapt.\ 6).
This expression can be simplified by the unitary transformation
$Q \sp \a \to U Q \sp \a U \sp {-1}$ with
$$ l n ~ U = -\ha\int ( \ha C\sp\a C\sb\a ) \fder{B} ' \quad .\eqno(8.3.2)$$
We then have the OSp(1,1$|$2) (from (3.3.7)):
$$ \li{ J \sb {-\a} & = \int C \sb \a \left( - i \ha\Hat P \sp 2 + i +
C \sp \b ' \fder{C \sp \b} + B ' \fder B \right) -B \fder {C
\sp \a} \quad , \cr
J \sb {\a\b} & = \int C \sb {(\a} \fder{C \sp {\b)}} \quad , \quad
J \sb {+\a} = 2 \int C \sb \a \fder B \quad , \cr
J\sb {-+} & = \int 2 B \fder B + C \sp \a \fder{C \sp\a} \quad .&(8.3.3) \cr}$$

A gauge-fixed kinetic operator for string field theory which is
invariant under the full OSp(1,1$|$2) can be derived, 
$$ K = \ha\left\{ J\sb -\sp\a , \left[ J\sb{-\a} , \int i\fder B\right]\right\}
= \int \ha \Hat P \sp 2 -1 + C \sp \a ' i \fder{C \sp \a} + B ' i \fder B
= \ha ( p \sp 2 + M \sp 2 ) \quad , \eqno(8.3.4)$$
\vskip.1in\noindent 
as the zero-mode of the generators $\Hat \cg (\s )$ from (3.3.10):
$$ \li{ \Hat \cg ( \s ) &= 
-\ha\left\{ J\sb - \sp \a , \left[ J \sb{-\a} , \fder B \right] \right\} \cr
&= -i(\ha\Hat P \sp 2 -1) + C \sp \a ' \fder{C \sp \a} + \left(
C \sp \a \fder{C \sp \a} \right)^\prime + B ' \fder B + \left( B \fder B
\right)^\prime \; .& (8.3.5) \cr}$$
The analog in the usual OSp(1,1$|$2) formalism is 
$$ \ha \{ J \sb - \sp \a , [ J \sb {-\a} , p \sb + \sp 2 ] \} =
p \sp A p \sb A \equiv 2 p \sb + p \sb - + p \sp \a p \sb \a = 
\bo - M \sp 2 \quad . \eqno(8.3.6)$$
This differs from the usual light-cone gauge-fixed hamiltonian $p \sb -$, 
which is not OSp(1,1$|$2) invariant.  Unlike the ordinary BRST case (but like 
the light-cone formalism), this operator is invertible, since it's of the
standard form $K = \ha p \sp 2 + \cdots$.  This was made possible by the
appearance of an even number (2) of anticommuting zero-modes.
In ordinary BRST (sect.\ 4.4), the kinetic operator is fermionic: 
$ c ( \bo - M \sp 2 ) -2\der c M \sp +$ is not
invertible because $M \sp +$ is not invertible.

As usual, the propagator can be converted into a form
amenable to first-quantized path-integral techniques by first
introducing the Schwinger proper-time parameter:
$$ {1 \over K} = \int_0^\infty d \t \; e \sp {-\t K} \quad , \eqno(8.3.7)$$
\vskip.1in\noindent 
where $\t$ is identified with the (Wick-rotated) world-sheet time.  
At the free level, the analysis of this propagator corresponds to solving
the Schr\"odinger equation or, in the Heisenberg picture (or classical
mechanics of the string), to solving for the time dependence of the
coordinates which follow from treating $K$ as the hamiltonian:
$$ [ K , Z ] = i Z' \quad , \quad
\dt Z - [ K , Z ] = 0 \quad \to \quad Z = Z ( \s + i \t ) \eqno(8.3.8)$$
for $Z = P , C , B , \d / \d C , \d / \d B$.  Thus in the mode expansion
$Z = z \sb 0 + \su_1^\infty ( z \sb n \sp\dag e \sp {-in(\s +i\t )} +
z \sb n e \sp {in(\s +i\t )} )$ the positive-energy $z \sb n \sp\dag$'s
are creation operators while the negative-energy $z \sb n$'s are
annihilation operators.  (Remember active vs.\ passive transformations:
In the Schr\"odinger picture coordinates are constant while states have
time dependence $e\sp{-tH}$; in the Heisenberg picture states are
constant while coordinates have time dependence $e\sp{tH}(~)e\sp{-tH}$.)

When doing string field theory, in order to define real string fields we
identify complex conjugation of the fields as usual with proper-time
reversal in the mechanics, which, in order to preserve handedness in the
world sheet, means reversing $\s$ as well as $\t$.  As a result, all
reparametrization-covariant variables with an even number of 2D vector 
indices are interpreted as string-field coordinates, while those with an
odd number are momenta.  (See sect.\ 8.1.)  This means that $X$ is a 
coordinate, while $B$ and $C$ are momenta.  Therefore, we should define 
the string field as
$\F [ X ( \s ) , G ( \s ) , F \sb \a ( \s ) ]$, where 
$B=i\d /\d G$ and $C \sp \a = i \d / \d F \sb \a$.  
This field is real under a combined
complex conjugation and ``twist'' ($\s \to - \s$), and $Q \sp \a$ is odd
in the number of functional plus $\s$ derivatives.
(Note that the corresponding replacement of $B$ with $G$ and $C$ with
$F$ would not be required if the {\hbox{$\cg \sb i$'s}} had been associated 
with a Yang-Mills symmetry rather than general coordinate
transformations, since in that case $B$ and $C$ carry no vector indices.)

This OSp(1,1$|$2) algebra can also be derived from the classical mechanics 
action.  The 2D general coordinate transformations (6.1.2) generated by
$\d = \int {d \s \over 2\p} \; \e \sp{\bf m} ( \s ) \cg \sb{\bf m} ( \s )$
determine the BRST transformations by (3.3.2):
$$ \li{ Q \sp \a X & = C \sp {{\bf m}\a} \pa \sb{\bf m} X \quad , \cr
Q \sp \a \Sc g \sp {\bf mn} & = \pa \sb{\bf p} ( C \sp {{\bf p}\a} \Sc g
\sp {\bf mn} ) -\Sc g \sp {\bf p(m} \pa \sb{\bf p} C \sp {{\bf n})\a}\quad ,\cr
Q \sp \a C \sp {{\bf m}\b} & = \ha C \sp {{\bf n} (\a} \pa \sb{\bf n}
C \sp {{\bf m}\b )} - C \sp {\a\b} B \sp{\bf m} \quad , \cr
Q \sp \a B \sp{\bf m} & = \ha ( C \sp {{\bf n}\a} \pa \sb{\bf n} B \sp{\bf m}
- B \sp{\bf n} \pa \sb{\bf n} C \sp {{\bf m}\a} ) \cr
& - \frac18 \left[
2 C \sp {{\bf n}\b} ( \pa \sb{\bf n} C \sp {{\bf p}\a} ) \pa \sb{\bf p} 
C \sp{\bf m} \sb \b + C \sp {{\bf n}\b} C \sp{\bf p} \sb \b \pa \sb{\bf n} 
\pa \sb{\bf p} C \sp {{\bf m}\a} \right] \; . \cr & & (8.3.9) \cr}$$
We then redefine
$$ \Tilde B \sp{\bf m} = B \sp{\bf m} - \ha C \sp {{\bf n}\a} \pa \sb{\bf n} 
C \sp{\bf m} \sb \a \quad \to $$
$$ \li{ Q \sp \a C \sp {{\bf m}\b} & = C \sp {{\bf n}\a} \pa \sb{\bf n} 
C \sp {{\bf m}\b} - C \sp {\a\b} \Tilde B \sp{\bf m} \quad , \cr
Q \sp \a \Tilde B \sp{\bf m} & = C \sp {{\bf n}\a} \pa \sb{\bf n} 
\Tilde B \sp{\bf m} \quad . & (8.3.10) \cr}$$
The rest of the OSp(1,1$|$2) follows from (3.3.7):
$$ \li{ J\sb {+\a} ( X , \Sc g \sp {\bf mn} , C \sp {{\bf m}\b} ) & = 0 \quad ,
\quad J \sb {+\a} \Tilde B \sp{\bf m} = 2 C \sp{\bf m} \sb \a \quad ; \cr
J \sb {\a\b} ( X , \Sc g \sp {\bf mn} , \Tilde B \sp{\bf m} ) & = 0 \quad ,
\quad J \sb {\a\b} C \sp {{\bf m}\g} = \d \sb {(\a} \sp \g 
C \sp{\bf m} \sb {\b )} \quad ; \cr}$$
$$ J \sb {-+} ( X , \Sc g \sp {\bf mn} ) = 0 \quad , \quad
J \sb {-+} \Tilde B \sp{\bf m} = 2 \Tilde B \sp{\bf m} \quad , \quad 
J \sb {-+} C \sp {{\bf m}\a} = C \sp {{\bf m}\a} \quad . \eqno(8.3.11) $$

An ISp(2)-invariant gauge-fixing term is (dropping boundary terms)
$$ \cl \sb 1 = Q\sb\a \sp 2 \ha \h \sb {\bf pq} \Sc g \sp {\bf pq} =
- \h \sb {\bf pq} \left[ \Tilde B \sp{\bf p} \pa \sb{\bf m} \Sc g \sp {\bf qm}
+ \ha \Sc g \sp {\bf mn} ( \pa \sb{\bf m} C \sp {{\bf p}\a} ) 
( \pa \sb{\bf n} C \sp{\bf q} \sb \a ) \right] \quad , \eqno(8.3.12)$$
where $\h$ is the flat world-sheet metric.  This expression is the
analog of the gauge-fixing term $Q\sp 2 \ha A\sp 2$ for Lorentz gauges
in Yang-Mills [3.6,12].  Variation of $\Tilde B$ gives the condition for
harmonic coordinates.  The ghosts have the same form of lagrangian as 
$X$, but not the same boundary conditions:  At the boundary, any
variable with an even number of 2D vector indices with the value 1 has its
$\s$-derivative vanish, while any variable with an odd number vanishes itself.
These are the only boundary conditions consistent with Poincar\'e and
BRST invariance.  They are preserved by the redefinitions below.

Rather than this Landau-type harmonic gauge, we can also define more
general Lorentz-type gauges, such as Fermi-Feynman, by adding to (8.3.12) a
term proportional to $Q \sp 2 \ha \h \sb {\bf mn} C \sp {{\bf m}\a} 
C \sp{\bf n} \sb \a = -\ha \h \sb {\bf mn} \Tilde B \sp{\bf m} 
\Tilde B \sp{\bf n} + \cdots$.  We will not consider such terms further here.

Although the hamiltonian quantum mechanical form of $Q \sp \a$ (8.3.3) also
follows from (3.3.2) (with the functional derivatives now with respect to
functions of just $\s$ instead of both $\s$ and $\t$), the relation to
the lagrangian form follows only after some redefinitions, which we now
derive.  The hamiltonian form that follows directly from (8.3.9,10) can be
obtained by applying the Noether procedure to $\cl = \cl \sb 0 + \cl \sb 1$:  
The BRST current is
$$\li{J \sp {{\bf m}\a} = &( \Sc g \sp {\bf np} C \sp {{\bf m}\a} - 
\Sc g \sp {\bf m(n} C \sp {{\bf p})\a} ) 
\ha \left[ ( \pa \sb{\bf n} X ) \cdot ( \pa \sb{\bf p} X ) + 
( \pa \sb{\bf n} C \sp {{\bf q}\b}) ( \pa \sb{\bf p} C \sb {{\bf q}\b} )
\right] \cr
& - \Tilde B \sb{\bf n} \pa \sb{\bf p} ( \Sc g \sp {\bf n[m} C \sp {{\bf
p}]\a}- \Sc g \sp {\bf mp} C \sp {{\bf n}\a} ) \quad , & (8.3.13) \cr}$$
where (2D) vector indices have been raised and lowered with the flat metric.
Canonically quantizing, with
$$ {1 \over \a '} P \sp{\bf 0} = i \fder X \quad , \quad {1 \over \a '} 
\Tilde B \sb{\bf m} = -i \fder {\Sc g \sp {\bf 0m}} \quad , \quad 
{1 \over \a '} \P \sb {{\bf m}\a} = i \fder {C \sp {{\bf m}\a}} \quad , 
\eqno(8.3.14a)$$
we apply
$$ \dt X = - {1 \over \Sc g \sp {\bf 00}} ( P \sp{\bf 0} + \Sc g \sp
{\bf 01} X ' ) \quad ,
\quad \dt C \sp {{\bf m}\a} = - {1 \over \Sc g \sp {\bf 00}} ( \P \sp
{{\bf m}\a} + \Sc g \sp {\bf 01} C \sp {{\bf m}\a} ' ) \quad , \eqno(8.3.14b)$$
to the first term in (8.3.13) and
$$ \pa \sb{\bf m} \Sc g \sp {\bf mn} = 0 \quad , \quad
\Sc g \sp {\bf 0m} \pa \sb{\bf m} C \sp {{\bf n}\a} = 
- \P \sp {{\bf n}\a} \quad , \eqno(8.3.14c)$$
to the second to obtain
$$ \li{ J \sp {{\bf 0}\a} & = \left( - {1 \over \Sc g \sp {\bf 00}} 
C \sp {{\bf 0}\a} \right)
\ha \left[ ( P \sp {{\bf 0}2} + X ' \sp 2 ) + ( \P \sp {{\bf m}\b} 
\P \sb {{\bf m}\b} + C \sp {{\bf m}\b} ' C \sb {{\bf m}\b} ' ) \right] \cr
& + \left( C \sp {{\bf 1}\a} - {\Sc g \sp {\bf 01} \over \Sc g \sp {\bf 00}} 
C \sp {{\bf 0}\a} \right) ( X ' \cdot P \sp{\bf 0} + C \sp {{\bf m}\b} '
\P \sb {{\bf m}\b} )
- \Tilde B \sb{\bf m} \left[ \P \sp {{\bf m}\a} + ( \Sc g \sp {\bf m[0} 
C \sp {{\bf 1}]\a} )' \right] \; , \cr & & (8.3.15) \cr}$$
where $Q \sp \a \sim \int d\s \; J \sp {{\bf 0}\a}$.

By comparison with (8.3.3), an obvious simplification is to absorb
the $\Sc g$ factors into the $C$'s in the first two terms.  This is
equivalent to
$$ C \sp {{\bf m}\a} \; \to \; \d \sb{\bf 1} \sp{\bf m} C \sp {{\bf 1}\a} -
\Sc g \sp {\bf 0m} C \sp {{\bf 0}\a} \quad , \eqno(8.3.16)$$
and the corresponding redefinitions (unitary transformation) of $\P$ 
and $\Tilde B$.  This puts $\Sc g$-dependence into the $\Tilde B \P$ terms,
$$ \P \sb{\bf m} \Tilde B \sp{\bf m} \quad \to \quad
\P \sb{\bf 0} \left( - {1 \over \Sc g \sp{\bf 00}} \Tilde B \sp{\bf 0}\right) +
\P \sb{\bf 1} \left( \Tilde B \sp{\bf 1} - {\Sc g \sp {\bf 01} \over 
\Sc g\sp{\bf 00}} \Tilde B \sp{\bf 0}\right)  + \cdots \quad , \eqno(8.3.17)$$
unlike (8.3.3), so we remove it by the $\Sc g$ redefinition
$$ \Sc g \sp {\bf 01} = -g \sp{\bf 1} \quad , \quad \Sc g \sp {\bf 00} =
- \sqrt {1 + 2 g \sp{\bf 0} + (g \sp{\bf 1})\sp 2} \quad ; $$
$$ {1 \over \a '} B \sb{\bf m} = i \fder {g \sp{\bf m}} \quad . \eqno(8.3.18)$$
These redefinitions give
$$ \li{  J \sp {{\bf 0}\a} = & \left[ C \sp {{\bf 0}\a} \ha ( P \sp
{{\bf 0}2} + X ' \sp 2 ) + C \sp {{\bf 1}\a} X ' \cdot P \sp{\bf 0} \right] \cr
& + \left\{ C \sp {{\bf 0}\a} ( C \sp {{\bf 0}\b} \P \sb {{\bf 1}\b} ) ' +
C \sp {{\bf 1}\a} \left[ C \sp {{\bf m}\b} ' \P \sb {{\bf m}\b} + 
( C \sp {{\bf 0}\b} \P \sb {{\bf 0}\b} ) ' \right] \right\} \cr
& - \left\{ C \sp {{\bf 0}\a} ( B \sp{\bf 1} + g \sb{\bf 1} B \sp{\bf 0} ) ' 
+ C \sp {{\bf }1\a} \left[ -B \sp{\bf 0} ' + g \sb{\bf m} B \sp{\bf m} '
+ ( g \sb{\bf 0} B \sp{\bf 0} ) ' \right] \right\} \cr
& + C \sp {{\bf 0}\a} \left[ \ha C \sp {{\bf m}\b} ' C \sb {{\bf m}\b} '
+ ( g \sb{\bf m} C \sp {{\bf 0}\b} ) ' C \sp{\bf m} \sb \b ' \right] 
- \P \sb{\bf m} \sp \a B \sp{\bf m} \quad . & (8.3.19) \cr}$$
The quadratic terms $C B '$ don't appear in (8.3.3), and can
be removed by the unitary transformation
$$ Q \sp \a ' = U Q \sp \a U \sp {-1} \quad , \quad l n ~ U =
- i {1 \over \a '} \int C \sp {{\bf 0}\a} C \sp{\bf 1} \sb \a ' \quad , 
\eqno(8.3.20)$$
giving
$$ \li{ J \sp {{\bf 0}\a} = & \left[ C \sp {{\bf 0}\a} \ha ( P \sp {{\bf
0}2} + X ' \sp 2 ) + C \sp {{\bf 1}\a} X ' \cdot P \sp{\bf 0} \right] \cr
& + \left\{ C \sp {{\bf 0}\a} ( C \sp {{\bf 0}\b} \P \sb {{\bf 1}\b} ) ' +
C \sp {{\bf 1}\a} \left[ C \sp {{\bf m}\b} ' \P \sb {{\bf m}\b} + 
( C \sp {{\bf 0}\b} \P \sb {{\bf 0}\b} ) ' \right] \right\} \cr
& - \left\{ C \sp {{\bf 0}\a} ( g \sb{\bf 1} B \sp{\bf 0} ) ' + C \sp
{{\bf 1}\a} \left[ g \sb{\bf m} B \sp{\bf m} ' + ( g \sb{\bf 0} 
B \sp{\bf 0} ) ' \right] \right\} \cr
& + C \sp {{\bf 0}\a} ( g \sb{\bf m} C \sp {{\bf 0}\b} ) ' C \sp{\bf m}\sb \b '
- \P \sb{\bf m} \sp \a B \sp{\bf m} \quad . & (8.3.21) \cr}$$
Finally, the remaining terms can be fixed by the transformation
$$ l n ~ U = i {1 \over \a '} \int 
C\sp {{\bf 0}\a} ( g \sb{\bf 1} C \sp{\bf 0} \sb \a ' + g \sb{\bf 0} 
C\sp{\bf 1} \sb\a ' )\eqno(8.3.22)$$
to get (8.3.3),
after extending $\s$ to $[ - \p , \p ]$ by making the definitions
$$ \hat P = {1 \over \sqrt {2 \a '}} ( P \sp{\bf 0} + X ' ) \quad , $$
$$ C \sp \a = {1 \over \sqrt {2 \a '}} ( C \sp {{\bf 0}\a} + C \sp {{\bf
1}\a} ) \quad , \quad i \fder {C \sp \a} = {1 \over \sqrt {2 \a '}}
( \P \sb {{\bf 0}\a} + \P \sb {{\bf 1}\a} ) \quad , $$
$$ G = {1 \over \sqrt {2 \a '}} ( g \sb{\bf 0} + g \sb{\bf 1} ) \quad , \quad
i \fder G = {1 \over \sqrt {2 \a '}} ( B \sp{\bf 0} + B \sp{\bf 1} )
\quad ,\eqno(8.3.23)$$
where the previous coordinates, defined on $[ 0 , \p ]$, have been
extended to $[ - \p , \p ]$ as
$$ Z ( - \s ) = \pm Z ( \s ) \quad , \eqno(8.3.24)$$
with ``$+$'' if $Z$ has an even number of vector indices with the value
1, and ``$-$'' for an odd number, in accordance with the boundary
conditions, so that the new coordinates will be periodic in $\s$ with
period $2\p$.

To describe the closed string with the world-sheet metric, we again use
2 sets of open-string operators, as in (7.1.17), with each set of
open-string operators given as in (8.3.3,4), and  
the translations are now, in terms of the zero-modes $b$ and $c$ of $B$
and $C$,
$$ \li{ p\sb + &= \sqrt{ -2i \der b } \quad , \cr
p\sb\a &= {1\over p\sb +} i \der{c\sp\a} \quad , \cr
p\sb - &= - {1\over p\sb +} \left( K + p\sb\a\sp 2 \right)
\quad . & (8.3.25) \cr} $$
The OSp(1,1$|$2)
subgroup of the resulting IOSp(1,1$|$2), after the use of the 
constraints $\D p = 0$, reduces to what would be
obtained from applying the general result (3.3.2) to closed string
mechanics.  ((3.2.6a),
without the $B\pa /\pa\tilde C$ term, gives the usual BRST operator when
applied to closed-string mechanics, which is the same as the sum of two
open-string ones with the two sets of physical zero-modes identified.)

We now put the OSp(1,1$|$2) generators in a form analogous to those 
derived from the light cone [3.13].  Let's first consider the open string.
Separating out the dependence on the zero-modes $g$ and $f\sp\a$ in the
OSp(1,1$|$2) operators,
$$\li{J\sb{\a\b} &= f\sb{(\a} \der{f \sp{\b )}} + \Tilde M \sb {\a\b} \quad,\cr
J\sb{-+} &= -2 g \der g - f \sp \a \der{f\sp\a} + \Tilde M \sb {-+} \quad , \cr
J \sb {+\a} &= -2 g \der{f\sp\a} + \Tilde M \sb {+\a} \quad , \cr
J \sb {-\a} &= f\sb\a \der g - ig\cb \sb \a + K \der{f\sp\a}
+ i\cc \sb \a \sp \b f \sb \b + \Tilde M \sb {-\a} \quad , &(8.3.26)\cr}$$
where the $\Tilde M$'s are the parts of the $J$'s not containing these
zero-modes, $K$ is given in (8.3.4), and
$$\li{ \cb \sb \a &= \int BC \sb \a ' \quad , \cr
\cc\sb{\a\b} &= \int C\sb\a C\sb\b ' = \cc \sb {\b\a} \quad .&(8.3.27)\cr}$$
From (3.3.8) and (8.3.26) we find the commutators of $\Tilde M \sb {AB}$,
$\cb \sb \a$, $\cc \sb {\a\b}$, and $K$; the nonvanishing ones are:
$$ \li{ [ \Tilde M \sb {\a\b} , \Tilde M \sb {\g\d} ]&= 
	- C \sb {( \g ( \a} \Tilde M \sb {\b ) \d )} \quad , \cr
[ \Tilde M \sb {\a\b} , \Tilde M \sb {\pm\g} ] & = 
	- C \sb {\g ( \a } \Tilde M \sb {\pm\b )} \quad , \cr
\{ \Tilde M \sb {-\a} , \Tilde M \sb {+\b} \} & = 
	- C \sb {\a\b} \Tilde M \sb {-+} - \Tilde M \sb {\a\b} \quad , \cr
[ \Tilde M\sb{-+} , \Tilde M\sb{\pm\a} ] & = \mp\Tilde M\sb {\pm\a} \quad , \cr
\{\Tilde M\sb{-\a}, \Tilde M \sb {-\b} \} & = 2iK \cc \sb {\a\b} \quad , \cr}$$
$$ [ \Tilde M \sb {\a\b} , \cb \sb \g ] = - C \sb {\g ( \a} \cb \sb {\b )}
\quad , \quad [ \Tilde M \sb {\a\b} , \cc \sb {\g\d} ] = - C \sb {( \g ( \a} 
\cc \sb {\b ) \d )} \quad , $$
$$ [ \Tilde M \sb {-+} , \cb \sb \a ] = 3 \cb \sb \a \quad , \quad
[ \Tilde M \sb {-+} , \cc \sb {\a\b} ] = 2 \cc \sb {\a\b} \quad , $$
$$ \{ \Tilde M \sb {+\a} , \cb \sb \b \} = 2 \cc \sb {\a\b} \quad , \quad
[ \Tilde M \sb {-\a} , \cc \sb {\b\g} ] = - C \sb {\a ( \b} \cb \sb {\g )}
\quad . \eqno(8.3.28) $$
(To show $\{ \Tilde M \sb {- [ \a} , \cb \sb {\b ]} \} = 0$ requires 
explicit use of (8.3.3), but it won't be needed below.)

We then make the redefinition (see (3.6.8))
$$ g = \ha h \sp 2 \eqno(8.3.29)$$
as for the particle.  We next redefine the zero-modes as in (3.5.2) 
by first making the unitary transformation
$$ ln~U = (ln~h) \left( \ha \left[ \der{f\sb\a}, f\sb\a \right] 
-\Tilde M\sb{-+}\right)\eqno(8.3.30a)$$
to redefine $h$ and then
$$ ln~U = - f\sp\a \Tilde M \sb {+ \a} \eqno(8.3.30b)$$
to redefine $c\sp\a$.  The net result is that the transformed operators are
$$ J \sb {-+} = - \der h h \quad , \quad J \sb {+\a} = -h\der{f\sp\a} \quad ,
\quad J\sb{\a\b} = f \sb {(\a} \der{f \sp{\b )}} + \Tilde M \sb {\a\b} \quad,$$
$$ J \sb {-\a} = f\sb\a \der h + {1 \over h} \left[
\der{f \sp \a} ( K + f \sp 2 ) - \Tilde M \sb \a \sp \b f\sb\b + \Hat\cq
\sb \a \right] \quad , \eqno(8.3.31a)$$
where
$$ \Hat\cq\sb\a = \Tilde M \sb {-\a} - i \ha \cb \sb \a + 
	K \Tilde M \sb {+\a} \quad . \eqno(8.3.31b)$$
We also have
$$ p\sb + = h \quad , \quad p\sb\a = -f\sb\a \quad , \quad
p\sb - = -{1\over h} ( K + f\sp 2 ) \quad . \eqno(8.3.32) $$
These expressions have the canonical form (3.4.2a), with 
the identification
$$ \Tilde M\sb{\a\b} \iff M\sb{\a\b} \quad , \quad \Hat\cq \sb \a \iff
\cq\sb\a \quad . \eqno(8.3.33)$$
From (8.3.28) we then find
$$ \{ \Hat\cq\sb\a , \Hat\cq\sb\b \} = - 2K\Tilde M\sb{\a\b} \quad , 
\eqno(8.3.34)$$
consistent with the identification (8.3.33).  Thus the IOSp(1,1$|$2) 
algebra (8.3.31,32) takes the canonical form of chapt.\ 3.
This also allows the closed string formalism to be constructed.

For expanding the fields, it's more convenient to expand the coordinate in {\it
real} oscillators (preserving the reality condition of the string field) as
$$ \hat P = p + \su_1^\infty \sqrt n ( - i a \sb n \sp\dag e \sp {-in\s} +
i a \sb n e \sp {in\s} ) \quad , $$
$$ G = g + \su_1^\infty \sqrt n ( g \sb n \sp\dag e \sp {-in\s} +
g \sb n e \sp {in\s} ) \quad , $$
$$ B = b + \su_1^\infty {1\over\sqrt n}( - i b \sb n \sp\dag e \sp {-in\s} +
i b \sb n e \sp {in\s} ) \quad , $$
$$ F \sp \a = f \sp \a + \su_1^\infty \sqrt n ( f \sp \a \sb n \sp\dag 
e \sp {-in\s} + f \sp \a \sb n e \sp {in\s} ) \quad , $$
$$ C \sp \a = c \sp \a + \su_1^\infty {1\over\sqrt n} ( -ic\sp \a \sb n 
\sp\dag e \sp {-in\s} + i c \sp \a \sb n e \sp {in\s} ) \quad .\eqno(8.3.35)$$
(With our conventions, always $z\sb n\sp\a\dg \equiv (z\sb n\sp\a)\dg$, 
and thus $z\sb{n\a}\dg = - (z\sb{n\a})\dg$.)
The commutation relations are then
$$ [ a \sb m , a \sb n \dg ] = [ b \sb m , g \sb n \dg ] = [ g \sb m , b
\sb n \dg ] = \d \sb {mn} \quad , $$
$$ \{ c \sb {m\a} , f \sb {n\b} \dg \} = \{ f \sb {m\a} , c \sb {n\b} \dg \} = 
\d \sb {mn} C \sb {\a\b} \quad , \eqno(8.3.36)$$
We then have
$$ \Hat\cq \sb \a = \su_1^\infty \left[ ( \co \sb n \dg c \sb {n\a} - c \sb
{n\a} \dg \co \sb n ) - ( b \sb n \dg f \sb {n\a} - f \sb {n\a} \dg b
\sb n ) \right] \quad , $$
$$ \co \sb n = {1\over\sqrt n} \Tilde L \sb n + \ha b \sb n - 2 g \sb n K 
\quad , $$
$$\li{K = \Tilde L \sb 0 &= -\ha \bo - 1 + \su_1^\infty n ( a \sb n \dg \cdot a
\sb n + b \sb n \dg g \sb n + g \sb n \dg b \sb n + c \sb n \sp \a \dg f
\sb {n\a} + f \sb n \sp \a \dg c \sb {n\a} ) \cr
&= -\ha\bo -1+N \quad , \cr} $$
$$ \li{ \Tilde L \sb n = & - \sqrt n \pa \cdot a \sb n \cr
& + \su_{m=1}^\infty \sqrt{m(n+m)} ( a \sb m \dg \cdot a \sb {n+m}
+ b \sb m \dg g \sb {n+m} + c \sb m \sp \a \dg f \sb {n+m , \a} \cr
& \qquad\qquad +
g \sb m \dg b \sb {n+m} + f \sb m \sp \a \dg c \sb {n+m , \a} ) \cr
& + \su_{m=1}^{n-1} \sqrt{m(n-m)} ( -\ha a \sb m \cdot a \sb {n-m} +
b \sb m g \sb {n-m} + c \sb m \sp \a f \sb {n-m , \a} ) 
\; . & (8.3.37) \cr}$$

The Lorentz-gauge OSp(1,1$|$2) algebra can also be derived by the method
of sect.\ 3.6.  In the GL(1) case, we get the same algebra as above,
while in the U(1) case we get a different result using the same
coordinates, suggesting a similar first-quantization with the
world-sheet metric.  For the GL(1) case we define new coordinates at 
the GL(2$|$2) stage of reduction:
$$ X\sp A = \tilde x\sp A + \Tilde\cc\sp A \quad , \quad
P\sb A = \tilde p\sb A + \Tilde\cc'\sb A \quad , \eqno(8.3.38) $$
where $\cc$ is the generalization of $\cc\sp\a$ of (8.2.3) to arbitrary index.
(Remember that in GL(2$|$2) a lower primed index can be converted into
an upper unprimed index.)  $P\sb A$ is then canonically conjugate to
$X\sp A$.  We also have relations between the coordinates such as
$$ \Tilde P\sb A = P\sb A \quad , \quad
\Tilde P\sp A = \check p\sp A + X'\sp A \quad , $$
$$ \Tilde X\sp A = \check p\sp A\s + X\sp A \quad , \eqno(8.3.39) $$
where $\Hat P\sb\ca = (\Tilde P\sb A , \Tilde P\sb{A'})$, etc.
However, $\Tilde X\sb A$ can be expressed in terms of $P\sb A$
only with an inverse $\s$-derivative.  As a result, when reexpressed in
terms of these new coordinates, of all the IOSp(2,2$|$4) generators only
the IGL(2$|$2) ones have useful expressions.  Of the rest, $\tilde
J\sp{AB}$ is nonlocal in $\s$, while $\tilde J\sb{AB}$ and $\tilde p\sb
A$ have explicit separate terms containing $x\sp A$ and $p\sb A$.
Explicitly, the local generators are $p\sb A$, $\check p\sp A$, and
$$ \tilde J\sp A\sb B = i(-1)\sp{AB}\check x\sb B\check p\sp A
-i\int X\sp A P\sb B \quad , \quad \tilde J\sp{AB} = -ix\sp{[A}\check p\sp{B)}
-i\int X\sp A X'\sp B \quad . \eqno(8.3.40) $$
In these expressions we also use the light-cone constraint and gauge
condition as translated into the new coordinates:
$$ P\sb - = -{1\over 2\check p\sp -}\left[\Hat P\sb a\sp 2 -2 
+2(\check p\sp + +X'\sp +) P\sb + +2(\check p\sp\a +X'\sp\a) P\sb\a\right]
\quad , \quad X\sp - = 0 \quad . \eqno(8.3.41) $$
After following the procedure of sect.\ 3.6 (or simply comparing
expressions to determine the $M\sb{ij}$), the final $OSp(1,1|2)$
generators (3.6.7c) are (dropping the primes on the $J$'s)
$$ J\sb{+\a} = 2i\int X\sb\a P\sb + \quad , \quad
J\sb{-+} = -i\int (2X\sb - P\sb + +X\sp\a P\sb\a) \quad , \quad
J\sb{\a\b} = -i\int X\sb{(\a}P\sb{\b )} \quad , $$
$$ J\sb{-\a} = -i\int [ X\sb - P\sb\a + X\sb\a ( \ha\Hat P\sb a\sp 2 -1
+X'\sb - P\sb + +X'\sp\b P\sb\b ) ] \quad . \eqno(8.3.42) $$
This is just (8.3.3), with the identification $P\sb + = G$ 
and $X\sp\a = C\sp\a$.

In the U(1) case there is no such redefinition possible (which gives
expressions local in $\s$).  The generators are
$$ J\sb{+\a} = ix\sb\a p\sb + + i\int \hat X\sb{\a'}\Hat P\sb{+'} \quad , \quad
J\sb{-+} = -ix\sb - p\sb + -i\int \hat X\sb{-'}\Hat P\sb{+'} \quad , $$
$$ J\sb{\a\b} = -ix\sb{(\a}p\sb{\b )} -i\int^\prime\hat X\sb\a\Hat P\sb\b
-i\int\hat X\sb{\a'}\Hat P\sb{\b'} \quad , $$
$$ \li{ J\sb{-\a} = & -ix\sb - p\sb\a + ix\sb\a \int\Hat P\sb - 
+i\int^\prime\hat X\sb\a\Hat P\sb - \cr
& -i{1\over p\sb +}\left( p\sp a \int^\prime\hat X\sb\a\Hat P\sb a
-p\sp\b \int^\prime\hat X\sb\a\Hat P\sb\b \right)
-i\int\hat X\sb{-'}\Hat P\sb{\a'} \quad , \cr} $$
$$ -2p\sb +\Hat P\sb - = \Hat P\sb a\sp 2 -2 +\Hat P\sp\a\Hat P\sb\a
+\Hat P\sp{\a'}\Hat P\sb{\a'} +2\Hat P\sb{+'}\Hat P\sb{-'} \quad , 
\eqno(8.3.43)$$
where $x\sb{A'}=p\sb{A'}=0$.  After performing the unitary
transformations (3.6.13), they become
$$ J\sb{+\a} = ix\sb\a p\sb + \quad , \quad J\sb{-+} = -ix\sb - p\sb +
\quad , \quad
J\sb{\a\b} = -ix\sb{(\a}p\sb{\b )} -i\int^\prime\hat X\sb\a\Hat P\sb\b
-i\int\hat X\sb{\a'}\Hat P\sb{\b'} \quad , $$
$$ \li{ J\sb{-\a} = & -ix\sb - p\sb\a + ix\sb\a \int\Hat P\sb - 
+i\int^\prime\hat X\sb\a\Hat P\sb - \cr
& -i{1\over p\sb +}\Bigg[ p\sp a \int^\prime\hat X\sb\a\Hat P\sb a
-p\sp\b \left( \int^\prime\hat X\sb\a\Hat P\sb\b +\int\hat X\sb{\a'}\Hat
P\sb{\b'} \right) \cr
& +\int\hat X\sb{-'}\Hat P\sb{\a'}
+\ha\int\hat X\sb{+'}\Hat P\sb{\a'} \left(p\sb a\sp 2 -2p\sb
+\int^\prime\Hat P\sb -\right) \Bigg] \quad . &(8.3.44) \cr} $$
We can still interpret the new coordinates
as the world-sheet metric, but different redefinitions would be
necessary to obtain them from those in the mechanics action, and the
gauge choice will probably differ with respect to the zero-modes.

Introducing extra coordinates has also been considered in [8.10], but in
a way equivalent to using bosonized ghosts, and requiring imposing
$D=26$ by hand instead of deriving it from unitarity.
Adding 4+4 extra dimensions to describe bosonic strings with enlarged
BRST algebras $OSp(1,1|4)$ and $OSp(2,2|4)$ has been considered by
Aoyama [8.11].

The use of such extra coordinates may also prove useful in the study of
loop diagrams:  In particular, harmonic-type gauges can be well-defined
globally on the world sheet (unlike conformal gauges), and consequently
certain parameters (the world-sheet generalization of proper-time
parameters for the particle, (5.1.13)) appear automatically [8.12].
This suggests that such coordinates may be useful for closed string
field theory (or superstring field theory, sect.\ 7.3).

The gauge-invariant action, its component analysis, and its comparison
with that obtained from the other OSp(1,1$|$2) will be made in sect.\ 11.2.

\sect{Exercises}

\Item{(1)} Prove that the operators (8.1.9) satisfy commutation
relations like (7.1.7a).  Prove that they are conformally invariant.
\Item{(2)} Derive (8.1.20).  Verify that the usual fermionic 
anticommutation relations for (8.1.18) then follow from (8.1.19).
Derive (8.1.22).
\Item{(3)} Derive (8.1.15).  Derive (8.1.23bc).  Prove $Q\sp 2 =0$.
\Item{(4)} Derive (8.1.23ac) from the energy-momentum tensors of (8.1.27,28ac).
\Item{(5)} Find the commutation relations of $\ha\hat Dd\hat D$ of
(7.2.2), generalizing (8.1.2).  Find the BRST operator.
Derive the gauge-fixed Virasoro operators, and show the conformal
weights of $\hat\J$ is 1/2, and of its ghosts are $-$1/2 and +3/2.
Use (8.1.23) to show the anomaly cancels for $D=10$.
\Item{(6)} Find an alternate first-order form of (8.1.28b) by rewriting
the last term in terms of $F$ and $\o$, and show how (8.1.28ac) follow.
\Item{(7)} Show from the explicit expression for $\o\sb{\bf a}$ that
(8.1.27) can be made vielbein-independent by field redefinition in the
conformal gauge.
\Item{(8)} Explicitly prove the equivalence of the IGL(1)'s derived in 
sect.\ 8.1 and from the light cone, using the analysis of sect.\ 8.2.
\Item{(9)} Derive (8.3.1,3).  Derive (8.3.9).
\Item{(10)} Derive (8.3.28).

%
%

\chsc{9. GRAPHS}{9.1. External fields}4

One way to derive Feynman graphs is by considering a propagator in an 
external field:

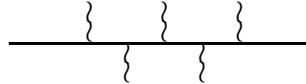
\begin{center}

\setlength{\unitlength}{1mm}
\begin{picture}(40,20)
\put(0,10){\line(1,0){40}}
\put(10,10.5){$\wr$}
\put(10,13.1){$\wr$}
\put(20,10.5){$\wr$}
\put(20,13.1){$\wr$}
\put(30,10.5){$\wr$}
\put(30,13.1){$\wr$}
\put(15,5.1){$\wr$}
\put(15,7.7){$\wr$}
\put(25,5.1){$\wr$}
\put(25,7.7){$\wr$}
\end{picture}

\end{center}

\noindent For example, for a scalar particle in an external scalar field,
$$ L = \ha \dot x \sp 2 - m \sp 2 - \f ( x ) \quad \to \eqno(9.1.1a)$$
$$ [ p \sp 2 + m \sp 2 + \f ( x ) ] \j ( x ) = 0 \quad \to \eqno(9.1.1b)$$
$$ propagator \quad {1 \over p \sp 2 + m \sp 2 + \f ( x )} = {1 \over 
p \sp 2 + m \sp 2} - {1 \over p \sp 2 + m \sp 2} \f {1 \over 
p \sp 2 + m \sp 2} + \cdots \eqno(9.1.1c)$$

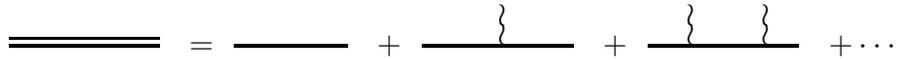
\setlength{\unitlength}{1mm}
\begin{picture}(140,15)
\put(20,5){\line(1,0){20}}
\put(20,6){\line(1,0){20}}
\put(44,4){$=$}
\put(50,5){\line(1,0){15}}
\put(69,4){$+$}
\put(75,5){\line(1,0){20}}
\put(85,5.5){$\wr$}
\put(85,8.1){$\wr$}
\put(99,4){$+$}
\put(105,5){\line(1,0){20}}
\put(110,5.5){$\wr$}
\put(110,8.1){$\wr$}
\put(120,5.5){$\wr$}
\put(120,8.1){$\wr$}
\put(129,4){$+ \cdots$}
\end{picture}

For the (open, bosonic) string, it's useful to use the ``one-handed'' 
version of $X ( \s )$ (as in (7.1.7b)) so that $\hat X$ and $\Hat P$ can be 
treated on an (almost) equal footing, so we will switch to that notation
hand and foot.  Then the generalization of (9.1.1b) (again jumping 
directly to the first-quantized form for convenience) is [7.5]
$$ \bra{\c} [ \ha \Hat P \sp 2 ( \s ) -1 + \cv ( \s ) ] \ket{\j} = 0 
\quad .\eqno(9.1.2)$$
(The OSp(1,1$|$2) algebra can be similarly generalized.)
However, for consistency of these equations of motion, they must 
satisfy the same algebra as $\ha \Hat P \sp 2 -1$ ((8.1.2)).  (In
general, this is only true including ghost contributions, which we will
ignore for the examples considered here.)  Expanding this 
condition order-by-order in $\cv$, we get the new relations
$$ \li{[ \ha \Hat P \sp 2 ( \s \sb 1 ) -1 , \cv ( \s \sb 2 )] &= 
2 \p i \d ' ( \s \sb 2 - \s \sb 1 ) \cv ( \s \sb 1 ) \cr
&= 2\p i\left[ \d (\s\sb 2 -\s\sb 1) \cv ' (\s\sb 2) + \d ' (\s\sb 2 -
\s\sb 1) \cv (\s\sb 2) \right] \; ,&(9.1.3a)\cr
[ \cv ( \s \sb 1 ) , \cv ( \s \sb 2 ) ] &= 0 \quad .&(9.1.3b)\cr}$$
The first condition gives the conformal transformation properties of 
$\cv$ (it transforms covariantly with conformal weight 1, like 
$\Hat P$), and the second condition is one of ``locality''.  A simple 
example of such a vertex is a photon field coupled to one end of the 
string:
$$ \cv ( \s ) = 2 \p \d ( \s ) A \left( X ( 0 ) \right) \cdot \Hat P \quad 
,\eqno(9.1.4)$$
where $X ( 0 ) = \hat X ( 0 )$.  Graphs are now given by expanding 
the propagator as the inverse of the hamiltonian
$$ H = \int {d \s \over 2 \p} \; ( \ha \Hat P \sp 2 -1 + \cv ) = 
{\bf L} \sb 0 + \int d \s \; \cv \equiv {\bf L} \sb 0 + V \quad .\eqno(9.1.5)$$

More general vertices can be found when normal ordering is carefully 
taken into account [1.3,4], and one finds that (9.1.3a) can be satisfied when 
the external field is on shell.  For example, consider the scalar vertex
$$ \cv (\s ) = - 2 \p \d ( \s ) \f \left( X ( 0 ) \right) \quad .\eqno(9.1.6)$$
Classically, scalar fields have the wrong conformal weight (zero):
$$ {1 \over 2 \p} \left[ \ha \Hat P \sp 2 ( \s \sb 1 ) -1 , \f \left( \hat X 
( \s \sb 2 ) \right) \right] = i \d ( \s \sb 2 - \s \sb 1 ) {\pa \over 
\pa \s \sb 2} \f \left( \hat X ( \s \sb 2 ) \right) \quad ;\eqno(9.1.7a)$$
but quantum mechanically they have weight ``$-\ha\bo$'':
$$ \li {{1 \over 2 \p} \left[ \ha \Hat P \sp 2 ( \s \sb 1 ) -1 , \f \left( 
	\hat X ( \s \sb 2 ) \right) \right] =& i \d ( \s \sb 2 - \s \sb 1 ) 
	{\pa \over \pa \s \sb 2} \f \left( \hat X ( \s \sb 2 ) \right) \cr
& + \left[ - \ha {\pa \sp 2 \over \pa \hat X \sp 2 
	( \s \sb 2)} \right] i \d ' ( \s \sb 2 - \s \sb 1 ) \f \left( 
	\hat X ( \s \sb 2 ) \right) \quad .\cr & &(9.1.7b)\cr}$$
Therefore $\f \left( \hat X ( \s \sb 2 ) \right)$, and thus $\cv$ of 
(9.1.6), satisfies (9.1.3a) if $\f$ is the on-shell ground state 
(tachyon):  $- \ha \bo \f = 1 \cdot \f$.  (Similar remarks apply quantum 
mechanically for the masslessness of the photon in the vertex (9.1.4).)

As an example of an S-matrix calculation, consider a string in an external 
plane-wave tachyon field, where the initial and final states of the
string are also tachyons:
$$ \f ( x ) = e \sp {- i k \cdot x} \quad \to \quad : e \sp {- i k 
\cdot X ( 0 )} : \eqno(9.1.8)$$
We then find

\setlength{\unitlength}{1mm}
\begin{picture}(60,25)
\put(0,14){$k\sb 1$}
\put(10,10){\line(1,0){40}}
\put(10,20){\line(1,0){40}}
\put(20,5.1){$\wr$}
\put(20,7.7){$\wr$}
\put(19,0){$k\sb 2$}
\put(28,5){$\cdots$}
\put(40,5.1){$\wr$}
\put(40,7.7){$\wr$}
\put(39,0){$k\sb{{\bf N}-1}$}
\put(55,14){$k\sb{\bf N}$}
\end{picture}

$$ \li{&= g\sp{{\bf N}-2} \sbra{k\sb N} V(k\sb{{\bf N}-1}) \cdots
\D (p) V (k\sb 3) \D (p) V (k\sb 2) \sket{k\sb 1} \cr
&= g \sp {{\bf N}-2} \left< 0 \left| \Tilde V ( k \sb {{\bf N}-1} ) \cdots 
\D ( k\sb 3 + k\sb 4 + \cdots + k\sb{\bf N} ) \Tilde V (k\sb 2) 
\right| 0 \right> \quad ,\cr}$$
$$ V ( k ) = ~: e \sp {-ik \cdot X ( 0 )} :~ = \Tilde V (k) e\sp{-ik\cdot x}
\quad , \quad
X( 0 ) = \su_1^\infty {1 \over \sqrt n}( a \sb n \dg + a \sb n ) \quad ,$$
$$ \D ( p ) = {1 \over \ha p \sp 2 + ( N - 1 )} \quad , \quad N = 
\su_1^\infty n a \sb n \dg \cdot a \sb n \quad ,\eqno(9.1.9)$$
where $g$ is the coupling constant, and we have pulled the $x$ pieces 
out of the $X$'s and pushed them 
to the right, causing shifts in the arguments of the $\D$'s (which 
were originally $p$, the momentum operator conjugate to $x$, not to be
confused with the constants $k\sb i$).  
We use Schwinger-like parametrizations (5.1.13) for the propagators:
$$ {1 \over \ha p \sp 2 + ( N - 1 )} = 
\int_0^\infty dt \; e\sp{-t[\ha p\sp 2 +(N-1)]} =
\int_0^1 {dx \over x} \; x \sp {\ha p \sp 2 + ( N - 1 )} \quad 
(x = e \sp {- t})\quad ,\eqno(9.1.10)$$
where we use $t\sb i$ for $\D (k\sb i + \cdots + k\sb{\bf N})$,
as the difference in proper time between $\Tilde V (k\sb{i-1})$ and
$\Tilde V (k\sb i)$.  Plugging (9.1.10) into (9.1.9), the amplitude is
$$ g\sp {{\bf N}-2} \left( \pr_{i=3}^{{\bf N}-1} \int_0^1 {dx \sb i \over x \sb
i} \; x\sb i\sp{\ha (k\sb i + \cdots + k\sb{\bf N})\sp 2 -1}
\right) \pr_n \left< 0 \left| \cdots x \sb 3 \sp {
n a \dg \cdot a} e \sp {-ik \sb 2 \cdot {1 \over \sqrt n} 
a \dg} \right| 0 \right> \quad .\eqno(9.1.11)$$
To evaluate matrix-elements of 
harmonic oscillators it's generally convenient to use coherent states:
$$ \left| z \right> \equiv e \sp {z a \dg} \left| 0 \right> \quad \to $$
$$ a \left| z \right> = z \left| z \right> \quad , \quad a\dg\ket{z} =
\der{z}\ket{z} \quad , \quad 
e \sp {z ' a \dg} \left| z \right> = \left| z + z ' \right> \quad , \quad 
x \sp {a \dg a} \left| z \right> = \left| xz \right> \quad ,$$
$$ \left< z \big| z ' \right> = e \sp {\bar z z '} \quad , \quad 1 = 
\int {d \sp 2 z \over \p} \; e \sp {- \left| z \right| \sp 2} \left| z 
\right> \left< z \right| \quad ,$$
$$ tr (\co ) = \int{d\sp 2z\over\p}\; e\sp{-|z|\sp 2}
\sbra{z}\co\sket{z} \quad .\eqno(9.1.12)$$
Using (9.1.12) and the identity $\pr_1^\infty e \sp{-cx \sp n / n} = 
( 1 - x ) \sp c$, (9.1.11) becomes
$$ g \sp {{\bf N}-2} \left( \pr \int {dx \sb i\over x\sb i} \; 
x\sb i \sp{\ha ( k \sb i + \cdots +
k \sb{\bf N} ) \sp 2 -1} \right) \pr_{2 \le i < j \le {\bf N}-1} \left( 1 - 
\pr_{k=i+1}^j x \sb k \right)^{k \sb i \cdot k \sb j} 
\quad.\eqno(9.1.13)$$
We next make the change of variables 
$$ \t\sb i = \su_{j=3}^i t\sb j \quad , \quad \t\sb 2 = 0 \quad ,
\quad \t\sb{\bf N} = \infty \quad\to\quad
0 = \t\sb 2 \le \t\sb 3 \le \cdots \le \t\sb{\bf N} = \infty \quad ,
\eqno(9.1.14a)$$
or
$$ z \sb i = \pr_{j=3}^i x \sb j \quad , \quad z \sb 2 = 1 \quad , 
\quad z \sb{\bf N} = 0 \quad \to \quad 1 = z \sb 2 \ge z \sb 3 \ge 
\cdots \ge z \sb{\bf N} = 0 \quad ,\eqno(9.1.14b)$$
with
$$ z \sb i = e \sp {-\t \sb i} \quad , \eqno(9.1.14c)$$
where $\t\sb i$ is the absolute proper time of the corresponding vertex.
Using the mass-shell condition $k \sb i \sp 2 = 2$, 
the final result is then [9.1]
$$ g \sp {{\bf N}-2} \left( \int \pr_{i=3}^{{\bf N}-1} dz \sb i \right) \pr_{2
\le i < j \le {\bf N}} ( z \sb i - z \sb j ) \sp {k \sb i \cdot k \sb j} 
\quad .\eqno(9.1.15)$$

The simplest case is the 4-point function ({\bf N}$=4$) [9.2]
$$ \li{& g \sp 2 \int _0^1 dz \; z \sp {- \ha s-2} ( 1 - z ) \sp{- \ha t-2}
	\qquad ( \; s = - ( k \sb 1 + k \sb 2 ) \sp 2 \quad , \quad t 
	= - ( k \sb 2 + k \sb 3 ) \sp 2 \; ) \cr
&= g \sp 2 B ( - \ha s-1 , - \ha t-1 ) = g \sp 2 {\G ( - \ha s-1) \G ( - 
	\ha t-1 ) \over \G ( -\ha s- \ha t-2 )} &(9.1.16a)\cr
&= g \sp 2 \su_{j=0}^\infty \left[ {( \ha t+1+j)( \ha t+j) \cdots ( \ha t+1) 
	\over j!} \right] {1 \over j-( \ha s+1)} &(9.1.16b)\cr
&= \left( in ~ \lim_{s \to - \infty \atop t ~ fixed} \right) \quad
	g \sp 2 \G ( - \ha t-1 ) ( - \ha s-1 ) \sp {\ha t+1} \quad .&(9.1.16c)
	\cr}$$
(9.1.16b) shows that the amplitude can be expressed as a sum of poles 
in the $s$ channel with squared masses $2(j-1)$, with maximum spin $j$ 
(represented by the coefficient with leading term $t \sp j$).  Since 
the amplitude is symmetric in $s$ and $t$, it can also be expressed as 
a sum of poles in the $t$ channel, and thus summing over poles in one 
channel generates poles in the other.  (It's not necessary to sum over 
both.)  This property is called ``duality''.  (9.1.16c) shows that the 
high-energy behavior goes like $s \sp {\ha t+1}$ instead of the usual 
fixed-power behavior $s \sp j$ due to the exchange of a spin $j$ 
particle, which can be interpreted as the exchange of a particle with 
effective spin $j= \ha t+1$.  This property is known as ``Regge behavior'', 
and $j(t)= \ha t+1$ is called the ``leading Regge trajectory'', which not only 
describes the high-energy behavior but also the (highest) spin at any 
given mass level (the mass levels being given by integral values of $j(t)$).

Instead of using operators to evaluate propagators in the presence of
external fields, we can also use the other approach to quantum
mechanics, the Feynman path integral formalism.  In particular, for the
calculation of purely tachyonic amplitudes considered above, we evaluate
(9.1.9) directly in terms of $V$ (rather than $\Tilde V$), after using
(9.1.10):
$$ g\sp{{\bf N}-2} \int_0^\infty dt\sb 3\cdots dt\sb{{\bf N}-1} \;
\left\langle k\sb{\bf N} \left| V(k\sb{{\bf N}-1}) \cdots
e\sp{-t\sb 3 (\ha p\sp 2 +N-1)}
V(k\sb 2) \right| k\sb 1 \right\rangle \quad . \eqno(9.1.17)$$
Using (9.1.14a), we can rewrite this as
$$ g\sp{{\bf N}-2} \left( \pr_{i=3}^{{\bf N}-1} \int_{\t\sb{i-1}}^{\t\sb{i+1}}
d\t\sb i\right) \; \sbra{k\sb{\bf N}}V(k\sb{{\bf N}-1},\t\sb{{\bf N}-1}) \cdots
V(k\sb 3 , \t\sb 3) V(k\sb 2 , \t\sb 2) \sket{k\sb 1} \quad , \eqno(9.1.18a)$$
where
$$ V(k,\t ) =~ : e\sp{-ik\cdot X (0,\t )} : \quad , \quad
X (0,\t ) = e\sp{\t (\ha p\sp 2 +N-1)} X(0) e\sp{-\t (\ha p\sp 2 +N-1)}
\quad , \eqno(9.1.18b)$$
is the vertex which has been (proper-)time-translated from $0$ to $\t$.
(Remember that in the Heisenberg picture operators have time dependence
$\co (t) = e\sp{tH}\co (0) e\sp{-tH}$, whereas in the Schr\"odinger
picture states have time dependence $\sket{\j (t)} = e\sp{-tH}\sket{\j
(0)}$, so that time-dependent matrix elements are the same in either
picture.  This is equivalent to the relation between first- and
second-quantized operators.)
External states can also be represented in terms of vertices:
$$ \sket{k} = \lim_{\t\to -\infty} V(k,\t ) e\sp{-\t} \sket{0} \quad ,
\quad \sbra{k} = \lim_{\t\to\infty} \sbra{0} e\sp\t V(k,\t ) \quad .
\eqno(9.1.19)$$
The amplitude can then be represented as, using (9.1.14c),
$$ g\sp{{\bf N}-2} \left( \pr_{i=3}^{{\bf N}-1} 
\int_{z\sb{i-1}}^{z\sb{i+1}} dz\sb i
\right) \lim_{z\sb 1\to\infty\atop
z\sb{\bf N}\to 0}\;(z\sb 1)^2\sbra{0} V'(k\sb{\bf N},
z\sb{\bf N}) \cdots V'(k\sb 1 , z\sb 1 ) \sket{0} \quad , \eqno(9.1.20a)$$
where
$$ V'(k,z) = \left(-{1\over z}\right) V\left( k,\t (z)\right) \eqno(9.1.20b)$$
according to (8.1.6), since vertices have weight $w=1$.
The amplitude with this form of the external lines can be evaluated by
the same method as the previous calculation.  (In fact, it directly
corresponds to the calculation with 2 extra external lines and vanishing
initial and final momenta.)  However, being a vacuum matrix
element, it is of the same form as those for which path integrals are
commonly used in field theory.  (Equivalently, it can also be
evaluated by the operator methods commonly used in field theory before
path integral methods became more popular there.)  More details will be
given in the following section, where such methods will be generalized
to arbitrary external states.

Coupling the superstring to external super-Yang-Mills is analogous to
the bosonic string and superparticle [2.6]:  
Covariantize $D \sb \a \to D \sb \a + \d ( \s ) \G \sb \a$,
$P \sb a \to P \sb a + \d ( \s ) \G \sb a$, $\O \sp \a \to \O \sp \a 
+ \d ( \s ) W \sp \a$.  Assuming $\int d \s \; \ca$ as kinetic 
operator (again ignoring ghosts), the vertex becomes
$$ V = W \sp \a D \sb \a + \G \sp a P \sb a - \G \sb \a \O \sp \a 
\eqno(9.1.21) $$
evaluated at $\s = 0$.  Solving the constraints (5.4.8) in a 
Wess-Zumino gauge, we find
$$ \li { W \sp \a &\approx \l \sp \a \quad ,\cr
\G \sp a &\approx A \sp a + 2 \g \sp a \sb {\a \b} \Q \sp \a \l \sp \b 
\quad ,\cr
\G \sb \a &\approx \g \sp a \sb {\a \b} \Q \sp \b A \sb a + \frac43
\g \sp a \sb {\a \b} \g \sb {a \g \d} \Q \sp \b \Q \sp \g \l \sp \d 
\quad , & (9.1.22) \cr}$$
evaluated at $\s = 0$, where ``$\approx$'' means dropping terms involving
$x$-derivatives of the physical fields $A \sb a$ and $\l \sp \a$.  Plugging 
(7.3.5) and (9.1.22) into (9.1.21) gives
$$ V \approx A \sp a \Hat P \sb a + \l \sp \a \left( {\d \over {\d 
\Q \sp \a}} - \g \sp a \sb {\a \b} \Hat P \sb a \Q \sp \b 
- \frac16 i \g \sp a \sb {\a \b} \g \sb {a \g \d} 
\Q \sp \b \Q \sp \g \Q ' \sp \d \right) \quad .\eqno(9.1.23) $$
Comparing with (7.3.6), we see that the vertices, in this approximation,
are the same as the integrands of the supersymmetry generators,
evaluated at $\s = 0$. (In the case of ordinary field theory, the 
vertices {\it are\/} just the supersymmetry generators $p \sb a$ and 
$q \sb \a$, to this order in $\q$.)  Exact expressions can be obtained by 
expansion of the superfields $\G \sb \a$, $\G \sb a$, and $W \sp \a$ 
in (9.1.21)
to all orders in $\Q$ [7.6].  In practice, superfield techniques should be used
even in the external field approach, so such explicit expansion (or even 
(9.1.22) and (9.1.23)) is unnecessary.  It's interesting to note that, 
if we generalize $D$, $P$, and $\O$ to gauge-covariant derivatives 
$\de \sb \a = D \sb \a + \G \sb \a$, $\de \sb a = D \sb a + \G \sb a$, 
$\de \sp \a = \O \sp \a + W \sp \a$, with $\G \sb \a$, $\G \sb a$, and 
$W \sp \a$ now functions of $\s$, describing the vector multiplets of
all masses, then the fact that the only mode of the $\de$'s missing is 
the zero-mode of $\O \sp \a$ ($\int d \s \; \O \sp \a = 0$) directly 
corresponds to the fact that the only gauge-invariant mode of the 
connections is the zero-mode of $W \sp \a$ (the massless spinor, the massive
spinors being Stueckelberg fields).

The external field approach has also been used in the string mechanics
lagrangian method to derive field theory lagrangians (rather than just 
S-matrices) for the lower mass levels (tachyons and massless 
particles) [9.3,1.16].  Since arbitrary external fields contain arbitrary 
functions of the coordinates, the string mechanics lagrangian is no 
longer free, and loop corrections give the field theory lagrangian 
including effective terms corresponding to eliminating the higher-mass 
fields by their classical field equations.  Thus, calculating all 
mechanics-loop corrections gives an effective field theory lagrangian whose 
S-matrix elements are the tree graphs of the string field theory with 
external lines corresponding to the lower mass levels.  Such effective 
lagrangians are useful for studying tree-level spontaneous breakdown 
due to these lower-mass fields (vacua where these fields are 
nontrivial).  Field-theory-loop corrections can be calculated by 
considering more general topologies for the string (mechanics-loops 
are summed for one given topology).

\sect{9.2. Trees} 

The external field approach is limited by the fact that it treats ordinary 
fields individually instead of treating the string field as a whole.  
In order to treat general string fields, a string graph can be 
treated as just a propagator with funny topology:  For example,

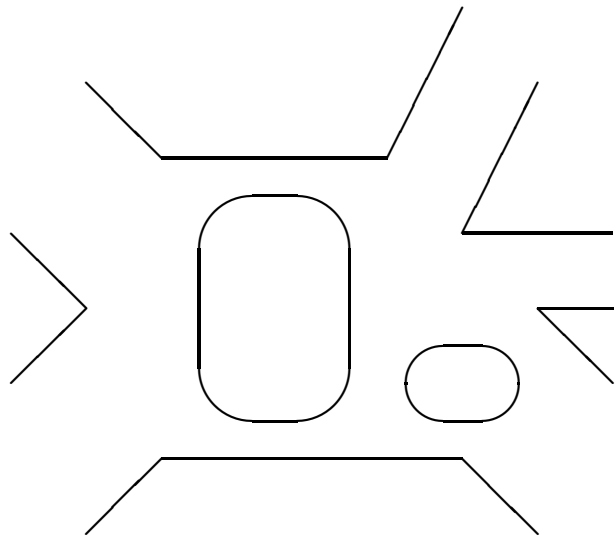
\begin{center}

\setlength{\unitlength}{1mm}
\begin{picture}(80,70)
\put(0,20){\line(1,1){10}}
\put(0,40){\line(1,-1){10}}
\put(10,0){\line(1,1){10}}
\put(10,60){\line(1,-1){10}}
\put(20,10){\line(1,0){40}}
\put(20,50){\line(1,0){30}}
\put(35,30){\oval(20,30)}
\put(60,20){\oval(15,10)}
\put(50,50){\line(1,2){10}}
\put(60,40){\line(1,2){10}}
\put(60,40){\line(1,0){20}}
\put(60,10){\line(1,-1){10}}
\put(70,30){\line(1,0){10}}
\put(70,30){\line(1,-1){10}}
\end{picture}

\end{center}

\noindent can be considered as a propagator where the initial and 
final ``one-string'' states just happen to be disconnected.  The holes 
in the world sheet represent loops.  When group theory indices are 
associated with the ends of the lines, the values of the indices are 
required to be the same along the entire line, which corresponds to 
tracing in the matrices associated with the string states.  (The ends 
of the strings can be interpreted as ``quarks'' which carry the 
``flavor'' quantum numbers, bound by a string of ``gluons'' which 
carry only ``color'' canceled by that of the quarks.)
Such an approach is limited to 
perturbation theory, since the string is necessarily gauge-fixed, and 
any one graph has a fixed number of external lines and loops, i.e., a 
fixed topology.  The advantage of this approach to graphs is that 
they can be evaluated by first-quantization, analogously to the free 
theory.  (Even the second-quantized coupling constant can be included 
in the first-quantized formalism by noting that the power of the 
coupling constant which appears in a graph, up to wave function 
normalizations, is just the Euler number, and then adding the 
corresponding curvature integral to the mechanics action.)

We first consider the light-cone formalism.  We Wick rotate the proper
time $\t\to i\t$ (see sects.\ 2.5-6), so now conformal transformations 
are arbitrary reparametrizations
of $\r = -\t + i \s$ (and the complex conjugate transformation on $\bar
\r$) instead of $\t + \s$ (and of $\t - \s$ independently: see
(6.2.7)), since the metric is now $d \s \sp 2 = d \r d \bar \r$.
There are three parts to the graph calculation: (1) 
expressing the S-matrix in terms of the Green function for the 2D 
Laplace equation, (2) finding an explicit expression for the Green 
function for the 2D surface for that particular graph, by conformally
transforming the $\r$ plane to the upper-half complex ($z$) plane 
where the Green function takes a simple form, and (3) finding the 
measure for the integration over the positions of the interaction 
points.

The first step is easy, and can be done using functional integration 
[9.4,1.4] or solving functional differential equations (the string analog 
of Feynman path integrals and the Schr\"odinger equation, respectively).  
Since all but the zero-mode (the usual spacetime coordinate) of the free
string is described by an infinite set of harmonic
oscillators, the most convenient basis is the ``number'' basis, where
the nonzero-modes are represented in terms of creation and annihilation
operators.  The basic idea is then to represent S-matrix elements as
$$ A = \langle ext | V \rangle = \sbra{ext} e\sp\D \sket{0} \quad , 
\eqno(9.2.1)$$
where $\sket{V}$ represents the interaction and $\sbra{ext}$ represents
all the states (initial and final) of the external strings, in the
interaction picture.  This is sort of a spacetime symmetric version of
the usual picture, where an initial state propagates into a final state:
Instead, the vacuum propagates into an ``external'' state.  The exponential 
$e \sp \D$ is then the analog of the S-matrix $exp(-H\sb{INT}t)$, which 
propagates the vacuum at time 0 to external states at time $t$.
It thus converts annihilation operators on its left (external, ``out'' 
states) into creation operators (for the ``in'' state, the vacuum, 
at ``time'' $x\sb + =0$).  $\D$ itself is then the ``connected''
S-matrix:  In this first-quantized picture, which looks like a free 2D
theory in a space with funny geometry, it corresponds directly to the
free propagator in this space.  Since we work in the interaction
picture, we subtract out terms corresponding to propagation in an
``ordinary'' geometry.

In the former approach, the amplitude can be evaluated as the Feynman 
path integral 

\font\godam=eufm10 scaled 1440
\def\SC#1{\hbox{\godam #1}}

\begin{large}
$$ \li{ A =& \int \left( \pr_{i=3}^{{\bf N}-1} d \t \sb i \right) \int 
	{\SC D} X\sp i ( \s , \t ) \cr
&\cdot \left[ \pr_r \int {\SC D} P \sb r ( \s ) \; \J [ P \sb r ] e 
	\sp {- i {1 \over \a '} \int {d \s \over 2 \p} \; P \sb r ( \s )
	\cdot X \sb r ( \s , \t \sb 1 )} \right] \cr
&\cdot e \sp {- \su p \sb {-r} \t \sb {1r} - {1 \over \a '} \int {d \sp 2 \s 
	\over 2 \p} \; \left[ \ha ( \dot X \sp 2 + X ' \sp 2 ) + constant 
	\right]} \quad ,&\mbox{\normalsize (9.2.2)}\cr}$$
\end{large}

\noindent corresponding to the picture (e.g., for {\bf N}$=5$)

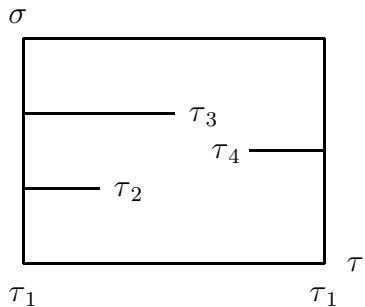
\begin{center}

\begin{picture}(60,50)
\put(10,10){\line(1,0){40}}
\put(10,40){\line(1,0){40}}
\put(10,10){\line(0,1){30}}
\put(50,10){\line(0,1){30}}
\put(10,20){\line(1,0){10}}
\put(10,30){\line(1,0){20}}
\put(40,25){\line(1,0){10}}
\put(8,5){$\t \sb 1$}
\put(8,42){$\s$}
\put(48,5){$\t \sb 1$}
\put(53,9){$\t$}
\put(22,19){$\t \sb 2$}
\put(32,29){$\t \sb 3$}
\put(35,24){$\t \sb 4$}
\end{picture}

\end{center}

\noindent where the $\t$'s have been Wick-rotated, $\t \sb {1r}$ is 
the end of the $r$th string (to be taken to $\pm \infty$), the $p \sb 
- \t \sb 1$ factor amputates external lines (converts from the
Schr\"odinger picture to the interaction picture), the factor in 
large brackets is the external-line wave function $\J [ X ]$ 
(or $\bar \J [ X ]$ for outgoing states) written as a Fourier transform,
and the constant corresponds to the usual normal-ordering constant in
the free hamiltonian.  For explicitness, we have 
written the integration over interaction points ($\pr d \t$) for the 
simple case of open-string tree graphs.  Planar graphs always appear as 
rectangles due to the string lengths being $2 \p \a ' p \sb +$, which is 
conserved.  The functional integral (9.2.2) is gaussian, so, making the 
definition 
$$ J ( \s , \t ) = i \d ( \t - \t \sb 1 ) {1\over\sqrt{\a '}}
P ( \s ) \quad ,\eqno(9.2.3)$$
we find
$$ -\int {d \sp 2 \s \over 2 \p} \; \left[ {1\over\a '} 
\ha ( \pa X ) \sp 2 + {1\over\sqrt{\a '}} J X \right] 
\quad \to \quad -\ha \int {d \sp 2 \s \over 2 \p} \; {d \sp 2 \s ' 
\over 2 \p} \; J ( \s ) G ( \s , \s ' ) J ( \s ' ) \quad ,$$
$$ \pa \sp 2 G ( \s , \s ' ) = 2 \p \d \sp 2 ( \s ' - \s ) \quad , \quad 
{\pa \over \pa n} G ( \s , \s ' ) = f ( \s ) \quad \left( \; \int d \sp 2 
\s \; J \sim \su p = 0 \; \right) \quad ,$$
$$ \li{ G =& \su (2-\d\sb{m0})(2-\d\sb{n0}) G\sp{rs}\sb{mn} \; 
cos \left( m {\s \sb r \over p \sb{+r}}\right) cos \left( n {\s 
\sb s ' \over p \sb{+s}}\right) e \sp {m \t \sb r / p \sb{+r} + n \t \sb 
s ' / p \sb{+s}} \cr
&+  G \sb {free} + (zero\hbox{-}mode)\sp 2~terms\cr}$$
$$\to \quad A = \int ( \pr d \t \sb i ) V ( \t ) \left< \J \left| e \sp 
\D \right| 0 \right> \quad , \quad \D = \frac14 \su G \sp {rs} \sb {mn} 
\a \sp r \sb m \cdot \a \sp s \sb n \quad,\eqno(9.2.4)$$
where $G$ is the 2D Green function for the kinetic operator (laplacian) 
$\pa \sp 2 / \pa \t \sp 2 + \pa \sp 2 / \pa \s \sp 2$ of that particular 
surface, and $V ( \t )$ comes from $det ( G )$.  We have used Neumann
boundary conditions (corresponding to (6.1.5)), where the ambiguity
contained in the arbitrary function $f$ (necessary in general to allow a
solution) is harmless because of the conservation of the current $J$
(i.e., the momentum $p$).  The $G\sb{free}$ term is dropped in
converting to the interaction picture:  In functional notation 
(see (9.2.2)), it produces the ground-state wave function.  The
(zero-mode)$\sp 2$ terms are due to boundary conditions at $\infty$, and
appear when the map to the upper-half plane is chosen so that the end of
one string goes to $\infty$, giving a divergence.  They correspond to the
factor $1/z\sb{\bf N}$ in the similar map used for (9.1.20a).  
The factors $(2-\d\sb{m0})$, which don't appear in the naive Green
function, are to correct for the fact
that the figure above is not quite the correct one:  The boundaries of
the initial and final strings should not go to $\pm\infty$ before the
source terms (9.2.3) (i.e., the wave functions) do, because of the
boundary conditions.  The net result is that nonzero-modes appear with an
extra factor of 2 due to reflections from the boundary.  However, these
relative factors of 2 are canceled in the $\s$-integration, since
$\int_0^\p{d\s\over 2\p}cos\sp 2 m\s =\frac14 (1+\d\sb{m0})$.  Explicitly,
after transforming the part of the $\r$ plane corresponding to the
string to the whole of the upper-half $z$ plane, the Green function becomes
$$ G ( z , z ' ) = ln | z - z ' | + ln | z - \bar z ' | \quad . \eqno(9.2.5)$$
The first term is the usual Green function without boundaries, whose use
in the $z$ plane (and not just the $\r$ plane) follows from the fact
that the Laplace equation is conformally invariant.  The second term,
which satisfies the homogeneous Laplace equation in the upper-half
plane, has been added according to the method of images in order to
satisfy the boundary conditions at the real axis, and gives the
reflections which contribute the factors of 2.

In the latter (Schr\"odinger equation/operator) approach, it all boils 
down to using the general expression
$$ \check \J ( z ) = \oint _z {{d z '} \over {2 \p i}} {1 \over {z ' - z}} 
\check \J ( z ' ) = - \su _r \oint _{z \sb r} {{d z '} \over {2 \p i}} {1 \over
{z ' - z}} \check \J ( z ' ) \quad , \eqno(9.2.6)$$
where $z \sb r$ are the points in the $z$-plane representing the ends 
of the strings (at $\r = \pm\infty$).  $\check\J (z)$ is an arbitrary
operator which has been conformally transformed to the $z$ plane:
$$ \check \J \sb r (z) = \left({\pa\r\over\pa z}\right)^w\tilde\J\sb r(\r )
\quad , \quad\tilde\J\sb r(\r ) = (p\sb{+r})\sp{-w}\J\sb r\left({\tilde\r \sb r
\over p \sb{+r}} \right) \quad , \quad \J \sb r ( \z ) = \su _{- \infty} ^ 
{\infty} \j \sb {rn} e \sp {-n \z} \quad ,\eqno(9.2.7a)$$
$$ \tilde\r\sb r = \r - i \p\su _{s=1} ^{r-1} p\sb{+s} \quad , \eqno(9.2.7b)$$
with $\J ( i \s ) = \hat \J ( \s )$ in terms of $\Hat 
P ( \s )$ (so the $\j \sb n$'s are the $\a \sb n$'s of (7.1.7a)), and 
the conformal transformations (9.2.7a) (cf.\ (8.1.6)) are
determined by the conformal weights $w$ ($=1$ for $\check P$).
The $\r\to z$ map is the map from the above figure to the upper-half
plane.  The $\z\to\r$ map is the map from the free-string $\s\in [0,\p ]$
to the $r$th interacting-string $\s\in [\p\su_{s=1}^{r-1}p\sb{+s} ,
\p\su_{s=1}^rp\sb{+s} ]$.
All $\s$ integrals from $- \p$ to $\p$ become contour integrals in the 
$z$ plane.  (The upper-half $z$ plane corresponds to $\s \in [ 0 , 
\p ]$, while the lower half is $\s \in [ - \p , 0 ]$.)  Since the
string (including its extension to $[-\p ,0]$) is mapped to the entire
$z$ plane without boundaries, (9.2.6) gets contributions from only the 
ends, represented by $z\sb r$.  We
work directly with $\Hat P(\s )$, rather than $X(\s )$, since $\Hat P$
depends only on $\r$, while $X$ depends on both $\r$ and $\bar\r$.
($\hat X$ has a cut in the $z$ plane, since $p\s$ isn't periodic in $\s$.)
The open string results can also be
applied directly to the closed string, which has separate operators
which depend on $\bar z$ instead of $z$ (i.e., $+$ and $-$ modes, in the
notation of sect.\ 6.2).  Actually, there is a bit of a cheat,
since $\Hat P (\s )$ doesn't contain the zero-mode $x$.  However, this
zero-mode needs special care in any method:  Extra factors of 2 appeared
in the path-integral approach because of the boundary conditions along
the real $z$-axis and at infinity.  In fact, we'll see that the lost
zero-mode terms can be found from the same calculation generally used in both
operator and path integral approaches to find the integration measure of
step 3 above, and thus requires no extra effort.

We therefore look directly for a propagator
$e\sp\D$ that gives 
$$ \check\J (z) = e\sp\D \check\J\sb r (z)e\sp{-\D} = \check\J\sb r (z)
+ [ \D , \check\J\sb r (z) ] + \cdots \quad , \eqno(9.2.8)$$
where $\check\J\sb r$ corresponds to a free
in-field for the $r$th string in the interaction picture, and $\check\J$
to the interacting field.  To do this, we first find a $\tilde\D$ for which 
$$ [ \tilde\D , \check\J\sb r ] = \check\J \quad .\eqno(9.2.9)$$
We next subtract out the free part of $\tilde\D$ (external-line amputation):  
$$ \tilde\D = \D + \D\sb{free} \quad , \quad
[ \D\sb{free} , \check\J\sb r ] = \check\J\sb r \quad .\eqno(9.2.10)$$
This gives a $\D$ which is quadratic in operators, but contains no 
annihilation operators (which are irrelevant anyway, since the $\sket{0}$ 
will kill them).  As a result, there are no terms with multiple
commutators in the expansion of the exponential.  We therefore obtain
(9.2.8).  When we subtract out free parts below, we will include the
parts of the external-line amputation which compensate for the fact that
$z$ and $z\sb r$ are not at the same time.

We first consider applying this method to operators of
arbitrary conformal weight, as in (8.1.23).  The desired form of $\D$
which gives (9.2.8) for both $f$ and $\d / \d f$ is
$$ \D \sb 0 = -\su_{r,s} \oint_{z\sb r} {dz\over 2\p i} \oint_{z\sb s}
{dz'\over 2\p i} \; {1\over z-z'} \check f\sb r (z) \fder{\check f\sb s
(z')} - free{\hbox{-}}string~terms \quad , \eqno(9.2.11a)$$
where
$$ \left[ \fder{\check f\sb r(z\sb 1)} , \check f\sb s(z\sb 2) \right\} =
2\p i \d\sb{rs} \d (z\sb 2 -z\sb 1) \quad\iff\quad
\left[ \fder{\hat f\sb r(\s\sb 1)} , \hat f\sb s(\s\sb 2) \right\} =
2\p \d\sb{rs} \d (\s\sb 2 -\s\sb 1) \quad , \eqno(9.2.11b)$$
as follows from the fact that the conformal transformations preserve 
the commutation relations of $f$ and $\d / \d f$.
The integration contours are oriented so that
$$ \oint _{\r \sb r} {{d \r} \over {2 \p i p \sb{+r}}} = \int _{- \p p 
\sb{+r}} ^{\p p \sb{+r}}{{d \s}\over{2 \p p \sb{+r}}} =1 \quad .\eqno(9.2.12)$$
(9.2.11) can easily be shown to satisfy (9.2.8).
The value of $\t\sb r$ ($\to\pm\infty$) for the 
integration contour is fixed, so the $\d$ in $\z$ in these commutation 
relations is really just a $\d$ in $\s$ of that contour.  The free-string 
terms are subtracted as explained above.  (In fact, they are poorly defined,
since the integration contours for $r=s$ fall on top of each other.)

Unfortunately, $\D$ will prove difficult to evaluate in
this form.  We therefore rewrite it by expressing the $1 / (z-z')$ as
the derivative with respect to either $z$ or $z'$ of a $ln$ and perform
an integration by parts.  The net result can be written as
$$\li{ \D ( \check\J\sb 1 , \check\J\sb 2 ) =& \su_{r,s} \left( \oint_{z\sb r} 
{{dz}\over{2\p i}} \oint_{z\sb s}{{dz '}\over{2\p i}} \right)^\prime \; 
ln ( z - z ' ) \check \J \sb {1r} ( z ) \check \J \sb {2s} ( z ' ) \cr
& -~free\hbox{-}string~terms\quad +~(zero\hbox{-}mode)\sp 2~terms
\; ,&(9.2.13a)\cr}$$
$$[ \check\J \sb {2r} ( z \sb 2 ) \; , \; \check\J \sb {1s} ( z \sb 1 )
\} = -2 \p i \d \sb {rs} \d ' ( z \sb 2 - z \sb 1 ) \quad ,\eqno(9.2.13b)$$
The $'$ on the contour integration is because the integration is poorly
defined due to the cut for the $ln$:  We therefore define it by
integration by parts with respect to either $z$ or $z'$, dropping
surface terms.  This also kills the constant part of the $ln$ which
contributes the (zero-mode)$\sp 2$ terms, which we therefore add back in.
Actually, (9.2.11) has no (zero-mode)$\sp 2$ terms, but
in the case $\J\sb 1 = \J\sb 2 = P$, these terms determine the evolution
of the zero-mode $x$, which doesn't appear in $\Hat P$, and thus could
not be determined by (9.2.8) anyway.  ($x$ does appear in $X$ and 
$\hat X$, but they're less convenient to work with, as explained above.)
These (quadratic-in-)zero-mode contributions are most easily calculated
separately by considering the case when all external states are ground
states of nonvanishing momentum (see below).
In order for the commutation relations (9.2.13b) to be preserved by the
conformal transformations, it's necessary that the conformal weights
$w\sb 1$ and $w\sb 2$ of $\J\sb 1$ and $\J\sb 2$ both be 1.  In that case,
$\check\J$ can be replaced with $\tilde\J$ in (9.2.13a) while replacing 
$dz$ with $d\r$.  However, one important use of this equation is for the
evaluation of vertices (S-matrices with no internal propagators).  Since
these vertices are just $\d$ functionals in the string field coordinates
(see below), and $\d$ functionals are independent of conformal weight
except for the $\z\to\r$ transformation (since that transformation
appears explicitly in the argument of the $\d$ functionals), we can
write this result, for the cases of S-matrices with $w\sb 1 = w\sb 2 =1$
or vertices (with $w\sb 1 + w\sb 2 =2$) as
$$\li{ \D ( \check\J\sb 1 , \check\J\sb 2 ) =&\su_{r,s} \left( \oint_{\r\sb r} 
{{d\r}\over{2\p i}} \oint_{\r\sb s}{{d\r '}\over{2\p i}} \right)^\prime \; 
ln ( z - z ' ) \check \J \sb {1r} ( \r ) \check \J \sb {2s} ( \r ' ) \cr
& -~free\hbox{-}string~terms\quad +~(zero\hbox{-}mode)\sp 2~terms
\; ,&(9.2.14a)\cr}$$
$$[ \J \sb {2r} ( \z \sb 2 ) \; , \; \J \sb {1s} ( \z \sb 1 )
\} = -2\p i\d \sb {rs} \d ' ( \z \sb 2 - \z \sb 1 ) \quad ,\eqno(9.2.14b)$$
where the appropriate $\D$ for $X$ ($w\sb 1 = w\sb 2 =1$) is
$$ \D = \ha \D ( \check P , \check P ) \quad .\eqno(9.2.15) $$
These $'$ and free-string corrections may seem awkward, but they will
automatically be fixed by the same method which gives a simple
evaluation of the contour integrals: i.e., the terms which are
difficult to evaluate are exactly those which we don't want.
(For non-vertex S-matrices with $w\sb 1 \ne 1 \ne w\sb 2$, (9.2.11) or 
(9.2.13) can be used, but their evaluated forms are much more complicated
in the general case.)
In general (for covariant quantization, supersymmetry, 
etc.) we also need extra factors which are evaluated at infinitesimal 
separation from the interaction (splitting) points, which follow from 
applying  the conformal transformation (9.2.7), and (9.2.6) with 
$z = z \sb {INT}$.  Only creation operators contribute.

The contour-integral form (9.2.14) can also be derived from the
path-integral form (9.2.4).  For the open string [9.5], 
these contour integrals can be obtained by either combining integrals over 
semicircles in the upper-half $z$ plane ($\s$ integrals from $0$ to $\p$)
with their reflections [9.6], or more directly by reformulating the 
open string as a closed string with modes of one handedness only, and 
with interactions associated with just the points $\s = 0 , \p$ rather 
than all $\s$.

For the second step, $G \sp {rs} \sb {mn}$ unfortunately is hard to 
calculate in general.  For open-string tree graphs, we perform the following 
conformal mapping to the upper-half complex plane [9.4], where $\D$ is
easy to calculate:
$$ \r = \su_{r=2}^{\bf N} p \sb{+r} ln ( z - z \sb r ) \quad . 
\eqno(9.2.16)$$
The boundary of the (interacting) string is now the real $z$ axis, 
and the interior is the upper half of the complex $z$-plane.
(The branches in the $ln$'s in (9.2.16) are thus chosen to run down into
the lower-half plane.  When we use the whole plane for contour integrals
below, we'll avoid integrals with contours with cuts inside them.)
As a result, operators such as $\Hat P$, which were periodic in $\s$,
are now meromorphic at $z\sb r$, so the contour integrals are easy to
evaluate.  Also, extending $\s$ from $[0,\p ]$ to $[-\p , \p ]$ extends
the upper-half plane to the whole complex plane, so there are no
boundary conditions to worry about.  To evaluate (9.2.14), we note that, 
since we are neglecting (zero-modes)$\sp 2$ and free string terms
($r=s$, $m=-n$), we can replace
$$ ln(z-z')e\sp{-m\r /p\sb{+r} - n\r' /p\sb{+s}} \quad\to\quad
{1\over{m\over p\sb{+r}}+{n\over p\sb{+s}}}\left[ \left( \der\r + \der{\r '}
\right) ln (z-z')\right] e\sp{-m\r /p\sb{+r} - n\r' /p\sb{+s}} \eqno(9.2.17)$$
by integration by parts.  We then use the identity, for the case of (9.2.16),
$$ \left( \der\r + \der{\r '} \right) ln (z-z') = 
\su_{r=2}^{\bf N} p\sb{+r} \left[ \der\r ln (z-z\sb r) \right]
\left[ \der{\r '} ln (z'-z\sb r) \right] \quad . \eqno(9.2.18)$$
Then we can trivially change completely to $z$ coordinates by using
$d\r \pa / \pa\r = dz\pa /\pa z$, and converting the $\r$ exponentials
into products of powers of $z$ monomials.  Differentiating the $ln$'s
gives (products of) single-variable contour integrals which can easily
be evaluated as multiple derivatives:
$$ \D = \su_{rsmn} \j\sb{1rm}\j\sb{2sn}(p\sb{+r})\sp{1-w\sb 1}
(p\sb{+s})\sp{1-w\sb 2}{1\over np\sb{+r} + mp\sb{+s}} \su_{t=2}^{\bf N}
p\sb{+t} A\sb{rtm}A\sb{stn} + (zero\hbox{-}mode)\sp 2 \quad , $$
$$ A\sb{rtm} = \oint_{z\sb r}{dz\over 2\p i} \; {1\over z-z\sb t}
\left[ (z-z\sb r) \pr_{s=2}^{r-1}(z\sb s-z)\sp{p\sb{+s} /p\sb{+r}}
\pr_{s=r+1}^{\bf N}(z-z\sb s)\sp{p\sb{+s} /p\sb{+r}} \right]^{-m}
\quad . \eqno(9.2.19) $$

For the third step, for open-string trees,
we also need the Jacobian from $\pr d \t \sb i \to 
( \pr d z \sb i ) V ( z )$, which for trees can easily be calculated by 
considering the graph where all external states are tachyons and all 
but 2 strings (one incoming and one outgoing) have infinitesimal length.
We can also restrict all transverse momenta to vanish, and determine
dependence on them at the end of the calculation by the requirement of
Lorentz covariance.  (Alternatively, we could complicate the calculation
by including transverse momenta, and get a calculation more similar to
that of (9.1.9).)  We then have the amplitude (from
nonrelativistic-style quantum mechanical arguments, or specializing (9.2.2))
$$ A = g\sp{{\bf N}-2} f (p\sb{+r} ) \int \left( \pr_{i=3}^{{\bf N}-1} 
d\t\sb i \right)
e\sp{-\su_{r=2}^{{\bf N}-1}p\sb{-r}\t\sb r} \quad , \eqno(9.2.20)$$
where $f$ is a function to be determined by Lorentz covariance, the
$\t$'s are the interaction points, the strings 1 and {\bf N} are those
whose length is not infinitesimal, and we also choose $z\sb 1 =\infty$,
$z\sb{\bf N}=0$ in the transformation (9.2.16).  We then solve for 
$\t\sb r$ ($=-Re(\r\sb r)$), in terms of $z\sb r$ (in this approximation
of infinitesimal lengths for all but 2 of the strings), as
the finite values of $\r$ where the boundary ``turns around'':
$$ \left. {\pa\r\over\pa z}\right|_{\r\sb r} = 0 \quad\to $$
$$ \r\sb r = p\sb{+\bf N} ln~z\sb r + p\sb{+r} \left[ ln \left(
-{p\sb{+r}\over p\sb{+\bf N}}z\sb r \right) -1\right] +
\su_{s=2,s\ne r}^{{\bf N}-1} p\sb{+s} ln(z\sb r -z\sb s) +
\co \left[\left({p\sb{+r}\over p\sb{+\bf N}}\right)^2\right] \quad .
\eqno(9.2.21)$$
We then find, using the mass-shell condition $p\sb - = 1/p\sb +$ for the
tachyon ($p\sb -$ is $-H$ in nonrelativistic-style calculations)
$$ A = g\sp{{\bf N}-2} \left[ f p\sb{+\bf N}\sp{-1} \left( 
\pr_{r=2}^{{\bf N}-1} {p\sb{+r}\over e} \right) \right] \int \left( 
\pr_{i=3}^{{\bf N}-1} dz\sb i \right) z\sb 2 \pr_{s>r=2}^{\bf N} 
( z\sb r - z\sb s )^{p\sb{+r}/ p\sb{+s} + p\sb{+s}/ p\sb{+r}} \cdot $$
$$ \cdot e\sp{\su p\sb{ir}n\sb{rs}(p\sb +,z)p\sb{is}} \quad , \eqno(9.2.22a)$$
where we have now included the transverse-momentum factor $n\sb{rs}$,
whose exponential form follows from previous arguments.  Its explicit
value, as well as that of $f$, can now be determined by the manifest
covariance of the tachyonic amplitude.  However, (9.2.22a) is also 
the correct measure for the $z$-integration to
be applied to (9.2.1) (or (9.2.4)), using $\D$ from (9.2.19) (which is 
expressed in terms of the same transformation (9.2.16)).
At this point we can see that Lorentz covariance determines $f$ to be 
such that the factor in brackets is a constant.  We then note that,
using $p\sb - = -H = -(\ha p\sb i\sp 2 +N-1)/p\sb +$, we have
$p\sb r \cdot p\sb s = p\sb{ir}p\sb{is}
-[(p\sb{+r}/p\sb{+s})(\ha p\sb{is}\sp 2+N-1) +r\iff s]$.  This determines the
choice of $n\sb{rs}$ which makes the amplitude manifestly covariant
for tachyons:
$$ \int dz\; V(z) = g\sp{{\bf N}-2} \int \left( \pr_{i=3}^{{\bf N}-1} 
dz\sb i \right)
z\sb 2 \pr_{s>r=2}^{\bf N} ( z\sb r - z\sb s )^{p\sb r\cdot p\sb s
-(p\sb{+r}/p\sb{+s})(\ha p\sb s\sp 2 -1) - (p\sb{+s}/p\sb{+r})
(\ha p\sb r\sp 2 -1)} . \eqno(9.2.22b)$$
Note that $p\sb -$ dependence cancels, so $p\sb r\cdot p\sb s$ and
$p\sb r\sp 2$ can be chosen to be the covariant ones.
Taking {\bf N}=1 to compare with the tachyonic particle 
theory, we see this agrees with the result (9.1.15) (after choosing 
also $z\sb 2 =1$).  It also gives the
(zero-mode)$\sp 2$ terms which were omitted in our evaluation of $\D$.
(That is, we have determined both of these factors by considering this
special case.)  For the case of the tachyon, we could have obtained the
covariant result (9.2.22b) more directly by using covariant amputation
factors, i.e., by using $p\sb -$ as an independent momentum instead of
as the hamiltonian (see sect.\ 2.5).  However, the result loses its
manifest covariance, even on shell, for excited states because of the
usual $1/p\sb +$ interactions which result in the light-cone formalism
after eliminating auxiliary fields.

As mentioned in sect.\ 8.1, there is an Sp(2) invariance of free string
theory.  In terms of the tree graphs, which were calculated by
performing a conformal map to the upper-half complex plane, it
corresponds to the fact that this is the subgroup of the conformal group
which takes the upper-half complex plane to itself.  Explicitly, the
transformation is $z \to (m\sb{11}z+m\sb{12})/(m\sb{21}z+m\sb{22})$, 
where the matrix $m\sb{ij}$ is real, and without loss of generality
can be chosen to have determinant 1.  This transformation also takes the
real line to itself, and when combined with (9.2.16) modifies it only by
adding a constant and changing the values of the $z\sb r$ (but not their
order).  In particular, the 3 arbitrary parameters allow arbitrary
values (subject to ordering) for $z\sb 1$, $z\sb 2$, and $z\sb{\bf N}$,
which were $\infty$, 1, and 0.  (This adds a term for $z\sb 1$ to
(9.2.16) which was previously dropped as an infinite constant.
$\r\to\infty$ as $z\to\infty$ in (9.2.16) corresponds to the end of the
first string.)  Because of the Sp(2) invariance, (9.2.22) can be
rewritten in a form with all $z$'s treated symmetrically:
The tachyonic amplitude is then
$$ A = g\sp{{\bf N}-2} \int {\pr_{r=1}^{\bf N}dz\sb r \over
dz\sb i dz\sb j dz\sb k}(z\sb i -z\sb j)(z\sb i -z\sb k)(z\sb j -z\sb k)
\pr_{1\le r<s\le{\bf N}}(z\sb r -z\sb s)\sp{p\sb r\cdot p\sb s}
\quad , \eqno(9.2.22c) $$
where $z\sb i$, $z\sb j$, $z\sb k$ are any 3 $z$'s, which are not
integrated over, and with all $z$'s cyclically ordered as in (9.1.14).

Closed-string trees are similar, but whereas open-string interaction 
points occur anywhere on the {\it boundary}, closed-string interaction 
points occur anywhere on the {\it surface}.  (Light-cone coordinates
are chosen so that these points always occur for those values of $\t$ 
where the strings split or join.)  Thus, for closed strings there are 
also integrations over the $\s$'s of the interaction points.
The amplitude corresponding to (9.1.15) or (9.2.22) for the 
closed string, since it has both clockwise and counterclockwise modes, 
has the product of the integrand for the open string (for one set of 
modes) with its complex conjugate (for the modes propagating in the 
opposite direction), and the integral is over both $z$'s and $\bar z$'s 
(i.e., both $\t$'s and $\s$'s).  (There are also additional factors of 
$1/4$ in the exponents due to the different normalization of the zero 
modes.)  However, whereas the integral in 
(9.1.15) for the open string is restricted by (9.1.14) so that the 
$z$'s (interaction points) lie on the boundary (the real axis), and 
are ordered, in the closed string case the $z$'s are anyplace on the 
surface (arbitrary complex).

We next consider the evaluation of the open-string 3-point 
function, which will be needed below as the vertex in the field theory 
action.  The 3-string vertex for the open string can be written 
in functional form as a $\d$-functional equating the coordinates of a 
string to those of the strings into which it splits.  In the case of 
general string coordinates {\bf Z}:
$$ S\sb{INT} = g \int d \t \; {\Sc D} \sp 3 {\bf Z} \; d \sp 3 p\sb + \;
\d\left(\su p\sb +\right)\d [\tilde{\bf Z}\sb 1 (\s ) -\tilde{\bf Z}\sb 3(\s )]
\d [ \tilde {\bf Z} \sb 2 ( \s ) - \tilde {\bf Z} \sb 3 ( \s ) ] 
\F [ 1 ] \F [ 2 ] \F [ 3 ] \quad ,\eqno(9.2.23)$$
with $\tilde {\bf Z}$ as in (9.2.7).
We now use (9.2.6-16) for {\bf N}$=3$, with $z \sb r = \infty , 1 , 0$ 
in (9.2.16), and $p\sb{+1}$ with the opposite sign to $p\sb{+2}$ and
$p\sb{+3}$.  The splitting point is 
$$ {\pa\r\over\pa z} =0 \quad\to\quad z = z \sb 0 = -{p\sb{+3}\over p\sb{+1}}
\quad , \quad \r = \t \sb 0 + i \p p \sb{+2} \quad , \eqno(9.2.24a)$$
$$ \t \sb 0 = \ha \su p\sb + \; ln \; ( p\sb +\sp 2 ) \quad . \eqno(9.2.24b) $$
For (9.2.19), we use the integral
$$ \oint_0 {dz\over 2\p i} \; z\sp{-n-1}(z+1)\sp u = 
{1\over n!}{d\sp n\over dz\sp n}(z+1)\sp u |\sb{z=0} =
{u(u-1)\cdots (u-n+1)\over n!} = \pmatrix{u\cr n\cr} \quad ,$$
$$ \pmatrix{-u+n-1\cr n\cr} = (-1)\sp n \pmatrix{u\cr n\cr} \eqno(9.2.25)$$
to evaluate
$$ \li{ m>0: \quad & A\sb{r2m} = p\sb{+3}\cn\sb{rm} \quad , \quad
A\sb{r3m} = -p\sb{+2} \cn\sb{rm} \quad ; \quad
\cn\sb{rm}={1\over p\sb{+,r+1}}\pmatrix{-m{p\sb{+,r+1}\over p\sb{+r}}\cr m\cr} 
\quad ; \cr
m=0: \quad & A\sb{rt0} = \d\sb{rt}-\d\sb{r1} \quad . &(9.2.26)\cr}$$
The result is then (see, e.g., [9.4]):
$$ \D (\check\J\sb 1 , \check\J\sb 2) = -\j\sb 1\Go N\j\sb 2 
- \Tilde\j\sb 1\cn{1\over n}\j\sb 2 - \j\sb 1{1\over n}\cn\Tilde\j\sb 2 
- \t\sb 0\su{{p\sp 2 + M\sp 2}\over 2p\sb +} \quad ,$$
$$ \Go N\sb{rsmn} = {p\sb{+1}p\sb{+2}p\sb{+3}\over np\sb{+r} + 
mp\sb{+s}}\cn\sb{rm}\cn\sb{sn} \quad , \quad
\Tilde\j = p\sb{+[r}(\j\sb 0)\sb{r+1]} \quad ,$$
$$ S = \int d\sp 3p\sb + \; d\sp 3\j \; \d\left(\su p\sb +\right) 
\d\left(\su\j\sb 0\right) \left< \F\sb 1\F\sb 2\F\sb 3 \left| e\sp\D \right| 
0 \right> \quad .\eqno(9.2.27)$$
(In some places we have used matrix notation with indices $r,s=1,2,3$ 
and $m,n=1,2,...,\infty$ implicit.)  The $\j$'s include the $p$'s.  
For simplicity, we have assumed the $\j$'s have $w=1$; otherwise, each 
$\j$ should be replaced with $p\sb +\sp{1-w}\j$.  The $\t\sb 0$ term comes 
from shifting the value of $\t$ at which the vertex is evaluated from 
$\t =0$ to the interaction time $\t = \t\sb 0$ (it gives just the 
propagator factor $e\sp{-\t\sb 0\su H\sb r}$, where $H\sb r$ is the 
free hamiltonian on each string).  In more general cases we'll also 
need to evaluate a regularized $\Hat\J$ at the splitting point, which 
is also expressed in terms of the mode expansion of $ln ( z - z \sb r )$
(actually its derivative $1 / ( z - z \sb r )$) which was used in 
(9.2.14) to obtain (9.2.27):
$$ \check\J (z\sb 0) \to {1\over\sqrt{p\sb{+1}p\sb{+2}p\sb{+3}}}\Tilde\j + 
\sqrt{p\sb{+1}p\sb{+2}p\sb{+3}}p\sb +\sp{-w}\cn\j \quad . \eqno(9.2.28)$$
(Again, matrix notation is used in the second term.)  We have
arbitrarily chosen a convenient normalization factor in the regularization.
(A factor which diverges as 
$z \to z \sb 0$ must be divided out anyway.)  The vertex is cyclically 
symmetric in the 3 strings (even though some strings have $p\sb + < 0$).
Besides the conservation law $\su p = 0$, we also have $\su p\sb + x = 0$, 
which is actually the conservation law for angular momentum $J \sb 
{+i}$.  These are special cases of the $\j\sb 0$ conservation law
indicated above by the $\d$ function, after including the $p\sb +\sp{1-w}$.
(Remember that a coordinate of weight $w$ is conjugate to one with weight
$1-w$.)  This conservation law makes the definition of $\tilde\j$ above
$r$-independent.  However, such 
conservation laws may be violated by additional vertex factors (9.2.28).

The 3-string vertex for the closed string in operator form is 
essentially just the product of open-string vertices for the clockwise 
and counterclockwise modes, since the $\d$ functionals can be written 
as such a product, except for the zero modes.  However, whereas 
open strings must join at their ends, closed strings may join 
anywhere, and the $\s$ parametrizing this joining is then integrated 
over.  Equivalently, the vertex may include projection operators 
$\d \sb {\D N , 0} = \int {d \s \over 2 \p} \; e \sp {i \s \D N}$
which perform a $\s$ translation equivalent to the integration.  (The 
former interpretation is more convenient for a first-quantized 
approach, whereas the latter is more convenient in the operator 
formalism.)  These projection operators are redundant in a ``Landau 
gauge,'' where the residual $\s \to \s + constant$ gauge invariance is 
fixed by introducing such projectors into the propagator.

In the covariant first-quantized formalism one can consider more 
general gauges for the $\s$-$\t$ reparametrization invariance and 
local scale invariance than $g \sb {\bf mn} = \h \sb {\bf mn}$.
Changing the gauge has the effect of ``stretching'' the surface in 
$\s$-$\t$ space.  Since the 2D metric can always be chosen to be flat 
in any small region of the surface, it's clear that the only invariant 
quantities are global.  These are topological quantities (some 
integers describing the type of surface) and certain proper-length 
parameters (such as the proper-length of the propagator in the case of 
the particle, as in (5.1.13)).  In particular, this applies to the light-cone 
formalism, which is just a covariant gauge with stronger gauge conditions (and 
some variables removed by their equations of motion).  Thus, the 
planar light-cone tree graph above is essentially a flat disc, and the 
proper-length parameters are the $\t \sb i$, $i = 3, \dots , {\bf N}-1$.
However, there are more general covariant gauges even for such 
surfaces with just straight-line boundaries:  For example, we can identify

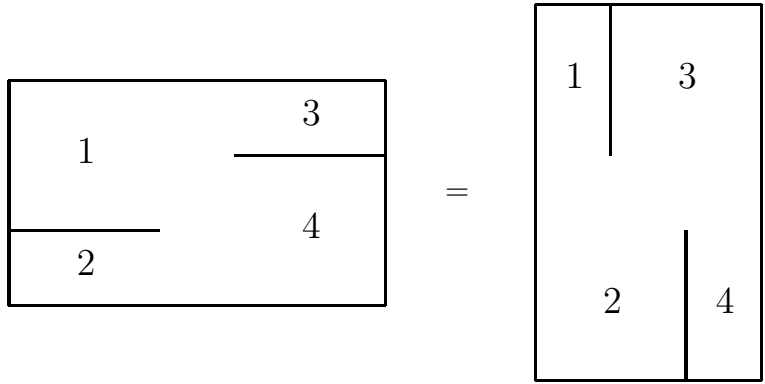
\phantom M

\phantom M

\begin{center}

\begin{picture}(100,55)
\put(0,10){\line(0,1){30}}
\put(50,10){\line(0,1){30}}
\put(70,0){\line(0,1){50}}
\put(100,0){\line(0,1){50}}
\put(90,0){\line(0,1){20}}
\put(80,30){\line(0,1){20}}
\put(0,10){\line(1,0){50}}
\put(0,40){\line(1,0){50}}
\put(70,0){\line(1,0){30}}
\put(70,50){\line(1,0){30}}
\put(0,20){\line(1,0){20}}
\put(30,30){\line(1,0){20}}
\put(58,24){$=$}
\put(9,14){\large 2}
\put(9,29){\large 1}
\put(39,19){\large 4}
\put(39,34){\large 3}
\put(79,9){\large 2}
\put(74,39){\large 1}
\put(94,9){\large 4}
\put(89,39){\large 3}
\end{picture}

\end{center}

\noindent with the proper-length parameter being the relative position 
of the 2 splitting points (horizontal or vertical displacement, 
respectively for the 2 graphs, with the value of the parameter being 
positive or negative).  More generally, the only invariants in 
this graph are the proper length distances measured along the boundary
between the {\it end}points (the points associated with the external 
particles), less 3 which can be eliminated by remaining projective 
invariance (consider, e.g., the surface as a disc, with the endpoints 
on the circular boundary).  Thus, we can keep 
the splitting points in the positions in the figures and vary the 
proper-length parameters by moving the ends instead.  If this is 
interpreted in terms of ordinary Feynman graphs, the first graph seems 
to have intermediate states formed by the collision of particles 1 and 
2, while the second one is from 1 and 3.  The identity of these 2 
graphs means that the same result can be obtained by summing over 
intermediate states in only 1 of these 2 channels as in the other, as 
we saw for the case of external tachyons in (9.1.16).  Thus, duality 
is just a manifestation of $\s$-$\t$ reparametrization invariance and 
local scale invariance.

\sect{9.3. Loops}

Here we will only outline the procedure and results of loop 
calculations (for details see [0.1,1.3-5,9.7-10,5.4] and the shelf of 
this week's 
preprints in your library).  In the first-quantized approach to loops
the only essential difference from trees is that the topology is
different.  This means that:  (1) It's no longer possible to conformally
map to the upper-half plane, although one can map to the upper-half
plane with certain lines identified (e.g., for the planar 1-loop graph,
which is topologically a cylinder, we can choose the region between 2
concentric semicircles, with the semicircles identified).  (2) The
integration variables include not only the $\t$'s of the interaction
points which define the position of the loop in the string, but also the
$\s$'s, which are just the $p \sb +$'s of the loop.  In covariant
gauges it's also necessary to take the ghost coordinates into account.
In the second-quantized approach the loop graphs follow directly from
the field theory action, as in ordinary field theory.  However, for
explicit calculation, the second-quantized expressions need to be
translated into first-quantized form, as for the trees.  1-loop graphs
can also be calculated in the external field approach by ``sewing''
together the 2 ends of the string propagator, converting the matrix
element in (9.1.9) into a trace, using the trace operator in (9.1.12).

An interesting feature of open string theories is that closed strings 
are generated as bound states.  This comes from stretching the one-loop 
graph with intermediate states of 2 180$\sp \circ$-twisted open strings:

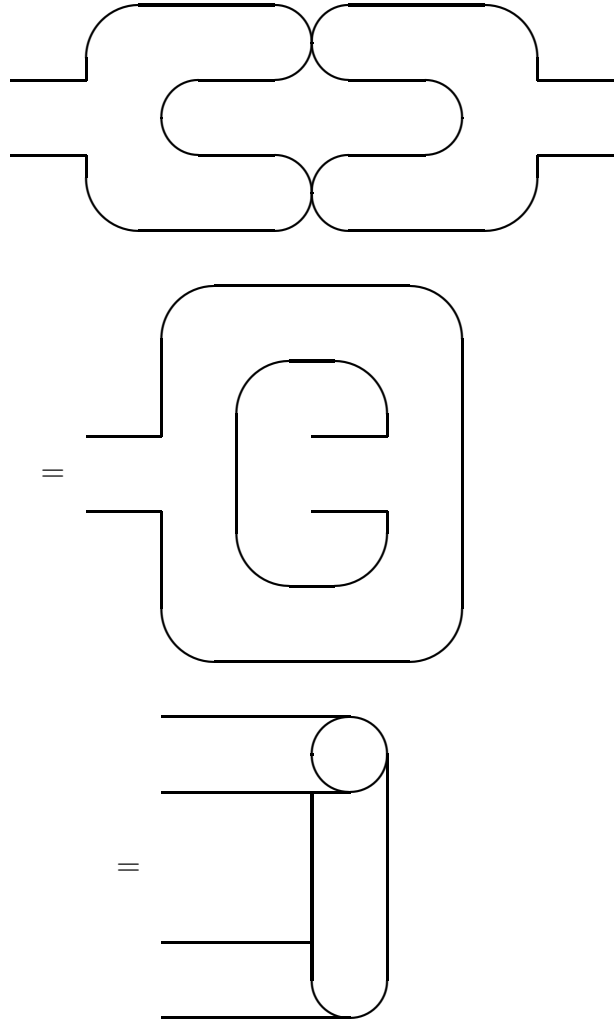
\begin{picture}(110,35)
\put(30,20){\line(1,0){10}}
\put(30,10){\line(1,0){10}}
\put(100,20){\line(1,0){10}}
\put(100,10){\line(1,0){10}}
\put(60,20){\oval(40,20)[tl]}
\put(60,10){\oval(40,20)[bl]}
\put(80,20){\oval(40,20)[tr]}
\put(80,10){\oval(40,20)[br]}
\put(60,15){\oval(20,10)[l]}
\put(80,15){\oval(20,10)[r]}
\put(60,25){\oval(20,10)[r]}
\put(60,5){\oval(20,10)[r]}
\put(80,25){\oval(20,10)[l]}
\put(80,5){\oval(20,10)[l]}
\end{picture}

\begin{picture}(90,55)
\put(34,24){=}
\put(40,30){\line(1,0){10}}
\put(40,20){\line(1,0){10}}
\put(70,30){\line(1,0){10}}
\put(70,20){\line(1,0){10}}
\put(60,20){\line(0,1){10}}
\put(90,20){\line(0,1){10}}
\put(70,30){\oval(40,40)[t]}
\put(70,20){\oval(40,40)[b]}
\put(70,30){\oval(20,20)[t]}
\put(70,20){\oval(20,20)[b]}
\end{picture}

\begin{picture}(80,45)
\put(44,19){=}
\put(50,40){\line(1,0){25}}
\put(50,30){\line(1,0){25}}
\put(50,10){\line(1,0){20}}
\put(50,0){\line(1,0){25}}
\put(70,5){\line(0,1){25}}
\put(80,5){\line(0,1){30}}
\put(75,35){\oval(10,10)}
\put(75,5){\oval(10,10)[b]}
\end{picture}

Thus, a closed string is a bound state of 2 open strings.  The 
closed-string coupling can then be related to the open-string coupling, 
either by examining more general graphs, or by noticing that the 
Gauss-Bonnet theorem says that twice the number of ``handles'' 
(closed-string loops) plus the number of ``windows'' (open 
string loops) is a topological invariant (the Euler number, up to a 
constant), and thus 1 closed-string loop can be converted into 2 
open-string loops.  Specifically,
$$ \hbar g \sb {closed} = ( \hbar g \sb {open} ) \sp 2 \quad \to \quad
g \sb {closed} = \hbar g \sb {open} \sp 2 \quad .\eqno(9.3.1)$$
Thus, for consistent $\hbar$ counting the open strings must be thought 
of as fundamental is such a theory (which so far means just the SO(32) 
superstring), and the closed strings as bound states.  Since (known) closed 
strings always contain gravitons, this makes the SO(32) superstring 
the only known example of a theory where the graviton appears as a 
bound state.  The graviton propagator is the result of ultraviolet 
divergences due to particles of arbitrarily high spin which sum to 
diverge only at the pole:
$$ \int_0^\infty dk \sp 2 \; ( 1 - \a ' \sp 2 p \sp 2 k \sp 2 + \ha \a ' 
\sp 4 p \sp 4 k \sp 4 - \cdots ) = \int dk \sp 2 \; e \sp { - \a ' \sp 
2 p \sp 2 k \sp 2} = {1 \over \a ' \sp 2 p \sp 2} \quad .\eqno(9.3.2)$$
(In general, $p \sp 2 + M \sp 2$ appears instead of $p \sp 2$, so the 
entire closed-string spectrum is generated.)

As mentioned in the introduction, the topology of a 2D surface is 
defined by a few integers, corresponding to, e.g., the number of holes. 
By choosing the coordinates of the surface appropriately 
(``stretching'' it in various ways), the surface takes the form of a 
string tree graph with one-loop insertions, each one loop insertion 
having the value 1 of one of the topological invariants (e.g., 1 
hole).  (Actually, some of these insertions are 1-loop closed-string 
insertions, and therefore are counted as 2-loop in an open-string 
theory due to (9.3.1).)  For example, a hole in an open-string sheet 
may be pushed around so that it represents a loop as in a box graph, a 
propagator correction, a tadpole, or an external line correction.  Such duality 
transformations can also be represented in Feynman graph notation as a 
consequence of the duality properties of simpler graphs such as the 
4-point tree graph (9.1.16):

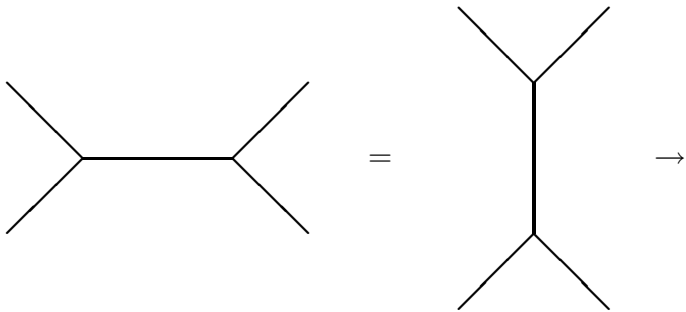
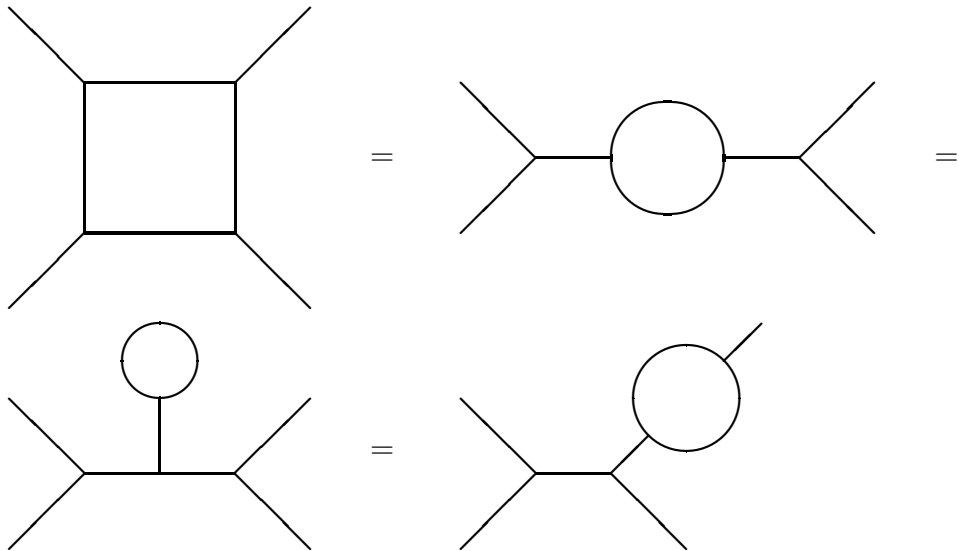
\begin{picture}(100,50)
\put(10,20){\line(1,0){20}}
\put(10,20){\line(-1,-1){10}}
\put(10,20){\line(-1,1){10}}
\put(30,20){\line(1,-1){10}}
\put(30,20){\line(1,1){10}}
\put(70,10){\line(0,1){20}}
\put(70,10){\line(-1,-1){10}}
\put(70,10){\line(1,-1){10}}
\put(70,30){\line(-1,1){10}}
\put(70,30){\line(1,1){10}}
\put(48,19){$=$}
\put(86,19){$\to$}
\end{picture}

\begin{picture}(130,50)
\put(10,10){\line(-1,-1){10}}
\put(10,10){\line(0,1){20}}
\put(10,10){\line(1,0){20}}
\put(30,30){\line(1,1){10}}
\put(30,30){\line(0,-1){20}}
\put(30,30){\line(-1,0){20}}
\put(10,30){\line(-1,1){10}}
\put(30,10){\line(1,-1){10}}
\put(48,19){$=$}
\put(70,20){\line(-1,-1){10}}
\put(70,20){\line(-1,1){10}}
\put(70,20){\line(1,0){10}}
\put(87.5,20){\oval(15,15)}
\put(105,20){\line(1,-1){10}}
\put(105,20){\line(1,1){10}}
\put(105,20){\line(-1,0){10}}
\put(123,19){$=$}
\end{picture}

\begin{picture}(110,35)
\put(10,15){\line(-1,-1){10}}
\put(10,15){\line(-1,1){10}}
\put(10,15){\line(1,0){20}}
\put(30,15){\line(1,-1){10}}
\put(30,15){\line(1,1){10}}
\put(20,15){\line(0,1){10}}
\put(20,30){\oval(10,10)}
\put(48,17){$=$}
\put(70,15){\line(-1,-1){10}}
\put(70,15){\line(-1,1){10}}
\put(70,15){\line(1,0){10}}
\put(80,15){\line(1,-1){10}}
\put(80,15){\line(1,1){5}}
\put(80,15){\line(-1,0){10}}
\put(90,25){\oval(14.1,14.1)}
\put(95,30){\line(1,1){5}}
\end{picture}

\noindent By doing stretching of such planar graphs out of the plane, 
these loops can even be turned into closed-string tadpoles:

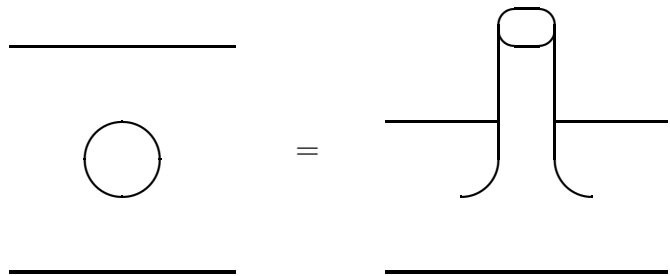
\begin{center}

\begin{picture}(90,50)
\put(0,5){\line(1,0){30}}
\put(0,35){\line(1,0){30}}
\put(15,20){\oval(10,10)}
\put(38,20){$=$}
\put(50,5){\line(1,0){37.5}}
\put(50,25){\line(1,0){15}}
\put(72.5,25){\line(1,0){15}}
\put(65,20){\line(0,1){17.5}}
\put(72.5,20){\line(0,1){17.5}}
\put(60,20){\oval(10,10)[br]}
\put(77.5,20){\oval(10,10)[bl]}
\put(67.5,35){\line(1,0){2.5}}
\put(67.5,40){\line(1,0){2.5}}
\put(67.5,37.5){\oval(5,5)[l]}
\put(70,37.5){\oval(5,5)[r]}
\end{picture}

\end{center}

Stretching represents continuous world-sheet coordinate transformations.
However, there are some coordinate transformations which can't be
obtained by combining infinitesimal transformations, and thus must be
considered separately in analyzing gauge fixing and anomalies [9.11].  The
simplest example is for a closed-string loop (vacuum bubble), which is a
torus topologically.  The group of general coordinate transformations
has as a subgroup conformal transformations (which can be obtained as a
residual gauge invariance upon covariant gauge fixing, sect.\ 6.2).
Conformal transformations, in turn, have as a subgroup the (complex)
projective group Sp(2,C):  The defining representation of this group is
given by 2$\times$2 complex matrices with determinant 1, so the
corresponding representation space consists of pairs of complex numbers.
If we consider the transformation property of a complex variable which
is the ratio of the 2 numbers of the pair, we then find:
$$ \pmatrix{z\sb 1 \cr z\sb 2 \cr}^\prime =
\pmatrix{a & b \cr c & d \cr}\pmatrix{z\sb 1 \cr z\sb 2 \cr}
\quad , \quad z\sb 0 = {z\sb 1 \over z\sb 2} \eqno(9.3.3a)$$
$$ \to\quad z\sb 0' = {az\sb 0+b\over cz\sb 0+d} \quad , \eqno(9.3.3b)$$
where $ad-bc=1$.  Finally, the projective group has as a discrete 
subgroup the ``modular'' group Sp(2,Z), where $a,b,c,d$ are (real)
integers (still satisfying $ad-bc=1$).  To see how this relates to the
torus, define the torus as the complex plane with the identification of
points 
$$ z \to z + n\sp 1 z\sb 1 + n\sp 2 z\sb 2 \eqno(9.3.4)$$
for any integers $n\sp 1 , n\sp 2$, for 2 particular complex numbers
$z\sb 1 , z\sb 2$ which point in different directions in the complex plane.  
We can then think of the torus as the parallelogram
with corners $0, z\sb 1, z\sb 2, z\sb 1 + z\sb 2$, with opposite sides
identified, and the complex plane can be divided up into an infinite
number of equivalent copies, as implied by (9.3.4).  The conformal
structure of the torus can be completely described by specifying the
value of $z\sb 0 = z\sb 1 / z\sb 2$.  (E.g., $z\sb 1$ and $z\sb 2$ both
change under a complex scale transformation, but not their ratio.
Without loss of generality, we can choose the imaginary part of $z\sb 0$
to be positive by ordering $z\sb 1$ and $z\sb 2$ appropriately.)
However, if we transform $(z\sb 1 , z\sb 2)$ under the modular group as in
(9.3.3a), then (9.3.4) becomes
$$ z \to z + n\sp i z'\sb i \quad , \quad z'\sb i = g\sb i\sp j z\sb j
\quad , \eqno(9.3.5a)$$
where $g\sb i\sp j$ is the Sp(2,Z) matrix, or equivalently
$$ z \to z + n'\sp i z\sb i \quad , \quad n'\sp i = n\sp j g\sb j\sp i
\quad . \eqno(9.3.5b)$$
In other words, an Sp(2,Z) transformation gives back the same torus,
since the identification of points in the complex plane (9.3.4) and
(9.3.5b) is the same (since it holds are all pairs of integers $n\sp
i$).  We therefore define the torus by the complex parameter $z\sb 0$,
modulo equivalence under the Sp(2,Z) transformation (9.3.3b).  It turns
out that the modular group can be generated by just the 2
transformations
$$ z\sb 0 \to -{1\over z\sb 0} \quad and \quad z\sb 0 \to z\sb 0 + 1 
\quad . \eqno(9.3.6)$$
The relevance of the modular group to the 1-loop closed-string diagram
is that the functional integral over all surfaces reduces (for tori) 
to an integral over $z\sb 0$.  Gauge fixing for Sp(2,Z) then means
picking just one of the infinite number of equivalent regions of the
complex plane (under (9.3.3b)).  However, for a closed string in less
than its critical dimension, there is an anomaly in the modular
invariance, and the theory is inconsistent.  Modular invariance also
restricts what types of compactification are allowed.

If the 2D general coordinate invariance is not violated by 
anomalies, it's then sufficient to consider these 1-loop objects to 
understand the divergence structure of the quantum string theory.  
However, while the string can be 
stretched to separate any two 1-loop divergences, we know from field 
theory that overlapping divergences can't be factored into 1-loop 
divergences.  This suggests that any 1-loop divergences, since they 
would lead to overlapping divergences, would violate the 2D 
reparametrization invariance which would allow the 1-loop divergences 
to be disentangled.  Hence, it seems that a string theory must be 
finite in order to avoid such anomalies.  Conversely, we expect that 
finiteness at 1 loop implies finiteness at all loops.  Some direct 
evidence of this is given by the fact that all known string theories 
with fermions have 1-loop anomalies in the usual gauge invariances of 
the massless particles if and only if they also have 1-loop 
divergences.  After the restrictions placed by tree-level duality 
(which determines the ground-state mass and restricts the open-string 
gauge groups to U(N), USp(N), and SO(N)), supersymmetry in the 
presence of massless spin-3/2 particles, and 1-loop modular invariance,
this last anomaly restriction allows only SO(32) as an 
open-string gauge group (although it doesn't restrict the 
closed-string theories) [1.11].  

Of the finite theories, the closed-string theories are finite 
graph-by-graph, whereas the open-string theory requires cancellation 
between pairs of 1-loop graphs, with the exception of the nonplanar 
loop discussed above.  The 1-loop closed-string graphs 
(corresponding to 2-loop graphs in the open-string theory) are (1) the 
torus (``handle'') and (2) the Klein bottle, with external lines attached.
The pairs of 1-loop open-string graphs are (1) the annulus (planar 
loop, or ``window'') + M\"obius strip (nonorientable loop) with 
external open and/or closed strings, and (2) the disk + a graph 
with the topology of RP$\sb 2$ (a disk with opposite points identified) 
with external closed strings.  The Klein bottle is allowed only for 
nonoriented closed strings, and the M\"obius strip and RP$\sb 2$ are 
allowed only for nonoriented open strings.  

It should be possible to simplify calculations and give simple proofs 
of finiteness by the use of background field methods similar to those 
which in gravity and supersymmetry made higher-loop calculations 
tractable and allowed simple derivations of no-renormalization 
theorems [1.2].  However, the use of arbitrary background string fields 
will require the development of gauge-covariant string field theory, 
the present status of which is discussed in the following chapters.

\sect{Exercises}

\Item{(1)} Generalize (9.1.3) to the spinning string, using (7.2.2)
instead of $\ha\Hat P\sp 2$.  Writing $\ha\hat Dd\hat D(\s ) \to
\ha\hat Dd\hat D(\s ) + \d (\s ) \cv$, $\cv = W + \q V$, show that $V$
is determined explicitly by $W$.
\Item{(2)} Derive (9.1.7b).
\Item{(3)} Fill in all the steps needed to obtain (9.1.15) from
(9.1.6,8).  Derive all parts of (9.1.16).  Derive (9.1.19).
\Item{(4)} Evaluate (9.1.20) by using the result (9.1.15) with ${\bf
N}\to{\bf N}+2$ (but dropping 2 $d\t$ integrations) and letting 
$k\sb 0 = k\sb{{\bf N}+1} =0$.
\Item{(5)} Derive (9.2.18).
\Item{(6)} Generalize (9.2.16) to arbitrary $z\sb 1$, $z\sb 2$,
$z\sb{\bf N}$ and derive (9.2.22c) by the method of (9.2.20,21).
Take the infinitesimal form of the Sp(2) transformation and show that
it's generated by ${\bf L}\sb 0$, ${\bf L}\sb{\pm 1}$ with the
correspondence (8.1.3), where $z=e\sp{i\s}$.
\Item{(7)} Derive (9.2.27).  Evaluate $G\sb{rsmn}$ of (9.2.4) using
(9.2.19,26,27). 

%
%

\chap{10. LIGHT-CONE FIELD THEORY}1

In this chapter we extend the discussion of sect.\ 2.1 to the string
and consider interacting contributions to the Poincar\'e algebra of sect.\
7.1 along the lines of the Yang-Mills case treated in sect.\ 2.3.

For the string [10.1,9.5,10.2], it's convenient to use a field 
$\F [ X \sb i ( \s ) , 
p \sb + , \t ]$, since $p \sb +$ is the length of the string.  This 
$X$ for $\s \in [ 0 , \p p\sb + ]$ is related to that in (7.1.7) for 
$p\sb + = 1$ by $X ( \s , p\sb + ) = X ( \s / p\sb + , 1 )$. 
The {\it hermiticity} condition on the (open-string) field is
$$ \F [ X \sb i ( \s ) , p \sb + , \t ] = \F \sp \dag [ X \sb i 
( \p p \sb + - \s ) , - p \sb + , \t ] \quad .\eqno(10.1a) $$
The same relation holds for the closed string, but we may replace 
$\p p \sb + - \s $ with just $- \s$, since the closed string 
has the residual gauge invariance $\s \to \s + constant$, which is 
fixed by the constraint (or gauge choice) $\D N \F = 0$.  (See (7.1.12).
In loops, this gauge choice can be 
implemented either by projection operators or by Faddeev-Popov ghosts.)
As described in sect.\ 5.1, this charge-conjugation condition 
corresponds to a combination of ordinary complex conjugation ($\t$ reversal)
with a twist (matrix transposition combined with $\s$ reversal).
The twist effectively acts as a charge-conjugation matrix
in $\s$ space, in the sense that expressions involving $tr \; \F 
\sp \dag \F$ acquire such a factor if reexpressed in terms of just 
$\F$ and not $\F \sp \dag$ (and (10.1a) looks like a reality 
condition for a group for which the twist is the group metric).
Here $\F$ is an N$\times$N matrix, and the odd mass levels of 
the string (including the massless Yang-Mills sector) are in the 
adjoint representation of U(N), SO(N), or USp(N) (for even N) [10.3], where 
in the latter 2 cases the field also satisfies the {\it reality} condition
$$ \h \F = ( \h \F ) \conj \quad ,\eqno(10.1b) $$
where $\h$ is the group metric (symmetric for SO(N), antisymmetric for 
USp(N)).  The fact that the latter cases use the operation of $\t$ 
reversal separately, or, by combining with (10.1a), the twist 
separately, means that they describe nonoriented strings:  The string 
field is constrained to be invariant under a twist.  The same is 
true for closed strings (although closed strings have no group theory, 
so the choice of oriented vs.\ nonoriented is arbitrary, and $\h = 1$ 
in (10.1b)).  The twist operator can be defined similarly for
superstrings, including heterotic strings.
For general open strings the twist is most simply written in terms of
the hatted operators, on which it acts as $\hat\co (\s ) \to
\hat\co ( \s - \p )$ (i.e., as $e\sp{i\p N}$).  For general closed
strings, it takes $\hat\co\sp{(\pm )}(\s ) \to \hat\co\sp{(\pm )}(-\s )$.
(For closed strings, $\s\to\s -\p$ is irrelevant, since $\D N=0$.)

Light-cone superstring fields [10.2] also satisfy the reality 
condition (in place of 
(10.1b), generalizing (5.4.32)) that the Fourier transform with respect to 
$\Q \sp {\bf a}$ is equal to the complex conjugate (which is the 
analog of a certain condition on covariant superfields)
$$ \int \Sc D \Q \; e \sp {\int \Q \sp {\bf a} \P \sb {\bf a} 
p \sb + /2} \h \F [ \Q \sp {\bf a} ] = \left( \h \F [ \bar \P \sp {\bf a}
] \right) \conj \quad .\eqno(10.1c)$$
$ \Q \sp {\bf a}$ has a mode expansion like that of the ghost $\Hat C$ 
or the spinning string's $\hat \J$ (of the fermionic sector).  The 
ground-state of the open superstring is described by the light-cone 
superfield of (5.4.35), which is a function of the zero-modes of all the above 
coordinates.  Thus, the lowest-mass (massless) sector of the open 
superstring is supersymmetric Yang-Mills.

The free action of the bosonic open string is [10.1,9.5]
$$ S \sb 0 = - \int \Sc D X \sb i \int _{-\infty}^\infty d p \sb +
\int_{-\infty}^\infty d \t \; tr \; \F \sp \dag p \sb + \left( i 
{\pa \over \pa \t} + H \right) \F \quad ,$$
$$ H = \int_0^{2 \p \a ' p \sb +} {d \s \over 2 \p} \; 
\left[ \ha \left( - \a ' {\d \sp 2 \over \d X \sb i \sp 2} + {1 \over 
\a '} X \sb i ' \sp 2 \right) - 1 \right] = \int_{- \p p\sb +}^{\p p\sb +} {d 
\s \over 2 \p} ( \ha \Hat P \sb i \sp 2 - 1 ) = {p \sb i \sp 2 + M \sp 
2 \over 2 p \sb +} \quad .\eqno(10.2)$$
The free field equation is therefore just the quantum mechanical 
Schr\"odinger equation.  (The $p \sb +$ integral can also be written 
as $2 \int_0^\infty$, due to the hermiticity condition.  This form 
also holds for closed strings, with $H$ the sum of 2 open-string ones, 
as described in sec.\ 7.1.)  Similar remarks apply to superstrings
(using (7.3.13)).

As in the first-quantized approach, interactions are described by 
splitting and joining of strings, but now the graph gets chopped up 
into propagators and vertices:

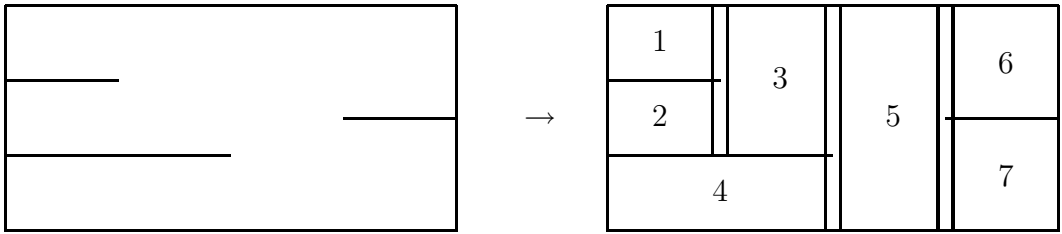
\setlength{\unitlength}{1mm}
\begin{picture}(140,45)
\put(0,5){\line(1,0){60}}
\put(0,35){\line(1,0){60}}
\put(0,15){\line(1,0){30}}
\put(0,25){\line(1,0){15}}
\put(45,20){\line(1,0){15}}
\put(0,5){\line(0,1){30}}
\put(60,5){\line(0,1){30}}
\put(69,19){$\to$}
\put(80,5){\line(1,0){60}}
\put(80,35){\line(1,0){60}}
\put(80,15){\line(1,0){30}}
\put(80,25){\line(1,0){15}}
\put(125,20){\line(1,0){15}}
\put(80,5){\line(0,1){30}}
\put(140,5){\line(0,1){30}}
\put(94,15){\line(0,1){20}}
\put(96,15){\line(0,1){20}}
\put(109,5){\line(0,1){30}}
\put(111,5){\line(0,1){30}}
\put(124,5){\line(0,1){30}}
\put(126,5){\line(0,1){30}}
\put(86,29){1}
\put(86,19){2}
\put(102,24){3}
\put(94,9){4}
\put(117,19){5}
\put(132,26){6}
\put(132,11){7}
\end{picture}

\noindent The 3-open-string vertex is then just a $\d$ functional setting 
1 string equal to 2 others, represented by an infinitesimal strip in the world 
sheet.  The interaction term in the action is given by (9.2.23) for $Z=X$.
The 3-closed-string vertex is a similar $\d$ functional for 3 closed 
strings, which can be represented as the product of 2 open-string $\d$ 
functionals, since the closed-string coordinates can be represented as 
the sum of 2 open-string coordinates (one clockwise and one 
counterclockwise, but with the same zero-modes).  This vertex 
generally requires an integration over the $\s$ of the integration 
point (since closed strings can join anywhere, not having any ends, 
corresponding to the gauge invariance $\s \to \s + constant$), but the 
equivalent operation of projection onto $\D N = 0$ can be absorbed 
into the propagators.

General vertices can be obtained by considering similar slicings of 
surfaces with general global topologies [10.2,9.7].  There are 2 of order $g$, 
corresponding {\it locally} to a splitting or joining:

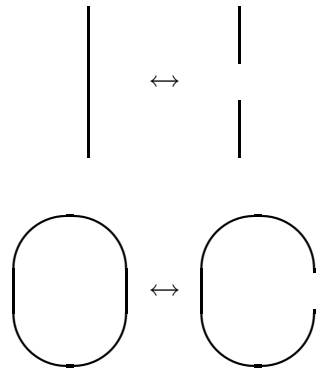
\setlength{\unitlength}{.5mm}
\begin{picture}(160,50)
\put(120,0){\line(0,1){40}}
\put(136,18){$\iff$}
\put(160,0){\line(0,1){15}}
\put(160,25){\line(0,1){15}}
\end{picture}

\begin{picture}(180,50)
\put(115,15){\oval(30,30)[b]}
\put(115,25){\oval(30,30)[t]}
\put(165,15){\oval(30,30)[b]}
\put(165,25){\oval(30,30)[t]}
\put(100,15){\line(0,1){10}}
\put(130,15){\line(0,1){10}}
\put(150,15){\line(0,1){10}}
\put(136,18){$\iff$}
\end{picture}

\noindent The former is the 3-open-string vertex.  The existence of 
the latter is implied by the former via the nonplanar loop graph (see 
sect.\ 9.3).  The rest are order 
$g \sp 2$, and correspond locally to 2 strings touching their middles 
and switching halves:

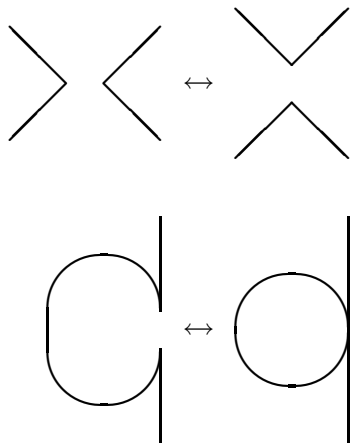
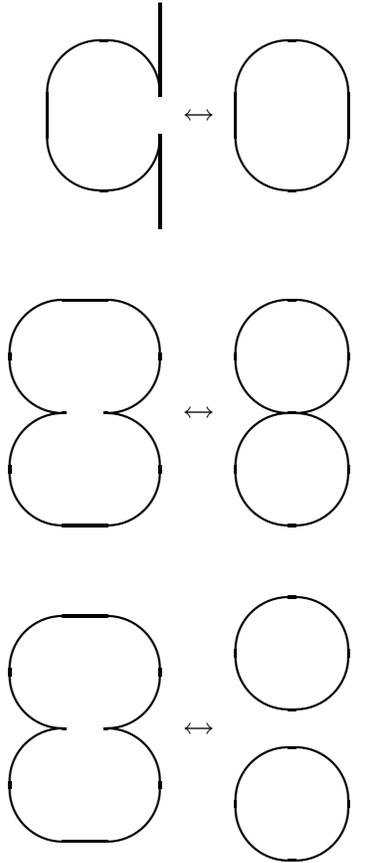
\begin{picture}(180,50)
\put(105,20){\line(-1,-1){15}}
\put(105,20){\line(-1,1){15}}
\put(115,20){\line(1,-1){15}}
\put(115,20){\line(1,1){15}}
\put(136,18){$\iff$}
\put(165,15){\line(-1,-1){15}}
\put(165,25){\line(-1,1){15}}
\put(165,15){\line(1,-1){15}}
\put(165,25){\line(1,1){15}}
\end{picture}

\begin{picture}(180,70)
\put(115,25){\oval(30,30)[b]}
\put(115,35){\oval(30,30)[t]}
\put(165,30){\oval(30,30)}
\put(100,25){\line(0,1){10}}
\put(130,0){\line(0,1){25}}
\put(130,35){\line(0,1){25}}
\put(180,0){\line(0,1){60}}
\put(136,28){$\iff$}
\end{picture}

\begin{picture}(190,70)
\put(115,25){\oval(30,30)[b]}
\put(115,35){\oval(30,30)[t]}
\put(165,25){\oval(30,30)[b]}
\put(165,35){\oval(30,30)[t]}
\put(100,25){\line(0,1){10}}
\put(150,25){\line(0,1){10}}
\put(180,25){\line(0,1){10}}
\put(130,0){\line(0,1){25}}
\put(130,35){\line(0,1){25}}
\put(190,0){\line(0,1){60}}
\put(136,28){$\iff$}
\end{picture}

\begin{picture}(180,70)
\put(105,15){\oval(30,30)[l]}
\put(105,45){\oval(30,30)[l]}
\put(115,15){\oval(30,30)[r]}
\put(115,45){\oval(30,30)[r]}
\put(165,15){\oval(30,30)}
\put(165,45){\oval(30,30)}
\put(105,0){\line(1,0){10}}
\put(105,60){\line(1,0){10}}
\put(136,28){$\iff$}
\end{picture}

\begin{picture}(180,80)
\put(105,20){\oval(30,30)[l]}
\put(105,50){\oval(30,30)[l]}
\put(115,20){\oval(30,30)[r]}
\put(115,50){\oval(30,30)[r]}
\put(165,15){\oval(30,30)}
\put(165,55){\oval(30,30)}
\put(105,5){\line(1,0){10}}
\put(105,65){\line(1,0){10}}
\put(136,33){$\iff$}
\end{picture}

\noindent The type-I (SO(32) open-closed) theory has all these 
vertices, but the type-IIAB and heterotic theories have only the last 
one, since they have only closed strings, and they are oriented 
(clockwise modes are distinguishable from counterclockwise).
If the type-I theory is treated as a theory of fundamental open 
strings (with closed strings as bound states, so $\hbar$ can be 
defined), then we have only the first of the order-$g$ vertices and 
the first of the order-$g \sp 2$.

The light-cone quantization of the spinning string follows directly 
from the corresponding bosonic formalism by the 1D supersymmetrization 
described in sect.\ 7.2.  In particular, in such a formalism the vertices 
require no factors besides the $\d$ functionals [10.4].  (In converting to a 
non-superfield formalism, integration of the vertex over $\q$ produces 
a vertex factor.)  However, the projection (7.2.5) must be put in by hand.
Also, the fact that boundary conditions can be either of 2 types (for
bosons vs.\ fermions) must be kept in mind.

The interactions of the light-cone superstring [10.2] are done as for the 
light-cone bosonic string, but there are extra factors.  For example, 
for the 3-open-string vertex (the interacting contribution to $p\sb -$), 
we have the usual $\d$-functionals times
$$ V(p\sb -) = P\sb L +\ha\bar P\sp{\bf ab}{\d\over\d\Q\sp{\bf a}} 
{\d\over\d\Q\sp{\bf b}} + \frac1{24}\bar P\sb L C\sp{\bf abcd}
{\d \over \d \Q \sp {\bf a}} {\d \over \d \Q \sp {\bf b}} 
{\d \over \d \Q \sp {\bf c}} {\d \over \d \Q \sp {\bf d}}\quad ,\eqno(10.3a)$$
evaluated at the splitting point.  The interacting contributions to
$q\sb -$ are given by the same overlap $\d$-functionals times the vertex
factors
$$ V(q\sb{-{\bf a}}) = \fder{\Q\sp{\bf a}} \quad , \quad
V(\bar q\sb -\sp{\bf a}) = \frac16 C\sp{\bf abcd}
\fder{\Q\sp{\bf b}}\fder{\Q\sp{\bf c}}\fder{\Q\sp{\bf d}} \quad .
\eqno(10.3b)$$ 
(The euphoric notation for $q$ is as in sect.\ 5.4 for $d$.)
These are evaluated as in (9.2.28), where $P$ and $\d / \d \Q$ have weight 
$w=1$.  Their form is determined by requiring that the supersymmetry
algebra be maintained.
The $\d$ functional part is given as in (9.2.27), but now the $\D$ 
of (9.2.14) instead of just (9.2.15) is
$$ \D = \ha \D ( \check P , \check P ) - \D \left( \check \Q ' , {\d 
\over \d \check \Q} \right) \quad .\eqno(10.4)$$

As for the bosonic string, the closed-superstring vertex 
is the product of 2 open-string ones (including 2 factors of the 
form (10.3a) for type I or II but just 1 for the heterotic, and 
integrated over $\s$).  For the general interactions above, all
order-$g$ interactions have a single open-string vertex factor, while all
order-$g \sp 2$ have 2, since the interactions of each order are locally
all the same.  The vertex factor is either $1$ or (10.3a), depending on
whether the corresponding set of modes is bosonic or supersymmetric. 
When these superstring theories are truncated to their ground states, 
the factor (10.3a) keeps only the zero-mode contributions, which is 
the usual light-cone, 3-point vertex for supersymmetric Yang-Mills,
and the product of 2 such factors (for closed strings) is the usual 
vertex for supergravity.

The second-quantized interacting Poincar\'e algebra for the light-cone string
can be obtained perturbatively.  For example,
$$ [ p \sb - , J \sb {-i} ] = 0 \quad \to $$
$$ [ p \sp {(2)} \sb - , J \sp {(2)} \sb {-i} ] = 0 \quad , \eqno(10.5a)$$
$$ [ p \sp {(3)} \sb - , J \sp {(2)} \sb {-i} ] +
[ p \sp {(2)} \sb - , J \sp {(3)} \sb {-i} ] = 0 \quad , \eqno(10.5b)$$
etc., where $(n)$ indicates the order in fields (see sect.\ 2.4).  
The solution to (10.5a) is known from the free theory.  The
solution to (10.5b) can be obtained from known results for the
first-quantized theory [1.4,10.5]:  The first term represents the invariance
of the 3-point interaction of the hamiltonian under free Lorentz
transformations.  The fact that this invariance holds only on-shell
is an expected consequence of the fact that the second term in (10.5b) is
simply the commutator of the free hamiltonian with the interaction
correction to the Lorentz generator.  Thus, the algebra of the complete
interacting generators closes off shell as well as on, and the explicit
form of $J \sp {(3)} \sb {-i}$ follows from the expression for
nonclosure given in [1.4,10.5]:  
$$ \cj\sp{(3)}\sb -\sp i (1,2,3) = -2ig X\sb{r}\sp i (\sigma_{int})
\d \left( \su p_+ \right) \D [X^i]\quad ,\eqno(10.6)$$
where ``$r$'' denotes any of the three strings and $\D$ represents the 
usual overlap-integral $\d$-functionals
with splitting point $\s \sb {INT}$.  This is the analog of the
generalization of (2.3.5) to the interacting scalar particle, where
$p \sb - \f \to  - (1 / 2 p \sb + ) ( p \sb i \sp 2 \f + \f \sp 2 )$.
Since $p \sb -$ also contains a 4-point interaction, there is a similar
contribution to $J \sp {(2)} \sb {-i}$ (i.e., (10.6) with $\D$
replaced by the corresponding 3-string product derived from the 4-string
light-cone vertex in $p\sb -$, and $g$ replaced with $g\sp 2$, but
otherwise the same normalization).  Explicit second-quantized operator
calculations show that this closes the algebra [10.6].  Similar
constructions apply to superstrings [10.7].

Covariant string rules can be obtained from the light-cone formalism in 
the same way as in sect.\ 2.6, and $p\sb +$ now also represents the string 
length [2.7].  Thus, from (10.2) we get the free action, in terms of a field 
$\F [ X \sp a ( \s ) , X\sp\a ( \s ) , p\sb + , \t ]$,
$$ S \sb 0 = - \int\Sc D X\sp a \;\Sc DX\sp\a
\int _{-\infty}^\infty d p\sb + \int_{-\infty}^\infty d \t \; tr \; 
\F \sp \dag p\sb + \left( i {\pa \over \pa \t} + H \right) \F \quad ,$$
$$ H = \int_0^{\p p\sb +} {d \s \over 2 \p} \; \left\{ \ha 
\left[ - \a ' \left( {\d \sp 2 \over \d X \sp {a2}} +\fder{X\sb\a}\fder{X\sp\a}
\right) + {1 \over \a '} \left( X \sp a ' \sp 2 + X\sp\a ' X\sb\a '
\right) \right] - 1 \right\} \quad .\eqno(10.7)$$
The vertex is again a $\d$ functional, in all variables.

The light-cone formalism for heterotic string field theory has also been
developed, and can be extended to further types of compactifications [10.8].

Unfortunately, the interacting light-cone formalism is not completely
understood, even for the bosonic string.  There are certain kinds of
contact terms which must be added to the superstring action and
supersymmetry generators to insure lower-boundedness of the energy
(supersymmetry implies positivity of the energy) and cancel divergences in
scattering amplitudes due to coincidence of vertex operator factors
[10.9], and some of these terms have been found.  
(Similar problems have appeared in the covariant spinning string
formulation of the superstring:  see sect.\ 12.2.)  This problem is 
particularly evident for closed strings, which were thought to have only
cubic interaction terms, which are insufficient to bound the potential
in a formalism with only physical polarizations.  Furthermore, the
closed-string bound states which have been found to follow at one loop from
open-string theories by explicitly applying unitarity to tree graphs do
not seem to follow from the light-cone field theory rules [10.1].  Since
unitarity requires that 1-loop corrections are uniquely determined by
tree graphs, the implication is that the present light-cone field
theory action is incomplete, or that the rules following from it have
not been correctly applied.  It is interesting to note that the type of
graph needed to give the correct closed-string poles resembles the
so-called Z-graph of ordinary light-cone field theory [10.10], which contains a
line backward-moving in $x\sb +$ when the light-cone formalism is
obtained as an ultrarelativistic limit, becoming an instantaneous line
when the limit is reached.

\sect{Exercises}

I can't think of any.

%
%

\chsc{11. BRST FIELD THEORY}{11.1. Closed strings}3
 
Since the gauge-invariant actions for free open strings follow directly 
from the methods of sects.\ 3.4-5 using the algebras of chapt.\ 8 (for
the bosonic and fermionic cases, using either OSp(1,1$|$2) or IGL(1) 
algebras), we will consider here just closed strings, after a few
general remarks.

Other string actions 
have been proposed which lack the complete set of Stueckelberg fields [11.1], 
and as a result they are expected to suffer from problems similar to those of 
covariant ``unitary'' gauges in spontaneously broken gauge theories: 
no simple Klein-Gordon-type propagator, nonmanifest renormalizability, 
and singularity of semiclassical solutions, including those representing 
spontaneous breakdown.  Further attempts with nonlocal, higher-derivative, 
or incomplete actions appeared in [11.2].  Equivalent gauge-invariant 
actions for the free Ramond string have been obtained by several groups 
[4.13-15,11.3].  The action of [11.3] is related to the rest by a
unitary transformation:  It has factors involving coordinates which are
evaluated at the midpoint of the string, whereas the others involve 
corresponding zero-modes.  

For open or closed strings, the hermiticity condition (10.1a) now 
requires that the ghost coordinates also be twisted:  In the IGL(1)
formalism  (where the ghost coordinates are momenta)
$$ \F [ X ( \s ) , C ( \s ) , \Tilde C ( \s ) ] = 
\F \sp \dag [ X ( \p - \s ) , -C ( \p - \s ) , \Tilde C ( \p - \s ) ]
\eqno(11.1.1a)$$
($\Tilde C$ gets an extra ``$-$'' because the twist is $\s$ reversal, and 
$\Tilde C$ carries a $\s$ index in the mechanics action), and in the
OSp(1,1$|$2) formalism we just extend (10.1a):
$$ \F [ X\sp a ( \s ) , X\sp\a (\s ) , p\sb + ] =
\F\dg [ X\sp a (\p p\sb + -\s ) , X\sp\a (\p p\sb + - \s ) , -p\sb + ]
\quad . \eqno(11.1.1b)$$
As in the light-cone formalism (see sect.\ 10.1), for discussing free
theories, we scale $\s$ by $p\sb +$ in (11.1.1b), so the twist then
takes $\s\to\p -\s$.  (For the closed string the twist is $\s\to -\s$,
so in both (11.1.1a) and (11.1.1b) the arguments of the coordinates are 
just $-\s$ on the right-hand side.)

In the rest of this section we will consider closed strings only.
First we show how to extend the OSp(1,1$|$2) formalism to closed strings
[4.10].  By analogy to (4.1.1), the kinetic operator for the 
closed string is a $\d$ function in IOSp(1,1$|$2):
$$ S = \int d\sp Dxd\sp 2x\sb\a dx\sb - d\sp 2\D p\sb\a d\D p\sb + \;\;
\F\dg \; p\sb + \sp 2 \d (J\sb{AB}) \d (\D p\sb A) \; \F \quad , $$
$$ \d \F = \ha J\sp{AB} \L\sb{BA} + \D p\sp A \L\sb A \quad ,\eqno(11.1.2)$$
where, as in (7.1.17), the Poincar\'e generators $J\sb{AB}$ and $p\sb A$
are given as sums, and $\D p\sb A$ as differences, of the left-handed
and right-handed versions of the open-string generators of (7.1.14).  
The Hilbert-space metric necessary for hermiticity is now $p\sb +\sp 2$ 
(or equivalently $p\sb +
\sp{(+)} p\sb +\sp{(-)}$), since a factor of $p\sb +$ is needed for the
open-string modes of each handedness.  For
simplicity, we do not take the physical momenta $p\sb a$ to be doubled
here, since the IOSp(1,1$|$2) algebra closes regardless, but
they can be doubled if the corresponding $\d$ functions and integrations
are included in (11.1.2).  More explicitly, the $\d$ function in the
Poincar\'e group is given by
$$ \d (J\sb{AB}) \d (\D p\sb A) = 
\d ( J \sb {\a\b} \sp 2 ) i \d ( J\sb{-+} ) \d\sp 2 ( J\sb{+\a} )
\d\sp 2 ( J \sb{-\a} ) \d ( \D p \sb - ) \d\sp 2 ( \D p\sb\a )
\d ( \D p \sb + ) \quad . \eqno(11.1.3) $$
To establish the invariance of (11.1.2), the fact that (4.1.1) is
invariant indicates that it's sufficient to show that $\d ( J\sb{AB} )$
commutes with $\d ( \D p \sb A )$.  This follows from the fact that each
of the $J\sb{AB}$'s commutes with $\d ( \D p\sb A )$.  
We interpret $\d ( \D p\sb - ) = p\sb + \d ( \D N )$ in the
presence of the other $\d ( \D p )$'s, where $\d ( \D N )$ is a Kronecker $\d$,
and the other $\d ( \D p )$'s are Dirac $\d$'s.

All the nontrivial terms are contained in the $\d \sp 2 ( J\sb{-\a} )$.
As in the open-string case, dependence on the gauge coordinates $x\sb\a$
and $x\sb -$ is eliminated by the $\d \sp 2 ( J\sb{+\a} )$ and $\d (
J\sb {-+} )$ on the left, and further terms are killed by $\d ( J\sb 
{\a\b}\sp 2 )$.  Similarly, dependence on $\D p\sb\a$ and $\D p\sb +$ 
is eliminated by the corresponding $\d$ functions on the right, and 
further terms are killed by $\d ( \D p \sb - )$.  For convenience, the 
latter elimination should be done before the former.
After making the redefinition $\F \to {1\over p\sb +}\F$ and integrating
out the unphysical
zero-modes, the action is similar to the OSp(1,1$|$2) case:
$$ S = \int d \sp D x \; \ha \f\dg \; \d (\D N) \d ( M \sb {\a\b} \sp 2 )
\left[ \bo - M \sp 2 + ( M \sb \a \sp a 
p \sb a + M \sb {\a m} M ) \sp 2 \right] \; \f \quad , $$
$$ \d \f = ( M \sp {\a a} p \sb a + M \sp \a \sb m M ) \L \sb \a 
+ \ha M \sp {\a\b} \L \sb {\a\b} + \D N\L \quad . \eqno(11.1.4) $$
This is the minimal form of the closed-string action.

The nonminimal form is obtained by analogy to the IGL(1) formalism, in
the same way OSp(1,1$|$2) was extended to IOSp(1,1$|$2):  Using a $\d$
function in the closed-string group GL(1$|$1) of sect.\ 8.2 (found from
sums and differences of the expressions in (8.2.6), as in (7.1.17)),
we obtain the action (with $\D p\sb - \to \D N$)
$$ S = \int d\sp Dxd c d\D p\sp{\tilde c} \; \F\dg \; i Q \d ( J\sp 3 ) 
\d ( \D p\sp{\tilde c} ) \d ( \D N ) \; \F \quad , $$
$$ \d \F = Q \L + J\sp 3 \grave\L + \D N \tilde\L + \D p\sp{\tilde c}
\check\L \quad , \eqno(11.1.5) $$
or, after integrating out $\D p\sp{\tilde c}$, with $\F = \f + \D
p\sp{\tilde c} \j$,
$$ S = \int d\sp Dxd c \; \f\dg \; i \Hat Q \d ( \Hat J\sp 3 ) 
\d ( \D N ) \; \f \quad , $$
$$ \d \f = \Hat Q \L + \Hat J\sp 3 \grave\L + \D N \tilde\L \quad , 
\eqno(11.1.6) $$
where the $\Hat{\phantom M}$'s indicate that all terms involving $\D
p\sp{\tilde c}$ and its canonical conjugate have been dropped.
The field $\f = \varphi +  c \c$ is commuting.

For the gauge-fixing in the GL(1$|$1) formalism above (or the 
equivalent one from first-quantization), we now choose [4.5]
$$\co = - 2 \D p\sp{\tilde c} \left[ c , {\pa \over {\pa c}} \right]
	\quad \to \quad K = \D p\sp{\tilde c} \left[ c ( \bo - M\sp 2 )
	-4 M\sp + \der c \right] -2 \D N \left[ c , {\pa \over {\pa c}} 
	\right] \quad ,\eqno(11.1.7) $$
where we have used
$$ Q = -i \frac14 c ( p\sp 2 + M\sp 2 ) +i\ha M\sp + \der c -i \D N
\der{\D p\sp{\tilde c}} +i \ha \D M\sp + \D p\sp{\tilde c}+\cq\sp + 
\quad .\eqno(11.1.8)$$
Expanding the string field over the ghost zero-modes,
$$\F =	( \f + ic \f ) + i \D p\sp{\tilde c} ( \hat \j + c\hat \f )
\quad ,\eqno(11.1.9)$$
we substitute into the lagrangian $L = \ha \F \sp \dag K \F$ and integrate
over the ghost zero modes:
$$ {\pa \over {\pa c}} {\pa \over {\pa \D p \sp{\tilde c}}} L =  \ha \f 
	\sp \dag ( \bo - M\sp 2 ) \f + 2 \j \sp \dag M \sp + \j + 
	4i ( \hat \f \sp \dag \D N \f + i \hat \j \sp \dag \D N \j )
	\quad .\eqno(11.1.10) $$
$\f$ contains propagating fields, $\j$ contains BRST auxiliary fields, 
and $\hat \f$ and $\hat \j$ contain lagrange 
multipliers which constrain $\D N = 0$ for the other fields.
Although the propagating fields are completely gauge-fixed, the BRST 
auxiliary fields again have the gauge transformations
$$ \d \j = \l \quad , \quad M \sp + \l = 0 \quad ,\eqno(11.1.11a)$$
and the lagrange multipliers have the gauge transformations
$$ \d\hat\f = \hat\l\sb\f \quad , \quad \d\hat\j = \hat\l\sb\j \quad ;
\quad \D N \hat\l\sb\f = \D N \hat\l\sb\j = 0 \quad . \eqno(11.1.11b)$$

\sect{11.2. Components}

To get a better understanding of the gauge-invariant string action in
terms of more familiar particle actions, we now expand the string action
over some of the lower-mass component fields, using the algebras of
chapt.\ 8 in the formalism of sect.\ 4.  All of these results can also
be derived by simply identifying the reducible representations which
appear in the light-cone, and then using the component methods of sect.\
4.1.  However, here we'll work directly with string oscillators, and not
decompose the reducible representations, for purposes of comparison.

As an example of how components appear in the IGL(1) quantization, 
the massless level of the open string is given by (cf.\ (4.4.6))
$$ \F = \left[ ( A\sp a a\sb a\dg + {\bf C}\sp\a a\sb \a\dg ) +
icBa\sp{\tilde c}\dg \right] \sket{0} \quad , \eqno(11.2.1)$$
where $a\sp\a = ( a\sp c , a\sp{\tilde c} )$, all oscillators are for 
the first mode, and we have used (4.4.5).  The lagrangian and BRST 
transformations then agree with (3.2.8,11) for $\z = 1$.  In order to
obtain particle actions directly without having to eliminate BRST auxiliary
fields, from now on we work with only the OSp(1,1$|$2) formalism.  (By
the arguments of sect.\ 4.2, the IGL(1) formalism gives the same actions
after elimination of BRST auxiliary fields.)

As described in sect.\ 4.1, auxiliary fields which come from the ghost
sector are crucial for writing local gauge-invariant actions.  (These
auxiliary fields have the same dimension as the physical fields, unlike the
BRST auxiliary fields, which are 1 unit higher in dimension and have
algebraic field equations.)  Let's consider
the counting of these auxiliary fields.  This requires finding
the number of Sp(2) singlets that can be constructed out of the ghost
oscillators at each mass level.  The Sp(2) singlet constructed from
two isospinor creation operators is $a\sb m\sp{\a\dag}a\sb{n\a}\dg$, 
which we denote as $(mn)$.  A general auxiliary field is obtained by 
applying to the vacuum some nonzero number of these pairs together with 
an arbitrary number of bosonic creation operators.  The first few independent
products of pairs of fermionic operators, listed by eigenvalue of the number
operator $N$, are:
$$\li{0 \; : \quad & I\cr
1 \; : \quad &  -\cr
2 \; : \quad & ( 1 1 )\cr
3 \; : \quad & ( 1 2 )\cr
4 \; : \quad & ( 1 3 ) \; , \; ( 2 2 )\cr
5 \; : \quad & ( 1 4 ) \; , \; ( 2 3 )\cr
6 \; : \quad & ( 1 5 ) \; , \; ( 2 4 ) \; , \; ( 3 3 ) \; , \; ( 1 1 )
( 2 2 )&(11.2.2)\cr}$$
where $I$ is the identity and no operator exists at level 1.  The
number of independent products of singlets at each level is given by
the partition function 
$${1 \over { \pr _{n=2}^{\infty} ( 1 - x \sp n )}}$$ $$ = 1 + x \sp 2 + x \sp 3
+ 2 x \sp 4 + 2 x \sp 5 + 4 x \sp 6 + 4 x \sp 7 + 7 x \sp 8 + 8 x \sp 9 +
12 x \sp {10} + 14 x \sp {11} + 21 x \sp {12} + \cdots \quad 
,\eqno(11.2.3)$$
corresponding to the states generated by a single bosonic coordinate
missing its zeroth and first modes, as described in sect.\ 8.1.

We now expand the open string up to the third mass level (containing a
massive, symmetric, rank-2 tensor) and the closed string up to the second
mass level (containing the graviton) [4.1].  The mode expansions of the 
relevant operators are given by (8.2.1).  Since the $\d (M\sb{\a\b}\sp 2)$
projector keeps only the Sp(2)-singlet terms, we find that up to 
the third mass level the expansion of $\f$ is
$$\li{\f =& [ \f \sb 0 + A \sp a a \sp \dag \sb {1a}\cr
&+ \ha h \sp {ab} a \sp \dag \sb {1a} a \sp \dag \sb {1b} + B \sp a a
\sp \dag \sb {2a} + \h (a\sb{1\a}\dg )\sp 2 ]\sket{0} \quad .&(11.2.4)\cr}$$
Here $\f \sb 0$
is the tachyon, $A \sp a$ is the massless vector, and $( h \sp {ab} , B
\sp a , \h )$ describe the massive, symmetric, rank-two tensor.  It's now
straightforward to use (8.2.1) to evaluate the action (4.1.6):
$$\cl = \cl \sb {-1} + \cl \sb 0 + \cl \sb 1 \quad ;\eqno(11.2.5)$$
$$\li{\cl \sb {-1} =& \ha \f \sb 0 ( \bo + 2 ) \f \sb 0 \quad,&(11.2.6a)\cr
\cl \sb 0 =& \ha A \cdot \bo A + \ha ( \pa \cdot A ) \sp 2 = - \frac14
F \sp 2 \quad ,&(11.2.6b)\cr
\cl \sb 1 =& \frac14 h \sp {ab} ( \bo - 2 ) h \sb {ab} + \ha B \cdot (
\bo - 2 ) B - \ha \h ( \bo - 2 ) \h \cr
&+ \ha ( \pa \sp b h \sb {ab} + \pa \sb a \h - B \sb a ) \sp 2 + \ha (
\frac14 h\sp a\sb a + \frac32 \h + \pa \cdot B ) \sp 2 \quad .&(11.2.6c)\cr}$$

The gauge transformations are obtained by expanding (4.1.6).  The
pieces involving $\L\sb{\a\b}$ are trivial in the component viewpoint,
since they are the ones that reduce the components of $\f$ to the
Sp(2) singlets given in (11.2.4).  Since then only the Sp(2) singlet
part of $\cq\sb\a\L\sp\a$ can contribute, only the ($M\sb{\a\b}$) isospinor
sector of $\L\sp\a$ is relevant.  We therefore take
$$\L\sb\a = (\x a\sb{1\a}\dg + \e\sp a a\sb{1a}\dg a\sb{1\a}\dg +
\e a\sb{2\a}\dg )\sket{0} \quad . \eqno(11.2.7)$$
Then the invariances are found to be
$$\li{\d A \sb a &= \pa \sb a \x \quad ;&(11.2.8a)\cr
&\cr
\d h \sb {ab} &= \pa \sb {(a} \e \sb {b)} - {1 \over \sqrt 2} \h \sb {ab} \e 
\quad ,\cr
\d B \sb a &= \pa \sb a \e + \sqrt 2 \e \sb a \quad ,\cr
\d \h &= - \pa \cdot \e + {3 \over \sqrt 2} \e \quad .&(11.2.8b)\cr}$$
(11.2.6b) and (11.2.8a) are the usual action and gauge invariance for a free
photon; however, (11.2.6c) and (11.2.8b) are not in the standard form for
massive, symmetric rank two.  Letting
$$h \sb {ab} = \Hat h \sb {ab} + \frac{1}{10} \h \sb {ab} \Hat \h
\quad , \quad \h = - \ha \Hat h \sp a \sb a - \frac{3}{10} \Hat \h
\quad ;\eqno(11.2.9)$$
one finds
$$\d \Hat h \sb {ab} = \pa \sb {(a} \e \sb {b)} \quad , \quad \d B \sb
a = \pa \sb a \e + \sqrt 2 \e \sb a \quad , \quad \d \Hat \h = - 5 
\sqrt 2 \e \quad .\eqno(11.2.10)$$
In this form it's clear that $\Hat \h$ and $B \sb a$ are Stueckelberg
fields that can be gauged away by $\e$ and $\e \sb a$.  (This was not
possible for $\h$, since the presence of the $\pa \cdot \e$ term in
its transformation law (11.2.8b) would require propagating Faddeev-Popov
ghosts.)  In this gauge $\cl \sb 1$ reduces to the Fierz-Pauli
lagrangian for a massive, symmetric, rank-two tensor:
$$\cl = \frac14 \Hat h \sp {ab} \bo \Hat h \sb {ab} + \ha ( \pa
\sb b \Hat h \sb {ab} ) \sp 2 - \ha ( \pa \sb b \Hat h \sp {ab} ) \pa
\sb a \Hat h \sp c \sb c - \frac14 \Hat h \sp a \sb a \bo \Hat h \sp b
\sb b - \frac14 ( \Hat h \sp {ab} 
\Hat h \sb {ab} - \Hat h \sp a \sb a \sp 2 ) \quad .\eqno(11.2.11)$$

The closed string is treated similarly, so we'll consider only the
tachyon and massless levels.  The expansion of the physical closed-string 
field to the second mass level is
$$ \f = (\f \sb 0 + h \sp {ab} a \sp +\sb{1a}\dg a\sp -\sb{1b}\dg +
A \sp {ab} a \sp +\sb{1a}\dg a\sp -\sb{1b}\dg +
\h a\sp +\sb 1\sp\a\dg a\sp -\sb{1\a}\dg )\sket{0} \quad ,\eqno(11.2.12)$$
where $\f \sb 0$ is the tachyon and $h \sb {ab}$, $A \sb {ab}$, and $\h$ 
describe the massless sector, consisting of the graviton, an
antisymmetric tensor, and the dilaton.  We have also dropped fields which
are killed by the projection operator for $\D N$.  We then find for the action 
(11.1.4):
$$\li{\cl =& \cl \sb {-2} + \cl \sb 0 \quad;\cr
\cl \sb {-2} =& \ha \f \sb 0 ( \bo + 4 ) \f \sb 0 \quad,\cr
\cl \sb 0 =& \frac14 h \sp {ab} \bo h \sb {ab} 
+ \frac14 A \sp {ab} \bo A \sb {ab} - \ha \h \bo \h \cr
&+ \ha ( \pa \sp b h \sb {ab} + \pa \sb a \h ) \sp 2 + \ha (
\pa \sp b A \sb {ab} ) \sp 2 \quad .&(11.2.13)\cr}$$

The nontrivial gauge transformations are found from (11.1.4):
$$\d h \sb {ab} = \pa \sb {(a} \e \sb {b)} \quad , \quad \d A \sb {ab}
= \pa\sb{[a}\z\sb{b]} \quad , \quad \d\h = \pa\cdot\e \quad .\eqno(11.2.14)$$
These lead to the field redefinitions
$$h \sb {ab} = \Hat h \sb {ab} + \h \sb {ab} \Hat \h \quad , \quad \h
= \Hat \h + \ha \Hat h \sp a \sb a \quad ;\eqno(11.2.15)$$
which result in the improved gauge transformations
$$\d \Hat h \sb {ab} = \pa \sb {(a} \e \sb {b)} \quad , \quad \d A \sb {ab}
= \pa \sb {[a} \z \sb {b]} \quad , \quad \d \Hat \h = 0 \quad .\eqno(11.2.16)$$
Substituting back into (11.2.13), we find the covariant action for a tachyon,
linearized Einstein gravity, an antisymmetric tensor, and a dilaton [4.10].

The formulation with the world-sheet metric (sect.\ 8.3)
uses more gauge and auxiliary degrees of freedom than even
the IGL(1) formulation.  We begin with the open string [3.13].
If we evaluate the kinetic operator for the gauge-invariant action
(4.1.6) between states without fermionic oscillators, we find
$ \Hat\cq\sp 2 \to \su_1^\infty ( b\sb n\dg \co\sb n + \co\sb n\dg b\sb n ) $,
so the kinetic operator reduces to
$$ \d ( M \sb {\a\b} \sp 2 ) ( -2K + \Hat\cq \sp 2 ) \quad \to \quad
-2K ( 1 - N \sb{GB} ) + \su_1^\infty \left[ - b \sb n \dg b \sb n
+ {1\over\sqrt n} ( b \sb n \dg \Tilde L \sb n + \Tilde L \sb n \dg b \sb n ) 
\right] \quad ,$$
$$ N \sb {GB} = \su_1^\infty ( b \sb n \dg g \sb n + g \sb n \dg b \sb
n ) \quad , \eqno(11.2.17)$$
dropping the $f$ and $c$ terms in $K$ and $\Tilde L \sb n$.
The operator $N \sb {GB}$ counts the number of $g \dg$'s plus $b \dg$'s
in a state (without factors of $n$).  We now evaluate the first few
component levels.  The evaluation of the tachyon action is trivial:
$$ \f = \varphi ( x ) \left| 0 \right> \quad\to\quad \cl =
\ha \varphi ( \bo + 2 ) \varphi \quad , \eqno(11.2.18)$$
where $S = \int d \sp {26} x \; \cl$.  For the photon, expanding in only
Sp(2) singlets,
$$ \f = ( A \cdot a \sb 1 \dg + \Sf B g \sb 1 \dg + \Sf G b \sb 1 \dg )
\left| 0 \right> \quad \to $$
$$ \cl = \ha A \cdot \bo A - \ha \Sf B \sp 2 - \Sf B \pa \cdot A =
- \frac14 F \sb {ab} \sp 2 - \ha ( \Sf B + \pa \cdot A ) \sp 2
\quad . \eqno(11.2.19)$$
The disappearance of {\sf G} follows from the gauge transformations
$$ \L \sb \a = ( f \sb {1\a} \dg \l \sb c + c \sb {1\a} \dg \l \sb f )
\left| 0 \right> \quad\to$$
$$ \d ( A , \Sf B , \Sf G ) = ( \pa \l \sb c , - \bo \l \sb c , \l \sb
f - \ha \l \sb c ) \quad . \eqno(11.2.20)$$
Since {\sf G} is the only field gauged by $\l\sb f$, no gauge-invariant
can be constructed from it, so it must drop out of the action.

For the next level, we consider the gauge transformations first in order
to determine which fields will drop out of the action so that its
calculation will be simplified.  The Sp(2) singlet part of the field is
$$ \li{ \f =& ( \ha h\sp{ab} a\sb{1a}\dg a\sb{1b}\dg + h\sp a a\sb{2a}\dg +
\Sf B\sp a g\sb 1\dg a\sb{1a}\dg + \Sf G\sp a b\sb 1\dg a\sb{1a}\dg \cr
&+ h g\sb 1\dg b\sb 1\dg + \widehat {\Sf B} g\sb 1\dg\sp 2 + \Sf G
b\sb 1\dg\sp 2 + \Sf B g\sb 2\dg + \widehat {\Sf G} b\sb 2\dg +
\h\sb + f\sb 1\dg\sp 2 + \h\sb - c\sb 1\dg\sp 2 + \h\sb 0 f\sb 1\sp\a\dg
c\sb {1\a}\dg ) \left| 0 \right> \quad .\cr & & (11.2.21) \cr}$$
The only terms in $\L \sb \a$ which
contribute to the transformation of the Sp(2) singlets are
$$ \li{ \L\sb\a = & \bigg[ ( \l\sb{cp}\cdot a\dg\sb 1 + \l\sb{cb}g\dg\sb 1
+\l\sb{cg}b\dg\sb 1 ) f\dg\sb{1\a} +
( \l\sb{fp}\cdot a\dg\sb 1 + \l\sb{fb}g\dg\sb 1
+\l\sb{fg}b\dg\sb 1 ) c\dg\sb{1\a} \cr
& + \l\sb c f\dg\sb{2\a} + \l\sb f c\dg\sb{2\a} \bigg] \ket{0}
\quad . & (11.2.22) \cr}$$
The gauge transformations of the components are then
$$ \li{ \d h\sb{ab} &= \pa\sb{(a}\l\sp{cp}\sb{b)} -\h\sb{ab}
	{1\over\sqrt 2} \l\sb c\quad , \cr
\d h\sb a &= \sqrt 2 \l\sp{cp}\sb a + \pa\sb a \l\sb c \quad , \cr
\d \Sf B\sb a &= \pa\sb a \l\sb{cb} + 2K \l\sp{cp}\sb a \quad , \cr
\d \Sf G\sb a &= -\ha\l\sp{cp}\sb a+\l\sp{fp}\sb a+\pa\sb a\l\sb{cg}\quad , \cr
\d h &= {1\over\sqrt 2} \l\sb c -\ha\l\sb{cb} +\l\sb{fb} +2K \l\sb{cg} 
	\quad , \cr
\d \widehat{\Sf B} &= 2K \l\sb{cb} \quad , \cr
\d \Sf G &= -\ha\l\sb{cg} +\l\sb{fg} \quad , \cr
\d \Sf B &= \sqrt 2 \l\sb{cb} +2K \l\sb c \quad , \cr
\d \widehat{\Sf G} &= -\ha\l\sb c +\l\sb f +\sqrt 2\l\sb{cg} \quad , \cr
\d \h\sb + &= \l\sb{cb} \quad , \cr
\d \h\sb - &= \sqrt 2\l\sb f -\ha\l\sb{fb} -\pa\cdot\l\sb{fp} +2K\l\sb{fg} 
	\quad , \cr
\d \h\sb 0 &= \sqrt 2\l\sb c -\frac14\l\sb{cb} +\ha\l\sb{fb}
	-\ha\pa\cdot\l\sb{cp} +K\l\sb{cg} \quad . & (11.2.23) \cr}$$
We then gauge away
$$ \li{ \Sf G \sb a &= 0 \quad\to\quad \l\sp{fp}\sb a = \ha\l\sp{cp}\sb
	a - \pa\sb a \l\sb{cg} \quad , \cr
\Sf G &= 0 \quad\to\quad \l\sb{fg} = \ha\l\sb{cg} \quad , \cr
\h\sb + &= 0 \quad\to\quad \l\sb{cb} = 0 \quad , \cr
\h\sb - &= 0 \quad\to\quad \l\sb f = {1\over 2\sqrt 2} \l\sb{fb} +
	{1\over 2\sqrt 2} \pa\cdot\l\sb{cp} - {1\over\sqrt 2} (\bo +K)\l\sb{cg}
	\quad , \cr
\h\sb 0 &= 0 \quad\to\quad \l\sb{fb} = -2\sqrt 2\l\sb c +\pa\cdot\l\sb{cp} 
	-2K\l\sb{cg} \quad . & (11.2.24) \cr}$$
The transformation laws of the remaining fields are
$$ \li{ \d h\sb{ab} &= \pa\sb{(a}\l\sb{b)} - \h\sb{ab}\l \quad , \cr
\d h\sb a &= \sqrt 2 (\l\sb a + \pa\sb a\l ) \quad , \cr
\d h &= -3\l + \pa\cdot\l \quad , \cr
\d \Sf B\sb a &= - (\bo -2) \l\sb a \quad , \cr
\d \Sf B &= - \sqrt 2 (\bo -2) \l \quad , \cr
\d \widehat{\Sf B} &= 0 \quad , \cr
\d \widehat{\Sf G} &= {1\over\sqrt 2}(- 3\l + \pa\cdot\l ) 
	\quad , &(11.2.25) \cr}$$
where $\l\sb a = \l\sp{cp}\sb a$ and $\l = \l\sb c / \sqrt 2$.  
$\l\sb{cg}$ drops out of the transformation law as a result of the gauge
invariance for gauge invariance
$$ \d\f = \Hat\cq\sp\a \L\sb\a \quad\to\quad \d\L\sb\a = 
\Hat\cq\sp\b\L\sb{(\a\b )} \quad . \eqno(11.2.26)$$
The lagrangian can then be computed from (4.1.6) and (11.2.17) to be
$$ \li{ \cl =& \frac14 h\sp{ab}(\bo -2)h\sb{ab} + \ha h\sp a(\bo -2)h\sb
	a - \ha h(\bo -2)h \cr
& -\ha\Sf B\sb a\sp 2-\ha\Sf B\sp 2 + 2\Hat{\Sf B}(\sqrt 2\Hat{\Sf G}-h) \cr
& + \Sf B\sp a(-\pa\sp b h\sb{ab}+\pa\sb a h+\sqrt 2 h\sb a) +
	\Sf B(- \pa\cdot h+{3\over\sqrt 2}h-{1\over 2\sqrt 2} h\sp a\sb a) 
	\, . &(11.2.27)\cr}$$

The lagrangian of (11.2.19) is equivalent to that obtained from the
IGL(1) action, whereas the lagrangian of
(11.2.27) is like the IGL(1) one but contains in addition the 2
gauge-invariant auxiliary fields $\widehat{\Sf B}$ and $\sqrt 2\Hat{\Sf G}-h$.
Both reduce to the OSp(1,1$|$2) lagrangians after elimination of
auxiliary fields.

The results for the closed string are similar.  Here we consider just
the massless level of the nonoriented closed string (which is symmetric
under interchange of $+$ and $-$ modes).  The Sp(2) and $\D N$ invariant
components are
$$ \li{ \f =& ( h\sp{ab}a\dg\sb{+a}a\dg\sb{-a} +B\sp ag\dg\sb{(+}
a\dg\sb{-)a} +G\sp ab\dg\sb{(+}a\dg\sb{-)a} +hg\dg\sb{(+}b\dg\sb{-)}
+Bg\dg\sb +g\dg\sb - +Gb\dg\sb +b\dg\sb - \cr
&+ \h\sb +f\dg\sb +\sp\a f\dg\sb{-\a} +\h\sb -c\dg\sb +\sp\a c\dg\sb{-\a}
+\h\sb 0f\dg\sb{(+}\sp\a c\dg\sb{-)\a} ) \left| 0 \right> \quad ,
&(11.2.28)\cr}$$
where all oscillators are from the first mode ($n=1$), and $h\sp{ab}$ is
symmetric.  The gauge parameters are
$$ \L\sb\a = \left[ ( \l\sb{cp}\cdot a\dg + \l\sb{cb}g\dg + \l\sb{cg}b\dg )
\sb{(+}f\dg\sb{-)\a} + ( \l\sb{fp}\cdot a\dg + \l\sb{fb}g\dg + \l\sb{fg}b\dg )
\sb{(+}c\dg\sb{-)\a} \right] \left| 0 \right> \quad . \eqno(11.2.29) $$
The component transformations are then
$$ \li{ \d h\sb{ab} &= \pa\sb{(a}\l\sp{cp}\sb{b)} \quad ,\cr
\d B\sb a &= \pa\sb a \l\sb{cb} - \ha\bo\l\sp{cp}\sb a \quad ,\cr
\d G\sb a &= 2\l\sp{fp}\sb a - \l\sp{cp}\sb a + \pa\sb a \l\sb{cg} \quad ,\cr
\d h &= 2\l\sb{fb} - \l\sb{cb} -\ha\bo\l\sb{cg} \quad , \cr
\d B &= -\bo\l\sb{cb} \quad , \cr
\d G &= 4\l\sb{fg} -2\l\sb{cg} \quad ,\cr
\d \h\sb + &= 2\l\sb{cb} \quad , \cr
\d \h\sb - &= -\l\sb{fb} -\pa\cdot\l\sb{fp} -\ha\bo\l\sb{fg} \quad ,\cr
\d \h\sb 0 &= \l\sb{fb} -\ha\l\sb{cb} -\ha\pa\cdot\l\sb{cp} -\frac14 
	\bo\l\sb{cg} \quad .&(11.2.30)\cr}$$
We choose the gauges
$$\li{ G\sb a &= 0 \quad\to\quad \l\sp{fp}\sb a = \ha\l\sp{cp}\sb a
	-\ha\pa\sb a \l\sb{cg} \quad , \cr
G &= 0 \quad\to\quad \l\sb{fg} = \ha\l\sb{cg} \quad , \cr
\h\sb + &= 0 \quad\to\quad \l\sb{cb} = 0 \quad , \cr
\h\sb 0 &= 0 \quad\to\quad \l\sb{fb} = \ha\pa\cdot\l\sb{cp}
	+\frac14\bo\l\sb{cg} \quad . &(11.2.31)\cr}$$
The remaining fields transform as
$$ \li{ \d h\sb{ab} &= \pa\sb{(a}\l\sb{b)} \quad , \cr
\d B\sb a &= -\ha\bo\l\sb a \quad , \cr
\d h &= \pa\cdot\l \quad , \cr
\d B &= 0 \quad , \cr
\d \h\sb - &= -\pa\cdot\l \quad , &(11.2.32)\cr}$$
where $\l\sb a = \l\sp{cp}\sb a$, and $\l\sb{cg}$ drops out.  Finally,
the lagrangian is
$$ \cl = \ha h\sp{ab}\bo h\sb{ab} - h\bo h - 4 B\sb a\sp 2 - 8B(\h\sb -
+ h) -4B\sp a (\pa\sp b h\sb{ab} -\pa\sb a h) \quad . \eqno(11.2.33)$$
Again we find the auxiliary fields $B$ and $\h\sb - +h$ in addition to
the usual nonminimal ones.  After elimination of auxiliary fields, this
lagrangian reduces to that of (11.2.13) (for the nonoriented sector,
up to normalization of the fields).

\sect{Exercises}

\Item{(1)} Derive the gauge-invariant actions (IGL(1) and OSp(1,1$|$2))
for free open strings.  Do the same for the Neveu-Schwarz string.
Derive the OSp(1,1$|$2) action for the Ramond string.
\Item{(2)} Find all the Sp(2)-singlet fields at the levels indicated in
(11.2.2).  Separate them into sets corresponding to irreducible
representations of the Poincar\'e group (including their Stueckelberg
and auxiliary fields).
\Item{(3)} Derive the action for the massless level of the open string
using the bosonized ghosts of sect.\ 8.1.
\Item{(4)} Derive the action for the next mass level of the closed
string after those in (11.2.13).  Do the same for (11.2.33).
\Item{(5)} Rederive the actions of sect.\ 11.2 for levels which include
spin 2 by first decomposing the corresponding light-cone representations
into irreducible representations, and then using the Hilbert-space
constructions of sect.\ 4.1 for each irreducible representation.

%
%

\Chsc{12. }{GAUGE-INVARIANT }{INTERACTIONS}{12.1. Introduction}3

The gauge-invariant forms of the interacting actions for string field
theories are far from understood.  Interacting closed string field
theory does not yet exist.  (Although a proposal has been made [12.1],
it is not even at the point where component actions can be examined.)
An open-string formulation exists [4.8] (see the following section), but it
does not seem to relate to the light-cone formulation (chapt.\ 10), and
has more complicated vertices.  Furthermore, all these formulations are
in the IGL(1) formalism, so relation to particles is less direct because
of the need to eliminate BRST auxiliary fields.  Most importantly, the
concept of conformal invariance is not clear in these formulations.  If
a formulation could be found which incorporated the world-sheet metric
as coordinates, as the free theory of sect.\ 8.3, it might be possible
to restore conformal transformations as a larger gauge invariance which
allowed the derivation of the other formulations as (partial) gauge choices.

In this section we will mostly discuss the status of the derivation of
an interacting gauge-covariant string theory from the light cone, with
interactions similar to those of the light-cone string field theory.  
The derivation follows the corresponding derivation for Yang-Mills
described in sects.\ 3.4 and 4.2 [3.14], but the important step (3.4.17) of
eliminating $p\sb +$ dependence has not yet been performed.

As performed for Yang-Mills in sect.\ 3.4, the transformation (3.4.3a)
with $\F\to U\sp{-1}\F$ is the first step in deriving an IGL(1) 
formalism for the interacting string from the light cone.  Since 
the transformation is nonunitary, the factor of $p_+$ in (2.4.9) is 
canceled.  In (2.4.7), using integration by parts, (3.4.3a) induces 
the transformation of the vertex function 
$$ \cv^{(n)} \quad\to\quad {1\over p_{+1}\cdots p_{+n}}U(1)\cdots U(n)
\cv^{(n)} \quad , \eqno(12.1.1) $$
where we work in momentum space with respect to $p_+$.
The lowest-order interacting contribution to $Q$ then follows from
applying this transformation to the OSp-extended form of the light-cone
vertex (10.6):
$$ \cj\sp{(3)}\sb -\sp c = -2ig {p_{+r}\over p_{+1}p_{+2}p_{+3}}
X\sb r\sp{ c} (\sigma_{int})\,\d \left( \su p_+ \right) \D [X^a]
\,\D [p_+ X^{ c}]\,\D [ p\sb +\sp{-1} X^{\tilde c}]\quad ,\eqno(12.1.2)$$
effectively giving conformal weight -1 to $X^{ c}$ and conformal
weight 1 to $X^{\tilde c}$.  A $\d$-functional that
matches a coordinate must also match the $\sigma$-derivative
of the coordinate, and with no zero-modes one must have
$$\D \left[ p\sb +\sp{-1} X^{\tilde c}\left({\s\over p_+}\right)
\bigg|_{\tilde c =0} \right] =
\D \left[ \pa_\s p\sb +\sp{-1} X^{\tilde c}\left({\s\over p_+}\right) \right]
= \D \left[ p\sb +\sp{-2} X^{\tilde c}'\left({\s\over p_+}\right) \right]
\quad .\eqno(12.1.3)$$
(Even the normalization is unambiguous, since without zero-modes $\D$
can be normalized to 1 between ground states.)
We now recognize $X^{ c}$ and
$X^{\tilde c}'$ to be just the usual Faddeev-Popov ghost $C(\s )$ of
$\t$-reparametrizations and Faddeev-Popov {\it anti\/}ghost $\d
/\d\Tilde C (\s )$ of $\s$-reparametrizations (as in (8.1.13)), of 
conformal weights -1 and 2, respectively, which is equivalent to the 
relation (8.2.2,3) (as seen by using (7.1.7b)).
Finally, we can (functionally) Fourier transform the antighost so that
it is replaced with the canonically conjugate ghost.
Our final vertex function is therefore
$$ \cj\sp{(3)}\sb -\sp c =  -2ig {p_{+r}\over p_{+1}p_{+2}p_{+3}}
C\sb r({\sigma}_{int}) \d \left( \su p_+ \right) \D [X^a]\,\D [ p\sb
+ C ]\, \D [ p\sb + \Tilde C ]\quad , \eqno(12.1.4a)$$
or in terms of momenta
$$ \Tilde\cj\sp{(3)}\sb -\sp c =  -2ig p_{+r} C\sb r({\sigma}_{int})
\d \left( \su p_+ \right) \D [p\sb +\sp{-1}P^a]\,\D \left[ p\sb +\sp{-2}\fder
C \right]\, \D \left[ p\sb +\sp{-2}\fder{\Tilde C}\right]\quad .
\eqno(12.1.4b)$$
The extra $p_+$'s disappear due to Fourier transformation of the 
zero-modes $c$:
$$ {1\over p_+}\int d c\; e\sp{-c p^{\tilde c}}f(p_+ c ) =
\int d c\; e\sp{-c p^{\tilde c} /p_+} f(c) = 
\tilde f\left( {p^{\tilde c}\over p_+} \right) \quad . \eqno(12.1.5) $$
Equivalently, the exponent of $U$ by (3.4.3a) is $c \pa /\pa c
+M^3$, so the zero-mode part just scales $c$, but $c \pa /\pa c =
1-(\pa /\pa c ) c$, so besides scaling $\pa /\pa c$ there is an extra factor of
$p_+$ for each zero-mode, canceling those in (12.1.1,2).
There is no effect on the normalization with respect
to nonzero-modes because of the above-mentioned normalization in the
definition of $\D$ with respect to the creation and annihilation
operators.  A similar analysis applies to closed strings [12.2].

Hata, Itoh, Kugo, Kunitomo, and Ogawa [12.3] proposed an interacting 
BRST operator equivalent to this one, and corresponding gauge-fixed 
and gauge-invariant actions, 
but with $p\sb +$ treated as an extra coordinate as in [2.7].
(A similar earlier attempt appeared in [4.9,7.5], with $p\sb +$ fixed, as
a consequence of which the BRST algebra didn't close to all orders.
Similar attempts appeared in [12.4].)
However, as explained in [2.7], such a formalism 
requires also an additional anticommuting coordinate in order for the 
loops to work (as easily checked for the planar 1-loop graph with 
external tachyons [2.7]), and can lead to problems with infrared behavior 
[4.4].

The usual four-point vertex of Yang-Mills (and even-higher-point vertices 
of gravity for the closed string) will be obtained only after 
field redefinitions of the massive fields.  This corresponds to the 
fact that it shows up in the zero-slope limit of the S-matrix only after 
massive propagators have been included and reduced to points.  In 
terms of the Lagrangian, for arbitrary massive fields $\m$ and 
massless fields $\n$, the terms, for example, 
$$ L = \ha \m [ \bo + M \sp 2 + M \sp 2 U ( \n ) ] \m + M \sp 2 \m V (\n ) 
\eqno(12.1.6a)$$
(where $U ( \n )$ and $V ( \n )$ represent some interaction terms)
become, in the limit $M \sp 2 \rightarrow \infty$, 
$$ L = M \sp 2 [ \ha ( 1 + U ) \m \sp 2 + V \m ] \quad .\eqno(12.1.6b)$$
The corresponding field redefinition is 
$$\m = \tilde \m - {V \over {1 + U}} \quad ,\eqno(12.1.6c)$$
which modifies the Lagrangian to 
$$ L = \ha \tilde \m ( \bo + M \sp 2 + M \sp 2 U ) \tilde \m + \ha M \sp 2 
{{V \sp 2}\over{1 + U}} + \co ( M \sp 0 ) \quad .\eqno(12.1.6d)$$
The redefined massive fields $\tilde \m$ no longer contribute in the 
zero-slope limit, and can be dropped from the Lagrangian before taking 
the limit.  However, the redefinition has introduced the new interaction term
$\ha M \sp 2 {{V \sp 2}\over{1 + U}}$ into the $\n$-part of the Lagrangian.

\sect{12.2. Midpoint interaction}

Witten has proposed an extension of the IGL(1) gauge-invariant open bosonic 
string action to the interacting case [4.8].  
Although there may be certain limitations with his construction, it shares 
certain general properties with the light-cone (and covariantized light-cone)
formalism, and thus we expect these properties will be common to any 
future approaches.  The construction is based on the use of a vertex 
which consists mainly of $\d$-functionals, as in the light-cone 
formalism.  Although the geometry of the infinitesimal surface 
corresponding to these $\d$-functionals differs from that of the light-cone 
case, they have certain algebraic features in common.  In particular, 
by considering a structure for which the $\d$-functional (times certain vertex 
factors) is identified with the product operation of a certain algebra, 
the associativity of the product follows from the usual properties of 
$\d$-functionals.  This is sufficient to define an interacting, nilpotent 
BRST operator (or Lorentz generators with $[ J \sb {-i} , J \sb {-j} ] =0$), 
which in turn gives an interacting gauge-invariant (or 
Lorentz-invariant) action.  

The string fields are
elements of an algebra: a vector space with an outer \hbox{product $*$.}
We write an explicit vector index on the string field $\F \sb i$,
where $i = {\bf Z} ( \s )$ is the coordinates ($X, C, \Tilde C$ for the 
covariant formalism and $X \sb T , x \sb \pm$ for the light cone),
and excludes group-theory indices.  Then the product can 
be written in terms of a rank-3 matrix
$$ ( \F * \J ) \sb i = f \sb i \sp {jk} \F \sb k \J\sb j \quad .\eqno(12.2.1)$$
In order to construct actions, and because of the relation of a field 
to a first-quantized wave function, we require, in addition to the 
operations necessary to define an algebra, a Hilbert-space inner product
$$ \left< \F | \J \right> = \int \Sc D Z \; tr \; 
\F \sp \dag \J = tr \; \F \sp i \sp \dag \J \sb i \quad ,\eqno(12.2.2)$$
where $tr$ is the group-theory trace.  Furthermore, in order to give 
the hermiticity condition on the field we require an indefinite, 
symmetric charge-conjugation matrix $\O$ on this space:
$$\F \sb i = \O \sb {ij} \F \sp j \sp \dag \quad , \quad ( \O \F ) [ X ( \s ) , 
C ( \s ) , \Tilde C ( \s ) ] \equiv \F [ X ( \p - \s ) , C ( \p - \s ) , 
- \Tilde C ( \p - \s ) ] \quad .\eqno(12.2.3a)$$
$\O$ is the ``twist'' of (11.1.1).  To allow a reality condition or, 
combining with (12.2.3a), a symmetry condition for real group 
representations (for SO(N) or USp(N)), we also require that indices 
can be freely raised and lowered:
$$ ( \h \F ) \sb i = \O \sb {ij} \d \sp {jk} ( \h \F ) \sp t \sb k 
\quad ,\eqno(12.2.3b)$$
where $\h$ is the group metric and $t$ is the group-index transpose.  
In order to perform the usual graphical manipulations implied by 
duality, the twist must have the usual effect on vertices, and thus on 
the inner product:
$$ \O ( \F * \J ) = ( \O \J ) * ( \O \F ) \quad .\eqno(12.2.4)$$
Further properties 
satisfied by the product follow from the nilpotence of the BRST 
operator and integrability of the field equations $Q \F = 0$:  Defining
$$ Q \F = Q \sb 0 \F + \F * \F \quad \to \quad ( Q \F ) \sb i = Q \sb {0i} \sp j
\F \sb j + f \sb i \sp {jk} \F \sb k \F \sb j \quad ,\eqno(12.2.5)$$
$$ S = \int \Sc D {\bf Z} \; tr \; \F \sp \dag ( \ha Q \sb 0 \F + 
\frac13 \F * \F ) = tr \; \left[ \ha ( \O Q \sb 0 ) \sp {ij} \F \sb j \F \sb i +
\frac13 (\O f ) \sp {ijk} \F \sb k \F \sb j \F \sb i \right] \quad ,
\eqno(12.2.6)$$
we find that hermiticity requires
$$ ( \O Q \sb 0 ) \sp {ij} = ( \bar Q \sb 0 \O ) \sp {ji} \quad , \quad 
( \O f ) \sp {ijk} = ( \bar f \O \O ) \sp {kji} \quad ,\eqno(12.2.7)$$
integrability requires antisymmetry of $Q \sb 0$ and cyclicity of $*$:
$$ ( \O Q \sb 0 ) \sp {ij} = - ( \O Q \sb 0 ) \sp {ji} \quad , \quad ( 
\O f ) \sp {ijk} = ( \O f ) \sp {jki} \eqno(12.2.8)$$
(where permutation of indices is in the ``graded'' sense, but we have 
omitted some signs: e.g., $( \O Q \sb 0 ) \sp {ij} \F \sb j \J \sb i = + 
( \O Q \sb 0 ) \sp {ij} \J \sb j \F \sb i$ when $\F [{\bf Z}]$ 
and $\J [ {\bf Z} ]$ are {\it anticommuting}, but $\F \sb i$ and $\J \sb i$ 
include components of {\it either} statistics), and nilpotence requires, 
besides $Q \sb 0 \sp 2 = 0$, that $*$ is BRST invariant (i.e., $Q \sb 
0$ is distributive over $*$) and associative:
$$ Q \sb 0 ( \F * \J ) = ( Q \sb 0 \F ) * \J + \F * ( Q \sb 0 \J ) $$
$$ \iff \quad Q \sb {0i} \sp l ( f \O \O ) \sb {ljk} + Q \sb {0j} \sp l 
( f \O \O ) \sb {ilk} + Q \sb {0k} \sp l ( f \O \O ) \sb {ijl} = 0 \quad ,
\eqno(12.2.9)$$
$$ \F * ( \J * \U ) = ( \F * \J ) * \U 
\quad \iff \quad ( \O f ) \sp {ijm} f \sb m \sp {kl} = ( \O f ) \sp 
{jkm} f \sb m \sp {li} \eqno(12.2.10)$$
(where we have again ignored some signs due to grading).  $*$ should 
also be invariant under all transformations under conserved 
quantities, and thus the operators $\pa / \pa {\bf z} \sim \int d \s \; \pa / 
\pa${\bf Z} must also be distributive over $*$.

At this point we have much more structure than in an ordinary algebra, 
and only one more thing needs to be introduced in order to obtain a
matrix algebra: an identity element for the outer product
$$ \F * I = I * \F = \F \quad \iff \quad f \sb i \sp {jk} I \sb k = \d 
\sb i \sp j \quad . \eqno(12.2.11)$$
It's not clear why string field theory must have such an object, but both the 
light-cone approach and Witten's approach have one.  In the light-cone 
approach the identity element is the ground state (tachyon) at 
vanishing momentum (including the string length, $2 \p \a ' p \sb +$), 
which is related to the fact that the vertex for an external tachyon 
takes the simple form $: e \sp {-ik \cdot X(0)} :$.  We now consider 
the fields as being 
matrices in {\bf Z}$( \s )$-space as well as in group space, although
the vector space on which such matrices act might not be (explicitly) defined. 
(Such a formalism might be a consequence of the same duality properties 
that require general matrix structure, as opposed to just adjoint 
representation, in the group space.)  $*$ is now the matrix product.
(12.2.7,10) then express just the usual hermiticity and associativity 
properties of the matrix product.  The trace operation $Tr$ of these
matrices is implied by the Hilbert-space inner product (12.2.2):
$$ \left< \F | \J \right> = Tr \; \F \sp \dag \J
\quad \iff \quad Tr \; \F = \left< I | \F \right> \quad .\eqno(12.2.12)$$
(12.2.8) states the usual cyclicity of the trace.  Finally, the twist 
metric (12.2.3) is identified with the matrix transpose, in 
addition to transposition in the group space, as implied by (12.2.4).  
Using this transposition in combination with the usual hermitian
conjugation to define the matrix complex conjugate, the hermiticity 
condition (12.2.3a) becomes just hermiticity in the group space:
Denoting the group-space matrix indices as $\F \sb {\bf a} \sp {\bf b}$,
$$ \F \sb {\bf a} \sp {\bf b} = \F \sb {\bf b} \sp {\bf a} \conj 
\quad .\eqno(12.2.13)$$
As a result of
$$ I \sp t = I \quad \iff \quad \O \sb {ij} I \sb j = I \sb i \eqno(12.2.14)$$
and the fact that $Q \sb 0$ and $\pa / \pa${\bf z} are distributive as well 
as being ``antisymmetric'' (odd under simultaneous twisting and 
integration by parts), we find
$$ Q \sb 0 I = {\pa \over \pa {\bf z}} I = 0 \quad .\eqno(12.2.15)$$

Given one $*$ product, it's possible to define other associative 
products by combining it with some operators $d$ which are distributive 
over it.  Thus, the condition of associativity of $\star$ and $*$ implies
$$ A \star B = A * dB \quad \to \quad d \sp 2 = 0 $$
$$ A \star B = ( dA ) * ( dB ) \quad \to \quad d \sp 2 = d ~ or ~ d 
\sp 2 = 0 \quad .\eqno(12.2.16)$$
The former allows the introduction of conserved anticommuting factors 
(as for the BRST open-string vertex), 
while the latter allows the introduction of projection operators (as 
expected for closed strings with respect to $\D N$).

The gauge transformations and action come directly from the 
interacting BRST operator:  Using the analysis of (4.2.17-21),
$$ \d \F = \left[ \left[ \int\L\F , Q \right]_c , \F \right]_c =
Q\sb 0 \L + [ \F ,* \L ] \quad ,\eqno(12.2.17a)$$
where the last bracket is the commutator with respect to the $*$ product, and
$$ S = -iQ = \int \ha\F\dg Q\sb 0 \F + \frac13 \F\dg ( \F * \F )
\quad ,\eqno(12.2.17b)$$
where for physical fields we restrict to
$$ J\sp 3 \F = 0  \quad , \quad J\sp 3 \L = - \L \quad .\eqno(12.2.17c)$$
Gauge invariance follows from $Q \sp 2 = \ha [ Q , Q ]\sb c =0$.
Actually, the projection onto $J\sp 3 = 0$ is somewhat 
redundant, since the other fields can be removed by gauge 
transformations or nondynamical field equations, at least at the 
classical level.  (See (3.4.18) for Yang-Mills.)

A possible candidate for a gauge-fixed action can be written in terms of $Q$ as
$$\li{S& = \left[ Q , \ha \int \F \sp \dag \co \F \right]_c + 
\frac16 \int \F \sp \dag ( Q \sb {INT} \F ) &(12.2.18a)\cr
&= - \ha \int ( Q \sb 0 \F ) \sp \dag \left[ c , {\pa \over {\pa c}} 
\right] \F - \ha \int ( Q \sb {INT} \F ) \sp \dag \left( \left[ 
c , {\pa \over {\pa c}} \right] - \frac13 \right) \F \; ,&(12.2.18b)\cr}$$
where $Q \sb 0$ and $Q \sb {INT}$ are the free and interaction terms 
of $Q$, and
$\co = \ha [ c , \pa / \pa c ]$.  Each term in (12.2.18a) is separately BRST 
invariant.  The BRST invariance of the second term follows from the 
associativity property of the $*$ product.
Due to the $-1/3$ in (12.2.18b) one can easily show that all 
$\f \sp 2 \j$ terms drop out, due just to the $c$ dependence of $\F$ and the 
cyclicity of $Q \sb {INT}$.  Such terms would contain auxiliary 
fields which drop out of the free action.  We would like these fields to
occur only in a way which could be eliminated by field redefinition, 
corresponding to maintaining a gauge invariance of the free action at the 
interacting level, so we could choose the gauge where these fields vanish.  
Unfortunately, this is not the case in (12.2.18), so allowing
this abelian gauge invariance would require adding some additional
cubic-interaction gauge-fixing term, each term of which contains
auxiliary-field factors, such that the undesired auxiliary fields are
eliminated from the action.

Whereas the $\d$-functionals used in the light-cone formalism 
correspond to a ``flat'' geometry (see chapt.\ 10), with all curvature 
in the boundary (specifically, the splitting point) rather than the 
surface itself, those used in Witten's covariant formalism correspond 
to the geometry (with the 3 external legs amputated)

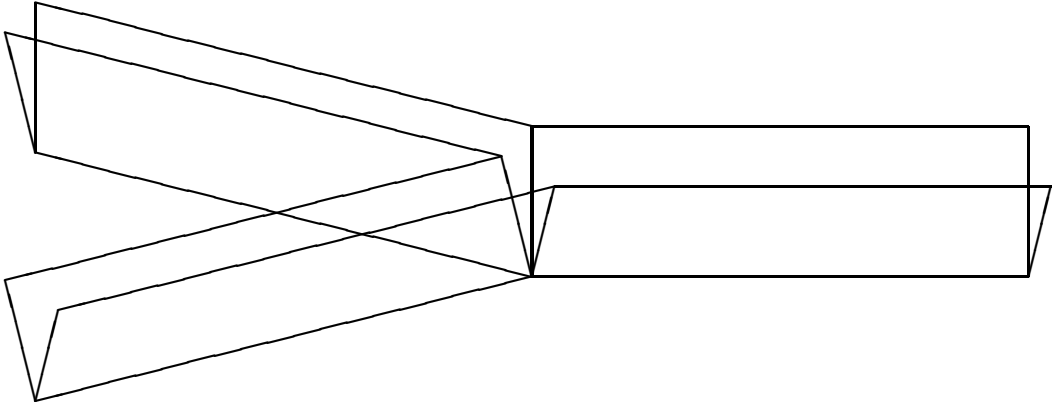
\setlength{\unitlength}{1mm}
\begin{picture}(140,65)
\put(70,20){\line(0,1){20}}
\put(4,36.5){\line(0,1){20}}
\put(136,20){\line(0,1){20}}
\put(70,20){\line(-1,4){4}}
\put(4,36.5){\line(-1,4){4}}
\put(4,3.5){\line(-1,4){4}}
\put(70,20){\line(1,4){3}}
\put(136,20){\line(1,4){3}}
\put(4,3.5){\line(1,4){3}}
\put(70,20){\line(-4,-1){66}}
\put(70,20){\line(-4,1){66}}
\put(70,20){\line(1,0){66}}
\put(70,40){\line(-4,1){66}}
\put(70,40){\line(1,0){66}}
\put(66,36){\line(-4,-1){66}}
\put(66,36){\line(-4,1){66}}
\put(73,32){\line(-4,-1){66}}
\put(73,32){\line(1,0){66}}
\end{picture}

\noindent (The boundaries of the strings are on top; the folds along 
the bottom don't affect the intrinsic geometry.)  All the curvature is 
concentrated in the cusp at the bottom (a small circle around it 
subtends an angle of $3 \p$), with no intrinsic curvature in the 
boundary (all parts of the boundary form angles of $\p$ with respect 
to the surface).  Each string is ``folded'' in the middle ($\p /2$),
and thus the vertex $\d$-functionals equate the coordinates ${\bf Z}
(\s )$ of one string with ${\bf Z}(\p -\s )$ for the next string for
$\s\in [ 0 , \p / 2 ]$ (and therefore with ${\bf Z}(\p -\s )$ for the
previous string for $\s\in [\p /2 ,\p ]$).
These $\d$ functions are easily seen to define an 
associative product:  Two successive $*$ products produce a 
configuration like the one above, but with 4 strings, and 
associativity is just the cyclicity of this 4-string object (see 
(12.2.10)).  (However, vertex factors can ruin this associativity because
of divergences of the coincidence of two such factors in the product of
two vertices, as in the superstring: see below.)

Similar remarks apply to the corresponding product 
implied by the $\d$ functionals of (9.2.23) of the light-cone 
formalism, but there associativity is violated by an amount which is
fixed by the light-cone 4-point interaction vertex.
This is due to the fact that in the
light-cone formalism there is a 4-point graph where a string has a
string split off from one side, followed by an incoming string joining
onto the same side.  If no conformal transformation is made, this graph 
is nonplanar, unlike the graph where this splitting and joining occur 
on opposite sides.  There is a similar graph where the splitting and
joining occur on the opposite side from the first graph, and these 2
graphs are continuously related by a graph with a 4-point vertex as
described in chapt.\ 10.  The $\s$-position of the interaction point in
the surface of the string varies from one end of the string to the
other, with the vertex having this point at an end being the same as the
limit of the 1 of the other 2 graphs where the propagator has vanishing
length.  On the other hand, in the covariant formalism the limits of the
2 corresponding graphs are the same, so no 4-point vertex is needed to
interpolate between them.

The appropriate vertex factor for the midpoint interaction
follows directly from the quantum
mechanics with bosonized ghosts.  From (8.1.28a), since $w=-1$ for $\Hat
C$, there is an $\int d\sp 2 \s \; {\rm e}\sp{-1} \frac32 R\c$ term
which contributes only where there is curvature.  Thus, the
total contribution of the curvature at the cusp to the path integral 
is an extra factor of an exponential whose exponent is 3/2 times
$\c$ evaluated at the cusp.  The BRST invariance of the 
first-quantized action then guarantees that the vertex conserves $Q 
\sb 0$, and the coefficient of the curvature term compensates for the 
anomaly in ghost-number conservation at the cusp.  (Similar remarks
apply for fermionized ghosts using the Lorentz connection term.)
Alternatively, the 
coefficient follows from considering the ghost number of the fields 
and what ghost number is required for the vertex to give the same 
matrix elements for physical polarizations as in the light-cone 
formalism:  In terms of bosonized-ghost coordinates, any physical state
must be $\sim e\sp{-\hat q /2}$ by (8.1.19,21a).  Since the $\d$
functional and functional integration in $\c$ have no such factors, and
the vertex factor can be only at the cusp (otherwise it destroys the
above properties of the $\d$ functionals), it must be $e\sp{3\c /2}$
evaluated at the cusp to cancel the $\hat q$-dependence of the 3 fields.
In terms of the original fermionic ghosts, we
use the latter argument, since the anomalous curvature term doesn't show
up in the classical mechanics lagrangian (although a similar argument
could be made by considering quantum mechanical corrections).  Then the
physical states have no dependence on $c$, while the vertex has a $dc$
integration for each of the 3 coordinates, and a single $\d$ function
for overall conservation of the ``momentum'' $c$.  (A similar argument
follows from working in terms of Fourier transformed fields which depend
instead on the ``coordinate'' $\pa / \pa c$.)  The appropriate vertex 
factor is thus
$$ C (\frac\p 2 ) \tilde C (\frac\p 2) \sim \hat C (\frac\p 2) \hat C
(-\frac\p 2) \quad , \eqno(12.2.19a)$$
or, in terms of bosonized ghosts (but still for fields with fermionic
ghost coordinates)
$$ e\sp{2\c ( \p / 2 )} \sim e\sp{\hat\c (\p /2)} e\sp{\hat\c (-\p /2)}
\quad ,\eqno(12.2.19b)$$
where $\p / 2$, the midpoint of each string, is the position of the cusp.
(The difference in the vertex factor for different coordinates is
analogous to the fact that the ``scalar'' $(-g)\sp{-1/2}\d\sp D (x-x')$
in general relativity has different expressions for $g$ in different
coordinate systems.)  The vertex, including the factor (12.2.19), can be
considered a Heisenberg-picture vacuum in the same way as in the
light-cone formalism, where the vertex $\sket{V}=e\sp\D\sket{0}$ in
(9.2.1) was the effect of acting on the interaction-picture vacuum with
the S-matrix (of the first-quantized theory).  However, in this case the
vacuum includes vertex factors because the appropriate vacuum is not the
tachyon one but rather the one left invariant by the Sp(2) subalgebra of
the Virasoro algebra [12.5] (see sect.\ 8.1).  This is a consequence of
the Sp(2) symmetry of the tree graphs.

Because of the midpoint form of the interaction there is a global
symmetry corresponding to conformal transformations which leave the
midpoint fixed.  In second-quantized notation, any operator which is the
bracket of (the interacting) $Q$ with something itself has a vanishing
bracket with $Q$, and is therefore simultaneously BRST invariant and
generates a global symmetry of the action (because $Q$ is the action).
In particular, we can consider
$$ \left[ Q , \int \F\dg \left( \fder{\Hat C (\s )} -
\fder{\Hat C (\p -\s )} \right) \F \right]_c \sim
\int \F\dg \left( \Hat\cg (\s ) - \Hat\cg (\p -\s ) \right) \F
\quad , \eqno(12.2.20)$$
where the 2 $\d / \d\Hat C$ terms cancel in the interaction term
because of the form of the overlap integral for the vertex (and the
location of the vertex factor (12.2.19) at the midpoint), and the
surviving free term comes from the first-quantized expression for
$\Hat\cg = \{ Q\sb 0 , \d / \d\Hat C \}$.  Thus, this subalgebra of the
Virasoro algebra remains a global invariance at the interacting level (without
becoming inhomogeneous in the fields).

The mode expansion of Witten's vertex can be evaluated [12.6,7] as in the 
light-cone case (sect.\ 9.2).  (Partial evaluations were given and BRST
invariance was also studied in [12.8].)  Now
$$ \D = \ha \D ( \check P , \check P ) - \D \left( \check C ' , {\d 
\over \d \check C} \right) \quad .\eqno(12.2.21)$$
($\Hat C$ has weight $w = 2$.)  The map from the $\r$ plane to the 
$z$ plane can be found from the following sequence of conformal
transformations:

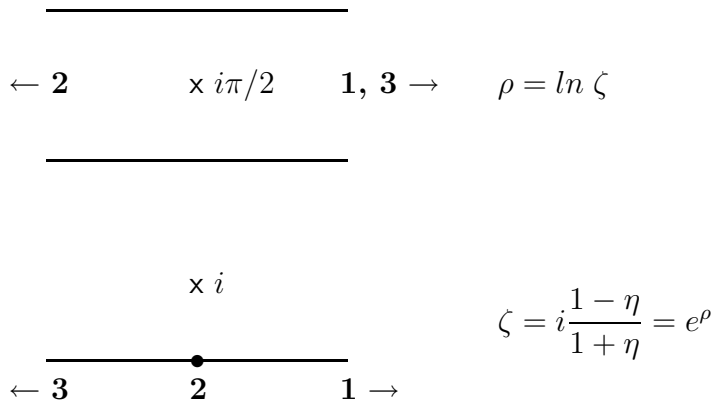
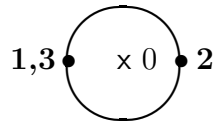
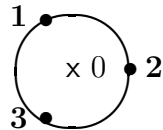
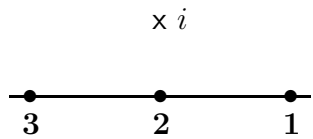
\begin{picture}(140,30)
\put(10,0){\line(1,0){40}}
\put(10,20){\line(1,0){40}}
\put(29,9){{\sf x}~$i \p / 2$}
\put(5,9){$\gets$ {\bf 2}}
\put(49,9){{\bf 1, 3} $\to$}
\put(70,9){$\r = ln \; \z$}
\end{picture}

\begin{picture}(140,30)
\put(10,5){\line(1,0){40}}
\put(29,14){{\sf x}~$i$}
\put(5,0){$\gets$ {\bf 3}}
\put(49,0){{\bf 1} $\to$}
\put(29,0){\bf 2}
\put(29,4){$\bullet$}
\put(70,9){$\z = i {\displaystyle{{1 - \h} \over {1 + \h}}} = e \sp \r$}
\end{picture}

\setlength{\unitlength}{.75mm}
\begin{picture}(165,30)
\put(40,10){\oval(20,20)}
\put(39,9){{\sf x}~0}
\put(29,9){$\bullet$}
\put(49,9){$\bullet$}
\put(53,9){\bf 2}
\put(20,9){\bf 1,3}
\put(93.3,9){$\h = \l \sp {3/2} = {\displaystyle{{i - \z} \over {i + \z}}}$}
\end{picture}

\begin{picture}(165,30)
\put(40,10){\oval(20,20)}
\put(39,9){{\sf x}~0}
\put(34,17.66){$\bullet$}
\put(34,.3){$\bullet$}
\put(49,9){$\bullet$}
\put(53,9){\bf 2}
\put(29,18){\bf 1}
\put(29,0){\bf 3}
\put(93.3,9){$\l = {\displaystyle{{i - z} \over {i + z}}} = \h \sp {2/3}$}
\end{picture}

\setlength{\unitlength}{1mm}
\begin{picture}(140,30)
\put(10,5){\line(1,0){40}}
\put(29,14){{\sf x}~$i$}
\put(29,0){\bf 2}
\put(29,4){$\bullet$}
\put(11.68,0){\bf 3}
\put(11.68,4){$\bullet$}
\put(46.32,0){\bf 1}
\put(46.32,4){$\bullet$}
\put(70,9){$z = i {\displaystyle{{1 - \l} \over {1 + \l}}}$}
\end{picture}

\noindent (The bold-face numbers label the ends of the strings.)
This maps the string from an infinite rectangle ($\r$) to the 
upper-half plane ($\z$) to the interior of the unit circle ($\h$) to a 
different circle with all three strings appearing on the same sheet 
($\l$) to the upper-half plane with all strings on one sheet ($z$).
If the cut for $\l ( \h )$ is chosen appropriately (the positive 
imaginary axis of the $\h$ plane), the cut under which the third 
string is hidden is along the part of the imaginary $\r$ axis below $i 
\p / 2$.  (More conveniently, if the cut is taken in the negative real 
direction in the $\h$ plane, then it's in the positive real direction 
in the $\r$ plane, with halves of 2 strings hidden under the cut.)
Since the last transformation is projective, we can drop it.
(Projective transformations don't affect equations like (9.2.14).)

Unfortunately, although the calculation can still be performed [12.7],
there is now no simple analog to (9.2.18).  It's easier to
use instead a map similar to (9.2.16) by considering a 6-string $\d$
functional with pairs of strings identified [12.6]:  Specifically, we replace
the last 2 maps above with

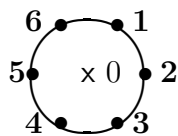
\setlength{\unitlength}{.75mm}
\begin{picture}(165,30)
\put(40,10){\oval(20,20)}
\put(39,9){{\sf x}~0}
\put(34,17.66){$\bullet$}
\put(29,18){\bf 6}
\put(34,.3){$\bullet$}
\put(29,0){\bf 4}
\put(49,9){$\bullet$}
\put(53,9){\bf 2}
\put(44,17.66){$\bullet$}
\put(48,18){\bf 1}
\put(44,.3){$\bullet$}
\put(48,0){\bf 3}
\put(29,9){$\bullet$}
\put(26,9){\bf 5}
\put(93.3,9){$\hat\l ={\displaystyle{{i-\hat z}\over{i+\hat z}}}=\h\sp{1/3}$}
\end{picture}

\setlength{\unitlength}{1mm}
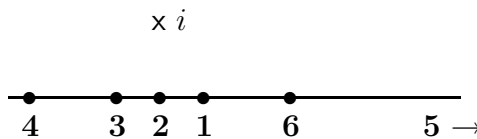
\begin{picture}(140,30)
\put(5,5){\line(1,0){60}}
\put(24,14){{\sf x}~$i$}
\put(24,0){\bf 2}
\put(24,4){$\bullet$}
\put(6.68,0){\bf 4}
\put(6.68,4){$\bullet$}
\put(41.32,0){\bf 6}
\put(41.32,4){$\bullet$}
\put(18.23,0){\bf 3}
\put(18.23,4){$\bullet$}
\put(29.77,0){\bf 1}
\put(29.77,4){$\bullet$}
\put(60,0){{\bf 5} $\to$}
\put(70,9){$\hat z = i {\displaystyle{{1 - \hat\l} \over {1 + \hat\l}}}$}
\end{picture}

\noindent and thus, relabeling $r\to r+1$ (or performing an equivalent
rotation of the $\hat\l$ circle)
$$ \r = \su_{r=1}^5 p\sb{+r} ln (\hat z-z\sb r) - ln~3 \quad , $$
$$ p\sb{+r} = (-1)\sp{r+1} \quad , \quad z\sb r = 
( \sqrt 3 , {1\over{\sqrt 3}} , 0 , -{1\over{\sqrt 3}} , -\sqrt 3 , \infty ) 
\quad . \eqno(12.2.22)$$
We can then use the same procedure as the light cone.  However, it turns
out to be more convenient to evaluate the contour integrals in terms of
$\z$ rather than $\hat z$.  Also, instead of applying (9.2.18) to
(12.2.22), we apply it to the corresponding expression for $\hat\l$:
$$ \r = \su_{r=1}^5 \a\sb r ln (\hat\l -\l\sb r) - \frac14 i\p \quad , $$
$$ p\sb{+r} = (-1)\sp r \quad , \quad \l\sb r = e\sp{-i\p (r-2)/3}
\quad . \eqno(12.2.23)$$
Reexpressing (9.2.18) in terms of $\hat\l$, we find
$$ \left( \der\r +\der{\r '} \right) ln (\hat\l -\hat\l ' ) =
\frac16 \left[ (\hat\l\sp 3 + \hat\l'\sp 3 ) + 
(\hat\l\hat\l'\sp 2 +\hat\l\sp 2\hat\l' ) +\left({1\over \hat\l\hat\l'\sp 2}
+{1\over \hat\l\sp 2\hat\l'}\right)\right] \quad .
\eqno(12.2.24)$$
Using the conservation laws, the first set of terms can be dropped.
Since it's actually $\l=\hat\l\sp 2$ (or $z$), and not $\hat\l$, for 
which the string
is mapped to the complex plane ($\hat\l$ describes a 6-string vertex,
and thus double counts), the $ln$ we actually want to evaluate is
$$ ln (\l -\l ' ) = ln (\hat\l -\hat\l ' ) + ln (\hat\l +\hat\l ' )
\quad . \eqno(12.2.25)$$
This just says that the general coefficients $N\sb{rs}$ in $\D$
multiplying oscillators from string $r$ times those from string $s$ is
related to the corresponding fictitious 6-string coefficients $\Tilde
N\sb{rs}$ by
$$ N\sb{rs} = \Tilde N\sb{rs} + \Tilde N\sb{r,s+3} \quad . \eqno(12.2.26)$$
The contour integrals can now be evaluated over $\z$ in terms of
$$ \left({1+x\over 1-x}\right)^{1/3} = \su_0^\infty a\sb n x\sp n \quad ,
\quad \left({1+x\over 1-x}\right)^{2/3} = \su_0^\infty b\sb n x\sp n \quad ,
\eqno(12.2.27)$$
These coefficients satisfy the recursion relations
$$ (n+1)a\sb{n+1} = \frac23 a\sb n + (n-1) a\sb{n-1} \quad , \quad
(n+1)b\sb{n+1} = \frac43 b\sb n + (n-1) b\sb{n-1} \quad , \eqno(12.2.28)$$
which can be derived by appropriate manipulations of the corresponding
contour integrals: e.g.,
$$ \li{ a\sb n &= \oint_0 {dx\over 2\p ix}{1\over x\sp n}
\left({1+x\over 1-x}\right)^{1/3} \cr
&= \oint_0 {dx\over 2\p ix}{1\over x\sp n}\frac32 (1-x\sp 2 )
\left[ \left({1+x\over 1-x}\right)^{1/3} \right]^\prime \quad .&(12.2.29)\cr}$$
Because of $i$'s relative to (12.2.27) appearing in the actual contour
integrals, we use instead the coefficients
$$ A\sb n = a\sb n \cdot \left\{ \matrix{ (-1)\sp{n/2} \hfill 
& ~(n~even) \hfill \cr
(-1)\sp{(n-1)/2} \hfill & ~(n~odd) \hfill \cr} \right. \quad , \eqno(12.2.30)$$
and similarly for $B\sb n$.  We finally obtain an expression similar 
to (9.2.27), except that we must use (12.2.26), and
$$ \Tilde N\sb{rsmn} = {1\over{m\over p\sb{+r}}+{n\over p\sb{+s}}}M\sb{rsmn}
\quad ,$$
$$ M\sb{r,r+t,mn} = \frac13 c\sb{mnt} \left[ A\sb m B\sb n +
(-1)\sp{m+n+t} B\sb m A\sb n \right] \quad , $$
$$ c\sb{mnt} = \left\{ \matrix{(-1)\sp m Re (e\sp{it2\p /3}) \hfill &
~(m+n~even) \hfill \cr Im (e\sp{it2\p /3}) \hfill
& ~(m+n~odd) \hfill \cr} \right. \quad . \eqno(12.2.31)$$
The terms for $n=m\ne 0$ or $n=0\ne m$ can be evaluated by taking the
appropriate limit ($n\to m$ or $n\to 0$).  $m=n=0$ can then be evaluated
separately, using (9.2.22b), (12.2.22), and (12.2.26).  The final result is
$$ \D ( \check\J\sb 1 , \check\J\sb 2 ) = -\su ' \j\sb 1 N \j\sb 2
-\frac14 ln \left( {3\sp 3\over 2\sp 4} \right) \su p\sp 2 \quad ,
\eqno(12.2.32)$$
where $\su '$ is over $r,s =1,2,3$ and $m,n =0,1,\dots ,\infty$ except
for the term $m=n=0$.  As for (9.2.27), $\j$ refers to all sets of
oscillators, with $\j$ replaced with $p\sb +\sp{1-w}\j$ for oscillators of
weight $w$.  In this case we use (12.2.21), and the $p\sb +$'s are all $\pm
1$, so for the ghosts there is an extra sign factor $p\sb{+r}p\sb{+s}$ for
$\Tilde N\sb{rsmn}$.

There are a number of problems to resolve for this formalism:  (1) 
In calculating S-matrix
elements, the 4-point function is considerably more difficult to
calculate than in the light-cone formalism [12.9], and
the conformal maps are so complicated that it's not yet known
how to derive even the 5-point function for tachyons, although arguments have 
been given for equivalence to the light-cone/external-field result [12.10].
(2) It doesn't seem possible to derive an external-field approach to 
interactions, since the string lengths are all fixed to be $\p$.  
In the light-cone
formalism the external-field approach follows from choosing the Lorentz frame
where all but 2 of the string lengths (i.e., $p \sb +$'s) vanish.  
(Thus, e.g., in the 3-string vertex 1 string reduces to a point on the 
boundary, reducing to a vertex as in sect.\ 9.1.)
This is related to the fact that $I$ of (12.2.11) is just the
harmonic oscillator ground state at vanishing momentum (and length) for the 
light-cone formalism, but for this formalism it's
$\sim\d [ X ( \s ) - X ( \p - \s ) ]$.  
(3) The fact that the gauge-invariant vertex is 
so different from the light-cone vertex indicates that gauge-fixing to 
the light-cone gauge should be difficult.  Furthermore, the light-cone
formalism requires a 4-point interaction in the action, whereas this
covariant formalism doesn't.  Perhaps a formalism with a 
larger gauge invariance exists such that these 2 formalisms are found 
by 2 different types of gauge choices.  (4) There is some difficulty 
in extending the 
discussion of sect.\ 11.1 for the closed string to the interacting case, 
since the usual form of the physical-state vertex requires that the 
vertex be related to the product of open-string vertices for the 
clockwise and counterclockwise states, multiplied by certain vertex 
factors which don't exist in this formalism (although they would in a 
formalism more similar to the light-cone one, since the 
light-cone formalism has more zero-mode conservation laws).
This is particularly confusing since open strings generate closed ones
at the 1-loop level.  However, some progress in understanding these closed
strings has been made [12.11].  Also, a general analysis has been made 
of some properties of the 3-point closed-string vertex required by
consistency of the 1-loop tadpole and 4-string tree graphs [12.12], using
techniques which are applicable to vertices more general than
$\d$-functionals [12.13].

The gauge-fixing of this formalism with a BRST algebra that closes on
shell has been studied [12.14].  It has been shown both in the formalism
of light-cone-like closed string theory [12.15] and for the
midpoint-interaction open string theory [12.16] that the kinetic term
can be obtained from an action with just the cubic term by giving an
appropriate vacuum value to the string field.  However, whereas in the
former case (barring difficulties in loops mentioned above) this vacuum
value is natural because of the vacuum value of the covariant metric
field for the graviton, in the latter case there is no classical
graviton in the open string theory, so the existence (or usefulness) of
such a mechanism is somewhat confusing.

The midpoint-interaction formulation of the open superstring (as a
truncated spinning string) has also been developed [11.3,12.17].
The supersymmetry algebra closes only on shell, and the action
apparently also needs (at least) 4-point interactions to cancel
divergences in 4-point amplitudes due to coincidence of vertex operator
factors (both of which occur at the midpoint) [12.18].  Such
interactions might be of the same type needed in the light-cone 
formulation (chapt.\ 10).

\sect{Exercises}

\Item{(1)} Check the BRST invariance of (12.2.18).
\Item{(2)} Find the transformation of $ln (z-z')$ under the projective
transformation $z\to{az+b\over cz+d}$ (and similarly for $z'$).
Use the conservation law $\su_r \j\sb{0r} =0$ to show that (9.2.14) is
unaffected. 
\Item{(3)} Derive the last term of (12.2.32).

%
%

\newpage

\def\chip#1{\vskip.3in\goodbreak\noindent{\bf Chapter #1}\par\nobreak
	\vskip.2in}

\refs

\parindent=27pt

\vskip.2in\noindent{\bf Preface}\par\vskip.2in

\Item{[0.1]}M.B. Green, J.H. Schwarz, and E. Witten, Superstring
	theory, v. 1-2 (Cambridge University, Cambridge, 1987).
\Item{[0.2]}V. Gates, E. Kangaroo, M. Roachcock, and W.C. Gall,
	The super G-string, {\it in} Unified string theories,
	eds. M. Green and D. Gross, Proc. of Santa Barbara Workshop,
	Jul. 29 - Aug. 16, 1985 (World Scientific, Singapore, 1986) p. 729;\\
	Super G-string field theory, {\it in} Superstrings, Cosmology
	and Composite Structures, eds. S.J. Gates, Jr. and R.N. Mohapatra,
	Proc. of International Workshop on Superstrings, Composite
	Structures, and Cosmology, College Park, MD, March 11-18, 1987
	(World Scientific, Singapore, 1987) p. 585.

\chip1

\Item{[1.1]}P. van Nieuwenhuizen, Phys. Rep. 68 (1981) 189.
\Item{[1.2]}M.T. Grisaru and W. Siegel, \NP 201 (1982) 292, B206 (1982) 496;\\
	N. Marcus and A. Sagnotti, \PL 135B (1984) 85;\\
	M.H. Goroff and A. Sagnotti, \PL 160B (1985) 81.
\Item{[1.3]}Dual theory, ed. M. Jacob (North-Holland, Amsterdam, 1974);\\
	J. Scherk, Rev. Mod. Phys. 47 (1975) 123;\\
	P.H. Frampton, Dual resonance models and string theories (World
	Scientific, Singapore, 1986).
\Item{[1.4]}S. Mandelstam, Phys. Rep. 13 (1974) 259.
\Item{[1.5]}J.H. Schwarz, Phys. Rep. 89 (1982) 223.
\Item{[1.6]}G. 't Hooft, \NP 72 (1974) 461.
\Item{[1.7]}S. Weinberg, Critical phenomena for field theorists, {\it 
	in} Understanding the fundamental constituents of matter, 
	proc. of the International School of Subnuclear Physics, Erice, 1976, 
	ed. A. Zichichi (Plenum, New York, 1978) p. 1;\\
	Ultraviolet divergences in quantum theories of gravitation, 
	{\it in} General relativity: an Einstein centenary survey, 
	eds. S.W. Hawking and W. Israel (Cambridge University, 
	Cambridge, 1979) p. 790.
\Item{[1.8]}G. 't Hooft, \PL 109B (1982) 474.
\Item{[1.9]}T.H.R. Skyrme, Proc. Roy. Soc. A260 (1961) 227;\\
	E. Witten, \NP 223 (1983) 433.
\Item{[1.10]}D.J. Gross, A. Neveu, J. Scherk, and J.H. Schwarz, \PL 31B (1970) 
	592.
\Item{[1.11]}M.B. Green and J.H. Schwarz, \PL 149B (1984) 117.
\Item{[1.12]}M.B. Green and J.H.Schwarz, \PL 109B (1982) 444.
\Item{[1.13]}D.J. Gross, J.A. Harvey, E. Martinec, and R. Rohm, \PR 54 (1985) 
	502, \NP 256 (1985) 253, \NP 267 (1986) 75.
\Item{[1.14]}L.J. Dixon and J.A. Harvey, \NP 274 (1986) 93;\\
	L. Alvarez-Gaum\'e, P. Ginsparg, G. Moore, and C. Vafa,
	\PL 171B (1986) 155;\\
	H. Itoyama and T.R. Taylor, \PL 186B (1987) 129.
\Item{[1.15]}E. Witten, \PL 149B (1984) 351.
\Item{[1.16]}P. Candelas, G.T. Horowitz, A. Strominger, and E. Witten, \NP 
	258 (1985) 46;\\
	L. Dixon, J.A. Harvey, C. Vafa, and E. Witten, \NP 261 (1985)
	678, 274 (1986) 285;\\
	K.S. Narain, \PL 169B (1986) 41;\\
	H. Kawai, D.C. Lewellen, and S.-H.H. Tye, \NP 288 (1987) 1;\\
	I. Antoniadis, C. Bachas, and C. Kounnas, \NP 289 (1987) 87;\\
	R. Bluhm, L. Dolan, and P. Goddard, \NP 289 (1987) 364,
	Unitarity and modular invariance as constraints on
	four-dimensional superstrings, Rockefeller preprint
	RU87/B1/25 DAMTP 88/9 (Jan. 1988).
\Item{[1.17]}S.J. Gates, Jr., M.T. Grisaru, M. Ro\v cek, and W. Siegel,
	Superspace, {\it or} One thousand and one lessons in supersymmetry
	(Benjamin/Cummings, Reading, 1983);\\
	J. Wess and J. Bagger, Supersymmetry and supergravity
	(Princeton University, Princeton, 1983);\\
	R.N. Mohapatra, Unification and supersymmetry: the frontiers of
	quark-lepton physics (Springer-Verlag, New York, 1986);\\
	P. West, Introduction to supersymmetry and supergravity
	(World Scientific, Singapore, 1986).

\chip2

\Item{[2.1]}S. Weinberg, Phys. Rev. 150 (1966) 1313;\\
	Kogut and D.E. Soper, \PRD 1 (1970) 2901.
\Item{[2.2]}K . Bardak\ced ci and M.B. Halpern, Phys. Rev. 176 (1968) 1686.
\Item{[2.3]}W. Siegel and B. Zwiebach, \NP 282 (1987) 125.
\Item{[2.4]}S.J. Gates, Jr., M.T. Grisaru, M. Ro\v cek, and W. Siegel,
	Superspace, {\it or} One thousand and one lessons in supersymmetry
	(Benjamin/Cummings, Reading, 1983) p. 74.
\Item{[2.5]}A. J. Bracken, Lett. Nuo. Cim. 2 (1971) 574;\\
	A.J. Bracken and B. Jessup, J. Math. Phys. 23 (1982) 1925.
\Item{[2.6]}W. Siegel, \NP 263 (1986) 93.
\Item{[2.7]}W. Siegel, \PL 142B (1984) 276.
\Item{[2.8]}G. Parisi and N. Sourlas, \PR 43 (1979) 744.
\Item{[2.9]}R. Delbourgo and P.D. Jarvis, J. Phys. A15 (1982) 611;\\
	J. Thierry-Mieg, \NP 261 (1985) 55;\\
	J.A. Henderson and P.D. Jarvis, Class. and Quant. Grav. 3
	(1986) L61.

\chip3

\Item{[3.1]}P.A.M. Dirac, Proc. Roy. Soc. A246 (1958) 326;\\
	L.D. Faddeev, Theo. Math. Phys. 1 (1969) 1.
\Item{[3.2]}W. Siegel, \NP 238 (1984) 307.
\Item{[3.3]}C. Becchi, A. Rouet, and R. Stora, \PL 52B (1974) 344;\\
	I.V. Tyutin, Gauge invariance in field theory and in statistical 
	physics in the operator formulation, Lebedev preprint FIAN No. 39 
	(1975), in Russian, unpublished;\\
	T. Kugo and I. Ojima, \PL 73B (1978) 459;\\
	L. Baulieu, Phys. Rep. 129 (1985) 1.
\Item{[3.4]}M. Kato and K. Ogawa, \NP 212 (1983) 443;\\
	S. Hwang, Phys.\ Rev.\ D28 (1983) 2614;\\
	K. Fujikawa, \PRD 25 (1982) 2584.
\Item{[3.5]}N. Nakanishi, Prog. Theor. Phys. 35 (1966) 1111;\\
	B. Lautrup, K. Dan. Vidensk. Selsk. Mat. Fys. Medd. 34 (1967)
	No. 11, 1.
\Item{[3.6]}G. Curci and R. Ferrari, Nuo. Cim. 32A (1976) 151,
	\PL 63B (1976) 91;\\
	I. Ojima, Prog. Theo. Phys. 64 (1980) 625;\\
	L. Baulieu and J. Thierry-Mieg, \NP 197 (1982) 477.
\Item{[3.7]}L. Baulieu, W. Siegel, and B. Zwiebach, \NP 287 (1987) 93.
\Item{[3.8]}S. Ferrara, O. Piguet, and M. Schweda, \NP 119 (1977) 493.
\Item{[3.9]}J. Thierry-Mieg, J. Math. Phys. 21 (1980) 2834.
\Item{[3.10]}E.S. Fradkin and G.A. Vilkovisky, \PL 55B (1975) 224;\\
	I.A. Batalin and G.A. Vilkovisky, \PL 69B (1977) 309;\\
	E.S. Fradkin and T.E. Fradkina, \PL 72B (1978) 343;\\
	M. Henneaux, Phys. Rep. 126 (1985) 1.
\Item{[3.11]}M. Quir\'os, F.J. de Urries, J. Hoyos, M.L. Maz\'on, and
	E. Rodriguez, J. Math. Phys. 22 (1981) 1767;\\
	L. Bonora and M. Tonin, \PL 98B (1981) 48.
\Item{[3.12]}S. Hwang, \NP 231 (1984) 386;\\
	F.R. Ore, Jr. and P. van Nieuwenhuizen, \NP 204 (1982) 317.
\Item{[3.13]}W. Siegel and B. Zwiebach, \NP 288 (1987) 332.
\Item{[3.14]}W. Siegel and B. Zwiebach, \NP 299 (1988) 206.
\Item{[3.15]}W. Siegel, \NP 284 (1987) 632.
\Item{[3.16]}W. Siegel, Universal supersymmetry by adding 4+4 dimensions
	to the light cone, Maryland preprint UMDEPP 88-231 (May 1988).

\chip4

\Item{[4.1]}W. Siegel and B. Zwiebach, \NP 263 (1986) 105.
\Item{[4.2]}T. Banks and M.E. Peskin, \NP 264 (1986) 513;\\
	K. Itoh, T. Kugo, H. Kunitomo, and H. Ooguri, Prog. Theo. 
	Phys. 75 (1986) 162.
\Item{[4.3]}M. Fierz and W. Pauli, Proc. Roy. Soc. A173 (1939) 211;\\
	S.J. Chang, Phys. Rev. 161 (1967) 1308;\\
	L.P.S. Singh and C.R. Hagen, \PRD 9 (1974) 898;\\
	C. Fronsdal, \PRD 18 (1978) 3624;\\
	T. Curtright, \PL 85B (1979) 219;\\
	B. deWit and D.Z. Freedman, \PRD 21 (1980) 358;\\
	T. Curtright and P.G.O. Freund, \NP 172 (1980) 413;\\
	T. Curtright, \PL 165B (1985) 304.
\Item{[4.4]}W. Siegel, \PL 149B (1984) 157, 151B (1985) 391.
\Item{[4.5]}W. Siegel, \PL 149B (1984) 162; 151B (1985) 396.
\Item{[4.6]}E.S. Fradkin and V.I. Vilkovisky, \PL 73B (1978) 209.
\Item{[4.7]}W. Siegel and S.J. Gates, Jr., \NP 147 (1979) 77;\\
	S.J. Gates, Jr., M.T. Grisaru, M. Ro\v cek, and W. Siegel, Superspace,
	{\it or} One thousand and one lessons in supersymmetry
	(Benjamin/Cummings, Reading, 1983) p. 242.
\Item{[4.8]}E. Witten, \NP 268 (1986) 253.
\Item{[4.9]}A. Neveu, H. Nicolai, and P.C. West, \PL 167B (1986) 307.
\Item{[4.10]}W. Siegel and B. Zwiebach, \PL 184B (1987) 325.
\Item{[4.11]}M. Scheunert, W. Nahm, and V. Rittenberg, J. Math. Phys. 18
	(1977) 155.
\Item{[4.12]}A. Galperin, E. Ivanov, S. Kalitzin, V. Ogievetsky, and E.
	Sokatchev, JETP Lett. 40 (1984) 912, Class. Quant. Grav. 1
	(1984) 469, 2 (1985) 601;\\
	W. Siegel, Class. Quant. Grav. 2 (1985) 439;\\
	E. Nissimov, S. Pacheva, and S. Solomon, \NP 296 (1988) 462, 
	297 (1988) 349;\\
	R.E. Kallosh and M.A. Rahmanov, Covariant quantization of the
	Green-Schwarz superstring, Lebedev preprint (March 1988).
\Item{[4.13]}J.P. Yamron, \PL 174B (1986) 69.
\Item{[4.14]}G.D. Dat\'e, M. G\"unaydin, M. Pernici, K. Pilch, and P. van
	Nieuwenhuizen, \PL 171B (1986) 182.
\Item{[4.15]}H. Terao and S. Uehara, \PL 173B (1986) 134;\\
	T. Banks, M.E. Peskin, C.R. Preitschopf, D. Friedan, and E.
	Martinec, \NP 274 (1986) 71;\\
	A. LeClair and J. Distler, \NP 273 (1986) 552.

\chip5

\Item{[5.1]}A. Barducci, R. Casalbuoni, and L. Lusanna, Nuo. Cim. 35A 
	(1976) 377;\\
	L. Brink, S. Deser, B. Zumino, P. DiVecchia, and P. Howe, \PL
	64B (1976) 435;\\
	P.A. Collins and R.W. Tucker, \NP 121 (1977) 307.
\Item{[5.2]}S. Deser and B. Zumino, \PL 62B (1976) 335.
\Item{[5.3]}E.C.G. Stueckelberg, Helv. Phys. Acta 15 (1942) 23;\\
	R.P. Feynman, Phys. Rev. 80 (1950) 440.
\Item{[5.4]}J. Polchinski, Commun. Math. Phys. 104 (1986) 37.
\Item{[5.5]}J. Schwinger, Phys. Rev. 82 (1951) 664.
\Item{[5.6]}R.J.Eden, P.V. Landshoff, D.I. Olive, and J.C. Polkinghorne,
	The analytic S-matrix (Cambridge University, Cambridge, 1966) p. 152.
\Item{[5.7]}{\it ibid.}, p. 88;\\
	J.D. Bjorken and S.D. Drell, Relativistic quantum fields
	(McGraw-Hill, New York, 1965) p. 225.
\Item{[5.8]}J. Iliopoulos, G. Itzykson, and A. Martin, Rev. Mod. Phys.
	47 (1975) 165.
\Item{[5.9]}M.B. Halpern and W. Siegel, \PRD 16 (1977) 2486.
\Item{[5.10]}F.A. Berezin and M.S. Marinov, Ann. Phys. (NY) 104 (1977) 336;\\
	A. Barducci, R. Casalbuoni, and L. Lusanna, \NP 124 (1977) 93;\\
	A.P. Balachandran, P. Salomonson, B.-S. Skagerstam, and J.-O. 
	Winnberg, \PRD 15 (1977) 2308.
\Item{[5.11]}H. Georgi, Lie algebras in particle physics: from isospin
	to unified theories (Benjamin/Cummings, Reading,1982) p. 192.
\Item{[5.12]}S.J. Gates, Jr., M.T. Grisaru, M. Ro\v cek, and W. Siegel,
	Superspace, {\it or\/} One thousand and one lessons in supersymmetry
	(Benjamin/Cummings, Reading, 1983) pp. 70, 88.
\Item{[5.13]}L. Brink, J.H. Schwarz, and J. Scherk, \NP 121 (1977) 77;\\
	W. Siegel, \PL 80B (1979) 220.
\Item{[5.14]}M.T. Grisaru, W. Siegel, and M. Ro\v cek, \NP 159 (1979) 429;\\
	M.T. Grisaru and W. Siegel, \NP 201 (1982) 292, B206 (1982) 496;\\
	S.J. Gates, Jr., M.T. Grisaru, M. Ro\v cek, and W. Siegel, 
	Superspace, pp. 25, 382, 383.
\Item{[5.15]}W. Siegel, \NP 156 (1979) 135.
\Item{[5.16]}J. Koller, \NP 222 (1983) 319;\\
	P.S. Howe, G. Sierra, and P.K. Townsend, \NP 221 (1983) 331;\\
	J.P. Yamron and W. Siegel, \NP 263 (1986) 70.
\Item{[5.17]}J. Wess and B. Zumino, \NP 70 (1974) 39;\\
	R. Haag, J.T. \Sll opusza\'nski, and M. Sohnius, \NP 88 (1975) 257;\\
	W. Nahm, \NP 135 (1978) 149.
\Item{[5.18]}P. Ramond, Physica 15D (1985) 25.
\Item{[5.19]}W. Siegel, Free field equations for everything, {\it in}
	Strings, Cosmology, Composite
	Structures, March 11-18, 1987, College Park, Maryland, eds.
	S.J. Gates, Jr. and R.N. Mohapatra (World Scientific, Singapore, 1987).
\Item{[5.20]}A. Salam and J. Strathdee, \NP 76 (1974) 477;\\
	S.J. Gates, Jr., M.T. Grisaru, M. Ro\v cek, and W. Siegel, 
	Superspace, pp. 72-73;\\
	R. Finkelstein and M. Villasante, J. Math. Phys. 27 (1986) 1595.
\Item{[5.21]}W. Siegel, \PL 203B (1988) 79.
\Item{[5.22]}J. Strathdee, Inter. J. Mod. Phys. A2 (1987) 273.
\Item{[5.23]}W. Siegel, Class. Quantum Grav. 2 (1985) L95.
\Item{[5.24]}W. Siegel, \PL 128B (1983) 397.
\Item{[5.25]}W. Siegel and S.J. Gates, Jr., \NP 189 (1981) 295.
\Item{[5.26]}S. Mandelstam, \NP 213 (1983) 149.
\Item{[5.27]}L. Brink, O. Lindgren, and B.E.W. Nilsson, \NP 212 (1983) 401.
\Item{[5.28]}R. Casalbuoni, \PL 62B (1976) 49;\\
	L. Brink and J.H. Schwarz, \PL 100B (1981) 310.
\Item{[5.29]}A.R. Mikovi\'c and W. Siegel, On-shell equivalence of 
	superstrings, Maryland preprint 88-218 (May 1988).
\Item{[5.30]}K. Kamimura and M. Tatewaki, \PL 205B (1988) 257.
\Item{[5.31]}R. Penrose, J. Math. Phys. 8 (1967) 345,
	Int. J. Theor. Phys. 1 (1968) 61 ;\\
	M.A.H. MacCallum and R. Penrose, Phys. Rep. 6C (1973) 241.
\Item{[5.32]}A. Ferber, \NP 132 (1978) 55.
\Item{[5.33]}T. Kugo and P. Townsend, \NP 221 (1983) 357;\\
	A. Sudbery, J. Phys. A17 (1984) 939;\\
	K.-W. Chung and A. Sudbery, \PL 198B (1987) 161.
\Item{[5.34]}Z. Hasiewicz and J. Lukierski, \PL 145B (1984) 65;\\
	I. Bengtsson and M. Cederwall, Particles, twistors, and the
	division algebras, Ecole Normale Sup\'erieure preprint
	LPTENS-87/20 (May 1987).
\Item{[5.35]}D.B. Fairlie and C.A. Manogue, \PRD 34 (1986) 1832,
	36 (1987) 475;\\
	J.M. Evans, \NP 298 (1988) 92.

\chip6

\Item{[6.1]}P.A. Collins and R.W. Tucker, \PL 64B (1976) 207;\\
	L. Brink, P. Di Vecchia, and P. Howe, \PL 65B (1976) 471;\\
	S. Deser and B. Zumino, \PL 65B (1976) 369.
\Item{[6.2]}Y. Nambu, lectures at Copenhagen Symposium, 1970 (unpublished);\\
	O. Hara, Prog. Theo. Phys. 46 (1971) 1549;\\
	T. Goto, Prog. Theo. Phys. 46 (1971) 1560.
\Item{[6.3]}S.J. Gates, Jr., R. Brooks, and F. Muhammad, \PL 194B (1987) 35;\\
	C. Imbimbo and A. Schwimmer, \PL 193B (1987) 455;\\
	J.M.F. Labastida and M. Pernici, \NP 297 (1988) 557;\\
	L. Mezincescu and R.I. Nepomechie, Critical dimensions for
	chiral bosons, Miami preprint UMTG-140 (Aug. 1987);\\
	Y. Frishman and J. Sonnenschein, Gauging of chiral bosonized
	actions, Weizmann preprint WIS-87/65/Sept.-PH (Sept. 1987);\\
	R. Floreanini and R. Jackiw, \PR 59 (1987) 1873;\\
	S. Bellucci, R. Brooks, and J. Sonnenschein, Supersymmetric
	chiral bosons, SLAC preprint SLAC-PUB-4458 MIT-CTP-1548
	UCD-87-35 (Oct. 1987);\\
	S.J. Gates, Jr. and W. Siegel, Leftons, rightons, nonlinear
	$\s$-models, and superstrings, Maryland preprint UMDEPP 88-113
	(Nov. 1987);\\
	M. Henneaux and C. Teitelboim, Dynamics of chiral (self-dual)
	p-forms, Austin preprint (Dec. 1987);\\
	M. Bernstein and J. Sonnenschein, A comment on the quantization
	of chiral bosons, SLAC preprint SLAC-PUB-4523 (Jan. 1988);\\
	L. Faddeev and R. Jackiw, Hamiltonian reduction of unconstrained
	and constrained systems, MIT preprint CTP\#1567 (Feb. 1988);\\
	J. Sonnenschein, Chiral bosons, SLAC preprint SLAC-PUB-4570
	(March 1988).
\Item{[6.4]}C.M. Hull, Covariant quantization of chiral bosons and
	anomaly cancellation, London preprint Imperial/TP/87/88/9
	(March 1988).
\Item{[6.5]}P. Goddard, J. Goldstone, C. Rebbi, and C.B. Thorn, \NP 56 
	(1973) 109.

\chip7

\Item{[7.1]}P. Ramond, \PRD 3 (1971) 86;\\
	A. Neveu and J.H. Schwarz, \NP 31 (1971) 86, \PRD 4 (1971) 1109.
\Item{[7.2]}D.B. Fairlie and D. Martin, Nuo. Cim. 18A (1973) 373, 21A 
	(1974) 647;\\
	L. Brink and J.-O. Winnberg, \NP 103 (1976) 445.
\Item{[7.3]}F. Gliozzi, J. Scherk, and D.I. Olive, \PL 65B (1976) 282, 
	\NP 122 (1977) 253.
\Item{[7.4]}Y. Iwasaki and K. Kikkawa, \PRD 8 (1973) 440.
\Item{[7.5]}W. Siegel, Covariant approach to superstrings, {\it in} 
	Symposium on anomalies, geometry and topology, eds. W.A. 
	Bardeen and A.R. White (World Scientific, Singapore, 1985) p. 348.
\Item{[7.6]}M.B. Green and J.H. Schwarz, \PL 136B (1984) 367, 
	\NP 243 (1984) 285.

\chip8

\Item{[8.1]}M. Virasoro, \PRD 1 (1970) 2933;\\
	I.M. Gelfand and D.B. Fuchs, Functs. Anal. Prilozhen 2 (1968) 92.
\Item{[8.2]}E. Del Giudice, P. Di Vecchia, and S. Fubini, Ann. Phys. 70
	(1972) 378.
\Item{[8.3]}F. Gliozzi, Nuovo Cim. Lett. 2 (1969) 846.
\Item{[8.4]}E. Witten, Some remarks about string field theory, {\it in}
	Marstrand Nobel Sympos. 1986, p. 70;\\
	I.B. Frenkel, H. Garland, and G.J. Zuckerman, Proc. Natl. Acad.
	Sci. USA 83 (1986) 8442.
\Item{[8.5]}P. Jordan, Z. Physik 93 (1935) 464;\\
	M. Born and N.S. Nagendra Nath, Proc. Ind. Acad. Sci. 3 (1936) 318;\\
	A. Sokolow, Phys. Z. der Sowj. 12 (1937) 148;\\
	S. Tomonaga, Prog. Theo. Phys. 5 (1950) 544;\\
	T.H.R. Skyrme, Proc. Roy. Soc. A262 (1961) 237;\\
	D. Mattis and E. Lieb, J. Math. Phys. 6 (1965) 304;\\
	B. Klaiber, {\it in} Lectures in theoretical physics, eds. A.O. Barut
	and W.E. Brittin (Gordon and Breach, New York, 1968) v. X-A, p. 141;\\
	R.F. Streater and I.F. Wilde, \NP 24 (1970) 561;\\
	J. Lowenstein and J. Swieca, Ann. Phys. (NY) 68 (1971) 172;\\
	K. Bardak\ced ci and M.B. Halpern, \PRD 3 (1971) 2493;\\
	G. Dell'Antonio, Y. Frishman, and D. Zwanziger, \PRD 6 (1972) 988;\\
	A. Casher, J. Kogut, and L. Susskind, \PR 31 (1973) 792, \PRD 10 (1974)
	732;\\
	A. Luther and I. Peschel, Phys. Rev. B9 (1974) 2911;\\
	A. Luther and V. Emery, \PR 33 (1974) 598;\\
	S. Coleman, \PRD 11 (1975) 2088;\\
	B. Schroer, Phys. Rep. 23 (1976) 314;\\
	S. Mandelstam, \PRD 11 (1975) 3026;\\
	J. Kogut and L. Susskind, \PRD 11 (1975) 3594.
\Item{[8.6]}M.B. Halpern, \PRD 12 (1975) 1684;\\
	I.B. Frenkel and V.G. Ka\v c, Inv. Math. 62 (1980) 23;\\
	I.B. Frenkel, J. Func. Anal. 44 (1981) 259;\\
	G. Segal, Comm. Math. Phys. 80 (1981) 301;\\
	P. Goddard and D. Olive, Vertex operators in mathematics and physics,
	eds. J. Lepowsky, S. Mandelstam, and I.M. Singer (Springer-Verlag,
	New York, 1985) p. 51.
\Item{[8.7]}R. Marnelius, \NP 211 (1983) 14.
\Item{[8.8]}W. Siegel, \PL 134B (1984) 318.
\Item{[8.9]}A.M. Polyakov, \PL 103B (1981) 207.
\Item{[8.10]}A.A. Tseytlin, \PL 168B (1986) 63;\\
	S.R. Das and M.A. Rubin, \PL 169B (1986) 182, Prog. Theor. Phys.
	Suppl. 86 (1986) 143, \PL 181B (1986) 81;\\
	T. Banks, D. Nemeschansky, and A. Sen, \NP 277 (1986) 67;\\
	G.T. Horowitz and A. Strominger, \PR 57 (1986) 519;\\
	G. M\"unster, Geometric string field theory, DESY preprint
	DESY 86-045 (Apr. 1986).
\Item{[8.11]}H. Aoyama, \NP 299 (1988) 379.
\Item{[8.12]}D.Z. Freedman, J.I. Latorre, and K. Pilch, Global aspects
	of the harmonic gauge in bosonic string theory, MIT preprint
	CTP\#1559 (Jan. 1988).

\chip9

\Item{[9.1]}K. Bardak\ced ci and H. Ruegg, Phys. Rev. 181 (1969) 1884;\\
	C.J. Goebel and B. Sakita, \PR 22 (1969) 257;\\
	Chan H.-M. and T.S. Tsun, \PL 28B (1969) 485;\\
	Z. Koba and H.B. Nielsen, \NP 10 (1969) 633, 12 (1969) 517.
\Item{[9.2]}G. Veneziano, Nuo. Cim. 57A (1968) 190.
\Item{[9.3]}C. Lovelace, \PL 135B (1984) 75;\\
	D. Friedan, Z. Qiu, and S. Shenker, proc. APS Div. of 
	Particles and Fields Conf., Santa Fe, eds. T. Goldman 
	and M. Nieto (World Scientific, Singapore, 1984);\\
	E. Fradkin and A. Tseytlin, \PL 158B (1985) 316;\\
	A. Sen, \PRD 32 (1985) 2102, \PR 55 (1985) 1846;\\
	C.G. Callan, E.J. Martinec, M.J. Perry, and D. Friedan, \NP 
	262 (1985) 593.
\Item{[9.4]}S. Mandelstam, \NP 64 (1973) 205;\\
	J.L. Torres-Hern\'andez, \PRD 11 (1975) 3565;\\
	H. Arfaei, \NP 85 (1975) 535, 112 (1976) 256.
\Item{[9.5]}E. Cremmer and J.-L. Gervais, \NP 90 (1975) 410.
\Item{[9.6]}W. Siegel, \NP 109 (1976) 244.
\Item{[9.7]}M.B. Green and J.H. Schwarz, \PL 151B (1985) 21.
\Item{[9.8]}P.H. Frampton, P. Moxhay, and Y.J. Ng, \PR 55 (1985) 2107,
	\NP 276 (1986) 599;\\
	L. Clavelli, \PRD 33 (1986) 1098, 34 (1986) 3262, Prog. Theor.
	Phys. Suppl. 86 (1986) 135.
\Item{[9.9]}O. Alvarez, \NP 216 (1983) 125;\\
	E. Martinec, \PRD 28 (1983) 2604;\\
	G. Moore and P. Nelson, \NP 266 (1986) 58;\\
	A. Cohen, G. Moore, P. Nelson, and J. Polchinski, \NP 267 (1986) 143;\\
	S. Mandelstam, The interacting-string picture and functional 
	integration, {\it in} Unified string theories, eds. M. Green and
	D. Gross (World Scientific, Singapore, 1986) pp. 46;\\
	E. D'Hoker and D.H. Phong, \NP 269 (1986) 205;\\
	M.A. Namazie and S. Rajeev, \NP 277 (1986) 332;\\
	D. Friedan, E. Martinec, and S. Shenker, \NP 271 (1986) 93;\\
	E. D'Hoker and D.H. Phong, Commun. Math. Phys. 104 (1986) 537,
	\PR 56 (1986) 912;\\
	E. Gava, R. Jengo, T. Jayaraman, and R. Ramachandran, 
	\PL 168B (1986) 207;\\
	G. Moore, P. Nelson, and J. Polchinski, \PL 169B (1986) 47;\\
	R. Catenacci, M. Cornalba, M. Martellini, C. Reina, \PL 172B
	(1986) 328;\\
	E. Martinec, \PL 171B (1986) 189;\\
	E. D'Hoker and D.H. Phong, \NP 278 (1986) 225;\\
	S.B. Giddings and S.A. Wolpert, Comm. Math. Phys. 109 (1987) 177;\\
	E. D'Hoker and S.B. Giddings, \NP 291 (1987) 90.
\Item{[9.10]}E. D'Hoker and D.H. Phong, The geometry of string
	perturbation theory, Princeton preprint PUPT-1039 (Feb. 1988),
	to appear in Rev. Mod. Phys. 60 (1988).
\Item{[9.11]}J. Shapiro, \PRD 5 (1972) 1945;\\
	E. Witten, Global anomalies in string theory, {\it in} 
	Symposium on anomalies, geometry and topology, eds. W.A. 
	Bardeen and A.R. White (World Scientific, Singapore, 1985) p. 348.

\chip{10}

\Item{[10.1]}M. Kaku and K. Kikkawa, \PRD 10 (1974) 1110, 1823,
	M. Kaku, \PRD 10 (1974) 3943;\\
	E. Cremmer and J.-L. Gervais, \NP 76 (1974) 209.
\Item{[10.2]}M.B. Green and J.H. Schwarz, \NP 243 (1984) 475.
\Item{[10.3]}Chan H.M. and J. Paton, \NP 10 (1969) 519;\\
	N. Marcus and A. Sagnotti, \PL 119B (1982) 97.
\Item{[10.4]}N. Berkovits, \NP 276 (1986) 650; Supersheet functional
	integration and the interacting Neveu-Schwarz string,
	Berkeley preprint UCB-PTH-87/31 LBL-23727 (July 1987);\\
	S.-J. Sin, Geometry of super lightcone diagrams and Lorentz
	invariance of lightcone string field theory, (II) closed
	Neveu-Schwarz string, Berkeley preprint LBL-25120 UCB-PTH-88/09
	(April 1988).
\Item{[10.5]}S. Mandelstam, \NP 83 (1974) 413.
\Item{[10.6]}T. Kugo, Prog. Theor. Phys. 78 (1987) 690;\\ 
	S. Uehara, \PL 196B (1987) 47;\\
	S.-J. Sin, Lorentz invariance of light cone string field
	theories, Berkeley preprint LBL-23715 (June 1987).
\Item{[10.7]}N. Linden, \NP 286 (1987) 429.
\Item{[10.8]}D.J. Gross and V. Periwal, \NP 287 (1987) 1.
\Item{[10.9]}J. Greensite and F.R. Klinkhamer, \NP 281 (1987) 269,
	291 (1987) 557; Superstring amplitudes and contact interactions,
	Berkeley preprint LBL-23830 NBI-HE-87-58 (Aug. 1987).
\Item{[10.10]}M. Saadi, Intermediate closed strings as open strings going
	backwards in time, MIT preprint CTP\#1575 (April 1988).

\chip{11}

\Item{[11.1]}C. Marshall and P. Ramond, \NP 85 (1975) 375;\\
	A. Neveu, H. Nicolai, and P.C. West, \NP 264 (1986) 573;\\
	A. Neveu, J.H. Schwarz, and P.C. West, \PL 164B (1985) 51,\\
	M. Kaku, \PL 162B (1985) 97;\\ 
	S. Raby, R. Slansky, and G. West, Toward a covariant string 
	field theory, {\it in} Lewes string theory workshop, eds.
	L. Clavelli and A. Halprin (World Scientific, Singapore, 1986).
\Item{[11.2]}T. Banks and M.E. Peskin, Gauge invariant functional field theory 
	for bosonic strings, {\it in} Symposium on anomalies, geometry, 
	topology, eds. W.A. Bardeen and A.R. White, Mar. 28-30, 1985 (World 
	Scientific, Singapore, 1985);\\
	M. Kaku, \NP 267 (1986) 125;\\
	D. Friedan, \NP 271 (1986) 540;\\
	A. Neveu and P.C. West, \NP 268 (1986) 125;\\
	K. Bardak\ced ci, \NP 271 (1986) 561;\\
	J.-L. Gervais, \NP 276 (1986) 339.
\Item{[11.3]}E. Witten, \NP 276 (1986) 291;\\
	T. Kugo and H. Terao, New gauge symmetries in Witten's Ramond
	string field theory, Kyoto preprint KUNS 911 HE(TH)88/01 (Feb. 1988).

\chip{12}

\Item{[12.1]}A. Strominger, \PR 58 (1987) 629.
\Item{[12.2]}C.-Y. Ye, Covariantized closed string fields from
	light-cone, MIT preprint CTP\#1566 (Feb. 1988).
\Item{[12.3]}H. Hata, K. Itoh, T. Kugo, H. Kunitomo, and K. Ogawa, \PL 172B
	(1986) 186, 195; \NP 283 (1987) 433;
	\PRD 34 (1986) 2360, 35 (1987) 1318, 1356; Prog. Theor. Phys. 77
	(1987) 443, 78 (1987) 453.
\Item{[12.4]}M.A. Awada, \PL 172B (1986) 32;\\
	A. Neveu and P. West, \PL 168B (1986) 192.
\Item{[12.5]}A. Hosoya and H. Itoyama, The vertex as a Bogoliubov
	transformed vacuum state in string field theory, Fermilab
	preprint FERMILAB-PUB-87/111-T OUHET-12 (Jan. 1988);\\
	T. Kugo, H. Kunitomo, and K. Suehiro, Prog. Theor. Phys. 78
	(1987) 923.
\Item{[12.6]}D.J. Gross and A. Jevicki, \NP 283 (1987) 1, 287 (1987) 225;\\
	Z. Hlousek and A. Jevicki, \NP 288 (1987) 131.
\Item{[12.7]}E. Cremmer, C.B. Thorn, and A. Schwimmer, \PL 179B (1986) 57;\\
	C.B. Thorn, The oscillator representation of Witten's three open
	string vertex function, {\it in} Proc. XXIII Int. Conf. on High
	Energy Physics, July 16-23, 1986, Berkeley, CA, ed. S.C. Loken
	(World Scientific, Singapore, 1987) v.1, p. 374.
\Item{[12.8]}S. Samuel, \PL 181B (1986) 255;\\
	N. Ohta, \PRD 34 (1986) 3785;\\
	K. Itoh, K. Ogawa, and K. Suehiro, \NP 289 (1987) 127;\\
	A. Eastaugh and J.G. McCarthy, \NP 294 (1987) 845.
\Item{[12.9]}S.B. Giddings, \NP 278 (1986) 242.
\Item{[12.10]}S.B. Giddings and E. Martinec, \NP 278 (1986) 91;\\
	S.B. Giddings, E. Martinec, and E. Witten, \PL 176B (1986) 362.
\Item{[12.11]}J.A. Shapiro and C.B. Thorn, \PRD 36 (1987) 432,
	\PL 194B (1987) 43;\\
	D.Z. Freedman, S.B. Giddings, J.A. Shapiro, and C.B. Thorn,
	\NP 298 (1988) 253.
\Item{[12.12]}B. Zwiebach, Constraints on covariant theories for closed
	string fields, MIT preprint CTP\#1583 (April 1988);
	A note on covariant Feynman rules for closed strings,
	MIT preprint CTP\#1598 (May 1988).
\Item{[12.13]}A. LeClair, M.E. Peskin, and C.R. Preitschopf,
	String field theory on the conformal plane,
	I. Kinematical principles, II. Generalized gluing, 
	SLAC preprints SLAC-PUB-4306, 4307 (Jan. 1988),
	C.R. Preitschopf, The gluing theorem in the operator formulation
	of string field theory, Maryland preprint UMDEPP-88-087
	(Oct. 1987);\\
	A. Eskin, Conformal transformations and string field
	redefinitions, MIT preprint CTP\#1560 (Feb. 1988).
\Item{[12.14]}M. Bochicchio, \PL 193B (1987) 31, 188B (1987) 330;\\
	C.B. Thorn, \NP 287 (1987) 61.
\Item{[12.15]}H. Hata, K. Itoh, T. Kugo, H. Kunitomo, and K. Ogawa, \PL 175B
	(1986) 138.
\Item{[12.16]}G.T. Horowitz, J. Lykken, R. Rohm, and A. Strominger, \PR 57
	(1986) 283.
\Item{[12.17]}D.J. Gross and A. Jevicki, \NP 293 (1987) 29;\\
	S. Samuel, \NP 296 (1988) 187.
\Item{[12.18]}M.B. Green and N. Seiberg, \NP 299 (1988) 559;\\
	C. Wendt, Scattering amplitudes and contact interactions in
	Witten's superstring field theory, SLAC preprint SLAC-PUB-4442
	(Nov. 1987).

%
%

\newpage
\def\lline#1#2{\hbox to\hsize{\noindent{#1}\hfill{#2}\quad}}
\def\iline#1#2{\hbox to\hsize{\indent{#1}\hfill{#2}\quad}}
\markboth{INDEX}{INDEX}
\bookmark0{Index}

\twocolumn[\hbox to\textwidth{\large\bf\hfill INDEX\hfill}\vskip.3in]
\footnotesize

\lline{Action}{}
	\iline{{\it see} BRST; Light cone; Yang-Mills}{}
\lline{Anomalies}{}
	\iline{{\it see} Graphs, anomalies}{}
\lline{Auxiliary fields}{2.1, 4.1}
			\vskip 12pt

\lline{Bosonic strings}{6.1, 7.1, 8.1}
\lline{Bosonization}{8.1}
\lline{Bound states}{9.3}
\lline{Boundary conditions}{6.1, 6.3}
\lline{Bracket, covariant}{3.4, 4.2}
\lline{BRST}{3.1, 4.1, 11.1, 12.1}
	\iline{action, gauge-invariant}{4.1, 11.1, 12.2}
	\iline{BRST1}{3.2, 3.4, 4.1}
	\iline{BRST2}{3.2, 3.3}
	\iline{closed strings}{11.1}
	\iline{extra modes}{3.2, 4.3}
	\iline{gauge fixing}{3.2, 4.4, 5.5, 12.2}
	\iline{GL(1$|$1)}{4.2, 8.2}
	\iline{GL(2$|$2)}{3.6, 4.1, 8.3}
	\iline{IGL(1)}{3.2, 4.2, 8.1, 12.1, 12.2}
	\iline{infinite tower of ghosts}{5.5}
	\iline{interacting}{2.4, 3.2, 3.3, 3.4, 4.2,}
		\iline{}{12.1, 12.2}
	\iline{IOSp(1,1$|$2)}{3.3, 11.1}
	\iline{IOSp(2,2$|$4)}{3.6, 4.1}
	\iline{IOSp(D,2$|$2)}{2.6, 3.4}
	\iline{IOSp(D+1,3$|$4)}{3.6, 4.1, 5.5, 8.3}
	\iline{Lorentz gauge}{3.2}
	\iline{Lorentz gauge, string}{8.3}
	\iline{open strings}{8.1, 8.3, 12.1, 12.2}
	\iline{OSp(1,1$|$2)}{3.3, 3.4, 3.5, 3.6, 4.1, 4.5,}
		\iline{}{8.2, 8.3}
	\iline{OSp(D,2$|$2)}{2.6, 3.4}
	\iline{particles}{5.2, 5.5}
	\iline{supersymmetry}{5.5, 7.3}
	\iline{temporal gauge}{3.1, 3.2}
	\iline{U(1,1$|$1,1)}{3.6, 4.1, 4.5, 5.5}
			\vskip 12pt

\lline{Canonical quantization}{}
	\iline{covariant}{3.4, 4.2}
	\iline{light cone}{2.4}
\lline{Chan-Paton factors}{}
	\iline{{\it see} Group theory indices}{}
\lline{Chiral scalars}{3.1, 6.2}
\lline{Classical mechanics}{}
	\iline{conformal gauge}{6.2}
	\iline{gauge covariant}{6.1}
	\iline{light cone}{6.3}
	\iline{particle}{5.1}
\lline{Closed strings}{1.1, 1.2, 6.1, 6.3, 7.1, 8.2,}
	\lline{}{10, 11.1, 11.2, 12.2}
\lline{Coherent states}{5.5, 9.1}
\lline{Compactification}{1.2, 1.3}
\lline{Component expansions}{11.2}
\lline{Conformal transformations}{}
	\iline{D=2}{6.2, 8.1, 9.1, 9.2}
	\iline{D$>$2}{2.2, 2.3}
	\iline{Sp(2)}{8.1, 9.2, 9.3, 12.2}
	\iline{superconformal}{5.4}
\lline{Conjugation, charge, complex, or hermitian}{}
	\iline{{\it see} Proper time reversal}{}
\lline{Constraints}{}
	\iline{{\it see} BRST}{}
\lline{Coset space methods}{2.2,5.4}
\lline{Covariantized light cone}{}
	\iline{{\it see} Light cone, covariantized}{}
\lline{Critical dimension}{1.2, 7.1, 8.1}
			\vskip 12pt

\lline{DDF operators}{8.1}
\lline{Dimension, spacetime}{}
	\iline{$D=1$}{1.1, 2.6, 5.1}
	\iline{$D=2$}{1.1, 1.3, 3.1, 6.1}
	\iline{$D=3$}{5.4, 7.3}
	\iline{$D=4$}{1.1, 1.3, 5.4, 7.3}
	\iline{$D=6$}{1.3, 5.4, 7.3}
	\iline{$D=10$}{1.2, 1.3, 5.4, 7.3}
	\iline{$D=26$}{1.2, 7.1, 8.1}
\lline{Dimensional reduction}{2.3, 5.4}
\lline{Divergences}{}
	\iline{{\it see} Graphs, loops}{}
\lline{Division algebras}{5.4}
\lline{Duality}{9.1}
			\vskip 12pt

\lline{External fields}{}
	\iline{{\it see} Graphs, external fields}{}
			\vskip 12pt

\lline{Fermionization}{8.1}
\lline{Fermions}{3.5, 4.5, 5.3}
\lline{Field, string}{10.1, 11.1, 12.1}
\lline{Field equations}{}
	\iline{field strengths}{2.2, 5.4, 7.3}
	\iline{gauge fields}{}
	\iline{{\it see} BRST; Yang-Mills}{}
\lline{Finiteness}{}
	\iline{{\it see} Loops}{}
\lline{4+4-extension}{}
	\iline{{\it see} BRST: GL(2$|$2), IOSp(2,2$|$4), U(1,1$|$1,1)}{}
\lline{Functional integrals}{9.2}
			\vskip 12pt

\lline{$\g$ matrices}{2.1, 3.5, 5.3, 5.4, 7.2, 7.3}
\lline{Gauge}{}
	\iline{{\it see} BRST; Light cone}{}
\lline{Gauge fixing}{}
	\iline{{\it see} BRST; Light cone}{}
\lline{Ghosts}{}
	\iline{{\it see} BRST}{}
\lline{Graphs}{}
	\iline{anomalies}{1.2, 8.1, 9.3}
	\iline{covariantized light cone}{2.6}
	\iline{external fields}{9.1}
	\iline{light cone}{2.5}
	\iline{loops}{1.1, 1.2, 9.3}
	\iline{trees}{9.2}
\lline{Gravity}{4.1, 11.2}
	\iline{super}{1.1}
\lline{Group theory indices}{7.1, 9.2, 10}
\lline{Gupta-Bleuler}{3.1, 3.2, 5.4, 7.3, 8.1}
			\vskip 12pt

\lline{Hadrons}{1.1}
\lline{Hamiltonian quantization}{3.1}
\lline{Heterotic strings}{}
	\iline{{\it see} Superstrings}{}
			\vskip 12pt

\lline{IGL(1)}{}
	\iline{{\it see} BRST}{}
\lline{Interacting string picture}{}
	\iline{{\it see} Graphs}{}
\lline{Interactions}{}
	\iline{{\it see} BRST; Light cone}{}
\lline{IOSp}{}
	\iline{{\it see} BRST; Light cone, covariantized}{}
			\vskip 12pt

\lline{Klein transformation}{7.2}
\lline{Koba-Nielsen amplitude}{9.1}
			\vskip 12pt

\lline{Lagrange multipliers}{3.1}
\lline{Length, string}{6.3}
\lline{Light cone}{2.1, 7.1, 10}
	\iline{actions}{2.1, 5.5, 10}
	\iline{bosonic particles}{5.1}
	\iline{bosonic strings}{6.3, 7.1, 10}
	\iline{covariantized}{2.6, 10}
	\iline{graphs}{2.5, 9.2}
	\iline{interactions}{2.1, 2.4, 9.2, 10}
	\iline{Poincar\'e algebra}{2.3, 7.1, 7.2, 10}
	\iline{spinning strings}{7.2}
	\iline{spinors}{5.3}
	\iline{superparticles}{5.4, 5.5}
	\iline{superstrings}{7.3}
	\iline{Yang-Mills}{2.1, 2.4}
\lline{Loops}{}
	\iline{{\it see} Graphs, loops}{}
			\vskip 12pt

\lline{Metric}{}
	\iline{$D=1$}{5.1}
	\iline{$D=2$}{6.2, 8.3}
	\iline{$D>2$}{}
	\iline{{\it see} Gravity}{}
	\iline{flat-space}{2.1, 6.2}
			\vskip 12pt

\lline{Neveu-Schwarz-Ramond model}{}
	\iline{{\it see} Spinning strings}{}
\lline{No-ghost theorem}{4.4, 4.5}
			\vskip 12pt

\lline{Octonions}{5.4}
\lline{Orientation}{}
	\iline{{\it see} Twists}{}
\lline{OSp}{}
	\iline{{\it see} BRST; Light cone, covariantized}{}
			\vskip 12pt

\lline{Particle}{1.1, 5.1}
	\iline{bosonic}{5.1, 5.2}
	\iline{isospinning}{5.3}
	\iline{spinning}{5.3}
	\iline{supersymmetric}{5.4}
\lline{Path integrals}{9.2}
\lline{Perturbation theory}{}
	\iline{{\it see} Graphs}{}
\lline{Poincar\'e algebra}{}
	\iline{{\it see} Light cone, Poincar\'e algebra}{}
\lline{Polyakov approach}{}
	\iline{{\it see} Graphs}{}
\lline{Projective group}{8.1, 9.3}
\lline{Proper time $\t$}{}
	\iline{{\it see} Dimension, spacetime, $D=1$}{}
\lline{Proper time reversal}{5.1, 6.3, 8.1, 8.3, 10}
			\vskip 12pt

\lline{QCD}{1.1}
\lline{Quantum mechanics}{}
	\iline{{\it see} BRST; Canonical quantization;}{}
	\iline{\phantom{see} Light cone; Path integrals}{}
\lline{Quaternions}{5.4}
			\vskip 12pt

\lline{Ramond-Neveu-Schwarz model}{}
	\iline{{\it see} Spinning strings}{}
\lline{Regge theory}{1.1, 9.1}
\lline{Reparametrization invariance}{}
	\iline{{\it see} Metric, $D=1$ and $D=2$}{}
			\vskip 12pt

\lline{Scale invariance, 2D local (Weyl)}{}
	\iline{{\it see} Metric, $D=2$}{}
\lline{$\s$ reversal}{}
	\iline{{\it see} Twists}{}
\lline{S-matrix}{}
	\iline{{\it see} Graphs}{}
\lline{Spectrum, mass}{7.1}
\lline{Spinning strings}{7.2}
\lline{Spinors}{}
	\iline{{\it see} Fermions; $\g$ matrices; Supersymmetry}{}
\lline{String}{}
	\iline{{\it see the whole book}}{}
\lline{Stueckelberg fields}{4.1}
\lline{Superconformal transformations}{5.4, 7.2}
\lline{Supergravity}{}
	\iline{{\it see} Gravity, super}{}
\lline{Superspin}{5.4, 5.5}
\lline{Superstrings (I, IIAB, heterotic)}{1.2, 7.3, 9.3,}
	\iline{}{10}
\lline{Supersymmetry}{}
	\iline{$D=1$}{5.3}
	\iline{$D=2$}{7.2}
	\iline{$D > 2$}{1.2, 1.3, 5.4, 5.5, 7.3}
			\vskip 12pt

\lline{Tension, string ($\a '$)}{6.1}
\lline{Trees}{}
	\iline{{\it see} Graphs, trees}{}
\lline{Twistors}{5.4}
\lline{Twists}{7.1, 10}
			\vskip 12pt

\lline{Veneziano model}{}
	\iline{{\it see} Bosonic string}{}
\lline{Vertices}{}
	\iline{{\it see} BRST; Light cone}{}
\lline{Virasoro operators}{}
	\iline{{\it see} Conformal transformations, 2D}{}
			\vskip 12pt

\lline{World sheet}{}
	\iline{{\it see} Dimension, spacetime, D=2}{}
			\vskip 12pt

\lline{Yang-Mills}{2.1, 2.4, 3.1, 3.2, 3.3, 3.4, 4.1,}
		\lline{}{4.2, 4.4, 4.6, 11.2}
	\iline{super}{5.4, 9.1}

\onecolumn
\end{document}